\documentclass[10pt,superscriptaddress,twocolumn,amsmath,amssymb,aps,prx,showpacs,longbibliography]{revtex4-2}
\usepackage{mathrsfs}
\usepackage{graphicx}
\usepackage{dcolumn}
\usepackage{bm}
\usepackage{amssymb}
\usepackage{amsmath}
\usepackage{paralist}
\usepackage{float}
\usepackage{mdframed}
\usepackage{booktabs}
\usepackage{mathtools} 
\usepackage{graphicx}
\usepackage[normalem]{ulem}
\usepackage[colorlinks,linkcolor=blue,anchorcolor=blue,citecolor=green]{hyperref} 
\usepackage{cleveref}
\usepackage{url}
\usepackage{color}

\usepackage{soul}
\usepackage{verbatim}

\newcommand{\dr}{\mathrm{dr}}
\newcommand{\Li}{\mathrm{Li}}
\newcommand{\eff}{\mathrm{eff}}

\begin{document}

\title{Universal Transport Properties of Continuous Quantum Gases}

\author{Zi-Yang Liu}
\affiliation{Innovation Academy for Precision Measurement Science and Technology, Chinese Academy of Sciences, Wuhan 430071, China}
\affiliation{University of Chinese Academy of Sciences, Beijing 100049, China.}

\author{Xiangguo Yin}
\affiliation{Institute of Theoretical Physics and State Key Laboratory of Quantum Optics and Quantum Optics Devices, Shanxi University, Taiyuan 030006, China}

\author{Yunbo Zhang}
\email[]{ybzhang@zstu.edu.cn}
\affiliation{Department of Physics and Key Laboratory of Optical Field Manipulation of Zhejiang Province, Zhejiang Sci-Tech University, Hangzhou 310018, China}

\author{Shizhong Zhang}
\email[]{shizhong@hku.hk}
\affiliation{Department of Physics and State Key Laboratory of Optical Quantum Materials, Hong Kong Institute of Quantum Science and Technology, The University of Hong Kong, Hong Kong, China}

\author{Xi-Wen Guan}
\email[]{xiwen.guan@anu.edu.au}
\affiliation{Innovation Academy for Precision Measurement Science and Technology, Chinese Academy of Sciences, Wuhan 430071, China}
\affiliation{Hefei National Laboratory, Hefei 230088, People’s Republic of China}
\affiliation{Department of Fundamental and Theoretical Physics, Research School of Physics, Australian National University, Canberra, ACT 0200, Australia}

\begin{abstract}
The Drude weight characterizes ballistic transport in quantum many-body systems. Although analytical calculations of Drude weights have been extensively studied in integrable models, their direct  connections to finite-temperature macroscopic state functions remain unestablished, especially for continuous multicomponent quantum gases.  In the present work, we use generalized hydrodynamics and the thermodynamic Bethe ans\"{a}tz to calculate exactly the Drude weights for one-dimensional continuous integrable systems, including the Lieb-Liniger and Bose-Fermi mixture models.  We derive universal exact relations between Drude weight matrix components and key thermodynamic densities (particle density, enthalpy, entropy). Analytic expressions for Drude weight are obtained across different physical regimes, i.e. strong- and weak-coupling regimes in addition to universal scaling laws near the quantum phase transitions.  To bridge theory and experiment, we simulate two experimental protocols, linear potential quench and bipartitioning quench, to enable reliable measurements of the Drude weights. Using these protocols, we calculate the charge and energy Drude weight for Lieb-Liniger gas and compare with recent measurements reported in [Science 391, 290 (2026)], showing excellent agreement with particle density and enthalpy, respectively, thus offering deeper physical insights into  experimental observations.  Our findings directly link ballistic transport properties to  thermodynamics, providing rigorous theoretical benchmarks for future ultracold atomic gas experiments.

\end{abstract}

\maketitle

\section{introduction}
\label{sec:intro}
Quantum transport is a powerful probe of the intrinsic properties of many-body systems and has been the focus of intense research for decades~\cite{J.Stat.Mech.2016CalabreseIntroduction,Rev.ModernPhys.2021BertiniFinitetemperature}. Central to this field is the Drude weight, a fundamental transport coefficient in condensed matter physics that quantifies the contributions from ballistic transport of charge, spin and energy. In clean systems, Drude weight acts as a definitive criterion that distinguishes between a metal for which it is finite and an insulator for which it is zero~\cite{Phys.Rev.1964KohnTheory}.

At finite temperature, Drude weights of interacting systems generally vanish due to incoherent collision, but integrable systems form a striking exception. Pioneering work using the Mazur-Suzuki inequality~\cite{Physica1969MazurNon-ergodicity,Physica1971SuzukiErgodicity} has established finite lower bounds for conductivity Drude weights. The preservation of ballistic transport arises from a macroscopic set of local conservation laws that constrain the transport dynamics, suppressing standard thermalization and full current decay~\cite{Phys.Rev.B1997ZotosTransport,Nucl.Phys.B2014ProsenQuasilocal,Phys.Rev.Lett.2011ProsenOpen,Phys.Rev.Lett.2013ProsenFamilies}.

In linear response theory, the Drude weight is related to current-current  correlation functions~\cite{J.Phys.Soc.Jpnkubo1954general,J.Phy.Soc.Jpa1957KuboStatistical-Mechanical,Springer1991KuboStatistical}. On the other hand, Kohn further showed the conductivity Drude weight can be related to the increments of ground state energy when an applied magnetic flux, or when twist boundary condition is imposed~\cite{Phys.Rev.1964KohnTheory,Phys.Rev.Lett.1995CastellaIntegrability,Phys.Rev.Lett.1990ShastryTwisted}. Combined with the Thermodynamic Bethe ans\"{a}tz (TBA), this formulation yields formal expressions for the Drude weight in the XXZ Heisenberg and Hubbard models~\cite{J.Phys.A:Math.Gen.Exact1998Fujimoto, J.Phys.Soc.Jpn.Finite2005Benz, Phys.Rev.B2011SirkerConservation, J.Stat.Mech:TheoryExp.2017ZotosTBA, Phys.Rev.Lett.1999ZotosFinite, arXiv2025LuoQuantum}. These lattice models exhibit extremely rich physical phenomena and have been comprehensively reviewed in Ref.~\cite{Rev.ModernPhys.2021BertiniFinitetemperature}.

A major recent breakthrough in the study of quantum transport in one-dimensional models is the development of Generalized Hydrodynamics (GHD) which furnishes a universal framework for describing the emergent large-scale dynamics of integrable systems. Intuitively, GHD describes the system dynamics as a set of stable quasi-particle excitations via collisionless Boltzmann equations~\cite{Phys.Rev.Lett.2016BertiniTransport,Phys.Rev.X2016Castro-AlvaredoEmergent}. It allows the calculation of transport coefficients using equilibrium physical quantities and correlation functions, particularly the rapidity distribution function and effective quasi-particle velocities~\cite{J.Math.Phys.1969YangThermodynamics}. So far, GHD has been successfully applied to derive Drude weights for canonical models, including the Lieb-Liniger gas, the XXZ spin chain and the Hubbard model~\cite{SciPostPhysics2017DoyonDrude,SciPostPhys.2019UrichukSpin,Phys.Rev.Lett.2017IlievskiMicroscopic,Phys.Rev.B2017IlievskiBallistic} with results in perfect agreement with those obtained via the Kohn formula and the Mazur-Suzuki inequality. Beyond transport coefficients, GHD has also demonstrated remarkable efficacy in describing entanglement dynamics, correlation spreading, and anomalous transport regimes~\cite{Phys.Rev.X2025DoyonGeneralized,J.Stat.Mech.2022BastianelloIntroduction}. Those theoretical advances have been paralleled by striking experimental progresses in ultracold atomic gas~\cite{Rep.Progr.Phys.2022GuanNew} which offer a uniquely controllable testbed for verifying GHD predictions~\cite{SciPostPhys.2019CauxHydrodynamics,Phys.Rev.Lett.2019SchemmerGeneralized,Phys.Rev.Lett.2019BastianelloGeneralizeda}. Most notably, recent experiments have successfully measured directly the charge Drude weight, $D_{nn}$, and simulated the corresponding transport dynamics~\cite{arXiv2024SchuttelkopfCharacterising,arXiv2025DuboisExperimental}.


For a Galilean invariant system, rigorous arguments establish the connection between certain elements of the Drude weight matrix such as $D_{nn}$ and the thermodynamical quantities of the system~\cite{Phys.Rev.Lett.2015VasseurExpansion,arXiv2025GohmannBallistic}. However, general relations for more complicated conserved currents, such as momentum and heat, and furthermore their analytic dependence on temperature,  have not been explored so far systematically. This theoretical gap is even more pronounced in multi-component systems, where particles with distinct statistical properties undergo complex dynamic coupling. Very recently, an important initial step was taken by Ref.~\cite{arXiv2025SohamBethe}, who analytically evaluated the zero-temperature Drude weights of an integrable Bose-Fermi mixture.

In this work, we aim to establish exact and universal relations that directly link the Drude weight matrix to thermodynamic functions such as compressibility, density, enthalpy and entropy. Using the GHD-TBA framework, we derive these analytical results for two paradigmatic continuous models: the Lieb-Liniger Bose gas~\cite{Phys.Rev.1963LiebExact} and the Bose-Fermi mixture~\cite{Phys.Rev.A2006ImambekovExactly}. Furthermore, we elucidate the evolution of ballistic transport across distinct parameter regimes, including both weak and strong coupling, low and high temperature regimes. At low temperature, we obtain analytic results that explicitly show the contribution from the thermal excitations in addition to that from the ground state.

For Lieb-Liniger model, we show how Drude weight behaves in three different statistical regimes. In the weak coupling regime dominated by Bose-Einstein statistics, the Drude weights reduce to that for a free Bose gas. In the strong coupling regime governed by Fermi-Dirac statistics (Tonks-Girardeau limit), the Drude weights are expressed in terms of Fermi-Dirac functions with finite interaction corrections. In the high temperature regime, the Drude weight reduces to that for a Maxwell-Boltzmann gas with quantum corrections that can be readily obtained via high temperature expansion. In the quantum critical regime near phase transitions, we establish universal scaling laws determined solely by critical exponents. We perform similar analysis for the Bose-Fermi mixture model.

 Finally, to bridge theory and experiment, we simulate two protocals, including linear potential quench and bipartitioning quench, used recently to measure the Drude weights~\cite{arXiv2024SchuttelkopfCharacterising}. We numerically simulate these protocols and show that the extracted Drude weights are consistent with experimental observations and more importantly, consistent with the thermodynamic identities established in this work. Particularly, the charge and energy Drude weights are in excellent agreement with the particle density and enthalpy, respectively.

 

 %

This paper is structured as follows. In section \ref{sec:integrableGHD}, we briefly review the conserved charges and currents of continuous integrable models and their expressions within GHD. Section \ref{sec:drude} presents the general formalism of the Drude weight and derives several universal identities. 
In Sections \ref{sec:LL} and \ref{sec:BF}, we apply this formalism to the Lieb-Liniger Bose gas and Bose-Fermi mixture, respectively, presenting formal expressions, universal features, and detailed analytical and numerical results across various parameter regimes. In section \ref{sec:expt}, we propose two experimental protocols and show that they provide a robust benchmark for measurements in ultracold atomic gases.

\section{Integrable systems and Generalized Hydrodynamics}
\label{sec:integrableGHD}
In many one-dimensional (1D) quantum systems, integrability entails the existence of an infinite set of local conserved quantities. These dynamical constraints suppress conventional thermalization, and a defining hallmark of such systems is that their quasiparticle excitations undergo purely elastic, factorizable scattering processes.
This is in stark contrast to non-integrable systems. Consequently, any many-body scattering event can be decomposed into a sequence of two-body interactions, each dictated by a scattering matrix obeying the Yang–Baxter equation. The properties of these stable quasi particles constitute the fundamental kinetic data. For Bethe ans\"{a}tz-solvable models, including the Lieb-Liniger model and Bose-Fermi mixture, eigenstates are uniquely characterized by rapidity sets $\{u_{\alpha,l}\}$. Here,  $\alpha$ labels the species of quasiparticle  dictated by the symmetries of the model, while $l$ indexes individual particles. Each rapidity corresponds to the quasi-momentum of the quasi particle. In the thermodynamic limit, solving the Bethe equations yield the rapidity density distributions  $\rho_{\alpha}(u)$ and their corresponding filling functions $\theta_{\alpha}(u)$. Together, they provide a full macroscopic characterization of the states of the system~\cite{Phys.Rev.Lett.1967YangExact,Cambridge1999TakahashiThermodynamics}.

Building on this quasi particle framework, GHD has emerged as a powerful theory for describing the large-scale, non-equilibrium dynamics of integrable systems. The theory is founded on the local continuity equation satisfied by each local conserved charge density $\hat{q}_{i}(x)$ and its associated current density $\hat{j}_{i}(x)$~\cite{Phys.Rev.X2016Castro-AlvaredoEmergent,Phys.Rev.Lett.2016BertiniTransport}
\begin{equation}
	\partial_{t}\hat{q}_{i}(x,t)+\partial_{x}\hat{j}_{i}(x,t)=0.
\end{equation}
The corresponding total charge $\hat{Q}_i$ and total current $\hat{J}_i$ are obtained by spatial integration: $\hat{Q}_{i}=\int dx\hat{q}_{i}(x)$ and $\hat{J}_{i}=\int dx\hat{j}_{i}(x)$.

GHD assumes a state of local equilibrium, described as a fluid of quasiparticles whose densities $\rho_{\alpha}(u, x, t)$ vary slowly in space and time. The local expectation values of the charge and current densities, $q_i = \langle\hat{q}_{i}\rangle$ and $j_i = \langle\hat{j}_{i}\rangle$, are then expressed in terms of these quasi particle densities
\begin{align}
	q_{i}(x,t) = & \sum_{\alpha}\int du~\rho_{\alpha}(u, x, t)h_{i,\alpha}(u), \\
	j_{i}(x,t) = & \sum_{\alpha}\int du~\rho_{\alpha}(u,x,t)v_{\alpha}^{\rm eff}(u)h_{i,\alpha}(u).
\end{align}
Here, $h_{i,\alpha}(u)$ is the single-particle eigenvalue of the total conserved charge $\hat{Q}_{i}$ for a quasiparticle of species $\alpha$ with rapidity $u$. The crucial quantity $v_{\alpha}^{\rm eff}(u)$ is the effective velocity of  quasiparticles. It is defined as the ratio of the dressed derivatives of the quasiparticle energy $e_{\alpha}(u)$ and momentum $p_{\alpha}(u)$ with respect to rapidity, namely, 
\begin{equation}
	v_{\alpha}^{\rm eff}(u)=\frac{(\partial_{u}e_{\alpha})^{\text{dr}}(u)}{(\partial_{u}p_{\alpha})^{\text{dr}}(u)}.
\end{equation}
The dressing operation, denoted $(\cdot)^{\text{dr}}$, renormalizes the bare velocity of a quasiparticle to account for the drag effect or momentum exchange from its interaction with the sea of all other quasiparticles, thus encapsulating the many-body effects on transport.

\section{DRUDE WEIGHTS: FORMALISMS AND DEFINITIONS}
\label{sec:drude}
To introduce Drude weight, let us take charge transport as an example and consider the conductivity matrix $\sigma_{ij}(\omega)$. In 1D, $i=j=z$ labels the only spatial direction. Its real part is decomposed as~\cite{Phys.Rev.1964KohnTheory}
\begin{equation}
	Re[\sigma(\omega)]=2\pi D\delta(\omega)+\sigma^{reg}(\omega),
	\label{eq:Resigmaij}
\end{equation}
where $\sigma^{reg}(\omega)$ is the regular part and the singular part proportional to $\delta(\omega)$ defines the Drude weight $D$. It quantifies the dissipationless contribution to conductivity and, in the case of superconductivity, is proportional to the superfluid density. The Drude weight can be conveniently expressed via the Kubo formula as the long-time average of the current-current correlation function~\cite{Springer1991KuboStatistical} (setting $k_B=1$):
\begin{equation}
	D=\lim_{t\rightarrow\infty}\frac{\beta}{tL}\int_{0}^{t}dt^{\prime}\int_{-\infty}^{\infty}dz\langle J(z,t^{\prime})J(0,0)\rangle_{c},
\end{equation}
where $L$ is the size of the system. $\beta=1/T$ and $\langle...\rangle_{c}$ denotes the connected correlation function. In integrable systems, the presence of an extensive number of conserved quantities $\{Q_{k}\}$ prevents the complete decay of current-current correlations. This gives rise to the Mazur-Suzuki bound~\cite{Physica1969MazurNon-ergodicity,Physica1971SuzukiErgodicity}:
\begin{equation}
	D\ge\sum_{k}\langle Q_{k}J\rangle^{2}/\langle Q_{k}^{2}\rangle,
\end{equation}
where any conserved charge $Q_k$ with $\langle Q_{k}J\rangle\ne0$ provides a lower bound for the time-averaged current-current correlation function. While a direct evaluation of the Kubo formula requires calculating exact real-time correlation functions, GHD offers a powerful alternative way to compute the Drude weight based on the hydrodynamic projection principle~\cite{Phys.Rev.X2025DoyonGeneralized,Phys.Rev.B2017IlievskiBallistic,JStatPhys2022DoyonDiffusionb}.

 The core physical insight is that the ballistic contribution to transport arises solely from the overlap between currents and conserved charges. 
 Mathematically, this is realized by projecting the current operator onto the space of conserved quantities. 
 This projection requires two static thermodynamic inputs: the charge-charge susceptibility matrix $C$, which acts as the metric in the charge space, and the charge-current correlation matrix (flux Jacobian) $B$. Consequently, the Drude weight matrix $D$ is exactly determined by~\cite{SciPostPhysics2017DoyonDrude,Phys.Rev.B2017IlievskiBallistic}
\begin{equation}
	D=\beta BC^{-1}B.
	\label{eq:Dij_GHD_M}
\end{equation}
Here, $C$ and $B$ are defined, respectively, as the static covariance matrices
\begin{align}
	C_{ij}=&\int dx\langle q_{i}(x)q_{j}(0)\rangle_{c}={-}\frac{\partial q_{i}}{\partial\beta_{j}}, \label{eq:Cij_M}\\
	B_{ij}=&\int dx\langle j_{i}(x)q_{j}(0)\rangle_{c}={-}\frac{\partial j_{i}}{\partial\beta_{j}}. \label{eq:Bij_M}
\end{align}
Here, $q_i$ and $j_i$ denote the local densities of the conserved charges and currents, while the set of Lagrange multipliers $\{\beta_j\}$ parameterizes the Generalized Gibbs Ensemble (GGE). Crucially, GHD allows these macroscopic correlation matrices to be recast as explicit functionals of the quasiparticle spectrum. By solving the dressing equations, $C$ and $B$ are determined by integrals over the rapidity distribution. In this work, we specialize this general framework to thermal equilibrium, where the GGE reduces to the standard Grand Canonical Ensemble governed solely by temperature $T$ and chemical potential $\mu$.

\section{Drude weight of the Lieb-Liniger model}
\label{sec:LL}
\subsection{Lieb-Liniger model and conserved charges}

We now apply the general framework to the canonical model of interacting bosons in one dimension: the Lieb-Liniger model~\cite{Phys.Rev.1963LiebExact}. The Hamiltonian in first quantized form is given by
\begin{equation}
\hat{H}=\frac{\hbar^{2}}{2m_{0}}\left[- \sum_{i=1}^{N}\frac{\partial^{2}}{\partial x_{i}^{2}}+2c\sum_{i<j}^{N}\delta(x_{i}-x_{j}) \right].
\end{equation}
The 1D $s$-wave scattering length $a_{1D}$ is related to the coupling constant $c$ by $a_{1D}=-c/2$. We shall adopt the units where $\hbar=m_{0}=1$ in the following calculations.



The model is integrable and can be solved exactly using the Bethe ans\"{a}tz. 
The eigenstates are parameterized by a set of quasi-momenta, or rapidities, $\{k_{j}\}_{j=1}^{N}$, which satisfy the Bethe ans\"{a}tz Equations (BAE) with periodic boundary conditions
\begin{equation}
e^{ik_{j}L}=\prod_{l\neq j}^{N}\frac{k_{j}-k_{l}+ic}{k_{j}-k_{l}-ic},\quad\text{for }j=1,...,N.
\end{equation}
In the thermodynamic limit ($N,L\rightarrow\infty$ with fixed density $n=N/L$), the state of the system is described by the density distribution of particle rapidities, $\rho(k)$. In thermal equilibrium, this distribution is determined by the TBA equations. $\rho(k)$ is a solution to the following integral equation
\begin{equation}
\rho(k)+\rho_{h}(k)=\frac{1}{2\pi}+\int_{-\infty}^{\infty}dk'\,\phi(k-k')\rho(k'),
\end{equation}
where $\rho_{h}(k)$ is the density of "holes" or unoccupied states, and the total density of states is $\rho_{t}(k)=\rho(k)+\rho_{h}(k)$.
The scattering kernel $\phi(k)$ is given by the derivative of the two-body scattering phase shift:
\begin{equation}
\phi(k)=\frac{1}{2\pi}\frac{2c}{c^{2}+k^{2}}.
\end{equation}
The macroscopic state is fully characterized by the filling function $\theta(k)=\rho(k)/\rho_{t}(k)$.

In thermal equilibrium, the infinite set of local conserved charge densities $q_{i}$ and their corresponding currents $j_{i}$ can be expressed within the GHD framework. 
The bare eigenvalues for the conserved quantities are $h_{n}(k)=1$ (particle number), $h_{k}(k)=k$ (momentum), $h_{e}(k)=k^2/2$ (energy), and $h_{\epsilon}(k)=k^2/2-\mu$ (thermal energy).
Specifically, $h_{\epsilon}(k)$ represents the single-particle eigenvalue corresponding to the charge $H - \mu N$. 
As was discussed in Ref.~\cite{Phys.Rev.X2016Castro-AlvaredoEmergent}, the energy current is physically associated with this charge. 
Accordingly, we refer to the charge $H - \mu N$ as thermal energy, and its associated flow as the thermal current.
In general, the charge densities and currents are given by~\cite{Phys.Rev.X2016Castro-AlvaredoEmergent,Phys.Rev.Lett.2016BertiniTransport}
\begin{align}
q_{i}=&\int dk \rho(k)h_{i}(k) ,\\
j_{i}=&\int dk \rho(k)v^{\rm eff}(k)h_{i}(k).
\end{align}
The effective velocity is given by the GHD formula as the ratio of dressed derivatives
\begin{equation}
v^{\eff}(k)=\frac{(d{h_e}/dk)^{\text{dr}}}{(d{h_k}/dk)^{\text{dr}}}=\frac{(k)^{\text{dr}}}{(1)^{\text{dr}}}.
\end{equation}
The dressing operation, denoted by $(\cdot)^{\text{dr}}$, transforms a bare quantity $f(k)$ by accounting for interactions with the medium. 
It is defined by the solution to the linear integral equation
\begin{equation}
f^{\dr}(k)=f(k)+\int dk'\,\phi(k-k')\theta(k')f^{\dr}(k').
\label{eq:droperat_LL}
\end{equation}
This equation can also be expressed in operator form as $f^{\dr}=(1-\tilde{T}\theta)^{-1}f$,
where $\tilde{T}$ is the integral operator with kernel $\phi(k-k')$, i.e. $(\tilde{T}\theta f)(k) = \int dk' \, \phi(k-k') \theta(k') f(k')$.

\subsection{Drude weight}
Let us consider a situation that is appropriate in actual experiments. Assuming that there exist in the system both the chemical potential gradient $\nabla\mu$ and temperature gradient $\nabla T/T$. In linear response theory, the particle current $j_{n}$, momentum current $j_{k}$, energy current $j_{e}$, and thermal current $j_{\epsilon}$ due to these two gradients are given by
\begin{equation}
	\left( \begin{array}{c} j_{n}\\ j_{k}\\ j_{e}\\ j_{\epsilon} \end{array}\right) =\left( \begin{array}{cc} \sigma_{nn} & \sigma_{ne}\\ \sigma_{kn} & \sigma_{ke}\\ \sigma_{en} & \sigma_{ee}\\ \sigma_{\epsilon n} & \sigma_{\epsilon e} \end{array}\right) \left( \begin{array}{c} \nabla\mu\\ -\nabla T/T \end{array}\right).
	\label{eq:J_sigma_t_LL}
\end{equation}
The Drude weights $D_{ij}$ represent the ballistic (zero-frequency) part of the corresponding Onsager coefficients $\sigma_{ij}$ (Eq.~\ref{eq:Resigmaij}). 
Here, the subscripts $i, j \in \{n, k, e, \epsilon\}$ label the relevant conserved charges. This maintains consistent notation with the bare eigenvalues $h_i(k)$ defined previously. These coefficients are constrained by Onsager reciprocity ($\sigma_{ij} = \sigma_{ji}$), which implies $D_{ij} = D_{ji}$. Specialize to chemical potential gradient, we now focus primarily on the Drude weights associated with both the chemical potential and temperature gradients, namely, $D_{nn}$, $D_{nk}$, $D_{ne}$, and $D_{n\epsilon}$ (particle, particle-momentum, particle-energy, and particle-thermal energy Drude weight).

Within the GHD approach, the components of the Drude weight matrix Eq.~\eqref{eq:J_sigma_t_LL} can be expressed in a compact and elegant form using the thermodynamic properties of the quasiparticles~\cite{SciPostPhysics2017DoyonDrude}
\begin{equation}
D_{ij}=\frac{1}{T}\int dk~\rho(k)\theta(k)[1-\theta(k)](v^{\text{eff}}(k))^{2}h_{i}^{\text{dr}}(k)h_{j}^{\text{dr}}(k).
\label{eq:DW_LL_MT}
\end{equation}
Here we recall $h_{i}^{\text{dr}}(k)$ are the dressed single-particle eigenvalues of the conserved charges, We have set the Boltzmann constant $k_B=1$. This formula is the explicit form of the general GHD expression $D=\beta B C^{-1} B$ [Eq.~\ref{eq:Dij_GHD_M}].



\begin{figure}[!htbp]
\includegraphics[width=1\columnwidth]{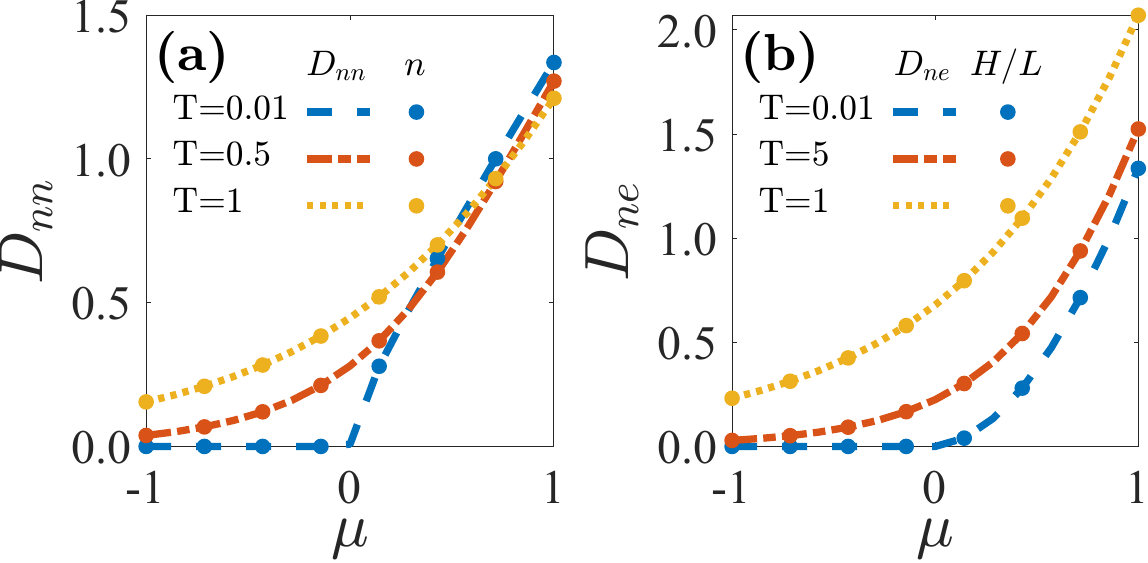}
\caption{
Numerical verification of the exact macroscopic identities $D_{nn}=n$ and $D_{ne}=H/L$ in the Lieb-Liniger model. (a) Particle Drude weight $D_{nn}$ and number density $n$, and (b) enthalpy Drude weight $D_{ne}$ and enthalpy density $H/L$ as functions of chemical potential $\mu$ at various temperatures $T$ ($c=1$). Solid lines represent the transport coefficients (Drude weights), while symbols denote the corresponding equilibrium thermodynamic densities. We set $\hbar = m_{0} = k_B = 1$ in all numerical evaluations.}
\label{fig:LL_phase}
\end{figure}

\subsection{Universal thermodynamic behavior}
%

The GHD integral formula Eq.~\eqref{eq:DW_LL_MT} not only facilitates numerical computation but also enables the exact analytical evaluation of the Drude weights for thermal equilibrium states. For example, we establish exact identities between the ballistic transport coefficients and macroscopic thermodynamic functions of the state (see Supplemental material \cite{SM} for detailed derivations). While partial results for the single-component particle Drude weight have been discussed~\cite{Phys.Rev.Lett.2015VasseurExpansion,arXiv2025GohmannBallistic}, extracting the complete Drude weight matrix remains technically challenging, especially for the transport coefficients associated with other higher order conserved quantities (such as $D_{nk}$, $D_{ne}$ etc.). By substituting the TBA equations into the GHD integral formula and performing complex analytical reductions, we successfully evaluate these previously inaccessible coefficients. Using this approach, we rigorously prove the following explicit identities
\begin{align}
	D_{nn} & =\frac{N}{L}=n, \label{eq:Dnn_LL_un} \\
	D_{nk} & =2\frac{P}{L}, \label{eq:Dnk_LL_un} \\
	D_{ne} & =\frac{E}{L}+p=\frac{H}{L}, \label{eq:Dne_LL_un} \\
	D_{n\epsilon} & =D_{ne}-\mu D_{nn}=T\frac{S}{L}. \label{eq:Dnep_LL_un} 
\end{align}
Here, $p$ is the thermodynamic pressure, and $n=N/L$, $P/L$, $H/L$, and $s=S/L$ are the densities of particle, momentum, enthalpy, and entropy, respectively. Note that in the rest frame of the thermal equilibrium state, the total momentum density vanishes ($P/L=0$). Given these, we can further derive their asymptotic expansions across distinct parameter regimes. In Fig.(\ref{fig:LL_phase}), we plot the numerically calculated Drude weights ($D_{nn}$ and $D_{ne}$) alongside their corresponding macroscopic thermodynamic densities (particle number density $n$ and enthalpy density $H/L$) as functions of chemical potential across various temperatures. The perfect agreement between the transport coefficients and thermodynamic quantities numerically confirms our exact analytical identities.

\subsection{Transport at zero temperatures}
In the low-temperature regime ($T\to 0$), the transport is governed by the low-energy particle-hole excitations near the (quasimomentum) Fermi surface. By applying the Sommerfeld expansion~\cite{Z.Physik1928SommerfeldZur} to the exact GHD formula (Eq.~\ref{eq:DW_LL_MT}), we derive the asymptotic behavior of the Drude weights
\begin{equation}
	D_{ij}=\frac{1}{\pi} \left[v^{\text{eff}}h_{i}^{\text{dr}}h_{j}^{\text{dr}}+\frac{\pi^{2}T^{2}}{6}\frac{\partial^{2}}{\partial\epsilon^{2}}\left(v^{\text{eff}}h_{i}^{\text{dr}}h_{j}^{\text{dr}}\right)\right]_{\epsilon=0}.
	\label{eq:DW_LL_LT}
\end{equation}
Here, all dressed quantities ($v^{\text{eff}}, h^{\text{dr}}$) are evaluated at the Fermi rapidity $k_F$ (defined by $\epsilon(k_F)=0$). The first term represents the contribution from the ground-state, solely determined by the Fermi surface properties, while the second term captures the leading quadratic thermal correction ($O(T^2)$).

Let us specialize to $T=0$, the momentum distribution $\rho(k)$ exhibits a sharp Fermi edge.  Consequently, the expression for the Drude weights in the ground state is determined solely by the properties at the Fermi surface
\begin{equation}
	D_{ij} = \frac{1}{\pi} v^{\text{eff}}(k_{F}) h_{i}^{\text{dr}}(k_{F}) h_{j}^{\text{dr}}(k_{F}).
	\label{eq:Dij_0T_LL_M}
\end{equation}
On the other hand, in the low-energy limit, the collective behavior of the system is effectively described by the Tomonaga-Luttinger liquid (TLL) theory. Within this framework, the compressibility is thermodynamically defined as $\kappa = \partial n / \partial \mu$. According to the TLL theory, it is fundamentally related to the Luttinger parameter $K$ and the sound velocity $v_s$ via the relation~\cite{Phys.Rev.Lett.Haldane1981Effective}
\begin{equation}
	\kappa = \frac{K}{\pi v_s}.
\end{equation}
These macroscopic TLL parameters are intrinsically linked to the exact Bethe ans\"{a}tz quantities evaluated at the Fermi surface. Specifically, the sound velocity corresponds to the effective velocity of the quasiparticles, $v_s = v^{\text{eff}}(k_F)$, while the Luttinger parameter is identified with the square of the dressed charge, $K = [h_{n}^{\text{dr}}(k_F)]^2$~\cite{Chin.Phys.B2015JiangUnderstanding}. Substituting these identifications into the Drude weight expression, we obtain the classic relation for the charge stiffness originally established by Haldane~\cite{J.Phys.C:SolidStatePhys.Haldane1981Luttinger, Phys.Rev.Lett.Haldane1981Effective}
\begin{equation}
	D_{nn} = v_s \frac{K}{\pi} = \kappa v_s^2.\label{dnn_kappa}
\end{equation}

Next, we derive the analytical results for the ground-state Drude weights in the strong-interaction limit. In the strong-coupling regime ($c \gg 1$), substituting the asymptotic expansions of the dressed charges into the general formula Eq.~\eqref{eq:Dij_0T_LL_M} yields explicit expressions for the ground-state transport coefficients up to order $1/\gamma^3$ ($\gamma = c/n$)~\cite{J.Phys.A:Math.Theor.2011GuanPolylogsa}
\begin{align}
	D_{nn} & =n, \label{eq:Dnn_LL_zTiC_M} \\
	D_{ne} & \approx \frac{\pi^{2}n^{3}}{2}\left(1-\frac{16}{3\gamma }+\frac{20}{\gamma^{2}}-\frac{64}{\gamma^{3}}+\frac{64\pi^{2}}{15\gamma^{3}}\right).
	\label{eq:Dne_LL_zTiC_M}
\end{align}

\begin{figure}[!htbp]
	\includegraphics[width=1\columnwidth]{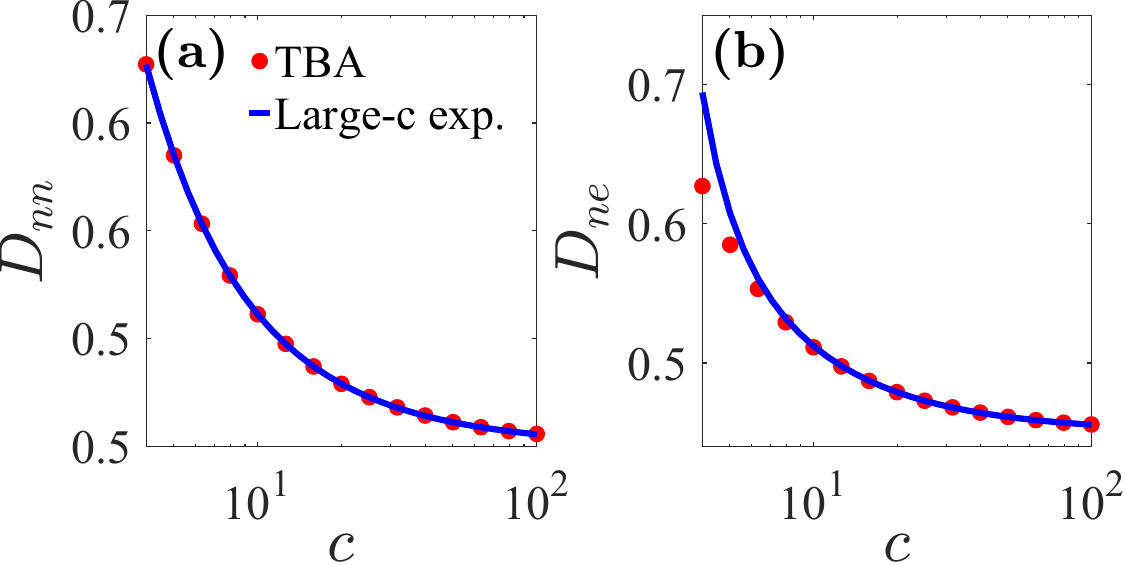}
	\caption{Ground-state (calculated with $T=0.001$) Drude weights ($D_{nn}$ in (a) and $D_{ne}$ in (b)) as functions of interaction strength $c$ for the Lieb-Liniger model at chemical potential $\mu=1$. Solid lines (black) are the exact numerical TBA solutions. Dashed lines (red) are the analytical large-$c$ expansions from Eq.~\eqref{eq:Dnn_LL_zTiC_M} and~\eqref{eq:Dne_LL_zTiC_M}.}
	\label{fig:LL_lowtep_strlim}
\end{figure}

These results represent the large-$c$ expansion of the exact ground state Drude weight. As shown in Fig.~\ref{fig:LL_lowtep_strlim}, this analytical expansions agree excellently with the numerical solution obtained from the Bethe ans\"{a}tz.

\subsection{Transport at finite temperature}

At finite temperature, the transport properties of the Lieb-Liniger gas undergo profound changes as the interaction strength changes. Here, we analytically derive the Drude weights in three distinct limits: the weak-coupling regime governed by Bose statistics, the strong-coupling regime characterized by fermionization at low temperature, and the high-temperature limit described by classical Maxwell-Boltzmann statistics.

Before detailing the explicit evaluations, it is necessary to make a few comments on the method of our derivation. The asymptotic expressions presented below, such as Eqs.~\eqref{eq:Dnn_LL_fTiC}-\eqref{eq:Dne_LL_fTiC} for the Lieb-Liniger gas and, similarly, Eqs.~\eqref{eq:Dnn_BF_fTiC_M}-\eqref{eq:Dme_BF_fTiC_M} for the Bose-Fermi mixture, are not obtained by substituting existing thermodynamic approximations into the macroscopic relations (Eqs.~\eqref{eq:Dnn_LL_un}-\eqref{eq:Dnep_LL_un}). Instead, they are  derived {\em ab initio} by directly expanding the dressed charges and root densities within the fundamental Drude weight integral equations (e.g., Eq.~\eqref{eq:DW_LL_MT} and Eq.~\eqref{eq:Dij_BF_M}; see Supplemental material \cite{SM} for detailed derivations). 

This independent approach demonstrates the robustness and self-consistency of the GHD-TBA framework. We note that the expansion of Eq.~\eqref{eq:Dnn_LL_fTiC} and Eq.~\eqref{eq:Dne_LL_fTiC} below consistently reproduce the results reported in Refs.~\cite{Chin.Phys.B2015JiangUnderstanding, J.Phys.A:Math.Theor.2011GuanPolylogsa} (see Supplemental material \cite{SM}). Beyond these specific consistencies, the vast majority of our analytical transport coefficients are entirely new. Although previous studies have evaluated various other thermodynamic quantities under specific limits, the explicit derivation of the enthalpy and heat Drude weights, as well as the comprehensive set of asymptotic transport coefficients for the multi-component Bose-Fermi mixture, has not been previously obtained.

In the weak-coupling regime ($\gamma \ll 1$), the system behaves as a degenerate Bose gas. 
Here, the scattering kernel approaches a delta function: $\phi(k) \to \delta(k)$, simplifying the dressing operation $f^{\text{dr}} \rightarrow f/(1-\theta)$. Consequently, the Drude weights reduce to the thermodynamic forms of free bosons
\begin{align}
	D_{nn} & = \frac{1}{\sqrt{2\pi}}T^{1/2}\text{Li}_{1/2}\left(e^{\mu/T}\right), \\
	D_{ne} & = \frac{3}{2\sqrt{2\pi}}T^{3/2}\text{Li}_{3/2}\left(e^{\mu/T}\right).
\end{align}
where $\Li_{s}(z)$ is the polylogarithm function. Notably,  the pressure is given by $p=2E/L$ (the equation of state for free bosons).

Conversely, in the strong-coupling limit ($\gamma \gg 1$) at finite temperatures $T \ll E_F$ where $E_F$ denotes the Fermi energy, the system enters the Tonks-Girardeau (TG) phase, where hard-core repulsion mimics the Pauli exclusion principle. Consequently, the thermodynamics maps onto that of an ideal Fermi gas. In this regime, the dispersion relation exhibits a parabolic form
 \begin{equation}
 	\epsilon(k)=\beta_{\eff}k^{2}-A,
 \end{equation} 
 where $\beta_{\eff}=1/2$, $A = \mu + 2p/c + (\sqrt{2}/\sqrt{\pi}c^{3}) T^{5/2}\mathrm{Li}_{5/2}(-e^{\mu/T})$. By performing a perturbative expansion of the scattering kernel in powers of $1/c^3$, we derive the Drude weights with finite interaction corrections:
\begin{align}
	D_{nn} & \approx \frac{1}{2\pi}\left(B_{1/2}+\frac{B_{1/2}^{2}}{\pi c}+\frac{B_{1/2}^{3}}{\pi^{2}c^{2}}+\frac{B_{1/2}^{4}}{\pi^{3}c^{3}}-\frac{2B_{1/2}B_{3/2}}{\pi c^{3}}\right),\label{eq:Dnn_LL_fTiC}\\ 
	D_{ne} & \approx \frac{1}{4\pi}\left(3B_{3/2} +\frac{B_{1/2}B_{3/2}}{\pi c}+\frac{B_{1/2}^{2}B_{3/2}}{\pi^{2}c^{2}}\right. \notag \\
	&\qquad \qquad \left.+\frac{B_{1/2}^{3}B_{3/2}}{\pi^{3}c^{3}}+\frac{3B_{3/2}^{2}}{\pi c^{3}}-\frac{B_{1/2}B_{5/2}}{\pi c^{3}}\right).
	\label{eq:Dne_LL_fTiC}
\end{align}
Crucially, the thermal integrals $B_m$ are expressed in terms of polylogarithms $\Li_s(-e^{A/T})$, a mathematical signature of Fermi-Dirac statistics ($B_{s}=\int k^{2s-1}\theta (k)dk=-\beta_{\eff}^{-s}T^{s}\Gamma(s)\text{Li}_{s}(-e^{A/T})$).

Finally, in the high-temperature limit ($T\to \infty$), the thermal de Broglie wavelength becomes smaller than the inter-particle spacing, $T\gg E_F$ and the Yang-Yang thermodynamic formalism naturally reduces to Maxwell-Boltzmann statistics. In this limit, since the fugacity $z = e^{\mu/T} \ll 1$ serves as a naturally small parameter, we employ a virial expansion to derive the transport coefficients rigorously up to second order ($\mathcal{O}(z^2)$):
\begin{align}
	D_{nn} & \approx \frac{T^{1/2}}{\sqrt{2\pi}}\left(z+2\sqrt{2}z^{2}G_{1}\right), \\	
	D_{ne} & \approx \frac{3T^{3/2}}{2\sqrt{2\pi}}\left(z+\frac{\sqrt{2}}{3}z^{2}G_{2}\right).
\end{align}
Here $G_{1}=\int dq'e^{-q'^{2}/T}\phi(2q')-1/2,  G_{2}=3\int_{-\infty}^{\infty}\phi(2q')e^{-q'^{2}/T}dq'+\frac{2}{T}\int_{-\infty}^{\infty}q'^{2}\phi(2q')e^{-q'^{2}/T}dq'-5/4$. The leading term ($\mathcal{O}(z)$) recovers the transport of an ideal classical gas, while the subleading term ($\mathcal{O}(z^2)$) captures the corrections arising from two-body interactions which included the effects of quantum statistics.

\begin{figure}[!htbp]
\includegraphics[width=1\linewidth]{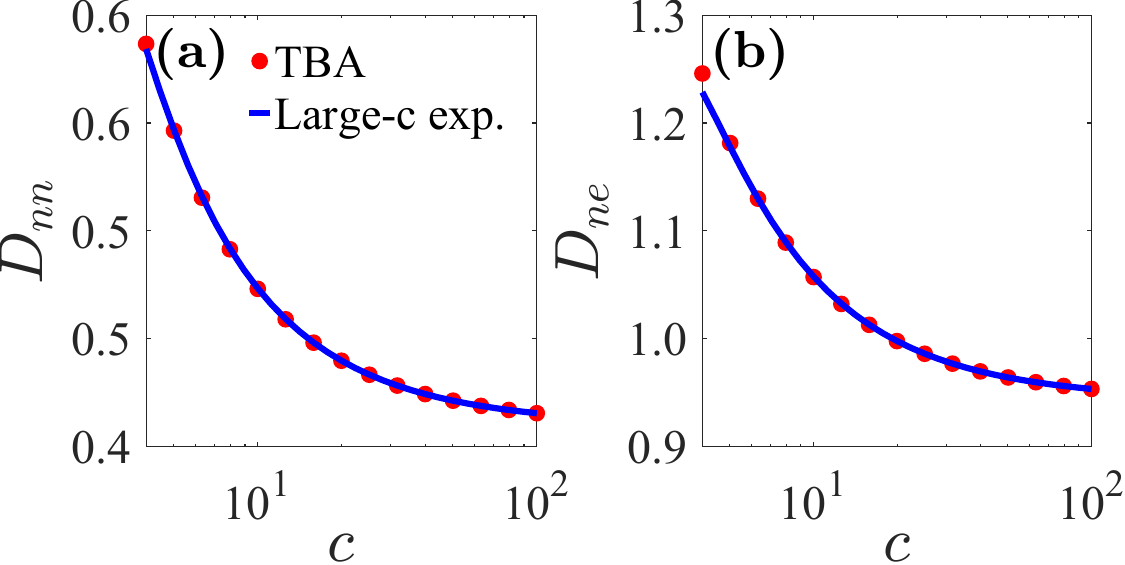}
\caption{Finite-temperature ($T=1$) Drude weights as functions of interaction strength $c$ for the Lieb-Liniger model at the chemical potential $\mu=1$. (a) $D_{nn}$ and (b) $D_{ne}$. Solid lines (black) are the exact numerical TBA solutions. Dashed lines (red) are the analytical large-$c$ expansions, Eqs.~\eqref{eq:Dnn_LL_fTiC} and~\eqref{eq:Dne_LL_fTiC}.}
\label{fig:LL_fintep_strlim}
\end{figure}

\subsection{Quantum criticality}
Finally, let us look at the quantum critical region (QCR), located near the continuous phase transition between the vacuum state and the Tomonaga-Luttinger liquid, which hosts a rich interplay between quantum and thermal fluctuations. In this regime, the physics is no longer governed by the specific microscopic details, but rather by the universal scaling properties of the dilute Bose gas universality class. The quantum phase transition occurs at the critical chemical potential $\mu_c=0$. In the vicinity of this critical point, the low-energy excitations are characterized by a quadratic dispersion $\varepsilon(k) \sim k^2$, which dictates a dynamical exponent $z=2$ and a correlation length exponent $\nu=1/2$.
Consequently, the Drude weights, which act as proxies for the conserved densities in this integrable setup, must obey the universal scaling hypothesis
\begin{align}
	D_{nn} & = n \approx T^{\frac{d}{z}+1-\frac{1}{\nu z}} \mathcal{F}_{1}\left(\frac{\mu-\mu_c}{T^{1/\nu z}}\right), \\
	D_{ne} & = H/L \approx T^{\frac{d}{z}+1} \mathcal{F}_{2}\left(\frac{\mu-\mu_c}{T^{1/\nu z}}\right),
\end{align}
where $d=1$.

To explicitly demonstrate that these transport coefficients are completely independent of the microscopic interaction $c$ in the QCR, we introduce the natural binding energy scale $\epsilon_{b} = \hbar^{2}c^2 / 2m$ (setting $\hbar=m=1$). By defining the dimensionless thermodynamic variables $\tilde{\mu} \equiv \mu/\epsilon_{b}$ and $\mathcal{T} \equiv T/\epsilon_{b}$, we rescale the Drude weights by their respective macroscopic dimensional factors as $\mathcal{D}_{nn} \equiv D_{nn}/c$ and $\mathcal{D}_{ne} \equiv D_{ne}/|c\epsilon_{b}|$. This gives rise to:
\begin{align}
	\mathcal{D}_{nn} &= -\frac{1}{2\sqrt{\pi}}\mathcal{T}^{1/2}\text{Li}_{1/2}\left(-e^{\frac{\Delta\tilde{\mu}}{\mathcal{T}}}\right), \label{eq:Dnn_LL_LJ_t}\\
	\mathcal{D}_{ne} &= -\frac{3}{4\sqrt{\pi}}\mathcal{T}^{3/2}\text{Li}_{3/2}\left(-e^{\frac{\Delta\tilde{\mu}}{\mathcal{T}}}\right), \label{eq:Dne_LL_LJ_t}
\end{align}
where $\Delta\tilde{\mu} \equiv (\mu - \mu_c)/\epsilon_b$.

\begin{figure}[!htbp]
\includegraphics[width=1\columnwidth]{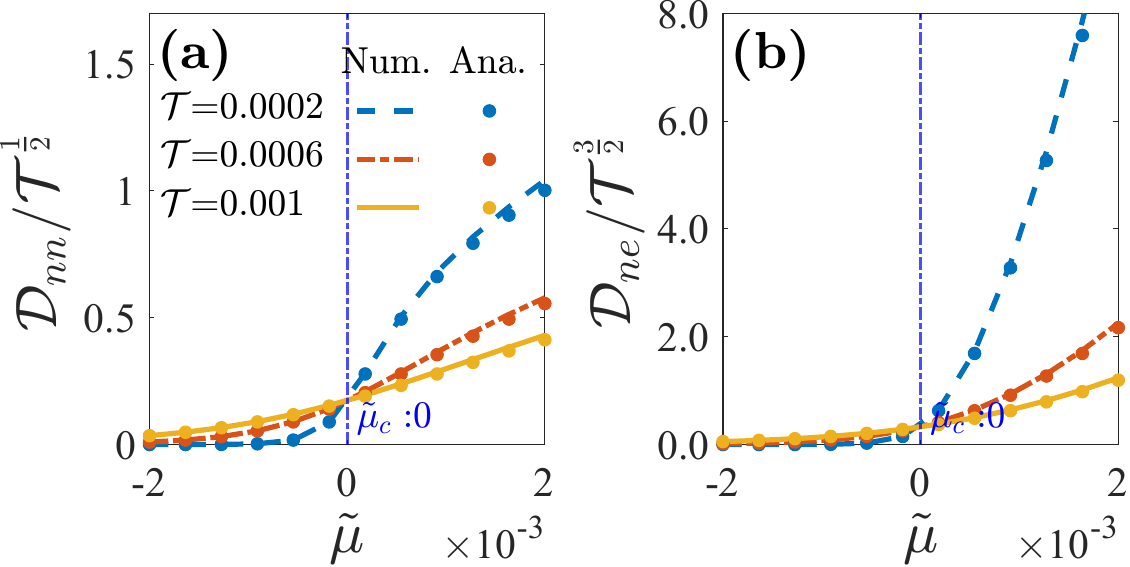}
\caption{The scaling behavior of (a) $D_{nn}$ and (b) $D_{ne}$ versus chemical potential $\mu$ for the phase transitoin with $c = \sqrt{10}$. The critical chemical potential for this transition is $\mu_{c} = 0$, as indicated by the dashed vertical line. The analytical results (symbols) based on Eqs.~\eqref{eq:Dnn_LL_LJ_t} and~\eqref{eq:Dne_LL_LJ_t} are in excellent agreement with numerical solutions (lines) of the TBA equations.}
\label{fig:LL_criticality}
\end{figure}

Here, the appearance of the polylogarithm functions $\text{Li}_s$ reveals the universal scaling function $\mathcal{F}(x)$; see Supplemental material \cite{SM}. The exponents of temperature, $\mathcal{T}^{1/2}$ for $D_{nn}$ and $\mathcal{T}^{3/2}$ for $D_{ne}$, are direct consequences of the $z=2$ dynamical scaling. In Fig.~\ref{fig:LL_criticality}, these analytical scaling laws show excellent agreement with the numerical TBA data, confirming that ballistic transport in the critical region is determined by the universal thermodynamics of the vacuum-to-liquid transition.


\section{DRUDE WEIGHT OF THE BOSE-FERMI MIXTURE MODEL}
\label{sec:BF}
\subsection{Bose-Fermi mixture model and conserved charges}

Having analyzed the single-component Bose gas, we now turn to a multicomponent system to investigate the interplay between different particle species and its effect on the Drude weight matrix. We consider a one-dimensional Bose-Fermi mixture of length $L$ with periodic boundary conditions, described by the second-quantized Hamiltonian~\cite{Phys.Rev.A1971LaiGroundstatea, Phys.Rev.A2006ImambekovExactly}
\begin{equation}
	\begin{split}
		\hat{H} &= \int_{0}^{L}dx \left(\frac{\hbar^{2}}{2m_{b}}\partial_{x}\Psi_{b}^{\dagger}\partial_{x}\Psi_{b}+\frac{\hbar^{2}}{2m_{f}}\partial_{x}\Psi_{f}^{\dagger}\partial_{x}\Psi_{f} \right. \\
		& \qquad \left. +{\frac{g_{bb}}{2}}\Psi_{b}^{\dagger}\Psi_{b}^{\dagger}\Psi_{b}\Psi_{b}+g_{bf}\Psi_{b}^{\dagger}\Psi_{f}^{\dagger}\Psi_{f}\Psi_{b} \right. \\
		& \qquad \left. -\mu_{{F}}\Psi_{f}^{\dagger}\Psi_{f}-\mu_{B}\Psi_{b}^{\dagger}\Psi_{b} \right).
		\label{eq:Hamiltonian_BF_2}
	\end{split}
\end{equation}
Here, $\Psi_{b(f)}$ are the field operators for bosons (fermions), and $g_{bb}$ and $g_{bf}$ are the boson-boson and boson-fermion interaction strengths, respectively. The fermions are spin fully-polarized, so the Pauli exclusion principle forbids $s$-wave interactions between them ($g_{ff}=0$). This general model becomes integrable under specific conditions originally discovered by Lai and Yang~\cite{Phys.Rev.A1971LaiGroundstatea} and further explored in recent years~\cite{Phys.Rev.A2006ImambekovExactly}. We focus on the integrable $SU(1|1)$ symmetric case where masses are equal ($m_b=m_f$) and $g_{bb}=g_{bf}=-(2\hbar^2/ma_{1D})$ (with $g_{ff}=0$). For convenience, we adopt units where $\hbar=m_b=m_f=m_0=1$. In this case,
the first-quantized Hamiltonian for this integrable model is written as
\begin{equation}
	\begin{split}
		\hat{H}=&-\sum_{i=1}^{N}{\frac{\hbar^{2}}{2m_0}}\frac{\partial^{2}}{\partial x_{i}^{2}}+{\frac{\hbar^{2}c}{m_{0}}}\sum_{i<j}^{N}\delta\left(x_{i}-x_{j}\right)\\
		&\qquad -\mu N-\frac{\mathcal{H}}{2}\left(N_{f}-M\right),
	\end{split}
\end{equation}
where $c=-2/a_{1D}$ and $N=N_{f}+M$ is the total number of particles, with $M$ bosons and $N_{f}$ spinless fermions. 
We have defined the average chemical potential $\mu=(\mu_{F}+\mu_{B})/2$ and the effective magnetic field $\mathcal{H}=\mu_{F}-\mu_{B}$, which controls the population imbalance.

The model is exactly solvable within the quantum inverse scattering framework using the nested Bethe ans\"{a}tz (NBA) originally developed by Lai and Yang~\cite{Phys.Rev.A1971LaiGroundstatea}, and has been employed recently to study emergent dynamics~\cite{J.Phys.A:Math.Theor.2020WangEmergent}. The NBA solution reveals that the elementary excitations are characterized by two distinct sets of rapidities: A set of $N$ charge rapidities $\left\{k_j\right\}_{j=1}^N$ (associated with species $\rho$), which carry the total momentum and describe the collective motion of the particle density. A set of $M$ auxiliary rapidities $\{\Lambda_\alpha\}_{\alpha=1}^M$ (associated with species $\sigma$), which correspond to the ``internal" bosonic degrees of freedom. In the thermodynamic limit, the many-body state is fully described by the continuous densities of particles and holes for these two branches. The macroscopic densities and the associated Yang-Yang thermodynamics are determined by~\cite{Phys.Rev.A1971LaiGroundstatea,J.Math.Phys.1974LaiThermodynamics, Phys.Rev.A2009YinYangyanga}:
\begin{align}
	n =& \int dk~\rho(k), \quad (\text{Particle density}),\\
	m =& \int d\Lambda~\sigma(\Lambda), \quad (\text{Boson density}),\\
	\frac{P}{L} =& \int dk~k\rho(k), \quad (\text{Momentum density}),\\
	\frac{E}{L} =& \int dk~\frac{k^{2}}{2}\rho(k). \quad (\text{Energy density}).
\end{align}
Within the GHD framework, these two species ($\rho, \sigma$) have the following bare single-particle eigenvalues ($h_i = \{h_{\rho,i}, h_{\sigma,i}\}$) for the conserved charges in thermal equilibrium~\cite{J.Phys.A:Math.Theor.2020WangEmergent}
\begin{align}
	h_{n}=& \{1,0\}, \quad (\text{Particle number}),\\
	h_{m}=& \{0,1\}, \quad (\text{Boson number}),\\
	h_{f}= & \{1,-1\}, \quad (\text{Fermion number}),\\
	h_{k}=& \{k,0\}, \quad (\text{Momentum}),\\
	h_{e}=& \{k^{2}/2,0\}, \quad (\text{Energy}),\\
	h_{\epsilon}=&\{k^{2}/2-(\mu+\mathcal{H}/2),\mathcal{H}\}, \quad (\text{Thermal energy}).
\end{align}
For notational brevity, we denote the components $h_{\rho,i}$ and $h_{\sigma,i}$ as $h_{\rho_{i}}$ and $h_{\sigma_{i}}$, respectively.
The total local conserved charge $q_i$ and current $j_i$ densities are summations over both species
\begin{align}
	q_{i}=&\int dk\rho(k)h_{\rho_{i}}(k)+\int d\Lambda\sigma(\Lambda)h_{\sigma_{i}}(\Lambda),\\
	j_{i}=&\int dk\rho(k)v_{\rho}^{\text{eff}}(k)h_{\rho_{i}}(k)+\int d\Lambda\sigma(\Lambda)v_{\sigma}^{\text{eff}}(\Lambda)h_{\sigma_{i}}(\Lambda).
\end{align}
Here interactions renormalize (dress) these bare quantities. 
The dressed charges $h_{i}^{\text{dr}}$ are obtained by solving a set of coupled linear integral equations, expressed in operator form as
\begin{equation}
	\left(\begin{array}{c} h_{\rho}^{\text{dr}}\\ h_{\sigma}^{\text{dr}} \end{array}\right)=(1-\widetilde{\boldsymbol{T}}_{\theta})^{-1}\left(\begin{array}{c} h_{\rho}\\ h_{\sigma} \end{array}\right),
\end{equation}
where $\widetilde{\boldsymbol{T}}_{\theta}$ is the matrix of integral operators
\begin{equation}
\tilde{\boldsymbol{T}}_{\theta}=\left[\begin{array}{cc}
	0 & \widetilde{T}_{\rho}\theta_{\sigma}\\
	\widetilde{T}_{\sigma}\theta_{\rho} & 0
\end{array}\right].
\end{equation}
The operators $\widetilde{T}_{\alpha}\theta_{\beta}$ represent the convolution with the scattering kernel weighted by the filling function of the other species, i.e. $(\widetilde{T}_{\rho}\theta_{\sigma} f)(k) = \int dq \, \phi(k-q) \theta_{\sigma}(q) f(q)$.

\subsection{Drude weight}
The linear transport properties of the Bose-Fermi mixture are encoded in the conductivity matrix $\boldsymbol{\sigma}$, which relates the system's currents to applied thermodynamic forces. For the BFM, the  six  primary currents ($j_n, j_f , j_m, j_k, j_e, j_\epsilon$) respond to the three fundamental thermodynamic forces: the gradients of average chemical potential ($\nabla\mu$), Bose chemical potential ($\nabla \mu_{B}$), and temperature ($-\nabla T/T$)
\begin{equation}
\left( \begin{array}{c}
j_{n}\\
j_{f} \\
j_{m}\\
j_{k}\\
j_{e}\\
j_{\epsilon}
\end{array}\right) =\left( \begin{array}{c}
\begin{array}{ccc}
\sigma_{nn} & \sigma_{nm} & \sigma_{ne}\\
\sigma_{fn} & \sigma_{fm}  & \sigma_{fe} \\
\sigma_{mn} & \sigma_{mm} & \sigma_{me}\\
\sigma_{kn} & \sigma_{km} & \sigma_{ke}\\
\sigma_{en} & \sigma_{em} & \sigma_{ee}\\
\sigma_{\epsilon n} & \sigma_{\epsilon m} & \sigma_{\epsilon e}
\end{array}\end{array}\right) \left( \begin{array}{c}
\nabla\mu\\
\nabla\mu_{B}\\
-\nabla T/T
\end{array}\right).
\end{equation}
The Drude weight matrix $D$ is the ballistic (zero-frequency) component of this conductivity matrix $\sigma$ and given by the general GHD projection formula (Eq.~\ref{eq:Dij_GHD_M}): $D=\beta BC^{-1}{B}$. Here again, $C$ (the static susceptibility matrix) and $B$ (the charge-current correlation matrix) are formally defined as thermodynamic derivatives in Eqs.~\ref{eq:Cij_M}-\ref{eq:Bij_M}. The GHD provides a powerful framework to compute these derivatives exactly, by expressing them as integrals over the quasiparticle distributions. The resulting expression for the components of the Drude weight matrix $D_{ij}$ in terms of the properties of the quasiparticle is (see \cite{SM} for derivation)
\begin{equation}
	\begin{split}
		D_{ij}=&\frac{1}{T}\sum_{\alpha=\rho,\sigma} \int du_\alpha~\rho_{\alpha}(u_\alpha)\theta_{\alpha}(u_\alpha)[1-\theta_{\alpha}(u_\alpha)] \\ & \times (v_{\alpha}^{\text{eff}}(u_\alpha))^{2}h_{\alpha_{i}}^{\text{dr}}(u_\alpha)h_{\alpha_{j}}^{\text{dr}}(u_\alpha),
	\end{split}
	\label{eq:Dij_BF_M}
\end{equation}
where the sum is over the two quasiparticle branches: particle ($\alpha=\rho$, $u_\alpha=k$) and boson ($\alpha=\sigma$, $u_\alpha=\Lambda$). The terms in the integrand are the density of states $\rho_{\alpha}$, the filling function factor $\theta_\alpha(1-\theta_\alpha)$, the squared effective velocity $v_{\alpha}^{\text{eff}}$, and the product of the dressed charges $h_{\alpha_{i}}^{\text{dr}}$. Obviously, Eq.(\ref{eq:Dij_BF_M}) satisfies the Onsager reciprocal relations, $D_{ij}=D_{ji}$. 
 
\begin{figure*}
\includegraphics[width=1\linewidth]{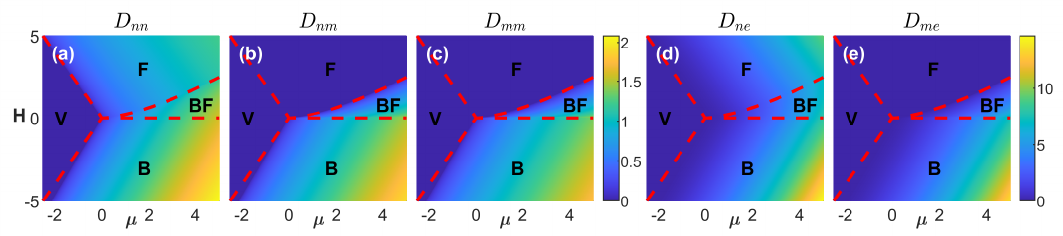}
\caption{Drude weight landscapes across quantum phases ($c=\sqrt{10}$) at low temperature $T=0.001$.
The panels display the values of (a) $D_{nn}$, (b) $D_{nm}$, (c) $D_{ne}$, (d) $D_{mm}$ and (e) $D_{me}$ in the chemical potential–effective field $(\mu, \mathcal{H})$ plane.
These contour plots  illustrate how the magnitude of each transport coefficient varies across the Vacuum, Fermi, Bose, and Bose-Fermi regions, identifying the distinct phase boundaries.}
\label{fig:BF_phase}
\end{figure*}

Depending on the parameters, there are quite a few different quantum phases as shown in Fig.~\ref{fig:BF_phase}. We will compute $D_{ij}$ as a function of the chemical potential $\mu$ and effective field $\mathcal{H}$, and the behavior of these transport coefficients provides a sensitive probe of the transitions between different quantum phases. For example, in the pure Fermi (F) phase, no bosons are present ($m=0$). In this limit, the $\sigma$-branch (boson) integrals in Eq.~\eqref{eq:Dij_BF_M} vanish (as $\rho_{\sigma}=0$), correctly yielding $D_{nm}=D_{mm}=D_{me}=0$, a feature naturally captured by our formalism.

\subsection{Universal thermodynamic behavior}
Using the similar procedure established for the single-component Lieb-Liniger gas, we proceed to evaluate explicitly the GHD integral expression (Eq.~\ref{eq:Dij_BF_M}) for the Bose-Fermi mixture. Strikingly, we find that the Drude weights associated with the conserved quantities of the total system (the $\rho$-branch) are exactly identical to those found in the Lieb-Liniger model
\begin{align}
	D_{nn} &= n,\\
    D_{nk} &=2\frac{P}{L},\\
	D_{ne} &=\frac{E}{L}+p=\frac{H}{L}, \\ 
	D_{n\epsilon} &=D_{ne}-\left(\mu+\frac{\mathcal{H}}{2}\right)D_{nn}+\mathcal{H}D_{nm}=T\frac{S}{L}.
\end{align}
This  structural equivalence reveals a crucial physical insight: despite the intricate internal dynamic coupling between bosons and fermions, the collective transport of the total fluid remains purely convective. The total mass and energy flows are protected by Galilean invariance and experience zero macroscopic drag, making them entirely dictated by the overall equilibrium state functions, as explained before. Furthermore, the methodology presented in Ref.~\cite{Phys.Rev.Lett.2015VasseurExpansion} is applicable to the BFM, leading to the result $D_{nn}=n$, which is consistent with the conclusions of the present work.

In addition, novel components of the Drude weight matrix emerge that are unique to the two-component mixture. For example, the primary cross-transport coefficient $D_{nm}$, which represents the coupling between the bosonic component and the entire system, is found to be precisely given by the boson density $m$
\begin{equation}
	D_{nm}=m. \label{eq:Dnm_BF_un}
\end{equation}
By Onsager reciprocity ($D_{nm}=D_{mn}$), this identity (Eq.~\ref{eq:Dnm_BF_un}) quantifies both the total particle current ($j_n$) driven by a gradient in the boson chemical potential ($\nabla \mu_{B}$), and the boson current ($j_m$) driven by a gradient in the total chemical potential ($\nabla\mu$). Similarly, the thermal cross term $D_{m\epsilon}$ is given by:
\begin{equation}
	D_{m\epsilon}= D_{me}-\left(\mu+\frac{\mathcal{H}}{2}\right)D_{mn}+\mathcal{H}D_{mm}.
\end{equation}

Now, let us address the transport properties of the fermions. Since the dressed charges satisfy the relation $h_{f}^{\text{dr}}=h_{n}^{\text{dr}}-h_{m}^{\text{dr}}$, it follows that the Drude weights involving the fermionic component obey\begin{equation}D_{fi}=D_{ni}-D_{mi},\end{equation}where the index $i$ runs over the set of conserved charges $\{n, m, e, \epsilon, \dots\}$. A prominent example is the coefficient $D_{fn}$, which simplifies to $D_{fn}=D_{nn}-D_{mn}=n_{f}$, where $n_{f}$, denotes the fermion density. We note that Ref.~\cite{arXiv2025SohamBethe}  recently derived the identities $D_{nm}=m$ and $D_{fn}=n_f$ at zero temperature using the generalized Kohn formula. Our GHD-based derivation demonstrates that these macroscopic correspondences  hold universally across all temperatures and interaction regimes.

However, in stark contrast to the perfectly convective flow of the total system, the boson-sector Drude weights $D_{mm}$ and $D_{me}$ cannot be expressed as simple macroscopic state functions. Instead, their complex forms essentially encode the interaction-dependent momentum exchange between the constituent species.

\subsection{Transport at zero temperature}
In the low-temperature limit ($T\to0$), the transport coefficients of the system can be analyzed by means of a Sommerfeld expansion~\cite{Z.Physik1928SommerfeldZur}, a method analogous to that used for the Lieb-Liniger model. 

This approach yields the low-temperature expression for $D_{ij}$ (see Supplemental material \cite{SM})
\begin{eqnarray}
		D_{ij}&=&\frac{1}{\pi} \sum_{{\alpha}=\rho, \sigma} \left[ v_{{\alpha}}^{\eff}h_{{\alpha}_{i}}^{\dr}h_{{\alpha}_{j}}^{\dr}
		+\frac{\pi^{2}T^{2}}{6}\frac{\partial^{2}}{\partial\epsilon^{2}}\left( v_{{\alpha}}^{\eff}h_{{\alpha}_{i}}^{\dr}h_{{\alpha}_{j}}^{\dr}\right) \right.\nonumber\\
		&&\left. 
		+\frac{7\pi^{2}T^{4}}{360}\frac{\partial^{4}}{\partial\epsilon^{4}}\left( v_{{\alpha}}^{\eff}h_{{\alpha}_{i}}^{\dr}h_{{\alpha}_{j}}^{\dr}\right) \right]_{\epsilon=0}.
	\label{eq:Dij_lT_BF_M}
\end{eqnarray}
The term $v_{{\alpha}}^{\text{eff}}$  with ${\alpha}= \rho, \sigma$ denotes the effective velocities  of the quasi-particles in total charge and boson degrees of freedom, $h_{{\alpha}_{i}}^{\text{dr}}$ denotes the corresponding dressed charge, and $u_{F}$ (i.e., $k_{F}$ or $\Lambda_{F}$) stand for the Fermi points for each branch. The leading term gives the ground-state ($T=0$) Drude weight, while the $T^{2}$ and higher terms in term of temperature describe thermal corrections.

At zero temperature, the system can be visualized as two separate Fermi seas for the particle ($\rho$) and boson ($\sigma$) rapidities. The Drude weights are determined by the properties at these Fermi surfaces:
\begin{eqnarray}
		D_{ij}&=&\frac{1}{\pi}v_{\rho}^{\eff}(k_{F})h_{\rho_{i}}^{\dr}(k_{F})h_{\rho_{j}}^{\dr}(k_{F}) \nonumber\\
		&&+\frac{1}{\pi}v_{\sigma}^{\eff}(\Lambda_{F})h_{\sigma_{i}}^{\dr}(\Lambda_{F})h_{\sigma_{j}}^{\dr}(\Lambda_{F}).
		\label{eq:Dij_0T_BF_M}
\end{eqnarray}
Following the matrix formulation in Refs.~\cite{Rep.Progr.Phys.2024LuoExact,Phys.Rev.B2012LeeThermodynamics}, we express the thermodynamic quantities and transport coefficients using the dressed charge matrix $\mathcal{Z}$. The elements $Z_{ij}$ are defined as the values of the dressed charges $h_{i}^{\text{dr}}$ evaluated at the respective Fermi rapidities of the $\alpha$-branch ($\alpha \in \{\rho, \sigma\}$):
\begin{equation}
Z_{i\rho} \equiv h_{i,\rho}^{\text{dr}}(k_F), \quad Z_{i\sigma} \equiv h_{i,\sigma}^{\text{dr}}(\Lambda_F),
\end{equation}
where the index $i$ runs over the conserved charges $\{n, m, f\}$. The dressed charges satisfy the relation $h_{f}^{\text{dr}}=h_{n}^{\text{dr}}-h_{m}^{\text{dr}}$. Consequently, the matrix elements satisfy:
\begin{equation}
	Z_{f\rho} = Z_{n\rho} - Z_{m\rho}, \quad Z_{f\sigma} = Z_{n\sigma} - Z_{m\sigma}.
\end{equation}
The compressibility matrix $\boldsymbol{\kappa}$ describes the system's thermodynamic response and is defined as the Jacobian of the charge densities $q_i$ with respect to the generalized chemical potentials $\mu_j$, i.e., $\kappa_{ij} \equiv \partial q_i / \partial \mu_j$. Explicitly, the components correspond to:
\begin{subequations}
\begin{align}
\kappa_{nn} &= \frac{\partial n}{\partial\mu}, \quad \kappa_{nm} = \frac{\partial m}{\partial\mu}, \quad \kappa_{nf} = \frac{\partial n_{f}}{\partial\mu}, \\
\kappa_{mm} &= \frac{\partial m}{\partial\mu_{b}}, \quad \kappa_{mf} = \frac{\partial n_{f}}{\partial\mu_{b}}, \quad \kappa_{ff} = \frac{\partial n_{f}}{\partial\mu_{f}}.
\end{align}
\end{subequations}
Analogous to the Drude weight matrix, the compressibility matrix inherently satisfies the Onsager reciprocal relations $\kappa_{ij} = \kappa_{ji}$.
	
Using the method outlined in Ref.~\cite{Rep.Progr.Phys.2024LuoExact}, these coefficients can be solved explicitly in terms of the dressed charges and effective velocities $v_\alpha$:
\begin{equation}
\kappa_{ij} = \frac{1}{\pi} \left( \frac{Z_{i\rho}Z_{j\rho}}{v_{\rho}} + \frac{Z_{i\sigma}Z_{j\sigma}}{v_{\sigma}} \right),
\end{equation}
which applies to all sectors $i,j \in \{n,m,f\}$. Here, the velocities are evaluated at the respective Fermi points: $v_{\rho}=v_{\rho}^{\text{eff}}(k_F)$ and $v_{\sigma}=v_{\sigma}^{\text{eff}}(\Lambda_F)$. By rewriting the dressed charges in Eq.~\eqref{eq:Dij_0T_BF_M} using the $\mathcal{Z}$ notation, we obtain the expression for the Drude weights:
\begin{equation}
D_{ij} = \frac{1}{\pi} \left( v_{\rho}Z_{i\rho}Z_{j\rho} + v_{\sigma}Z_{i\sigma}Z_{j\sigma} \right).
\end{equation}

These results can be synthesized into a compact and elegant matrix form. We define the dressed charge matrix $\mathcal{Z}$ (spanning the physical charge space and quasiparticle space) and the effective velocity matrix $\mathcal{V}$ as:
\begin{equation}
\mathcal{Z} = \begin{pmatrix}
			Z_{n\rho} & Z_{n\sigma} \\
			Z_{m\rho} & Z_{m\sigma} \\
			Z_{f\rho} & Z_{f\sigma}
\end{pmatrix}, \quad 
\mathcal{V} = \begin{pmatrix}
			v_{\rho} & 0 \\
			0 & v_{\sigma}
\end{pmatrix}.
\end{equation}
In this notation, the compressibility matrix $\boldsymbol{\kappa}$ and the Drude weight matrix $\boldsymbol{D}$ satisfy the dual relations:
\begin{equation}
\boldsymbol{\kappa} = \frac{1}{\pi} \mathcal{Z} \mathcal{V}^{-1} \mathcal{Z}^{T}, \quad \boldsymbol{D} = \frac{1}{\pi} \mathcal{Z} \mathcal{V} \mathcal{Z}^{T}.
\label{eq:Matrix_DK_Final}
\end{equation}
This formulation reveals a direct link between thermodynamics and transport in integrable systems. By introducing the effective velocity matrix in the physical basis, $\boldsymbol{V} \equiv \mathcal{Z} \mathcal{V} \mathcal{Z}^{-1}$, we recover a relation that connects $\boldsymbol{D}$ and $\boldsymbol{\kappa}$ via the square of the velocity matrix:
\begin{equation}
\boldsymbol{D} = \boldsymbol{V}^2 \boldsymbol{\kappa},
\label{eq:DVK}
\end{equation}
analogous to Eq.(\ref{dnn_kappa}) derived before. Our analytical expressions for the ground-state Drude weights, derived independently by taking the $T=0$ limit of the GHD-TBA integral equations, are in exact agreement with the calculations reported in Ref.~\cite{arXiv2025SohamBethe} following the Kohn approach.

Next, we calculate the analytical results for the ground-state Drude weights in the strong-interaction limit ($c \to \infty$). In the strong-coupling regime ($c \to \infty$), substituting the asymptotic expansions of the dressed charges and effective velocities into the zero-temperature formula Eq.~\eqref{eq:Dij_0T_BF_M} yields explicit expressions up to order $1/\gamma$:
\begin{align}
	D_{nn} & =n \label{eq:Dnn_BF_zTiC_M},\\
	D_{nm} & =m=n\alpha, \\
	D_{mm} & \approx\frac{m^{2}}{n}=n\alpha^{2}, \\
	D_{ne} & \approx\frac{\pi^{2}n^{3}}{2}\left(1-\frac{16}{3\gamma}\alpha\right), \label{eq:Dne_BF_zTiC_M}\\
	D_{me} & \approx\frac{\pi^{2}n^{3}}{2}\alpha\left(1-\frac{16}{3\gamma}\alpha\right), \label{eq:Dme_BF_zTiC_M}
\end{align}
where $\gamma=n/c$ is the dimensionless coupling strength and $\alpha=m/n$ is the boson fraction parameter.
These results reveal a profound underlying structure-a simple proportionality relation:
\begin{equation}
	\left\{ D_{nm},D_{mm},D_{me},D_{m\epsilon}\right\} =\alpha\left\{ D_{nn},D_{nm},D_{ne},D_{n\epsilon}\right\},
\end{equation}
This proportionality relation has a clear physical interpretation: in the TG limit, the fermionized bosons exhibit the same dynamical behavior as the other fermions, causing the entire system to behave as a two-component non-interacting Fermi gas. %
The  transport response of the bosonic component ($\sigma$-branch) is therefore directly proportional to the response of the entire system ($\rho$-branch), scaled by the boson fraction $\alpha$.

\begin{figure}
\includegraphics[width=1\linewidth]{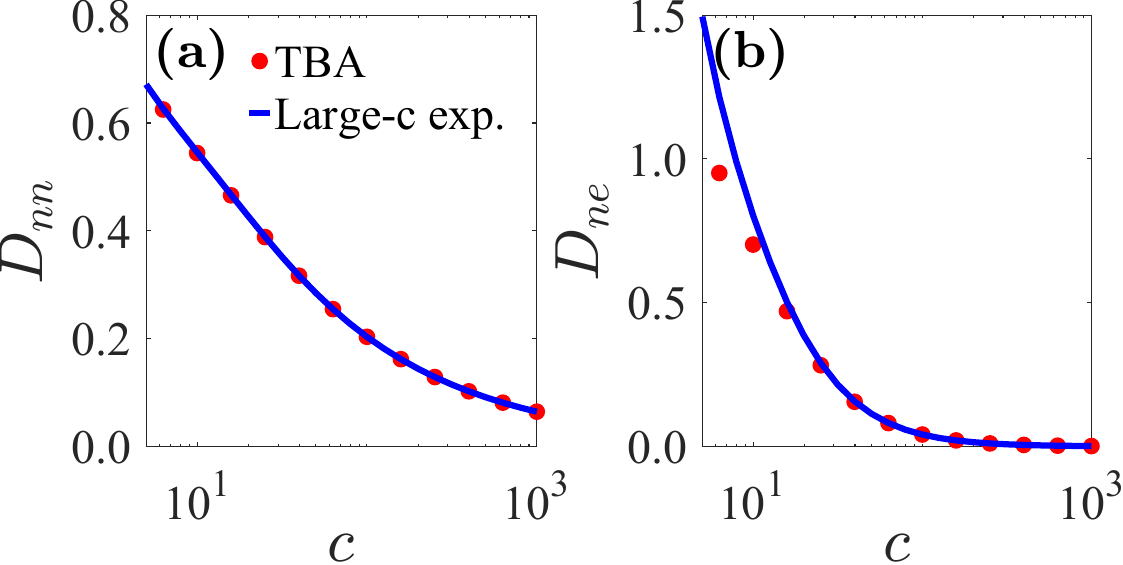}\\
\caption{
Drude weights $D_{nn}$ and $D_{ne}$ in the Bose-Fermi mixture for $T=0.001$, $\mu=1$, and $\mathcal{H}=-1$ (circles). The blue lines represent the analytical large-$c$ expansions [Eqs.~\eqref{eq:Dnn_BF_zTiC_M} and \eqref{eq:Dne_BF_zTiC_M}]. The remaining vanishing components ($D_{nm}$, $D_{mm}$, $D_{me}$) are provided in the Supplemental Material~\cite{SM}.
}
\label{fig:BF_lowtep_strlim}
\end{figure}

To validate these analytical predictions Eqs.~(\ref{eq:Dnn_BF_zTiC_M}-\ref{eq:Dme_BF_zTiC_M}), we compared them with direct numerical solutions from the Bethe ans\"{a}tz equations. As shown in Fig.~\ref{fig:BF_lowtep_strlim}, the agreement between our large-$c$ expansions (red dashed) and the numerical solutions (black solid) in the strong-coupling regime is excellent.

\subsection{Transport at finite temperature}

In the quantum degenerate regime where $T \ll E_F$, the transport signature of the Bose-Fermi mixture evolves dramatically as the interaction strength tunes the effective statistics of the constituents. We analyze two limiting regimes.

In the weak-interaction limit ($c\to0$), the system reduces to a non-interacting gas of bosons and fermions~\cite{Rep.Prog.Phys.2019SowinskiOnedimensional}. In this limit, the two quasiparticle species fully decouple. The analytical results confirm the expectations:
\begin{align}
	D_{nn} & = n =\frac{T^{1/2}}{\sqrt{2\pi}}\Li_{\frac{1}{2}}\left(e^{\mu_{B}/T}\right)+\frac{T^{1/2}}{\sqrt{2\pi}}\Li_{\frac{1}{2}}\left(e^{\mu_{F}/T}\right), \\
	D_{nm} & = m =\frac{T^{1/2}}{\sqrt{2\pi}}\Li_{\frac{1}{2}}\left(e^{\mu_{B}/T}\right), \\
	D_{ne} & = H=\frac{3T^{3/2}}{2\sqrt{2\pi}}\Li_{\frac{3}{2}}\left(e^{\mu_{B}/T}\right)+\frac{3T^{3/2}}{2\sqrt{2\pi}}\Li_{\frac{3}{2}}\left(e^{\mu_{F}/T}\right).
\end{align}
where $H$  is the enthalpy. The Drude weights for the entire system, $D_{nn}$ and $D_{ne}$, become the simple sum of the individual contributions from the free bosonic and free fermionic components.

The Drude weights for the boson sector also assumes their free-gas forms:
\begin{align}
D_{me} & =3\frac{E_{\sigma}}{L}=\frac{3}{2\sqrt{2\pi}}T^{3/2}\Li_{\frac{3}{2}}\left(e^{\mu_{B}/T}\right),\\
D_{mm} & =m=\frac{1}{\sqrt{2\pi}}T^{1/2}\Li_{\frac{1}{2}}\left(e^{\mu_{B}/T}\right),
\end{align}
where $E_{\sigma}/L$ is the energy density of the bosonic component; see Supplemental material \cite{SM}. Notably, in the weak interaction limit, the cross-transport coefficient $D_{nm}$ is exactly the boson density, which is precisely the value of the particle Drude weight $D_{nn}$ for Lieb-Liniger gas of bosons. Similarly, $D_{me}$ matches the particle-energy Drude weight $D_{ne}$ of Lieb-Liniger gas. 

\begin{figure}
\includegraphics[width=1\linewidth]{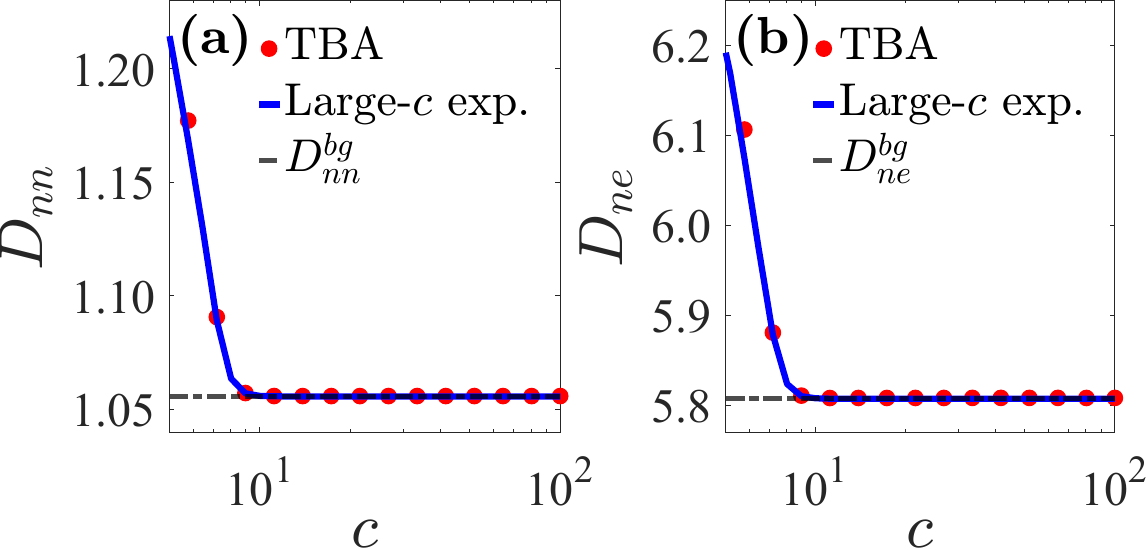}
\caption{
Drude weights $D_{nn}$ and $D_{ne}$ in the Bose-Fermi mixture for $T=0.1$, $\mu=5$, and $\mathcal{H}=1$ (circles). The blue lines represent the analytical large-$c$ expansions [Eqs.~\eqref{eq:Dnn_BF_fTiC_M} and \eqref{eq:Dne_BF_fTiC_M}]. Horizontal lines denote the non-zero free-fermion background limits $D_{nn}^{bg}$ and $D_{ne}^{bg}$. The remaining components ($D_{nm}$, $D_{mm}$, $D_{me}$) are provided in the Supplemental Material~\cite{SM}
}
\label{fig:BF_fintep_strlim}
\end{figure}

On the other hand, in the strong-interaction limit $c\to\infty$, the Bose-Fermi mixture exhibits a phenomenon known as fermionization. The infinite short-range repulsion between bosons prevents them from occupying the same position, forcing all particles to obey an exclusion principle akin to that of fermions. This regime, often called the Tonks-Girardeau (TG) limit, results in thermodynamics equivalent to that of two distinct species of non-interacting fermions~\cite{Rep.Prog.Phys.2019SowinskiOnedimensional}. Crucially, in the regime where thermal fluctuations are suppressed relative to the interaction energy scale ($T \ll c^2$), the system supports two independent types of low-energy excitations, corresponding to particle-hole pairs near the Fermi surfaces of the particle ($\rho$) and boson ($\sigma$) species. The dispersion relations for the particle and boson branches become parabolic, characteristic of free fermions
\begin{equation}
\epsilon(k)=\beta_{\epsilon}k^{2}-A_{\epsilon},\psi(\Lambda)=\beta_{\psi}\Lambda^{2}-A_{\psi}.
\end{equation}
The thermal contributions from these branches are described by the functions $B_{\alpha_s}$
\begin{equation}
	B_{\epsilon_{s}}=\int k^{2s-1}\theta_{\rho}(k)dk,\quad B_{\psi_{s}}=\int\Lambda^{2s-1}\theta_{\sigma}(\Lambda)d\Lambda,
\end{equation}
which can be evaluated to give (see see Supplemental material \cite{SM}):
\begin{align}
	B_{\psi_{s}} & =-\beta_{\psi}^{-s}T^{s}\Gamma(s)\text{Li}_{s}(-e^{A_{\psi}/T}),\\
	B_{\epsilon_{s}} & =-\beta_{\epsilon}^{-s}T^{s}\Gamma(s)\text{Li}_{s}(-e^{A_{\epsilon}/T}).
\end{align}
This allows for an expansion of the Drude weights up to order $1/c^3$:
\begin{align}
D_{nn}  &\approx  \frac{1}{2\pi}\left(B_{\epsilon_{\frac{1}{2}}}+\frac{1}{\pi^{2}c^{2}}B_{\psi_{\frac{1}{2}}}B_{\epsilon_{\frac{1}{2}}}^{2}\right), \label{eq:Dnn_BF_fTiC_M}\\\nonumber
D_{ne}& \approx  \frac{3}{4\pi}\left(B_{\epsilon_{\frac{3}{2}}}+\frac{1}{3\pi^{2}c^{2}}B_{\epsilon_{\frac{3}{2}}}B_{\psi_{\frac{1}{2}}}B_{\epsilon_{\frac{1}{2}}} \right. \label{eq:Dne_BF_fTiC_M} \\
& \left. + \frac{2}{3\pi^{2}c^{2}}B_{\epsilon_{1}}B_{\psi_{\frac{1}{2}}}B_{\epsilon_{1}}\right), \\
D_{mn} & \approx  \frac{1}{2\pi^{2}c}\left(B_{\psi_{\frac{1}{2}}}B_{\epsilon_{\frac{1}{2}}}-\frac{1}{c^{2}}B_{\psi_{\frac{3}{2}}}B_{\epsilon_{\frac{1}{2}}} \right. \notag \\
& \left.-\frac{2}{c^{2}}B_{\psi_{1}}B_{\epsilon_{1}}-\frac{1}{c^{2}}B_{\psi_{\frac{1}{2}}}B_{\epsilon_{\frac{3}{2}}}\right), \\
D_{mm} & \approx  \frac{1}{2\pi^{3}c^{2}}\left(B_{\psi_{\frac{1}{2}}}^{2}B_{\epsilon_{\frac{1}{2}}}+2\pi^{2}B_{\epsilon_{\frac{1}{2}}}^{-1}B_{\psi_{\frac{1}{2}}}B_{\epsilon_{\frac{3}{2}}}\right), \\
D_{me} & \approx  \frac{3}{4\pi^{2}c}\left(B_{\psi_{\frac{1}{2}}}B_{\epsilon_{\frac{3}{2}}}-\frac{2}{3c^{2}}B_{\epsilon_{\frac{1}{2}}}^{-1}B_{\psi_{1}}B_{\epsilon_{1}}B_{\epsilon_{\frac{3}{2}}} \right. \notag \\
& \left.+\frac{2}{3c^{2}}B_{\epsilon_{\frac{1}{2}}}^{-1}B_{\psi_{\frac{1}{2}}}B_{\epsilon_{\frac{3}{2}}}^{2}\right).\label{eq:Dme_BF_fTiC_M}
\end{align}
While these expressions look complex, they actually reveal the separate contributions from the two different thermal excitations to the transport properties, see Supplemental material \cite{SM}. Particularly, the leading-order behavior describes the transport characteristics of two independent quasi-particle gases, while the higher-order terms describe the influence of residual interactions on the transport properties. Notably, in the infinite interaction limit ($c \to \infty$), the cross-transport coefficients involving magnetization vanish, whereas the diagonal terms $D_{nn}$ and $D_{ne}$ do not vanish but instead converge to the finite background limits of non-interacting fermions:
\begin{align}
	D_{nn}^{bg} &\approx -\frac{T^{1/2}}{\sqrt{2\pi}}\text{Li}_{1/2}\left(-e^{(\mu+\mathcal{H}/2)/T}\right),\\
	D_{ne}^{bg} &\approx -\frac{3T^{3/2}}{2\sqrt{2\pi}}\text{Li}_{3/2}\left(-e^{(\mu+\mathcal{H}/2)/T}\right).
\end{align}
The excellent agreement between these analytical expansions Eqs.~(\ref{eq:Dnn_BF_fTiC_M}-\ref{eq:Dme_BF_fTiC_M}) and the numerical TBA solutions is demonstrated in Fig.~\ref{fig:BF_fintep_strlim}.

\subsection{Quantum criticality}

In the vicinity of a zero-temperature quantum phase transition (QPT), the interplay between thermal and quantum fluctuations becomes profound, giving rise to a quantum critical region.
This region, defined by the condition $T \gg |\mu - \mu_c|$, is governed by universal scaling laws where thermodynamic and transport properties are dictated solely by the dimensionality and symmetries of the system, and not by its microscopic details. Drude weights serve as a particularly powerful probe in this regime. 

As we have demonstrated (in Fig.~\ref{fig:BF_phase}), the Drude weight matrix provides sharp, unambiguous signatures to distinguish different quantum phases. 
Here, we delve into the universal scaling behavior of Drude weights across the two primary quantum phase transitions in our model. The control parameters for these transitions are the chemical potential $\mu$ and the effective field $\mathcal{H}$.

\subsubsection{The vacuum-to-Fermi (V-F) transition}
\begin{figure}
\includegraphics[width=1\linewidth]{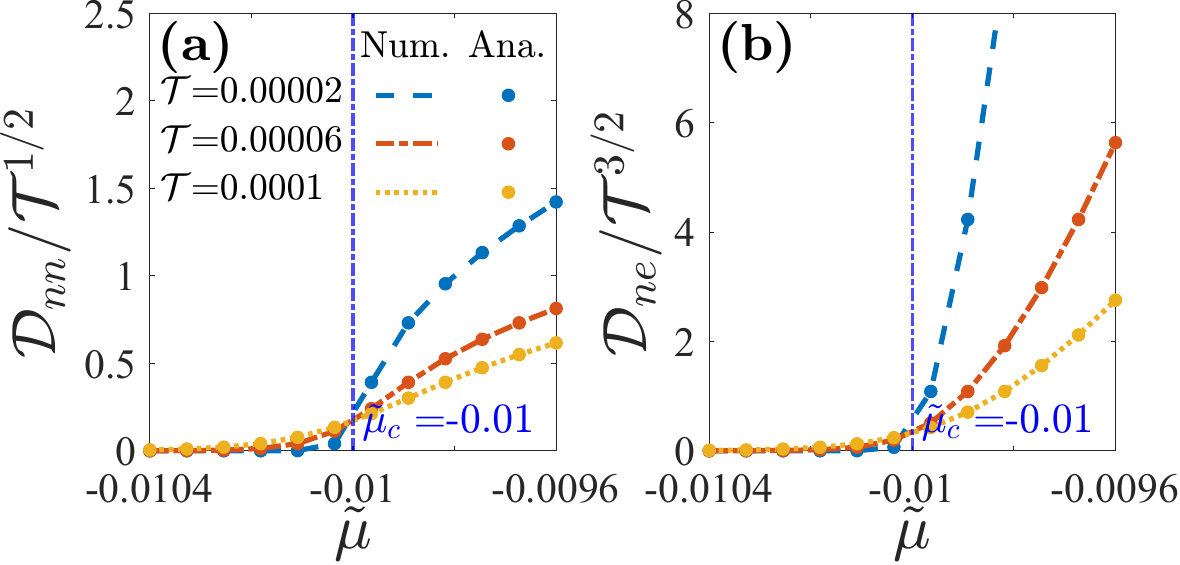}
\caption{
Scaling of Drude weights $D_{nn}$ and $D_{ne}$ versus chemical potential $\mu$ across the phase transition (solid lines). Solid circles represent the analytical results Eqs.~\eqref{eq:Dnn_BF_LJ1_dimless}-\eqref{eq:Dne_BF_LJ1_dimless}. The vertical dashed line marks the critical point $\tilde{\mu}_{c} = -0.01$. The remaining components are provided in the Supplemental Material~\cite{SM}. }
\label{fig:BF_criticality-1}
\end{figure}

The first quantum critical point occurs at $\mathcal{H}_{c}=-2\mu_{c}$ for $\mathcal{H}_{c}>0$, marking the transition from vacuum phase (V) to a pure Fermi (F) phase. Here, the absence of a background density implies that all transport is generated by critical thermal fluctuations. The elementary excitations obey a quadratic dispersion $\varepsilon(k) \sim k^2$, dictating a dynamical exponent $z=2$ and correlation length exponent $\nu=1/2$. 
In order to obtain dimensionless scaling functions, we introduce the natural binding energy scale $\epsilon_b = \hbar^{2}c^2/2m$ and define the dimensionless variables $\tilde{\mu} \equiv \mu/\epsilon_b$, $\tilde{\mathcal{H}} \equiv \mathcal{H}/\epsilon_b$, and $\mathcal{T} \equiv T/\epsilon_b$. By rescaling the Drude weights with their respective macroscopic dimensional factors as $\mathcal{D}_{nn} \equiv D_{nn}/c$ and $\mathcal{D}_{ne} \equiv D_{ne}/|c\epsilon_b|$,  the Drude weights exhibit pure power-law scaling identical to the single-component case
\begin{eqnarray}
		\mathcal{D}_{nn} & \approx& -\frac{1}{2\sqrt{\pi}}\mathcal{T}^{1/2}\text{Li}_{\frac{1}{2}}\left(-e^{\frac{\Delta\tilde{\mu}+\Delta \tilde{\mathcal{H}}/2}{\mathcal{T}}}\right), \label{eq:Dnn_BF_LJ1_dimless} \\
		\mathcal{D}_{ne} & \approx& -\frac{3}{4\sqrt{\pi}}\mathcal{T}^{3/2}\text{Li}_{\frac{3}{2}}\left(-e^{\frac{\Delta\tilde{\mu}+\Delta \tilde{\mathcal{H}}/2}{\mathcal{T}}}\right), \label{eq:Dne_BF_LJ1_dimless} \\
		\mathcal{D}_{nm} & = & \mathcal{D}_{mm} = \mathcal{D}_{me} =0,
	\end{eqnarray} 
where $\Delta\mu=\mu-\mu_{c}$ and $\Delta \mathcal{H}=\mathcal{H}-\mathcal{H}_{c}$. In the Fermi phase, the absence of a bosonic component implies that the associated Drude weights  ($D_{nm},D_{mm},D_{me}$) are trivially zero. The analytical results in Eqs.~\eqref{eq:Dnn_BF_LJ1_dimless}-\eqref{eq:Dne_BF_LJ1_dimless}  highlight the universal character of the vacuum-to-liquid transition. Strikingly, the Drude weights exhibit scaling forms identical to those of the single-component Bose gas, confirming that the physics is governed by the dilute gas fixed point where specific particle statistics become secondary. The observed power-law dependencies--$D_{nn} \propto T^{1/2}$ and $D_{ne} \propto T^{3/2}$--verify the dynamical scaling hypothesis for a system with quadratic dispersion $z=2$, demonstrating that the macroscopic transport is fully determined by the universal thermodynamics of the underlying quantum fluid. This scaling behavior is numerically verified in Fig.~\ref{fig:BF_criticality-1}.

\begin{figure}
\includegraphics[width=1\linewidth]{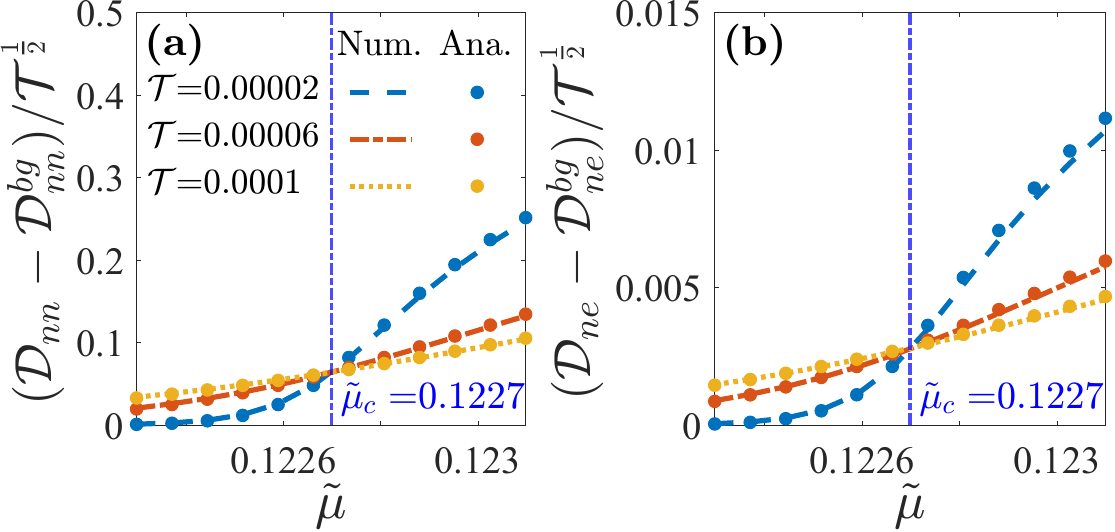}
\caption{
Scaling of Drude weights $D_{nn}$ and $D_{ne}$ versus chemical potential $\mu$ across the F-BF phase transition and $\mathcal{H}=1$ (lines). Circles represent the analytical results Eqs.~\eqref{eq:Dnn_BF_LJ2_dimless} and \eqref{eq:Dne_BF_LJ2_dimless}. The vertical dashed line marks the critical point $\tilde{\mu}_{c}=0.1227$. The remaining components are provided in the Supplemental Material~\cite{SM}. }
\label{fig:BF_criticality-2}
\end{figure}

\subsubsection{Fermi-to-Bose-Fermi (F-BF) transition}

The second and more intricate QPT occurs when the system transits from a pure Fermi (F) phase to a mixed Bose-Fermi (BF) phase. This critical point is located at $\mathcal{H}_{c} \approx \frac{4\sqrt{2}}{3\pi c}\left(\mu_{c}+\frac{\mathcal{H}_{c}}{2}\right)^{3/2},\mathcal{H}_{c}>0$ in the strong interaction limit. A distinct critical behavior emerges at phase boundaries where one species (or spin component) is already populated, forming a degenerate Fermi sea, while the other undergoes a quantum phase transition. This scenario is characterized by the coexistence of a temperature-independent background and singular critical fluctuations. 

The total Drude weight splits into a ``ground-state" contribution (dictated by the background Fermi surface) and a ``critical" correction scaling with temperature. 
After  rescaling  the Drude weights with their respective macroscopic dimensional factors as $\mathcal{D}_{nn} \equiv D_{nn}/c$, $\mathcal{D}_{ne} \equiv D_{ne}/|c\epsilon_b|$, $\mathcal{D}_{nm} \equiv D_{nm}/c$, $\mathcal{D}_{mm} \equiv D_{mm}/c$, and $\mathcal{D}_{me} \equiv D_{me}/|c\epsilon_b|$, we obtain 
\begin{align}
	\mathcal{D}_{nn} & \approx \frac{\left(\tilde{\mu}+\tilde{\mathcal{H}}/2\right)^{1/2}}{\pi} \nonumber \\
	& \quad \,\, -\frac{\sqrt{3}\mathcal{T}^{1/2}\left(\tilde{\mu}+\tilde{\mathcal{H}}/2\right)^{1/4}}{\pi^{2}}\text{Li}_{\frac{1}{2}}\left(-e^{\frac{\tilde{R}_{0}\Delta\tilde{\mu}+\tilde{S}_{0}\Delta\tilde{\mathcal{H}}}{\mathcal{T}}}\right), \label{eq:Dnn_BF_LJ2_dimless} \\
	\mathcal{D}_{ne} & \approx \frac{\left(\tilde{\mu}+\frac{\tilde{\mathcal{H}}}{2}\right)^{3/2}}{\pi} \nonumber \\
	& \quad \,\, -\frac{\mathcal{T}^{1/2}\left(\tilde{\mu}+\frac{\tilde{\mathcal{H}}}{2}\right)^{5/4}}{\sqrt{3}\pi^{2}}\text{Li}_{\frac{1}{2}}\left(-e^{\frac{\tilde{R}_{0}\Delta\tilde{\mu}+\tilde{S}_{0}\Delta\tilde{\mathcal{H}}}{\mathcal{T}}}\right), \label{eq:Dne_BF_LJ2_dimless} \\
	\mathcal{D}_{nm} & \approx -\frac{\sqrt{3}\mathcal{T}^{1/2}}{2\pi\left(\tilde{\mu}+\tilde{\mathcal{H}}/2\right)^{1/4}}\text{Li}_{\frac{1}{2}}\left(-e^{\frac{\tilde{R}_{0}\Delta\tilde{\mu}+\tilde{S}_{0}\Delta\tilde{\mathcal{H}}}{\mathcal{T}}}\right),\\
	\mathcal{D}_{mm} & \approx -\frac{\mathcal{T}^{1/2}\left(\tilde{\mu}+\tilde{\mathcal{H}}/2\right)^{1/4}}{2\sqrt{3}}\text{Li}_{\frac{1}{2}}\left(-e^{\frac{\tilde{R}_{0}\Delta\tilde{\mu}+\tilde{S}_{0}\Delta\tilde{\mathcal{H}}}{\mathcal{T}}}\right) , \\
	\mathcal{D}_{me} & \approx -\frac{\sqrt{3}\mathcal{T}^{1/2}\left(\tilde{\mu}+\tilde{\mathcal{H}}/2\right)^{3/4}}{2\pi}\text{Li}_{\frac{1}{2}}\left(-e^{\frac{\tilde{R}_{0}\Delta\tilde{\mu}+\tilde{S}_{0}\Delta\tilde{\mathcal{H}}}{\mathcal{T}}}\right). \label{eq:Dme_BF_LJ2_dimless}
\end{align}  
with $\tilde{R}_{0}=\frac{2\sqrt{2}}{\pi c}(\tilde{\mu}_{c}+\tilde{\mathcal{H}}_{c}/2)^{1/2}$ and $\tilde{S}_{0}=-1+\frac{\sqrt{2}}{\pi c}(\tilde{\mu}_{c}+\tilde{\mathcal{H}}_{c}/2)^{1/2}$. For details, see Supplemental material \cite{SM}. The analytical expressions in Eqs.~\eqref{eq:Dnn_BF_LJ2_dimless}-\eqref{eq:Dme_BF_LJ2_dimless} demonstrate that the total Drude weight is composed of a background contribution and a critical correction. The leading term represents the background contribution arising from the majority species ($D^{\text{bg}}$). 
Furthermore, in the low-temperature $\mathcal{D}_{nn}^{\text{bg}}=\left(\tilde{\mu}+\tilde{\mathcal{H}}/2\right)^{1/2}/\pi$ and $\mathcal{D}_{ne}^{\text{bg}}=\left(\tilde{\mu}+\tilde{\mathcal{H}}/2\right)^{3/2}/\pi$. This scaling behavior is numerically verified in Fig.~\ref{fig:BF_criticality-2}.

\section{EXPERIMENTAL PROTOCOLS AND NUMERICAL VALIDATION}
\label{sec:expt}
The Drude weights calculated in this work are fundamental transport coefficients, and are directly accessible to measurements in modern cold-atom experiments. To bridge our theoretical findings with actual experiment, we propose two concrete protocols and numerically evaluate their non-equilibrium dynamics. We use the GHD framework at the Euler scale to simulate the macroscopic evolution, starting from initial thermal equilibrium states strictly determined by the TBA. Specifically, we resort to a numerical solution of the GHD equations based on the method of characteristics. This numerical approach efficiently tracks the inhomogeneous translation of the quasiparticle distributions in phase space, allowing for a precise and direct evaluation of the local currents and density profiles.

\subsection{Protocol one: linear quench dynamics}

This protocol extracts the Drude weight by monitoring the ballistic acceleration of the system driven by a thermodynamic force. The system is initialized in global thermal equilibrium. At $t=0$, we suddenly quench the system by introducing a linear spatial gradient to the chemical potential, $\mu(x) = \mu_0 + \mathcal{F} x$. Physically, this chemical potential gradient $\nabla \mu = \mathcal{F}$ acts as a constant effective force on the quasiparticles. In a ballistic system without dissipation, such a force generates a constant acceleration, leading to a drift velocity that increases linearly with time ($v \propto t$)~\cite{SciPostPhys.2017DoyonNote}. Consequently, the induced macroscopic current $j(t)$ is expected to grow linearly~\cite{arXiv2024SchuttelkopfCharacterising}. The Drude weight $D$ is then extracted from the slope of this linear growth in the long-time limit~\cite{Commun.Math.Phys.2013IlievskiThermodyamic,Phys.Rev.Lett.2015VasseurExpansion}
\begin{equation}
	D_{nj}=\lim_{\mathcal{F}\to 0}\lim_{t\to\infty}\frac{j_{j}(t)}{\mathcal{F} \cdot t}.\label{eq:Dij_protocol1}
\end{equation}

\begin{figure}[!htbp]
	\includegraphics[width=1\linewidth]{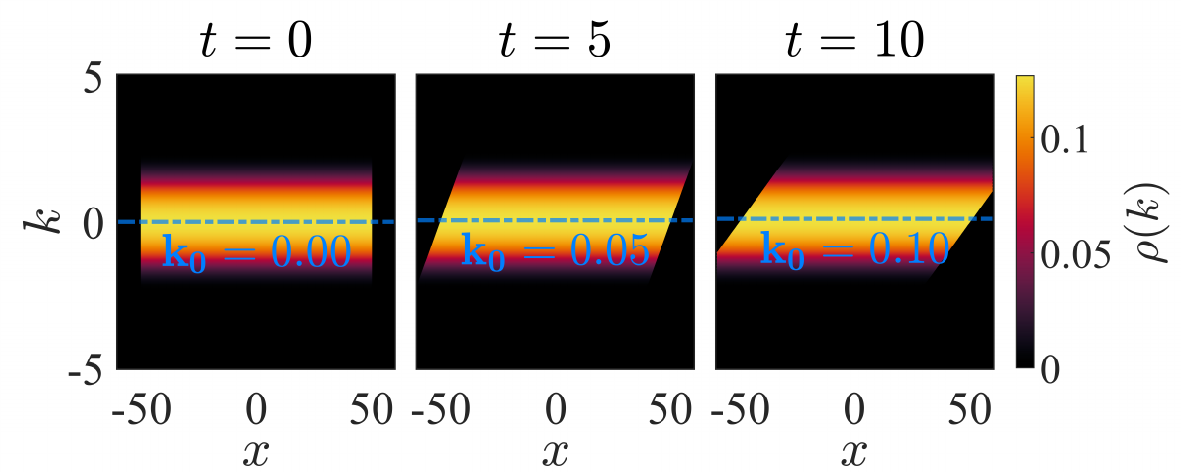}
	\caption{GHD simulation of the Lieb-Liniger gas under a linear potential quench (Protocol 1). The plot shows snapshots of the density functions of the momentum $\rho(k, x)$ in the $(k, x)$ plane at $t=0, 5, 10$. System parameters: $T=1, \mu=1, H=1, c=1$, force $\mathcal{F}=0.01$.}
	\label{fig:LL_n_k_x}
\end{figure}

\begin{figure}[!htbp]
\includegraphics[width=0.8\linewidth]{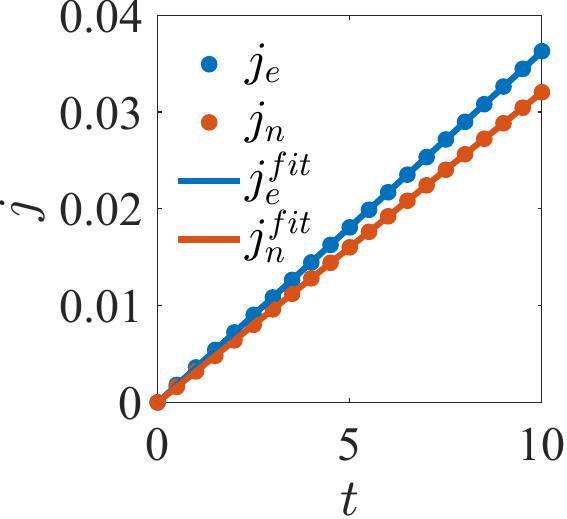}
\caption{Extraction of Drude weights from the linear growth of currents in the Lieb-Liniger gas (Protocol one). The time evolution of the particle current $j_n$ (red) and energy current $j_e$ (blue) is shown following the quench. The discrete symbols (dots) represent the exact numerical results obtained from the GHD evolution. The solid lines denote the linear fits to the data. The strictly linear increase of the currents with time ($j \propto t$) confirms the ballistic nature of the transport, allowing the Drude weights $D_{nn}$ and $D_{ne}$ to be extracted directly from the slopes via Eq.~\eqref{eq:Dij_protocol1}.}
\label{fig:LL_current_t}
\end{figure}

\noindent\textbf{Application to the Lieb-Liniger gas.} 
We first demonstrate this protocol in the single-component Bose gas. The GHD simulation of the phase-space dynamics is presented in Fig.~\ref{fig:LL_n_k_x}.  The snapshots clearly reveal the microscopic mechanism: driven by the chemical potential gradient, the entire momentum distribution $\rho(k, x)$ shifts continuously towards higher momenta. This indicates that the quasiparticles are being uniformly accelerated by the effective force. Macroscopically, this manifests as a strictly linear increase in the particle and energy currents, as shown in Fig.~\ref{fig:LL_current_t}. The linearity confirms the ballistic nature of the transport, allowing us to extract $D_{nn}$ and $D_{ne}$ accurately from the fitted slopes, see Supplemental material \cite{SM}.

\begin{figure}[!htbp]
\includegraphics[width=1\linewidth]{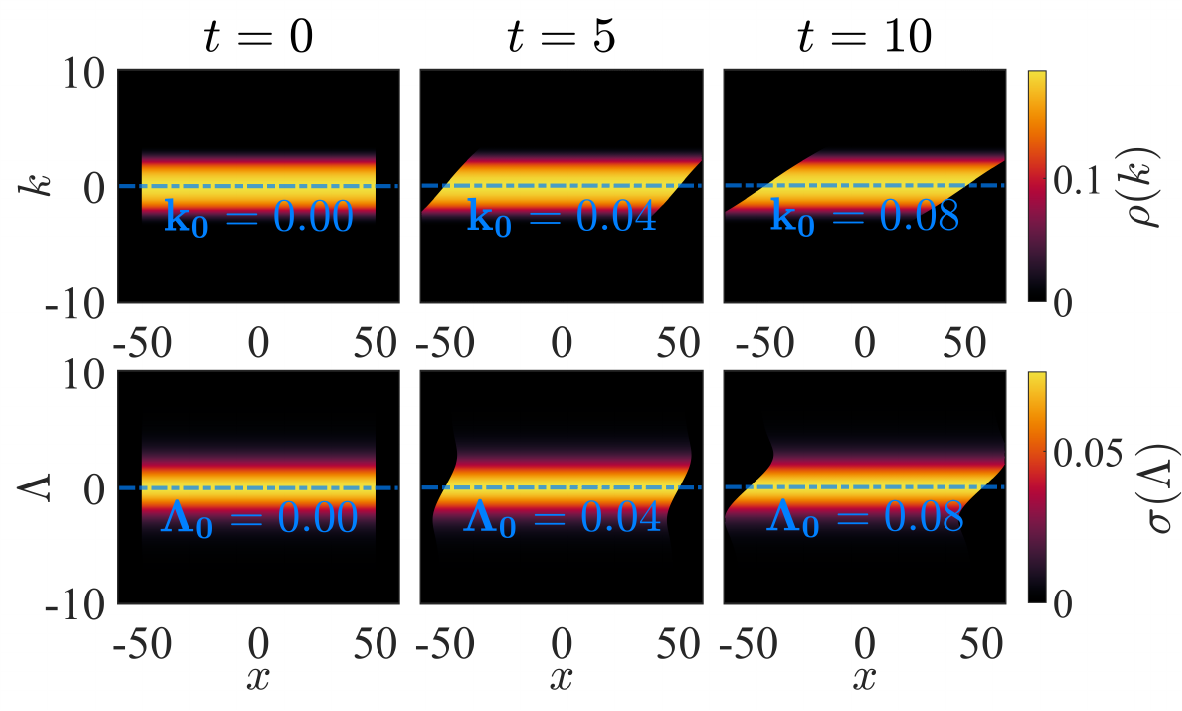}
\caption{GHD simulation of the Bose-Fermi mixture under a linear potential quench (Protocol 1). Top row: density functions of the particle momentum $\rho(k, x)$. Bottom row: density functions of the boson momentum $\sigma(\Lambda, x)$. Snapshots are at $t=0, 5, 10$. System parameters: $T=1, \mu=1, \mathcal{H}=1, c=\sqrt{10}$, force $\mathcal{F}=0.01$.}
\label{fig:BF_n_k_x}
\end{figure}

\begin{figure}[!htbp]
\includegraphics[width=0.8\linewidth]{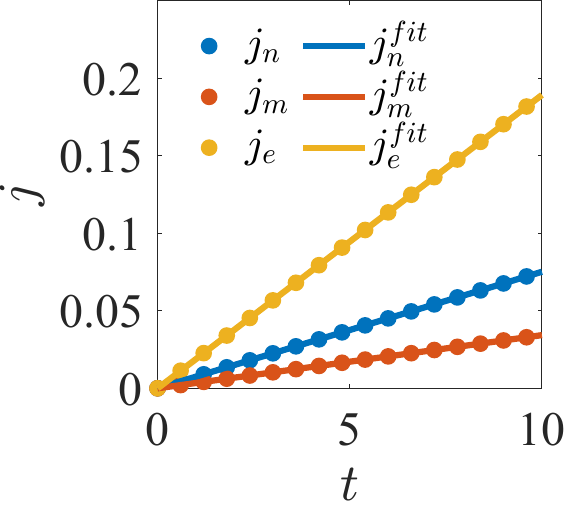}
\caption{Extraction of Drude weights from the linear growth of currents in the Bose-Fermi mixture (Protocol 1).The time evolution of the boson current $J_m(t)$ (red line), particle current $j_n$ (red) and energy current $j_e$ (blue) is shown following the quench. The discrete symbols (dots) represent the exact numerical results obtained from the GHD evolution. The solid lines denote the linear fits to the data. The strictly linear increase of the currents with time ($j \propto t$) confirms the ballistic nature of the transport, allowing the Drude weights $D_{nm}$, $D_{nn}$ and $D_{ne}$ to be extracted directly from the slopes via Eq.~\eqref{eq:Dij_protocol1}.}
\label{fig:BF_current_t}
\end{figure}

\noindent\textbf{Application to the Bose-Fermi mixture.} We further extend this protocol to the Bose-Fermi mixture. Here, the chemical potential gradient acts simultaneously on both species, driving a coupled non-equilibrium evolution. As illustrated in Fig.~\ref{fig:BF_n_k_x}, the momentum distributions for both the particle branch ($\rho$) and the boson branch ($\sigma$) exhibit a synchronized drift in rapidity space. This confirms that both species undergo coherent acceleration. Correspondingly, the total currents $j_n(t)$ and $j_e(t)$ in Fig.~\ref{fig:BF_current_t} maintain a robust linear growth. This demonstrates that Protocol 1 remains a reliable tool for characterizing ballistic transport even in multi-component systems, see Supplemental material \cite{SM}.


\subsection{Protocol two: bipartitioning dynamics}

A second method is the bipartitioning quench protocol. Two semi-infinite thermal subsystems ($x<0$ and $x>0$) are prepared at the same temperature $T$ but with a small imbalance in a thermodynamic potential, $\delta\beta_i$ (e.g., a chemical potential bias $\mu_L = \mu + \delta\mu/2$, $\mu_R = \mu - \delta\mu/2$). At $t=0$, they are joined. The system evolves according to the GHD Euler equations, relaxing to a local quasi-stationary state along each ray $\xi=x/t$. The Drude weight $D_{ij}$ is obtained by integrating the resulting steady-state current $j_j(\xi)$, Specifically, when the bias is applied to the chemical potential (corresponding to the particle sector, $i=n$), the derived quantity corresponds to the Drude weight component $D_{nj}$
\begin{equation}
D_{nj}=\lim_{\delta\mu\to0}\frac{1}{\delta\mu}\int d\xi\,j_{j}(\xi).
\label{eq:Dij_bias}
\end{equation}

The physical equivalence between this bipartitioning dynamics and the microscopic Drude weight can be established within the linear response regime. Consider the small chemical potential bias $\delta\mu$ across the interface as a constant thermodynamic driving force. In the long-time limit, the total transported charge associated with the $j$-th conserved quantity, $\mathcal{J}_j(t) = \int dx \langle\hat{j}_j(x,t)\rangle$, grows strictly linearly in a ballistic system, yielding $\mathcal{J}_j(t) \simeq D_{nj} \delta\mu t$~\cite{Phys.Rev.Lett.2015VasseurExpansion}. Simultaneously, within the Euler-scale GHD framework, the macroscopic current strictly follows the self-similar scaling $j_j(x,t) = j_j(\xi)$. By substituting $x = \xi t$, the spatial integral can be re-written as $\mathcal{J}_j(t) = t \int d\xi j_j(\xi)$. By equating these two expressions of $\mathcal{J}_j(t)$, one finds Eq.(\ref{eq:Dij_bias}). The proof establishing the equivalence with Eq.~(\ref{eq:Dij_BF_M}) is provided in Appendix D of the Supplemental Material \cite{SM}.

\begin{figure}[tb]
\includegraphics[width=1\linewidth]{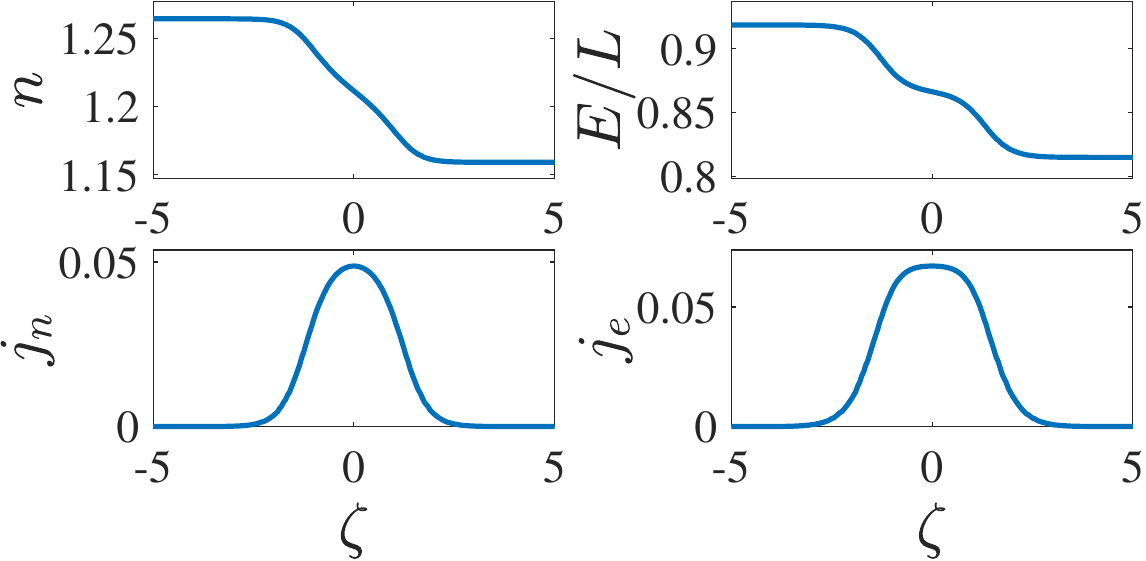}
\caption{GHD steady-state profiles for the Lieb-Liniger gas. The system is evolved using the bipartitioning protocol with interaction strength $c=1$, background chemical potential $\mu=1$, and temperature $T=1$, driven by a small bias $\delta\mu=0.1$. (Top row) Particle density $n$ and energy density $E/L$. (Bottom row) Transport currents for particles $j_n$ and energy $j_e$.}
\label{fig:LL_current_xi}
\end{figure}

\noindent\textbf{Application to the Lieb-Liniger model.}
We now apply the bipartitioning protocol to the single-component Bose gas. The resulting steady-state profiles are presented in Fig.~\ref{fig:LL_current_xi} as a function of the ray $\xi = x/t$. The top row illustrates the spatial variation of the thermodynamic quantities, specifically the particle density $n$ and energy density $E/L$, while the bottom row depicts the resulting transport currents for particles ($j_n$) and energy ($j_e$).
Crucially, the non-zero current signal is observed to be strictly confined within the light cone defined by the maximal quasiparticle velocity.
By integrating these current profiles over $\xi$ according to Eq.~\eqref{eq:Dij_bias}, we obtain a direct measurement of the Drude weights $D_{nn}$ and $D_{ne}$.

\begin{figure}[tb]
\includegraphics[width=1\linewidth]{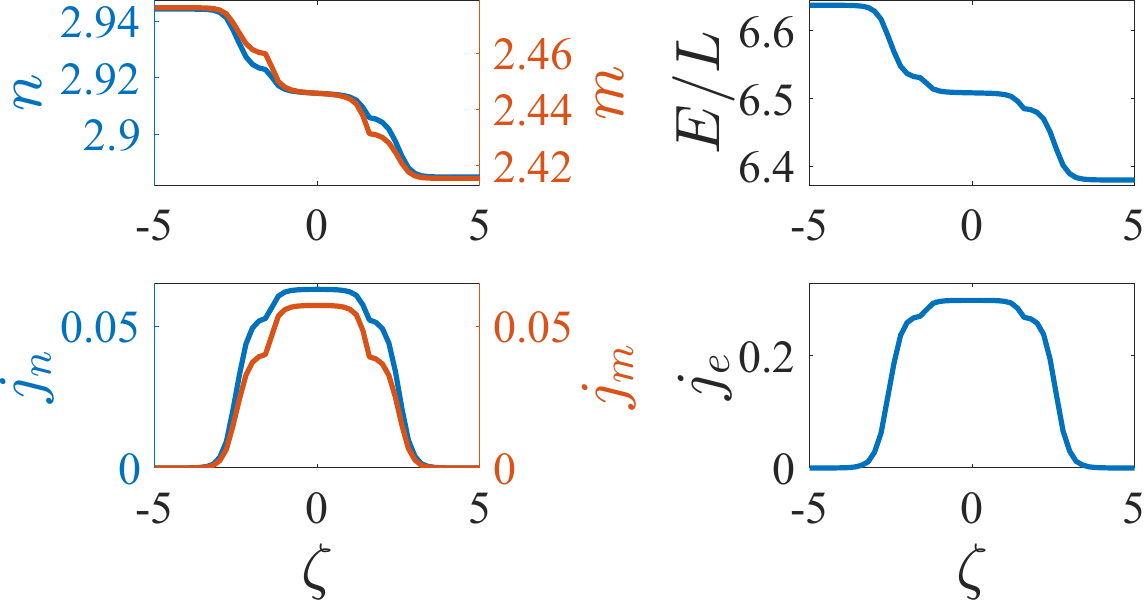}
\caption{GHD steady-state profiles for the Bose-Fermi mixture. Parameters are set to $c=1$, $\mu=1$, $\mathcal{H}=1$, and $T=1$. (Top row) Total particle density $n$, boson density $m$, and energy density $E/L$ plotted against the ray $\xi = x/t$.(Bottom row) Corresponding steady-state currents for the particle ($j_n$), boson ($j_m$), and energy ($j_e$) sectors.}
\label{fig:BF_current_xi}
\end{figure}

\noindent\textbf{Application to the Bose-Fermi mixture.}
This strategy is equally applicable to analyzing transport in the Bose-Fermi mixture.
Figure~\ref{fig:BF_current_xi} shows the steady-state profiles that develop after joining two mixture subsystems with a chemical potential imbalance.
The top panels show the relaxation of the macroscopic densities, including the total particle density $n$, the boson density $m$, and the energy density $E/L$. Correspondingly, the bottom panels reveal the induced currents in the particle ($j_n$), boson ($j_m$), and energy ($j_e$) sectors. Similar to the single-component case, the dynamics are fully captured by GHD, and the components of the Drude weight matrix are precisely determined by integrating these current profiles via Eq.~\eqref{eq:Dij_bias}.

\begin{figure}[tb]
\includegraphics[width=1\linewidth]{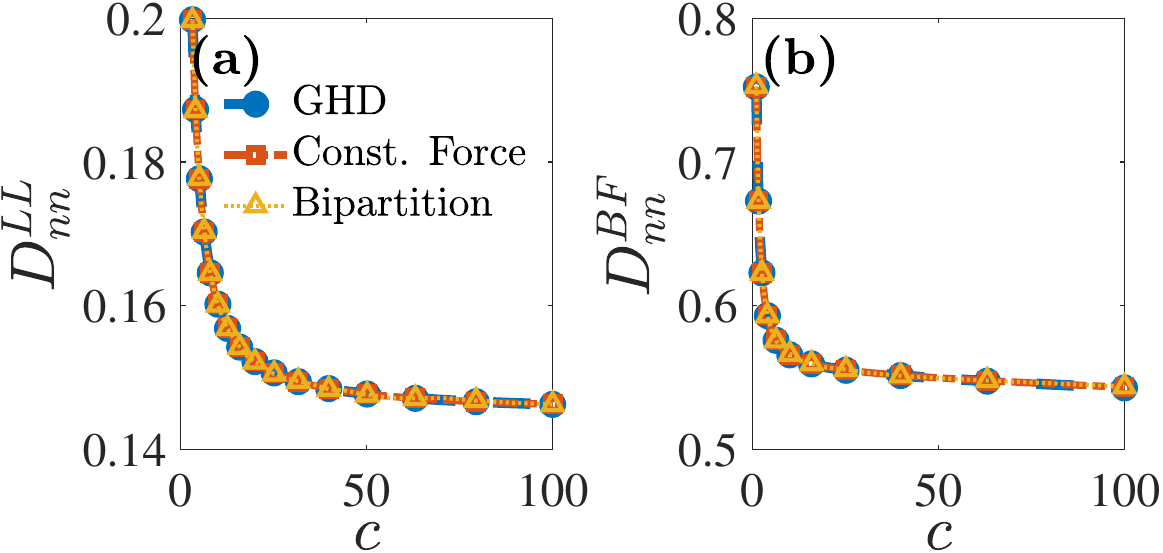}
\caption{Validation of Drude weight protocols.
The particle Drude weight $D_{nn}$ is plotted for (a) the Lieb-Liniger gas and (b) the Bose-Fermi mixture.
The analytical predictions (solid lines) are in perfect agreement with numerical results from the linear quench (circles) and bipartitioning (triangles) protocols.}
\label{fig:LL_compare}
\end{figure}

Finally, the ultimate validation of our theoretical framework lies in comparing these different approaches. In Fig.~\ref{fig:LL_compare}, we present a comprehensive comparison for both the Lieb-Liniger model and the Bose-Fermi mixture. We find perfect agreement between the Drude weights calculated via: (i) The direct analytical GHD integral formula Eq.~\eqref{eq:DW_LL_MT} and Eq.~\eqref{eq:Dij_BF_M}. (ii) The dynamical simulation of the linear potential quench protocol. (iii) The dynamical simulation of the bipartitioning protocol. This remarkable consistency demonstrates the robustness of the GHD framework and confirms that these experimental protocols provide reliable and direct access to the ballistic transport coefficients of one-dimensional integrable systems, see Supplemental material \cite{SM}.

\subsection{Comparison with recent experimental measurements}

Very recently, direct measurements of Drude weights in a 1D Bose gas have been achieved experimentally via ballistic currents generated by constant-force and bipartitioning protocols~\cite{arXiv2024SchuttelkopfCharacterising}. The extracted Drude weight is compared with the explicit analytic expressions based on GHD [our eq.(\ref{eq:DW_LL_MT})] and good agreements are found. However, its connection with the thermodynamics of the system is not emphasized and revealed. 

\begin{figure}[tb]
	\includegraphics[width=1\linewidth]{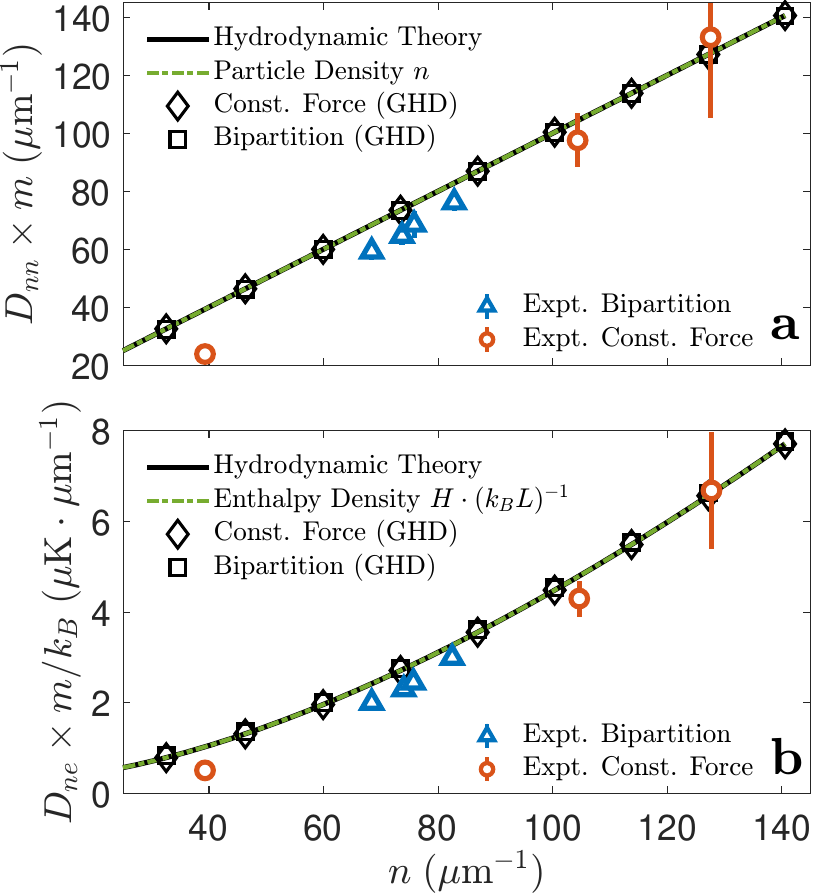}
	\caption{
Comparison of the measured Drude weights with theory  prediction  in the 1D Lieb-Liniger gas. (a) The particle Drude weight $D_{nn}$ and (b) the energy-current Drude weight $D_{ne}$ as functions of the mean atomic density $n$. The solid black lines represent the direct evaluation of our analytical GHD integral formulas [Eqs.~\eqref{eq:DW_LL_MT} and~\eqref{eq:Dij_BF_M}]. The green dashed lines plot the fundamental thermodynamic state variables: the particle density $n$ in (a) and the enthalpy density $H/L = {E}/{L} + p$ in (b). Their overall consistency  confirms  the thermodynamic identities $D_{nn} = n$ and $D_{ne} = H/L$. Black diamonds and squares stand for our numerical evaluations of the GHD equations using the bipartition and constant-force protocols, respectively. The colored symbols (red circles and blue triangles) display the experimental data for the $^{87}\text{Rb}$ atoms, extracted from Fig.~4 of Ref.~\cite{arXiv2024SchuttelkopfCharacterising}, in which the setting for temperature   $T=50\text{ nK}$, and the initial chemical potential imbalance for both the constant force and  bipartition protocols at $\delta\mu = 15\text{ nK}$. The constant force $\phi_n$ is extracted from the relation $\delta \mu=\phi_n L/2$ with $L=100\mu m$.  Here we  explicitly restored the physical constants in our calculation. }
	\label{fig:LL_ex}
\end{figure}

As shown in Fig.~\ref{fig:LL_ex}, we compare our GHD predictions (black solid lines) with experimental data of Ref.~\cite{arXiv2024SchuttelkopfCharacterising} (red circles and  blue triangles stand for constant force and bipartition protocols, respectively )~\cite{arXiv2024SchuttelkopfCharacterising} and our independent numerical solutions of the GHD equations (black diamonds, squares, accordingly). The solid black lines, computed directly from our GHD integral expressions Eqs.~\eqref{eq:DW_LL_MT} and~\eqref{eq:Dij_BF_M}, exhibit excellent agreement with both experimental (red circles, blue triangles) and numerical results (black diamonds, squares) over a broad range of mean atomic densities $n$.

On the other hand, the relations $D_{nn} = n$ and $D_{ne} = H/L$ that links the Drude weight with thermodynamic functions, shed further light on integrability-protected ballistic transport. It is plotted as dash-dotted line in Fig.~\ref{fig:LL_ex} and, as expected, exhibits excellent agreement with the experimental data. Such agreement provides an important independent check on these non-trivial relations and offer another consistent check on the experimental measurements. 


%

%

%

\section{Conclusions}
\label{sec:conclusion}

In summary, by combining GHD with the TBA, we have systematically investigated the ballistic transport properties of the Lieb-Liniger gas and the Bose-Fermi mixture. 
A central achievement of this work is the derivation of exact universal relations that directly link the Drude weight matrices to equilibrium thermodynamic quantities. 
Specifically, we have rigorously identified the Drude weight components corresponding to particle density ($D_{nn}=n$), momentum ($D_{nk}=2P/L$), enthalpy ($D_{ne}=H/L$), and heat ($D_{n\epsilon}=Ts$) beyond the known results at zero temperature \cite{arXiv2025SohamBethe}. Furthermore, we have obtained a comprehensive set of analytical formulas for the Drude weights across diverse interaction regimes. 
These include the strong-coupling limits for both the ground state and the thermally excited states, low-temperature thermal corrections, finite-temperature weak-coupling limits, high temperature quantum corrections to Maxwell-Boltzmann gases and the quantum critical regions. We have also accurately characterized the universal scaling laws governing the Drude weights in the vicinity of quantum phase transitions.

In addition, by numerically investigating two experimental protocols, it is shown that they fit exactly with our analytic derived expressions. This allows us to use the same computational framework for evaluating Drude weights associated with higher order conserved quantities. 
With these protocols, we have  rigorously computed the particle  and energy Drude weights under experimental setting \cite{arXiv2024SchuttelkopfCharacterising}, yielding excellent agreement with particle density and enthalpy, respectively. Our observation  sheds  light on universal relations between the Drude weight elements and thermodynamical functions. We expect our results and methods to have broader  application in other one-dimensional continuum models and can be verified in future cold atom experiments.



\section{Acknowledgements}

We  acknowledge support from the Natural Science Foundation of China (NSFC) key grants  No. 92365202, No. 12134015, U25D8013 and  Grants No.12461160324, No.12474492, and the Innovation Program for Quantum Science and Technology Grant No.  2021ZD0302000 and 2023ZD0300404.
They acknowledges partial support from the National Key R\&D Program of China under Grant No. 2022YFA1404104 and the Science Challenge Project  Grant No.TZ2025017. 
SZ acknowledges support from HK GRF (Grant No. 17306024), CRF (Grants No. C6009-20G, No. C7012-21G, No. C4050-23GF), CRS HKU701/24 and a RGC Fellowship Award No. HKU RFS2223-7S03.

\bibliography{Ref}

\clearpage\newpage
\setcounter{figure}{0}
\setcounter{table}{0}
\setcounter{equation}{0}
\def\thefigure{S\arabic{figure}}
\def\thetable{S\arabic{table}}
\def\theequation{S\arabic{equation}}
\setcounter{page}{1}
\pagestyle{plain}

\begin{widetext}

\section*{Supplementary Material}
\begin{center}
	{Zi-yang Liu, Xiangguo Yin,  Yunbo Zhang, Shizhong Zhang, Xi-Wen Guan}
\end{center}

In this Supplementary Material, we present detailed derivations of the key results reported in the main text.
It is organized as follows: Appendix A gives the analytical results for the Drude weight (DW) of the Lieb-Liniger model.
Appendix B details the derivation of the Generalized Gibbs Ensemble (GGE), dressed operators, and the universal laws governing the DWs with respect to distinct gradients of potential strengths in the Bose-Fermi mixture model.
In Appendix C, we use the general expression for the DW to provide an explicit derivation under thermal equilibrium conditions, from which the universal relations are obtained.
In particular, we calculate the DWs at various temperatures in the strongly interacting regime, including the scaling functions of the DWs for the quantum critical regions.
The results obtained herein establish an intrinsic connection to the Luttinger parameter of the model, thereby shedding light on ballistic transport in quantum systems with multiple degrees of freedom.
In Appendix D, we derive several insightful results for the DWs in a bipartite quench protocol under a homogeneous external potential. We further derive its dynamical evolution via the non-equilibrium hydrodynamic equations.

\section*{Appendix A: Drude weights in Lieb-Liniger model}

\def\thefigure{A\arabic{figure}}
\def\thetable{A\arabic{table}}
\def\theequation{A\arabic{equation}}

\subsection{Lieb-Liniger model}

\subsubsection*{Thermodynamic Bethe Ansatz}

This subsection summarizes the key definitions and standard transport equations
for the Lieb-Liniger model, which form the basis for the analytical
derivations in the following appendices.
The Lieb-Liniger model describes one-dimensional bosons with a repulsive
contact interaction\cite{Phys.Rev.1963LiebExact}. We adopt units
where $\hbar=m_{0}=1$, $m_{0}$ is the mass of the particles.
The first quantization form of the  Hamiltonian for $N$
particles on a ring of length $L$  reads
	\begin{equation}
		\hat{H}=\frac{\hbar^{2}}{2m_{0}}\left[- \sum_{i=1}^{N}\frac{\partial^{2}}{\partial x_{i}^{2}}+2c\sum_{i<j}^{N}\delta(x_{i}-x_{j}) \right].
	\end{equation}
Here $c$ is the interaction strength. 
The model is characterized by a single dimensionless interaction parameter
$c/n$. 

The model is integrable by means of  the Bethe Ansatz. 
The eigenstates are
parameterized by a set of quasi-momenta (rapidities) $\{k_{j}\}_{j=1}^{N}$,
which satisfy the Bethe Ansatz Equations (BAE):
\begin{equation}
	e^{ik_{j}L}=\prod_{l\neq j}^{N}\frac{k_{j}-k_{l}+ic}{k_{j}-k_{l}-ic}
\end{equation}
with $j=1,2,\ldots, N$.
In the thermodynamic limit ($N,L\to\infty$ with $n=N/L$ fixed),
the particle rapidity density $\rho(k)$ and the hole rapidity density
$\rho_{h}(k)$ satisfy the integral equation
\begin{equation}
	\rho(k)+\rho_{h}(k)=\frac{1}{2\pi}+\int_{-\infty}^{\infty}dk'\,\phi(k-k')\rho(k').
\end{equation}

The macroscopic states  at thermal equilibrium (temperature $T=1/\beta$
and chemical potential $\mu$) are  described by the dressed energy
$\epsilon(k)$ of the quasiparticle excitations. 
The dressed energy is the solution to the core TBA integral equation
\begin{equation}
	\epsilon(k)=\frac{k^{2}}{2}-\mu-T\int_{-\infty}^{\infty}dk'\phi(k-k')\ln(1+e^{-\epsilon(k')/T}). \label{Lieb-Liniger-TBA}
\end{equation}
In the above equation, the scattering kernel $\phi(k)$ is the derivative of the two-body
scattering phase shift, i.e. 
\begin{equation}
	\phi(k)=\frac{1}{2\pi}\frac{2c}{c^{2}+k^{2}}.
\end{equation}
The dressed energy (\ref{Lieb-Liniger-TBA}) determines the filling function $\theta(k)$, which
is the probability that a state with rapidity $k$ is occupied.
It is given by the distribution of the dressed energy
\begin{equation}
	\theta(k)=\frac{\rho(k)}{\rho(k)+\rho_{h}(k)}=\frac{1}{1+e^{\epsilon(k)/T}}. 
\end{equation}

\subsubsection*{Local conserved charges, currents and the Dressing Operation}

The GHD framework is built upon the local conserved charges $q_{i}$
and currents $j_{i}$. The charge and current densities are expressed
as projections of their bare eigenvalues $h_{i}(k)$ onto the particle
distribution:
\begin{align}
	q_{i} & =\int dk\,\rho_{p}(k)h_{i}(k), \\
	j_{i} & =\int dk\,\rho_{p}(k)v^{\eff}(k)h_{i}(k). 
\end{align}
The bare charges corresponding to particle number, momentum, energy, and thermal energy are defined as $h_{n}(k)=1$, $h_k(k)=k$, $h_{e}(k)=k^{2}/2$, and $h_{\epsilon}(k)=k^{2}/2-\mu$, respectively.
Specifically, $h_{\epsilon}(k)$ represents the single-particle eigenvalue the  the charge $H - \mu N$.
As being discussed in Ref.~\cite{Phys.Rev.X2016Castro-AlvaredoEmergent}, the energy current is physically associated with this charge. 
Accordingly, we refer the charge $H - \mu N$  to the thermal energy, , i.e. heat. 
Its associated flow is thus  regarded as  the thermal current.
For our convenience, in the following discussion, we denote the dressed charges of particle number, momentum, energy, and thermal energy as  $h_{n}^{\dr}$, $h_k^{\dr}$, and $h_{e}^{\dr}$ as $1^{\dr}$, $k^{\dr}$, and $e^{\dr}$, respectively. 
The interaction-renormalized quantities are obtained via the dressed  operation, $(\cdot)^{\dr}$. A bare function $f(k)$ is ``dressed" by solving the linear integral equation:
\begin{equation}
	f^{\dr}(k)=f(k)+\int dk'\,\phi(k-k')\theta (k')f^{\dr}(k'). 
\end{equation}
This can be expressed in operator form as $f^{\dr}=(1-\tilde{T}\theta )^{-1}f$,
where $\tilde{T}$ is the integral operator with kernel $\phi(k-k')$.

The effective velocity $v^{\eff}(k)$ of an excitation, which
determines the current, is given by the GHD formula as the ratio of
dressed derivatives
\begin{equation}
	v^{\eff}(k)=\frac{(k)^{\dr}}{(1)^{\dr}}. 
\end{equation}

We note that if  the convention  is taken  as $\hbar=2m_{0}=1$  Ref.~\cite{Chin.Phys.B2015JiangUnderstanding}, the obtained energy and Drude weights (DW) are twice  of the present  results.  Some of other  quantities also have different factors.

\subsection{Universal Relation}

In the Lieb-Liniger model, the formal expression of the Drude weights
(DWs) $D_{ij}$ is given by \cite{SciPostPhysics2017DoyonDrude}:

\begin{equation}
	D_{ij}=-\frac{1}{2\pi}\int-\frac{2\pi}{T}\rho(k)\left[1-\theta(k)\right]v^{\eff}(k)v^{\eff}(k)h_{i}^{\dr}(k)h_{j}^{\dr}(k) \,dk. \label{eq:Dij_LL}
\end{equation}
To facilitate the calculation, we first transform the prefactor of
this expression. We take the derivative of the thermal occupation
factor $\theta(k)$ with respect to $k$, noting that $\frac{d\epsilon}{dk}=k^{\dr}=v^{\eff}(h_k')^{\dr}$
and $(h_k')^{\dr}=2\pi\rho/\theta$
\begin{align}
	\frac{d\theta}{dk} & =\frac{d}{dk}\frac{1}{1+e^{\epsilon(k)/T}}\nonumber \\
	& =-\frac{1}{T}\frac{e^{\epsilon(k)/T}}{(1+e^{\epsilon(k)/T})^{2}}\frac{d}{dk}\epsilon(k)\nonumber \\
	& =-\frac{1}{T}\frac{1}{1+e^{\epsilon/T}}\left(1-\frac{1}{1+e^{\epsilon/T}}\right)k^{\dr}\nonumber \\
	& =-\frac{1}{T}\theta (1- \theta)v^{\eff}(h_k')^{\dr}\nonumber \\
	& =-\frac{1}{T} \theta (1-\theta )v^{\eff}\frac{2\pi\rho}{\theta }\nonumber \\
	& =-\frac{2\pi}{T}(1- \theta)v^{\eff}\rho.  \label{eq:dndk_LL}
\end{align}

Substituting this result (Eq.~\ref{eq:dndk_LL}) back into the original
DW expression (Eq.~\ref{eq:Dij_LL}) yields another form
\begin{align}
	D_{ij}= & -\frac{1}{2\pi}\int-\frac{2\pi}{T}\rho(k)\left[1-\theta(k)\right]v^{\eff}(k)v^{\eff}(k)h_{i}^{\dr}(k)h_{j}^{\dr}(k)\,dk\nonumber \\
	= & -\frac{1}{2\pi}\int\frac{d\theta(k)}{dk}v^{\eff}(k)h_{i}^{\dr}(k)h_{j}^{\dr}(k) \,dk,  \label{eq:Dij_dn_LL}
\end{align}
here $i,j$ denote the dressed charges, such as  particle number, momentum, energy, and thermal energy. 
This form (Eq. \ref{eq:Dij_dn_LL} ) is the starting point for our
subsequent calculations. The specific DW matrix elements are obtained
by substituting the corresponding dressed charges $h_{i}^{\dr}$.

\subsubsection*{Universal Relation for $D_{nn}$}

We first calculate the universal relation for $D_{nn}$ by substituting
the dressed charge for particle number, $h_{i}^{\dr}=h_{j}^{\dr}=1^{\dr}$,
into Eq. (\ref{eq:Dij_LL}). 
We also use the identity $v^{\eff}1^{\dr}=k^{\dr}$
\begin{align}
	D_{nn} & =-\frac{1}{2\pi}\int\frac{d\theta(k)}{dk}v^{\eff}(k)1_{\rho}^{\dr}(k)1_{\rho}^{\dr}(k)\,dk\nonumber \\
	& =-\frac{1}{2\pi}\int\frac{d\theta(k)}{dk}k_{\rho}^{\dr}(k)1_{\rho}^{\dr}(k)\,dk. \label{eq:Dnn_LL}
\end{align}
We find a relation for the term $\frac{d\theta}{dk}1^{\dr}$ by differentiating
the momentum density distribution $\rho(k)$:

\begin{align}
	\frac{\rho}{\theta} & =\frac{1}{2\pi}+T\rho, \\
	2\pi\rho & =\theta\left(1+2\pi \tilde{T}\rho\right), \\
	2\pi\frac{d\rho}{dk} & =\frac{d\theta}{dk}\left(1+2\pi \tilde{T}\rho\right)+2\pi \theta\frac{d}{dk}\tilde{T}\rho\nonumber \\
	& =\frac{d\theta}{dk}\left(1+2\pi \tilde{T}\rho\right)+2\pi \theta \tilde{T}\frac{d\rho}{dq} \nonumber \\
	& =\frac{d\theta}{dk}\frac{2\pi\rho}{\theta}+2\pi \theta \tilde{T}\frac{d\rho}{dq}.\label{relation-rho-theta}
\end{align}
This yields the relation $\frac{d\theta}{dk}\frac{2\pi\rho}{\theta}=\frac{d\theta}{dk}1^{\dr}=2\pi\frac{d\rho}{dk}-2\pi \theta \tilde{T}\frac{d\rho}{dq}$.
Substituting this back into the expression for $D_{nn}$ (Eq. \ref{eq:Dnn_LL})
gives:
\begin{align}
	D_{nn} & =-\frac{1}{2\pi}\int\frac{d\theta(k)}{dk}k^{\dr}(k)1^{\dr}(k)\,dk\nonumber \\
	& =-\frac{1}{2\pi}\int\left(2\pi\frac{d\rho}{dk}-2\pi \theta \tilde{T}\frac{d\rho}{dq}\right)k^{\dr}(k)\,dk\nonumber \\
	& =-\int\,dk\frac{d\rho}{dk}k^{\dr}(k)+\int\,dk\theta k^{\dr}(k)\tilde{T}\frac{d\rho}{dq}\nonumber \\
	& =-\int\,dk\frac{d\rho}{dk}k^{\dr}(k)+\int\,dk\theta k^{\dr}(k)\int dq\phi\left(k-q\right)\frac{d\rho}{dq}\nonumber \\
	& =-\int\,dk\frac{d\rho}{dk}k^{\dr}(k)+\int dk\frac{d\rho}{dk}\int\,dq\theta \phi\left(k-q\right)k^{\dr}(q)\nonumber \\
	& =-\int\,dk\frac{d\rho}{dk}\left[k^{\dr}(k)-\int\,dq\theta \phi\left(k-q\right)k^{\dr}(q)\right]\nonumber \\
	& =-\int\,dk\frac{d\rho}{dk}\left[k^{\dr}(k)-\tilde{T}\theta k^{\dr}(q)\right]\nonumber \\
	& =-\int\,dk\frac{d\rho}{dk}k\nonumber \\
	& =k\rho\big|_{-\infty}^{\infty}+\int\rho\,dk\nonumber \\
	& =\int\rho\,dk\nonumber \\
	& =\frac{N}{L}. \label{Dnn-calculation}
\end{align}

Through the calculation above, we rigorous prove  the universal relation $D_{nn}=N/L=n$,
which is the particle number density. 
This conclusion was previously
obtained in Refs. \cite{Phys.Rev.Lett.2015VasseurExpansion,arXiv2025LuoQuantum}.

\subsubsection*{Universal Relation for $D_{ne}$}

Next, we calculate the universal relation expression for $D_{ne}$
by substituting $h_{i}^{\dr}=1^{\dr}$ and $h_{j}^{\dr}=e^{\dr}$ into
Eq. \ref{eq:Dij_LL}:
\begin{align}
	D_{ne} & =-\frac{1}{2\pi}\int\frac{d\theta(k)}{dk}v^{\eff}(k)1^{\dr}(k)e^{\dr}(k)\,dk\nonumber \\
	& =-\frac{1}{2\pi}\int\frac{d\theta(k)}{dk}k^{\dr}(k)e^{\dr}(k)\,dk. \label{eq:Dne_LL}
\end{align}
Following a similar strategy to the $D_{nn}$ calculation by defining
an auxiliary function $u=e^{\dr}\theta=\theta\left(\frac{k^{2}}{2}+\tilde{T}u\right)$.
Differentiating $u$ with respect to $k$ yields:
\begin{align}
	\frac{du}{dk} & =\frac{d}{dk}\theta\left(\frac{k^{2}}{2}+\tilde{T}u\right)\nonumber \\
	& =\frac{d\theta}{dk}\left(\frac{k^{2}}{2}+\tilde{T}u\right)+\theta k+\theta \tilde{T}\frac{du}{dq}\nonumber \\
	& =\frac{d\theta}{dk}e^{\dr}+\theta k+\theta \tilde{T}\frac{du}{dq}.
\end{align}
This gives the relation $\frac{d\theta }{dk}e^{\dr}=\frac{du}{dk}-\theta k-\theta \tilde{T}\frac{du}{dq}$.
Substituting this relation back into Eq. (\ref{eq:Dne_LL}), then we have 
\begin{align}
	D_{ne} & =-\frac{1}{2\pi}\int\frac{d\theta(k)}{dk}k^{\dr}(k)e^{\dr}(k)\,dk\nonumber \\
	& =-\frac{1}{2\pi}\int\left(\frac{du}{dk}-\theta k-\theta \tilde{T}\frac{du}{dq}\right)k^{\dr}(k)\,dk\nonumber \\
	& =-\frac{1}{2\pi}\int\frac{du}{dk}k^{\dr}(k)\,dk+\frac{1}{2\pi}\int \theta kk^{\dr}(k)\,dk+\frac{1}{2\pi}\int \theta \tilde{T}\frac{du}{dq}k^{\dr}(k)\,dk. \label{eq:Dne_ann_LL}
\end{align}
The resulting expression consists of three terms. We first solve the
third term by swapping the integration variables and applying the
dressing operation:
\begin{align}
	\frac{1}{2\pi}\int \theta \tilde{T}\frac{du}{dq}k^{\dr}(k)\,dk= & \frac{1}{2\pi}\int\,dk\theta k^{\dr}(k)\int dq\phi\left(k-q\right)\frac{du}{dq}\nonumber \\
	= & \frac{1}{2\pi}\int dq\frac{du}{dq}\int\,dk\phi\left(k-q\right)\theta k^{\dr}(k)\nonumber \\
	= & \frac{1}{2\pi}\int dk\frac{du}{dk}\int\,dq\phi\left(q-k\right)\theta k^{\dr}(q)\nonumber \\
	= & \frac{1}{2\pi}\int dk\frac{du}{dk}\tilde{T}\theta k^{\dr}(q)
\end{align}
Substituting this back into Eq. (\ref{eq:Dne_ann_LL}) yields and recombining
the terms (using the integral equation for $k^{\dr}$) yields:
\begin{align*}
	D_{ne} & =-\frac{1}{2\pi}\int\frac{du}{dk}k^{\dr}(k)\,dk+\frac{1}{2\pi}\int \theta kk^{\dr}(k)\,dk+\frac{1}{2\pi}\int \theta \tilde{T}\frac{du}{dq}k^{\dr}(k)\,dk\\
	& =-\frac{1}{2\pi}\int\frac{du}{dk}k^{\dr}(k)\,dk+\frac{1}{2\pi}\int \theta kk^{\dr}(k)\,dk+\frac{1}{2\pi}\int dk\frac{du}{dk}\tilde{T}\theta k^{\dr}(q)\\
	& =-\frac{1}{2\pi}\int\,dk\frac{du}{dk}\left[k^{\dr}(k)-\tilde{T}\theta k^{\dr}(q)\right]+\frac{1}{2\pi}\int \theta kk^{\dr}(k)\,dk\\
	& =-\frac{1}{2\pi}\int\,dk\frac{du}{dk}k+\frac{1}{2\pi}\int \theta kk^{\dr}(k)\,dk\\
	&=D_{ne}\left(1\right)+ D_{ne}\left(2\right), \label{Dne}
\end{align*}
Where   by integration by part, the first term, $D_{ne}(1)$ is given by 
\begin{align}
	D_{ne}\left(1\right) & =-\frac{1}{2\pi}\int\,dk\frac{du}{dk}k\nonumber \\
	& =\frac{1}{2\pi}\int\,dke^{\dr}\theta  =\frac{1}{2\pi}\int\,dke\theta 1^{\dr}\nonumber \\
	& =\int dke\rho  =\frac{E}{L}. 
\end{align}
In the above equation, we  utilized  the relation$\int dkg^{\dr}\theta f=\int dkg\theta f^{\dr}$, $g$ stands for an arbitrary conserved charge. 
The proof of this relation  is as follows
\begin{align}
	\int dkg^{\dr}\theta f & =\int dkg^{\dr}\theta \left(f^{\dr}-\tilde{T}\theta f^{\dr}\right)\nonumber \\
	& =\int dkg^{\dr}\theta f^{\dr}-\int dkg^{\dr}\theta \int dq\phi\left(k-q\right)\theta f^{\dr}\nonumber \\
	& =\int dkg^{\dr}\theta f^{\dr}-\int dq\theta f^{\dr}\int dk\phi\left(k-q\right)g^{\dr}\theta \nonumber \\
	& =\int dkg^{\dr}\theta f^{\dr}-\int dk \theta f^{\dr}\int dq\phi\left(q-k\right)g^{\dr} \theta\nonumber \\
	& =\int dk\theta f^{\dr}\left[g^{\dr}-\int dq\phi\left(q-k\right)\theta g^{\dr}\right]\nonumber \\
	& =\int dkg\theta f^{\dr}. 
\end{align}

For the second term $D_{ne}(2)$ in Eq. (\ref{Dne}),  we  prove that it relates
to the thermodynamic pressure $p$. 
First, we compute  the integration by parts on the standard expression for pressure
\begin{align}
	p & =\frac{T}{2\pi}\int_{-\infty}^{\infty}\ln\left(1+e^{-\epsilon(k)/T}\right)dk\nonumber \\
	& =\frac{T}{2\pi}\left\{ k\ln\left[1+e^{-\epsilon(k)/T}\right]\right\} _{-\infty}^{\infty}-\frac{T}{2\pi}\int_{-\infty}^{\infty}kd\ln\left(1+e^{-\epsilon(k)/T}\right)\nonumber \\
	& =\frac{T}{2\pi}\int_{-\infty}^{\infty}k\frac{1}{1+e^{\epsilon(k)/T}}\frac{1}{T}\frac{d\epsilon(k)}{dk}dk\nonumber \\
	& =\frac{1}{2\pi}\int_{-\infty}^{\infty}\frac{k}{1+e^{\epsilon(k)/T}}\frac{d\epsilon(k)}{dk}dk\nonumber \\
	& =\frac{1}{2\pi}\int_{-\infty}^{\infty}\theta(k)k\frac{d\epsilon(k)}{dk}dk\nonumber \\
	& =\frac{1}{2\pi}\int_{-\infty}^{\infty}\theta(k)k\frac{d\epsilon(k)}{dk}dk.  \label{eq:p_n}
\end{align}
We compare the derivative of the dressed energy $\frac{d\epsilon(k)}{dk}$ 
with the dressed momentum $k^{\dr}$ 
\begin{align}
	\frac{d\epsilon(k)}{dk} & =k+\int\frac{1}{\pi}\frac{c}{c^{2}+(k-q)^{2}}\theta\frac{d\epsilon(q)}{dq}dq, \\
	k^{\dr} & =k+\int\frac{1}{\pi}\frac{c}{c^{2}+(k-q)^{2}}\theta k^{\dr}dq.
\end{align}
Thus the uniqueness of the solution implies
they are identical:
\begin{equation}
	\frac{d\epsilon(k)}{dk}=k^{\dr}
\end{equation}
Comparing this identity with expressions for $p$ (Eq. \ref{eq:p_n})
and $D_{ne}(2)$ in Eq. (\ref{Dne}), we obtain
\begin{equation}
	D_{ne}\left(2\right)=\frac{1}{2\pi}\int \theta kk^{\dr}(k)\,dk=p
\end{equation}
Therefore, we find the universal relation
for $D_{ne}$:
\begin{equation}
	D_{ne}=D_{ne}\left(1\right)+D_{ne}\left(2\right)=\frac{E}{L}+p=H/L
\end{equation}
Thus, we have conclude $D_{ne}=H/L$, which is the enthalpy density.
A similar conclusion was also presented  in Ref. \cite{Phys.Rev.Lett.2015VasseurExpansion}.

\subsubsection*{Universal Relation for $D_{nk}$}

We next calculate the universal relation expression for $D_{nk}$
by substituting $h_{i}^{\dr}=1^{\dr}$ and $h_{j}^{\dr}=k^{\dr}$ into
the $D_{ij}$ expression (Eq. \ref{eq:Dij_dn_LL})
\begin{align}
	D_{nk} & =-\frac{1}{2\pi}\int\frac{d\theta(k)}{dk}v^{\eff}(k)1^{\dr}(k)k^{\dr}(k)\,dk\nonumber \\
	& =-\frac{1}{2\pi}\int\frac{d\theta(k)}{dk}k^{\dr}(k)k^{\dr}(k)\,dk. \label{eq:Dnk_LL}
\end{align}
We again follow the same method, introducing an auxiliary function
$r=k^{\dr}\theta=\theta\left(k+\tilde{T}r\right)$. 
Differentiating $r$ with respect
to $k$ yields
\begin{align}
	\frac{\dr}{dk} & =\frac{d}{dk}\theta \left(k+\tilde{T}r\right)\nonumber \\
	& =\frac{dn}{dk}\left(k+\tilde{T}r\right)+\theta+\theta \tilde{T}\frac{\dr}{dq}\nonumber \\
	& =\frac{\dr}{dk}k^{\dr}+\theta+\theta \tilde{T}\frac{\dr}{dq}
\end{align}

This gives the relation $\frac{d\theta}{dk}k^{\dr}=\frac{\dr}{dk}-\theta-\theta \tilde{T}\frac{\dr}{dq}$.
Substituting this back into the original expression for $D_{nk}$
(Eq. \ref{eq:Dnk_LL}):
\begin{align}
	D_{nk} & =-\frac{1}{2\pi}\int\frac{d\theta(k)}{dk}k^{\dr}(k)k^{\dr}(k)\,dk\nonumber \\
	& =-\frac{1}{2\pi}\int\left(\frac{\dr}{dk}-1-\theta \tilde{T}\frac{\dr}{dq}\right)k^{\dr}(k)\,dk\nonumber \\
	& =-\frac{1}{2\pi}\int\frac{\dr}{dk}k^{\dr}(k)\,dk+\frac{1}{2\pi}\int \theta k^{\dr}(k)\,dk+\frac{1}{2\pi}\int \theta \tilde{T}\frac{\dr}{dq}k^{\dr}(k)\,dk. \label{eq:Dnk_ann_LL}
\end{align}
This expression has three terms. The third term is solved identically
to the $D_{ne}$ case:
\begin{align}
	\frac{1}{2\pi}\int \theta \tilde{T}\frac{\dr}{dq}k^{\dr}(k)\,dk= & \frac{1}{2\pi}\int\,dk\theta k^{\dr}(k)\int dq\phi\left(k-q\right)\frac{\dr}{dq}\nonumber \\
	= & \frac{1}{2\pi}\int dq\frac{\dr}{dq}\int\,dk\phi\left(k-q\right)\theta k^{\dr}(k)\nonumber \\
	= & \frac{1}{2\pi}\int dk\frac{\dr}{dk}\int\,dq\phi\left(q-k\right)\theta k^{\dr}(q)\nonumber \\
	= & \frac{1}{2\pi}\int dk\frac{\dr}{dk}\tilde{T}\theta k^{\dr}(q).
\end{align}

Substituting the above equation  into the Eq. (\ref{eq:Dnk_ann_LL}) yields 
\begin{align}
	D_{nk} & =-\frac{1}{2\pi}\int\frac{\dr}{dk}k^{\dr}(k)\,dk+\frac{1}{2\pi}\int \theta k^{\dr}(k)\,dk+\frac{1}{2\pi}\int \theta \tilde{T}\frac{\dr}{dq}k^{\dr}(k)\,dk\nonumber \\
	& =-\frac{1}{2\pi}\int\frac{\dr}{dk}k^{\dr}(k)\,dk+\frac{1}{2\pi}\int \theta k^{\dr}(k)\,dk+\frac{1}{2\pi}\int dk\frac{\dr}{dk}\tilde{T}\theta k^{\dr}(q)\nonumber \\
	& =-\frac{1}{2\pi}\int\,dk\frac{\dr}{dk}\left[k^{\dr}(k)-\tilde{T}\theta k^{\dr}(q)\right]+\frac{1}{2\pi}\int  \theta k^{\dr}(k)\,dk\nonumber \\
	& =-\frac{1}{2\pi}\int\frac{\dr}{dk}k\,dk+\frac{1}{2\pi}\int \theta k^{\dr}(k)\,dk\nonumber \\
	& =\frac{1}{2\pi}\int r\,dk+\frac{1}{2\pi}\int \theta k^{\dr}(k)\,dk\nonumber \\
	& =\frac{1}{2\pi}\int k^{\dr}\theta\,dk+\frac{1}{2\pi}\int \theta k^{\dr}(k)\,dk\nonumber \\
	& =\frac{1}{\pi}\int k^{\dr}\theta \,dk =\frac{1}{\pi}\int k\theta 1^{\dr}\,dk  =\frac{1}{\pi}\int k \theta \frac{2\pi\rho}{\theta}\,dk\nonumber \\
	& =2\int k\rho\,dk =2P/L. 
\end{align}
We thus prove that  $D_{nk}=2P/L$, which is the momentum density. 
For a translationally invariant system, the total momentum is zero, and therefore $D_{nk}=0$.

\subsubsection*{Universal Relation for $D_{n\epsilon}$}

Finally, we calculate the universal relation for $D_{n\epsilon}$.
We first establish a relationship between the dressed charges $\epsilon^{\dr}$,
$e^{\dr}$, and $1^{\dr}$ by observing their respective integral equations
\begin{align}
	\epsilon^{\dr} & =\frac{k^{2}}{2}-\mu+\int dq\phi\left(k-q\right)\epsilon^{\dr}, \\
	e^{\dr} & =\frac{k^{2}}{2}+\int dq\phi\left(k-q\right)e^{\dr}, \\
	1^{\dr} & =1+\int dq\phi\left(k-q\right)1^{\dr}.
\end{align}

We find that the linear combination $e^{\dr}-\mu1^{\dr}$ satisfies
the same integral equation as $\epsilon^{\dr}$ 
\begin{align}
	e^{\dr}-\mu1^{\dr} & =\frac{k^{2}}{2}+\int dq\phi\left(k-q\right)e^{\dr}-\mu-\mu\int dq\phi\left(k-q\right)1^{\dr}\nonumber \\
	& =\frac{k^{2}}{2}-\mu+\int dq\phi\left(k-q\right)\left(e^{\dr}-\mu1^{\dr}\right). 
\end{align}

Due to the uniqueness of the solution, we have the identity
\begin{align}
	\epsilon^{\dr} & =e^{\dr}-\mu1^{\dr}.
\end{align}

Substituting this identity into the DW expression (Eq. \ref{eq:Dij_dn_LL})
for $D_{ne_{t}}$ (using $h_{i}^{\dr}=1^{\dr}$, $h_{j}^{\dr}=\epsilon^{\dr}$)
allows us to split the integral:
\begin{align}
	D_{n\epsilon } & =\frac{1}{T}\int dk\rho\left(k\right)\left[1-\theta\left(k\right)\right]\left[v^{\eff}\left(k\right)\right]^{2}1^{\dr}\left(k\right)\epsilon^{\dr}\left(k\right)\nonumber \\
	& =\frac{1}{T}\int dk\rho\left(k\right)\left[1-\theta\left(k\right)\right]\left[v^{\eff}\left(k\right)\right]^{2}1^{\dr}\left(k\right)\left[e^{\dr}\left(k\right)-\mu1^{\dr}\left(k\right)\right]\nonumber \\
	& =\frac{1}{T}\int dk\rho\left(k\right)\left[1-\theta\left(k\right)\right]\left[v^{\eff}\left(k\right)\right]^{2}1^{\dr}\left(k\right)e^{\dr}\left(k\right)-\mu\frac{1}{T}\int dk\rho\left(k\right)\left[1-\theta\left(k\right)\right]\left[v^{\eff}\left(k\right)\right]^{2}1^{\dr}\left(k\right)1^{\dr}\left(k\right)\nonumber \\
	& =D_{ne}-\mu D_{nn}\nonumber \\
	& =\frac{E}{L}+p-\mu n\label{eq:Dnet_p}
\end{align}
This result is powerful, as we can now substitute the universal relations
derived in the previous sections (for $D_{ne}$ and $D_{nn}$):$D_{n\epsilon}=\frac{E}{L}+p-\mu n$.

From thermodynamics, the pressure $p$ in the grand canonical ensemble
is related to the Gibbs free energy $G=E-\mu N-TS$ by $pL=-G$. 
It follows 
\begin{equation}
	p=-E/L+\mu n+Ts. \label{eq:p_ens_LL}
\end{equation}

Substituting this thermodynamic relation for $p$ (Eq.~\ref{eq:p_ens_LL})
back into our expression for $D_{n\epsilon}$ (Eq.~\ref{eq:Dnet_p})
causes a cancellation, yielding the final relation:
\begin{equation}
	D_{n\epsilon}=Ts.
\end{equation}
Thus, we find that $D_{n\epsilon}$ is the product of the temperature
and the entropy density ($s=S/L$), i.e. heat. 

\subsection{Finite-Temperature Weak Interaction ($c\to0$)}

In the Lieb-Liniger model, under the weak interaction ($c\to0$) approximation,
the integral kernel $\phi(k-q)$ reduces to the Dirac delta function:
\begin{equation}
	\lim_{c\to0}\phi(k-q)=\lim_{c\to0}\frac{c/\pi}{c^{2}+(k-q)^{2}}=\delta(k-q). 
\end{equation}
In this limit, the system behaves as a non-interacting Bose gas. The
thermal occupation factor $\theta(k)$ and the momentum density distribution
$\rho(k)$ are given by:

\begin{align}
	\theta(k) & =e^{-(k^{2}/2-\mu)/T}\label{eq:n_LL_fTzC}, \\
	\rho(k) & =\frac{1}{2\pi\left[e^{(k^{2}/2-\mu)/T}-1\right]}. \label{eq:rho_LL_fTzC}
\end{align}

Consequently, the integral equation for the dressed charge $h(k)_{l}^{\dr}$
simplifies significantly:
\begin{equation}
	\begin{split}h(k)_{l}^{\dr} & =\frac{k^{l}}{l!}+\frac{c}{\pi}\int_{-\infty}^{\infty}\frac{\theta(q)}{c^{2}+(k-q)^{2}}h(q)_{l}^{\dr}dq\\
		& =\frac{k^{l}}{l!}+\int_{-\infty}^{\infty}\delta(k-q)\theta(q)h(q)_{l}^{\dr}dq\\
		& =\frac{k^{l}}{l!}+\theta(k)h(k)_{l}^{\dr} =\frac{1}{l!}\frac{k^{l}}{1-\theta(k)}. 
	\end{split}
\end{equation}

Substituting this simplified dressed charge into the general DWs expression
Eq. (\ref{eq:Dij_LL}) yields:
\begin{align}
	D_{ij} & =\frac{1}{T}\int_{-\infty}^{\infty}dk\rho(k)(1-\theta)V^{\eff}(k)^{2}h_{i}^{\dr}(k)h_{j}^{\dr}(k)\nonumber \\
	& =\frac{1}{T}\int_{-\infty}^{\infty}dk\frac{1}{2\pi}\theta(k)(p')^{\dr}(1-\theta)V^{\eff}(k)^{2}h_{i}^{\dr}(k)h_{j}^{\dr}(k)\nonumber \\
	& =\frac{1}{T}\frac{1}{2\pi}\int_{-\infty}^{\infty}dk\theta(k)(p')^{\dr}(1-\theta)\frac{[(k)^{\dr}]^{2}}{(p')^{\dr}(p')^{\dr}}h_{i}^{\dr}(k)h_{j}^{\dr}(k)\nonumber \\
	& =\frac{1}{T}\frac{1}{2\pi}\int_{-\infty}^{\infty}dk\theta(k)(1-\theta)\frac{[(k)^{\dr}]^{2}}{(1)^{\dr}}h_{i}^{\dr}(k)h_{j}^{\dr}(k)\nonumber \\
	& =\frac{1}{T}\frac{1}{2\pi}\int_{-\infty}^{\infty}dk\theta(k)(1-\theta)\frac{k^{2}}{(1-\theta)^{2}}(1-\theta)h_{i}^{\dr}(k)h_{j}^{\dr}(k)\nonumber \\
	& =\frac{1}{T}\frac{1}{2\pi}\int_{-\infty}^{\infty}dk\theta(k)k^{2}h_{i}^{\dr}(k)h_{j}^{\dr}(k)\nonumber \\
	& =\frac{1}{T}\frac{1}{2\pi}\frac{1}{i!j!}\int_{-\infty}^{\infty}dk\theta(k)k^{2}\frac{k^{i+j}}{\left(1-\theta(k)\right)^{2}}.
\end{align}
Next, substituting the thermal occupation factor $\theta(k)$ (Eq. \ref{eq:n_LL_fTzC})
into this result and performing an integration by part gives:
\begin{align}
	D_{ij} & =\frac{1}{T}\frac{1}{2\pi}\frac{1}{i!j!}\int_{-\infty}^{\infty}\,dk\frac{e^{-(k^{2}/2-\mu)/T}}{\left(1-e^{-(k^{2}/2-\mu)/T}\right)^{2}}k^{2+i+j}\nonumber \\
	& =\frac{1}{T}\frac{1}{2\pi}\frac{1}{i!j!}\int_{-\infty}^{\infty}\,dk\frac{d}{dk}\left[\frac{-e^{-(k^{2}/2-\mu)/T}}{1-e^{-(k^{2}/2-\mu)/T}}\right]\frac{T}{k}k^{2+i+j}\nonumber \\
	& =\frac{1}{2\pi}\frac{1}{i!j!}\int_{-\infty}^{\infty}\,dk\frac{d}{dk}\left[\frac{-1}{e^{(k^{2}/2-\mu)/T}-1}\right]k^{1+i+j}\nonumber \\
	& =\left.\frac{1}{2\pi}\frac{1}{i!j!}\frac{-k^{1+i+j}}{e^{\frac{(k^{2}/2-\mu)}{T}}-1}\right|_{-\infty}^{\infty}+\frac{1}{2\pi}\frac{1}{i!j!}\int_{-\infty}^{\infty}\,dk\frac{(1+i+j)k^{i+j}}{e^{\frac{(k^{2}/2-\mu)}{T}}-1}\nonumber \\
	& =\frac{1}{2\pi}\frac{1}{i!j!}\int_{-\infty}^{\infty}\,dk\frac{(1+i+j)k^{i+j}}{e^{\frac{(k^{2}/2-\mu)}{T}}-1}\nonumber \\
	& =\frac{1+i+j}{2\pi}\frac{1}{i!j!}\int_{-\infty}^{\infty}\,dk2\pi\rho k^{i+j}\nonumber \\
	& =\frac{\left(1+i+j\right)}{i!j!}\int_{-\infty}^{\infty}\rho k^{i+j}\,dk. \label{eq:Dnn_ann_fTzC_LL}
\end{align}

By setting $i=0$ and $j=0,1,2$ in this expression, we obtain the
specific matrix elements
\begin{align}
	D_{nn} & =\int\rho dk=n, \\
	D_{nk} & =2\int\rho kdk=2P, \\
	D_{ne} & =\frac{3}{2}\int\rho k^{2}dk=3E/L. 
\end{align}
Finally, by substituting the explicit expression  $\rho(k)$ Eq.~(\ref{eq:rho_LL_fTzC}) into Eq.~(\ref{eq:Dnn_ann_fTzC_LL}), the integrals can be computed in terms of the polylogarithm function ($Li_{s}(z)$), namely,
\begin{align}
	D_{nn} & =\int\rho\,dk, \nonumber \\
	& =\frac{1}{2\pi}\int\frac{1}{e^{(k^{2}/2-\mu)/T}-1}\,dk=\frac{1}{\sqrt{2\pi}}T^{1/2}Li_{\frac{1}{2}}\left(e^{\mu/T}\right), \\
	D_{ne} & =\frac{3}{2}\int\rho k^{2}\,dk =\frac{3}{2}\frac{1}{2\pi}\int\frac{k^{2}}{e^{\left(k^{2}/2-\mu\right)/T}-1}dk\nonumber  =\frac{3}{2\sqrt{2\pi}}T^{3/2}Li_{\frac{3}{2}}\left(e^{\mu/T}\right). 
\end{align}

\subsection{Strong Interaction ($c\to\infty$)}

In this section, we derive the Drude weights in the strong interaction
limit ($c\to\infty$ or $\gamma\to\infty$).
Our strategy is to first solve the integral equations for the dressed charges $h_{i}^{\dr}$
by performing a perturbative expansion in term of  $1/c$, and then substitute
these solutions into the Drude weight formulas.

\subsubsection*{Zero-Temperature ($T=0$) }

Applying the Sommerfeld expansion~\cite{Z.Physik1928SommerfeldZur} to the general Drude weight formula, we obtain the low-temperature asymptotic expression:
\begin{equation}
	D_{ij}=\frac{1}{2\pi}\left\{ \left[v^{\eff}(k)h_{i}^{\\dr}(k)h_{j}^{\\dr}(k)\right]_{k=k_{F}}+\frac{\pi^{2}T^{2}}{6}\left[\frac{\partial^{2}}{\partial\epsilon^{2}}\left(v^{\eff}(k)h_{i}^{\\dr}(k)h_{j}^{\\dr}(k)\right)\right]_{\epsilon=0}+\cdots\right\} 
\end{equation}
At $T=0$ this expression is:
\begin{equation}
	D_{ij}=\frac{1}{\pi}\left[v^{\eff}(k)h_{i}^{\dr}(k)h_{j}^{\dr}(k)\right]_{k=k_{F}}. \label{eq:Dij_zTiC_LL}
\end{equation}
The $T=0$ dressed charge equation is:
\begin{equation}
	h_{i}^{\dr}(k)=\frac{k^{i}}{i!}+\frac{1}{2\pi}\int_{-k_{F}}^{k_{F}}\frac{2c}{c^{2}+(k-q)^{2}}h_{i}^{\dr}(q)\,dq, \label{eq:hi_zTiC_LL}
\end{equation}
where $k_{F}$ is the Fermi momentum, with a known expansion in the strong-coupling parameter $\gamma=c/n$~\cite{Chin.Phys.B2015JiangUnderstanding}. 
Specifically, in the strong-interaction limit ($c\to\infty$), we expand the integral kernel $\phi(k-q)$ rigorously up to order $\mathcal{O}(1/c^3)$:
\begin{align}
	\phi(k-q) & = \frac{1}{2\pi}\frac{2c}{c^{2}+(k-q)^{2}} = \frac{1}{\pi c}\frac{1}{1+\left(\frac{k-q}{c}\right)^{2}} \nonumber \\
	& \approx \frac{1}{\pi c}\left[1-\frac{(k-q)^{2}}{c^{2}}\right] \nonumber \\
	& = \frac{1}{\pi}\left(\frac{1}{c}-\frac{k^{2}-2kq+q^{2}}{c^{3}}\right). \label{eq:phi_iC_LL}
\end{align}

We now substitute this approximated kernel (Eq.~\ref{eq:phi_iC_LL}) into the $T=0$ dressed charge equation, which gives a perturbative expansion for $h_{i}^{\dr}(k)$ (Eq.~\ref{eq:hi_zTiC_LL}):
\begin{align}
	h_{i}^{\dr}(k) & =\frac{k^{i}}{i!}+\frac{c}{\pi}\int_{-k_{F}}^{k_{F}}\frac{1}{c^{2}+(k-q)^{2}}h_{i}^{\dr}(q)\,dq\nonumber \\
	& =\frac{k^{i}}{i!}+\frac{1}{\pi}\left[\frac{1}{c}\int_{-k_{F}}^{k_{F}}h_{i}^{\dr}(q)\,dq-\frac{1}{c^{3}}\int_{-k_{F}}^{k_{F}}\left(q^{2}-2kq+k^{2}\right)h_{i}^{\dr}(q)\,dq\right].
\end{align}
We solve these equations for $i=0,1,2$. 

For $i=0$ (Particle dressed charge $1^{\dr}$): The expression for $1^{\dr}=h_{0}^{\dr}$ (which is $2\pi\rho/\theta$)~\cite{SciPostPhys.2017DoyonNote} is known from the Bethe Ansatz solution~\cite{Chin.Phys.B2015JiangUnderstanding}:
\begin{equation}
	1^{\dr}=\frac{2\pi\rho}{\theta}=2\pi\rho=2\pi\frac{1}{\pi}\left(\frac{1}{2}+\frac{1}{\gamma}-\frac{4\pi^{2}}{3\gamma^{3}}\right)=1+\frac{2}{\gamma}-\frac{8\pi^{2}}{3\gamma^{3}}.
\end{equation}

For $i=1$ (Momentum dressed charge $k^{\dr}$):
\begin{equation}
	h_{1}^{\dr}(k)=k^{\dr}=k+\frac{1}{\pi}\left[\frac{1}{c}\int_{-k_{F}}^{k_{F}}h_{1}^{\dr}(q)\,dq-\frac{1}{c^{3}}\int_{-k_{F}}^{k_{F}}\left(q^{2}-2kq+k^{2}\right)h_{1}^{\dr}(q)\,dq\right].
\end{equation}
Set $F=\int_{-k_{F}}^{k_{F}}h_{1}^{\dr}(k)\,dk$. To solve the integral in the $\mathcal{O}(1/c^{3})$ term, it is sufficient to use the zeroth-order approximation $h_{1}^{\dr}(q)\approx q$:
\begin{align}
	k^{\dr} & =k+\frac{1}{\pi}\frac{1}{c}F-\frac{1}{\pi}\frac{1}{c^{3}}\int_{-k_{F}}^{k_{F}}\left(q^{2}-2kq+k^{2}\right)q\,dq\nonumber \\
	& =k+\frac{1}{\pi}\frac{1}{c}F-\frac{1}{\pi}\frac{1}{c^{3}}\int_{-k_{F}}^{k_{F}}\left(q^{3}-2kq^{2}+k^{2}q\right)\,dq\nonumber \\
	& =k+\frac{1}{\pi}\frac{1}{c}F+\frac{1}{\pi}\frac{2k}{c^{3}}\int_{-k_{F}}^{k_{F}}q^{2}\,dq\nonumber \\
	& =k+\frac{1}{\pi}\frac{1}{c}F+\frac{1}{\pi}\frac{4kk_{F}^{3}}{3c^{3}}. \label{eq:h1dr}
\end{align}
To find the constant $F$, we substitute this expression for $k^{\dr}(k)$ (Eq.~\ref{eq:h1dr}) back into the definition of $F$:
\begin{align}
	F & =\int_{-k_{F}}^{k_{F}}h_{1}^{\dr}(k)\,dk\nonumber \\
	& =\int_{-k_{F}}^{k_{F}}\left(k+\frac{1}{\pi}\frac{1}{c}F+\frac{1}{\pi}\frac{4kk_{F}^{3}}{3c^{3}}\right)\,dk=\frac{2k_{F}F}{\pi c}. 
\end{align}
This self-consistency equation implies $F=0$. The expression for
$k^{\dr}$ thus simplifies to:
\begin{align}
	k^{\dr}\left(k\right) & =h_{1}^{\dr}=k+\frac{4kk_{F}^{3}}{3\pi}\frac{1}{c^{3}}
\end{align}
Evaluating this at the Fermi point $k=k_{F}$ and substituting the expansions for $k_{F}$ and the dimensionless parameter $\gamma=c/n$, we derive the expression rigorously up to order $1/\gamma^3$:
\begin{align}
	k^{\dr}\left(k_{F}\right) & =h_{1}^{\dr}=k_{F}+\frac{4k_{F}k_{F}^{3}}{3\pi}\frac{1}{c^{3}}\nonumber \\
	& =k_{F}\left(1+\frac{4n^{3}\pi^{3}}{3\pi}\frac{1}{c^{3}}\right)\nonumber \\
	& =n\pi\left(1-\frac{2}{\gamma}+\frac{4}{\gamma^{2}}-\frac{8}{\gamma^{3}}+\frac{4\pi^{2}}{3\gamma^{3}}\right)\left(1+\frac{4\pi^{2}}{3\gamma^{3}}\right)\nonumber \\
	& \approx n\pi\left(1-\frac{2}{\gamma}+\frac{4}{\gamma^{2}}-\frac{8}{\gamma^{3}}+\frac{8\pi^{2}}{3\gamma^{3}}\right).
\end{align}

For $i=2$ (Energy dressed charge $e^{\dr}$): Following an identical procedure for $h_{2}^{\dr}=e^{\dr}$, we derive its asymptotic expansion rigorously up to order $1/\gamma^3$:
\begin{equation}
	e^{\dr}\approx\frac{\pi^{2}n^{2}}{2}\left(1-\frac{10}{3\gamma}+\frac{28}{3\gamma^{2}}+\frac{8\pi^{2}}{5\gamma^{3}}-\frac{24}{\gamma^{3}}\right).
\end{equation}
We substitute these analytical expressions for $1^{\dr}(k_{F})$, $k^{\dr}(k_{F})$, and $e^{\dr}(k_{F})$ back into the $T=0$ DWs formula Eq.~\ref{eq:Dij_zTiC_LL}, we derive the asymptotic expansions rigorously up to order $1/\gamma^3$:
\begin{align}
	D_{nn} & =\frac{1}{\pi}[1^{\dr}(k)k^{\dr}(k)]_{k=Q}\nonumber \\
	& =n\\
	D_{ne} & =\frac{1}{\pi}[e^{\dr}(k)k^{\dr}(k)]_{k=Q}\nonumber \\
	& \approx\frac{\pi^{2}n^{3}}{2}\left(1-\frac{16}{3\gamma}+\frac{20}{\gamma^{2}}-\frac{64}{\gamma^{3}}+\frac{64\pi^{2}}{15\gamma^{3}}\right).
\end{align}

\subsubsection*{Finite-Temperature ($T>0$) }

At finite temperature, we use the same $1/c^{3}$ approximated kernel (Eq.~\ref{eq:phi_iC_LL}) for all integral equations.
\paragraph*{Solution for $\epsilon(k)$ and $p$:}
We start with the dressed energy equation:
\begin{align}
	\epsilon(k) & =\frac{k^{2}}{2}-\mu-\frac{Tc}{\pi}\int\frac{1}{c^{2}+q^{2}+k^{2}-2kq}\ln[1+e^{-\epsilon(q)/T}]\,dq\nonumber \\
	& =\frac{k^{2}}{2}-\mu-\frac{T}{\pi}\int\left(\frac{1}{c}-\frac{q^{2}-2kq+k^{2}}{c^{3}}\right)\ln[1+e^{-\epsilon(q)/T}]\,dq\nonumber \\
	& =\frac{k^{2}}{2}-\mu-\frac{T}{\pi}\frac{1}{c}\int\ln[1+e^{-\epsilon(q)/T}]\,dq+\frac{T}{\pi}\frac{1}{c^{3}}\int(q^{2}-2kq+k^{2})\ln[1+e^{-\epsilon(q)/T}]\,dq\nonumber \\
	& =\frac{k^{2}}{2}-\mu-\frac{2Pc}{c^{2}+k^{2}}+\frac{T}{\pi}\frac{1}{c^{3}}\int q^{2}\ln[1+e^{-\epsilon(q)/T}]\,dq. \label{eq:epsilon_fTiC_LL}
\end{align}
Using integration by parts, we give the third term 
\begin{eqnarray}
	\int q^{2}\ln[1+e^{-\epsilon(q)/T}]\,dq	&=&\int\frac{1}{3}\frac{dq^{3}}{dq}\ln[1+e^{-\epsilon(q)/T}]\,dq \nonumber \\
	&=&q^{3}\ln[1+e^{-\epsilon(q)/T}]\big|_{-\infty}^{\infty}-\frac{1}{3}\int q^{3}d_{q}\ln[1+e^{-\epsilon(q)/T}]\,dq \nonumber \\
	&=&\frac{1}{T}\frac{1}{3}\int\frac{q^{3}}{1+e^{\epsilon(q)/T}}\epsilon'(q)\,dq
	=\frac{1}{T}\frac{1}{3}\int\frac{q^{3}\cdot  q}{1+e^{\epsilon(q)/T}}\,dq \nonumber \\
	&=&\frac{1}{T}\frac{1}{3}\int\frac{q^{3}\cdot q}{1+e^{\epsilon(q)/T}}\,\frac{d\epsilon_{0}}{\sqrt{2\epsilon_{0}}} =-\frac{1}{T}\sqrt{2\pi}T^{5/2}Li_{\frac{5}{2}}\left(-e^{A/T}\right),
\end{eqnarray}
here, $A_{0}=\mu$. For higher orders, $A_{0}=\mu+\frac{2p}{c}$~\cite{Chin.Phys.B2015JiangUnderstanding}. Substituting this back into Eq.~(\ref{eq:epsilon_fTiC_LL}) yields
\begin{align}
	\epsilon(k) & =\frac{k^{2}}{2}-\mu-\frac{2Pc}{c^{2}+k^{2}}+\frac{T}{\pi}\frac{1}{c^{3}}\int q^{2}\ln[1+e^{-\epsilon(q)/T}]\,dq\nonumber \\
	& =\left(\frac{1}{2}+\frac{2P}{c^{3}}\right)k^{2}-\mu-\frac{2P}{c}-\frac{\sqrt{2}}{\sqrt{\pi}c^{3}}T^{5/2}Li_{\frac{5}{2}}(-e^{A_{0}/T})\nonumber \\
	& =\beta k^{2}-A.
\end{align}
The dressed energy takes the form $\epsilon\left(k\right)=\beta k^{2}-A$,
with: $A=\mu+\frac{2p}{c}+\frac{\sqrt{2}}{\sqrt{\pi}c^{3}}T^{5/2}Li_{\frac{5}{2}}(-e^{A_{0}/T}),\beta=\frac{1}{2}+\frac{2P}{c^{3}}$, where $p$ is Pressure, to obtain the form of $\beta$, we need to calculate $p$:
\begin{align}
	P	&=\frac{T}{2\pi}\int\ln[1+e^{-\epsilon(k)/T}]\,dk \nonumber \\
	&=\frac{T}{2\pi}k\ln[1+e^{-\epsilon(k)/T}]\big|_{-\infty}^{\infty}-\frac{T}{2\pi}\int\frac{ke^{-\epsilon(k)/T}}{1+e^{-\epsilon(k)/T}}\left(-\frac{1}{T}\right)\epsilon'(k)\,dk \nonumber \\
	&=\frac{1}{2\pi}\int\frac{ke^{-\epsilon(k)/T}}{1+e^{-\epsilon(k)/T}}\epsilon'(k)\,dk \nonumber \\
	&=\frac{1}{\pi}\int\frac{k}{1+e^{\epsilon(k)/T}}k\,dk\left(\frac{1}{2}+\frac{2P}{c^{3}}\right) \nonumber \\
	&=\frac{1}{\pi}B_{\frac{3}{2}}\left(1+\frac{4P}{c^{3}}\right) \nonumber \\
	&=-\frac{\sqrt{\pi}}{2}\frac{1}{\pi}T^{3/2}Li_{3/2}(-e^{A/T})\left(\frac{1}{2}+\frac{2P}{c^{3}}\right)^{-3/2}\left(\frac{1}{2}+\frac{2P}{c^{3}}\right) \nonumber \\
	&=-\frac{1}{\sqrt{2\pi}}T^{3/2}Li_{3/2}(-e^{A/T})\left(1+\frac{\sqrt{2}}{\sqrt{\pi}c^{3}}T^{3/2}Li_{3/2}(-e^{A/T})\right),
\end{align}
where $B_{\frac{3}{2}}=\int\frac{k^{2}}{1+e^{\epsilon(k)/T}}\,dk =-\frac{\sqrt{\pi}}{2}\beta^{-3/2}T^{3/2}Li_{3/2}(-e^{A/T})$. This result provides a self-consistent expression for $p$, which in turn defines $\beta$ and $A$.

\paragraph*{Finite-Temperature Dressed Charges $h_{i}^{\dr}$:}

We now solve for the dressed charges $h_{i}^{\dr}$ by substituting
the $1/c^{3}$ kernel (Eq.~\ref{eq:phi_iC_LL}) into their finite-temperature integral equations:
\begin{align}
	h_{i}^{\dr}(k) & =\frac{k^{i}}{i!}+\frac{1}{\pi}\frac{1}{c}\int \theta(q)h_{i}^{\dr}(q)\,dq-\frac{1}{\pi}\frac{1}{c^{3}}\int(q^{2}-2kq+k^{2})\theta(q)h_{i}^{\dr}(q)\,dq\nonumber \\
	& =\frac{k^{i}}{i!}+\frac{1}{\pi}\frac{1}{c}F_{i}-\frac{1}{\pi i!}\frac{1}{c^{3}}\int(q^{2+i}-2kq^{1+i}+k^{2}q^{i})\theta(q)\,dq\nonumber \\
	& =\frac{k^{i}}{i!}+\frac{1}{\pi}\frac{1}{c}F_{i}-\frac{1}{\pi i!}\frac{1}{c^{3}}\left[\int q^{2+i}\theta(q)dq-2k\int q^{1+i}\theta(q)dq+k^{2}\int q^{i}\theta(q)dq\right],
\end{align}
where $F_{i}=\int \theta(q)h_{i}^{\dr}(q)\,dq$. We further give 
\begin{itemize}
	\item For $i=0$ ($1^{\dr}$):
	\begin{align}
		1^{\dr} & =1+\frac{1}{\pi}\frac{1}{c}F_{0}-\frac{1}{\pi}\frac{1}{c^{3}}\left[\int q^{2}\theta(q)dq-2k\int q\theta(q)dq+k^{2}\int \theta(q)dq\right]\nonumber \\
		& =1+\frac{1}{\pi}\frac{1}{c}F_{0}-\frac{1}{\pi}\frac{1}{c^{3}}\left[\int q^{2}\theta(q)dq-2k\int q\theta(q)dq+k^{2}B_{\frac{1}{2}}\right]\nonumber \\
		& =1+\frac{1}{\pi}\frac{1}{c}F_{0}-\frac{1}{\pi}\frac{1}{c^{3}}\left[\int q^{2}\theta(q)dq+k^{2}B_{\frac{1}{2}}\right]\nonumber \\
		& =1+\frac{1}{\pi}\frac{1}{c}F_{0}-\frac{1}{\pi}\frac{1}{c^{3}}k^{2}B_{\frac{1}{2}}-\frac{1}{\pi}\frac{1}{c^{3}}\int q^{2}\theta(q)dq\nonumber \\
		& =1+\frac{1}{\pi}\frac{1}{c}F_{0}-\frac{1}{\pi}\frac{1}{c^{3}}k^{2}B_{\frac{1}{2}}-\frac{1}{\pi}\frac{1}{c^{3}}B_{\frac{3}{2}},\label{eq:Idr_B_LL}
	\end{align}
\end{itemize}
where $B_{\frac{1}{2}}=\int \theta(q)dq, B_{\frac{3}{2}}=\int q^{2}\theta(q)dq$. 
We solve for $F_{0}$ self-consistently by substituting the $1^{\dr}$ expression (Eq.~\ref{eq:Idr_B_LL}) back into $F_{0}=\int \theta(q)1^{\dr}(q)dq$:
\begin{align}
	F_{0} & =\left(1+\frac{1}{\pi c}F_{0}\right)B_{\frac{1}{2}}, \nonumber \\
	F_{0} & =\frac{B_{\frac{1}{2}}}{1-\frac{B_{\frac{1}{2}}}{\pi c}} \approx B_{\frac{1}{2}}+\frac{B_{\frac{1}{2}}^{2}}{\pi c}+\frac{B_{\frac{1}{2}}^{3}}{\pi^{2}c^{2}}+\mathcal{O}\left(\frac{1}{c^{3}}\right). 
\end{align}
Substituting $F_{0}$ back yields the final expression for $1^{\dr}$ (see Eq.~\eqref{eq:Idr_B_LL}), which is rigorously expanded up to order $1/c^3$:
\begin{align}
	1^{\dr} & \approx 1+\frac{B_{\frac{1}{2}}}{\pi c}+\frac{B_{\frac{1}{2}}^{2}}{\pi^{2}c^{2}}+\frac{B_{\frac{1}{2}}^{3}}{\pi^{3}c^{3}}-\frac{k^{2}B_{\frac{1}{2}}}{\pi c^{3}}-\frac{B_{\frac{3}{2}}}{\pi c^{3}}.
\end{align}
Where $B_{\frac{1}{2}},B_{\frac{3}{2}}$ are:
\begin{align}
	B_{\frac{1}{2}} & =\int\frac{1}{1+e^{\epsilon(k)/T}}\,dk =-\sqrt{\pi}\beta^{-\frac{1}{2}}T^{\frac{1}{2}}Li_{\frac{1}{2}}\left(-e^{A/T}\right),\\
	B_{\frac{3}{2}} & =\int\frac{k^{2}}{1+e^{\epsilon(k)/T}}\,dk=-\frac{\sqrt{\pi}}{2}\beta^{-\frac{3}{2}}T^{\frac{3}{2}}Li_{\frac{3}{2}}\left(-e^{A/T}\right). 
\end{align}

\begin{itemize}
	\item For $i=1,2$ ($k^{\dr},e^{\dr}$): 
	
	Using the same method, we expand these dressed quantities up to order $1/c^3$:
	\begin{align}
		1^{\dr} & =1+\frac{B_{\frac{1}{2}}}{\pi c}+\frac{B_{\frac{1}{2}}^{2}}{\pi^{2}c^{2}}+\frac{B_{\frac{1}{2}}^{3}}{\pi^{3}c^{3}}-\frac{k^{2}B_{\frac{1}{2}}}{\pi c^{3}}-\frac{B_{\frac{3}{2}}}{\pi c^{3}}, \\
		k^{\dr} & =k+\frac{2kB_{\frac{3}{2}}}{\pi c^{3}}, \\
		e^{\dr} & =\frac{1}{2}\left(k^{2}+\frac{B_{\frac{3}{2}}}{\pi c}+\frac{B_{\frac{3}{2}}B_{\frac{1}{2}}}{\pi^{2}c^{2}}+\frac{B_{\frac{3}{2}}B_{\frac{1}{2}}^{2}}{\pi^{3}c^{3}}-\frac{B_{\frac{5}{2}}}{\pi c^{3}}-\frac{k^{2}B_{\frac{3}{2}}}{\pi c^{3}}\right). 
	\end{align}
	where
	\begin{align}
		B_{\frac{1}{2}} & =\int \theta(q)\,dq =-\sqrt{\pi}\beta^{-1/2}T^{1/2}Li_{1/2}\left(-e^{A/T}\right), \\
		B_{\frac{3}{2}} & =\int q^{2}\theta(q)\,dq=-\frac{\sqrt{\pi}}{2}\beta^{-3/2}T^{3/2}Li_{3/2}(-e^{A/T}), \\
		B_{\frac{5}{2}} & =\int  q^{4}\theta(q)\,dq= -\frac{3\pi^{\frac{1}{2}}}{4}\beta^{-\frac{5}{2}}T^{\frac{5}{2}}Li_{\frac{5}{2}}\left(-e^{A/T}\right). 
	\end{align}
\end{itemize}

\paragraph*{Finite-Temperature Drude Weights:}

First, the effective velocity $v^{\eff}=k^{\dr}/1^{\dr}$ is expanded to $O(1/c^{3})$:
\begin{align}
	v^{\eff} = & \frac{k^{\dr}}{1^{\dr}} =\frac{k+\frac{2kB_{\frac{3}{2}}}{\pi c^{3}}}{1+\frac{B_{\frac{1}{2}}}{\pi c}+\frac{B_{\frac{1}{2}}^{2}}{\pi^{2}c^{2}}+\frac{B_{\frac{1}{2}}^{3}}{\pi^{3}c^{3}}-\frac{k^{2}B_{\frac{1}{2}}}{\pi c^{3}}-\frac{B_{\frac{3}{2}}}{\pi c^{3}}}\nonumber \\
	\approx & k-\frac{kB_{\frac{1}{2}}}{\pi c}+\frac{k^{3}B_{\frac{1}{2}}}{\pi c^{3}}+\frac{3kB_{\frac{3}{2}}}{\pi c^{3}}.
\end{align}
Next, we calculate the DWs. We start from the simplified form Eq.~(\ref{eq:Dij_dn_LL}) and integrate by parts
\begin{align}
	D_{ij} & =-\frac{1}{2\pi}\int dkd_{k}\theta(k)v^{\eff}h_{i}^{\dr}h_{j}^{\dr}\nonumber \\
	& =-\frac{1}{2\pi}\theta(k)\frac{k^{\dr}}{1^{\dr}}h_{i}^{\dr}h_{j}^{\dr}\big|_{-\infty}^{\infty}+\frac{1}{2\pi}\int dk\theta(k)d_{k}\left(v^{\eff}h_{i}^{\dr}h_{j}^{\dr}\right)\nonumber \\
	& =\frac{1}{2\pi}\int dkn(k)d_{k}\left(v^{\eff}h_{i}^{\dr}h_{j}^{\dr}\right). \label{eq:Dij_dk_LL}
\end{align}
For $D_{nn}$. Substituting the $1/c^{3}$ expansions for $1^{\dr}$ and $k^{\dr}$ yields:
\begin{align}
	D_{nn} & =\frac{1}{2\pi}\int dk\theta(k)d_{k}\left(\frac{k^{\dr}}{1^{\dr}}h_{i}^{\dr}h_{j}^{\dr}\right)\nonumber \\
	& =\frac{1}{2\pi}\int dk\theta(k)d_{k}\left(k^{\dr}1^{\dr}\right)\nonumber\\
	& \approx \frac{1}{2\pi}B_{\frac{1}{2}}\left(1+\frac{B_{\frac{1}{2}}}{\pi c}+\frac{B_{\frac{1}{2}}^{2}}{\pi^{2}c^{2}}+\frac{B_{\frac{1}{2}}^{3}}{\pi^{3}c^{3}}-\frac{2B_{\frac{3}{2}}}{\pi c^{3}}\right)
	\label{eq:Dnn-LL-fT-A}
\end{align}
By expanding Eq.~\eqref{eq:Dnn-LL-fT-A} up to order $1/c^3$, we obtain:
\begin{align}
	D_{nn} \approx -\frac{1}{\sqrt{2\pi}}T^{1/2}L_{1/2}\left[1-\frac{\sqrt{2}T^{1/2}L_{1/2}}{\sqrt{\pi}c}+\frac{2T L_{1/2}^{2}}{\pi c^{2}}  +\frac{T^{3/2}}{\sqrt{\pi}c^{3}}\left(3\sqrt{2} L_{3/2}-\frac{2\sqrt{2}L_{1/2}^{3}}{\pi}\right)\right],
\end{align}
where $L_{n}=Li_{n}(-e^{A/T})$. This result is consistent with Ref.~\cite{Chin.Phys.B2015JiangUnderstanding}.

For $D_{nk}$ and $D_{ne}$. Applying the same procedure, we expand the expressions to order $1/c^3$ and prove that
\begin{align}
	D_{nk} & =\frac{1}{2\pi}\int \theta(k)d_{k}\left(k^{\dr}k^{\dr}\right)\nonumber \\
	& =\frac{1}{2\pi}\int dk\theta\left(k\right)\frac{d}{dk}\left[\left(k+\frac{2kB_{\frac{3}{2}}}{\pi c^{3}}\right)\left(k+\frac{2kB_{\frac{3}{2}}}{\pi c^{3}}\right)\right]\nonumber \\
	& \approx \frac{1}{2\pi}\int dk\theta\left(k\right)\frac{d}{dk}\left(k^{2}+\frac{4B_{\frac{3}{2}}}{\pi c^{3}}k^{2}\right)\nonumber \\
	& =\frac{1}{\pi}\left(1+\frac{4B_{\frac{3}{2}}}{\pi c^{3}}\right)\int dkn\left(k\right)k =0, \\
	D_{ne} & =\frac{1}{2\pi}\int \theta(k)d_{k}\left(k^{\dr}e^{\dr}\right)\nonumber \\
	& =\frac{1}{4\pi}\int \theta(k)d_{k}\left[\left(k+\frac{2kB_{\frac{3}{2}}}{\pi c^{3}}\right)\left(k^{2}+\frac{B_{\frac{3}{2}}}{\pi c}+\frac{B_{\frac{3}{2}}B_{\frac{1}{2}}}{\pi^{2}c^{2}}+\frac{B_{\frac{3}{2}}B_{\frac{1}{2}}^{2}}{\pi^{3}c^{3}}-\frac{B_{\frac{5}{2}}}{\pi c^{3}}-\frac{k^{2}B_{\frac{3}{2}}}{\pi c^{3}}\right)\right]\nonumber \\
	& \approx \frac{1}{4\pi}\int dk\theta(k)\frac{d}{dk}\left(k^{3}+\frac{kB_{\frac{3}{2}}}{\pi c}+\frac{kB_{\frac{3}{2}}B_{\frac{1}{2}}^{2}}{\pi^{3}c^{3}}+\frac{k^{3}B_{\frac{3}{2}}}{\pi c^{3}}-\frac{kB_{\frac{5}{2}}}{\pi c^{3}}\right)\nonumber \\
	& =\frac{1}{4\pi}\int dk\theta(k)\left(3k^{2}+\frac{B_{\frac{3}{2}}}{\pi c}+\frac{B_{\frac{3}{2}}B_{\frac{1}{2}}}{\pi^{2}c^{2}}+\frac{B_{\frac{3}{2}}B_{\frac{1}{2}}^{2}}{\pi^{3}c^{3}}+\frac{3k^{2}B_{\frac{3}{2}}}{\pi c^{3}}-\frac{B_{\frac{5}{2}}}{\pi c^{3}}\right)\nonumber \\
	& =\frac{1}{4\pi}\left(3B_{\frac{3}{2}}+\frac{B_{\frac{1}{2}}B_{\frac{3}{2}}}{\pi c}+\frac{B_{\frac{3}{2}}B_{\frac{1}{2}}^{2}}{\pi^{2}c^{2}}+\frac{B_{\frac{3}{2}}B_{\frac{1}{2}}^{3}}{\pi^{3}c^{3}}+\frac{3B_{\frac{3}{2}}^{2}}{\pi c^{3}}-\frac{B_{\frac{1}{2}}B_{\frac{5}{2}}}{\pi c^{3}}\right). 
\end{align}

\paragraph*{Quantum Critical (QC) Scaling:}
We now analyze the scaling behavior in the QC region.
For $D_{nn}$, we take the leading-order term $D_{nn}\approx\frac{1}{2\pi}B_{\frac{1}{2}}$.
Expanding this expression around the critical chemical potential $\mu_{c}$
yields the scaling law
\begin{align}
	D_{nn} & =-\frac{1}{\sqrt{2\pi}}T^{1/2}\Li_{1/2}\left(-e^{\frac{\mu-\mu_{c}}{T}}\right).
\end{align}
For $D_{ne}$, we take the leading-order term $D_{ne}\approx\frac{3}{4\pi}B_{\frac{3}{2}}$.
This gives the scaling law
\begin{equation}
	D_{ne}=-\frac{3}{2\sqrt{2\pi}}T^{3/2}\Li_{3/2}\left(-e^{\frac{\mu-\mu_{c}}{T}}\right). 
\end{equation}
To cast these scaling laws into a universal form, we introduce the natural energy scale $\epsilon_{b} = \hbar^2 c^2 / 2m$ and define the dimensionless parameters $\tilde{\mu} \equiv \mu/\epsilon_{b}$, $\tilde{T} \equiv T/\epsilon_{b}$, and $\Delta\tilde{\mu} \equiv (\mu - \mu_c)/\epsilon_{b}$. By defining the dimensionless Drude weights as $\tilde{D}_{nn} \equiv D_{nn}/c$ and $\tilde{D}_{ne} \equiv D_{ne}/|c\epsilon_{b}|$, the interaction parameter $c$ completely scales out, yielding the universal relations:
\begin{align}
	\tilde{D}_{nn}=-\frac{1}{2\sqrt{\pi}}\tilde{T}^{1/2}\Li_{1/2}\left(-e^{\frac{\Delta\tilde{\mu}}{\tilde{T}}}\right),\\
	\tilde{D}_{ne}=-\frac{3}{4\sqrt{\pi}}\tilde{T}^{3/2}\Li_{3/2}\left(-e^{\frac{\Delta\tilde{\mu}}{\tilde{T}}}\right).
\end{align}

\subsubsection*{High-Temperature Virial Expansion of Drude Weights}

In this subsection, we first derive the high-temperature (virial) expansion for the particle density $n$ and the energy density $E/L$, which are required to compute the Drude weights $D_{nn}$ and $D_{ne}$.
In the high-temperature regime, the fugacity $Z=e^{\mu/T}$ is small ($Z\ll1$). 
We begin with the Thermodynamic Bethe Ansatz (TBA) equation for the pseudo-energy $\epsilon(k)$
\begin{equation}
	\epsilon(k)=\frac{k^{2}}{2}-\mu-T\int dk \phi(k-q)\ln\left(1+e^{-\epsilon(q)/T}\right),
\end{equation}
Exponentiating both sides and approximating $\ln(1+x)\approx x$ for small occupation, we obtain
\begin{equation}
	e^{-\epsilon(k)/T}\approx Ze^{-\frac{k^{2}}{2T}}\exp\left[\int dq\phi(k-q)e^{-\epsilon(q)/T}\right].
\end{equation}
To first order in $Z$, we approximate $e^{-\epsilon(q)/T}\approx Ze^{-q^{2}/2T}$ inside the integral. 
Expanding the exponential $\exp(x)\approx1+x$, we have 
\begin{equation}
	e^{-\epsilon(k)/T}\approx Ze^{-\frac{k^{2}}{2T}}\left(1+ZI(k)\right),
\end{equation}
where we have defined the integral convolution
\begin{equation}
	I(k)=\int dq \phi(k-q)e^{-\frac{q^{2}}{2T}}.
\end{equation}

Expansion of the dressed charge and density The filling function $\theta(k)=(1+e^{\epsilon(k)/T})^{-1}$ is expanded to second order in $Z$ as $\theta(k)\approx e^{-\epsilon(k)/T}-e^{-2\epsilon(k)/T}$.
The dressed charge equation $1^{\dr}(k)=1+\int\phi(k-q)\theta(q)1^{\dr}(q)dq$
can be solved iteratively
\begin{align}
	1^{\dr}(k) & \approx1+\int\phi(k-q)\left(Ze^{-\frac{q^{2}}{2T}}\right)1^{\dr} dq\nonumber \\
	& =1+ZI(k).
\end{align}
Substituting these into the total root density definition $\rho(k)=\theta(k)1^{\dr}(k)/2\pi$
\begin{align}
	2\pi\rho(k) & \approx\left[Ze^{-\frac{k^{2}}{2T}}(1+ZI(k))-Z^{2}e^{-\frac{k^{2}}{T}}\right](1+ZI(k))\nonumber \\
	& \approx Ze^{-\frac{k^{2}}{2T}}+2Z^{2}e^{-\frac{k^{2}}{2T}}I(k)-Z^{2}e^{-\frac{k^{2}}{T}}.
\end{align}
We give the calculation of Particle Drude Weight $D_{nn}$.  
The particle Drude weight corresponds to the particle density $D_{nn}=n=\int\rho(k)dk$. 
Integrating the expression for $\rho(k)$
\begin{equation}
	D_{nn}=\frac{1}{2\pi}\left[Z\int dke^{-\frac{k^{2}}{2T}}+2Z^{2}\int dke^{-\frac{k^{2}}{2T}}I(k)-Z^{2}\int dke^{-\frac{k^{2}}{T}}\right].
\end{equation}

Using the Gaussian integral $\int e^{-ax^{2}}dx=\sqrt{\pi/a}$, the first and third terms yield $Z\sqrt{2\pi T}and-Z^{2}\sqrt{\pi T}$, respectively. 
For the second term involving $I(k)$, we utilize the coordinate transformation
\begin{equation}
	k=k'+q',\quad q=k'-q'\implies dkdq=2dk'dq'.
\end{equation}
The quadratic form becomes $k^{2}+q^{2}=2{k'}^{2}+2{q'}^{2}$ and the kernel argument becomes $k-q=2q'$. 
The integral factorizes
\begin{align}
	\int dk e^{-\frac{k^{2}}{2T}}I(k) & =2\int dk'e^{-\frac{{k'}^{2}}{T}}\int dq'\phi(2q')e^{-\frac{{q'}^{2}}{T}}=2\sqrt{\pi T}J_{0},
\end{align}
where $J_{0}\equiv\int dq'\phi(2q')e^{-{q'}^{2}/T}$. 
Combining these results
\begin{align}
	D_{nn} & =\frac{T^{1/2}}{\sqrt{2\pi}}\left[Z+2\sqrt{2}Z^{2}J_{0}-\frac{\sqrt{2}}{2}Z^{2}\right]\nonumber \\
	& =\frac{T^{1/2}}{\sqrt{2\pi}}\left(Z+2\sqrt{2}Z^{2}G_{1}\right),
\end{align}
with the coefficient $G_{1}=J_{0}-1/2$.

We now present calculation of energy Drude Weight $D_{ne}$. 
The energy Drude weight is given by $D_{ne}=E/L+P$.
We first compute the energy density 
\begin{equation}
	\frac{E}{L}=E/L=\frac{1}{4\pi}\int k^{2}2\pi\rho(k)dk=\frac{1}{4\pi}\left[Z\int k^{2}e^{-\frac{k^{2}}{2T}}dk*{\sqrt{2\pi}T^{3/2}}+2Z^{2}\mathcal{K}-Z^{2}\int k^{2}e^{-\frac{k^{2}}{T}}dk*{\frac{1}{2}\sqrt{\pi}T^{3/2}}\right].
\end{equation}
The integral $\mathcal{K}=\int k^{2}e^{-k^{2}/2T}I(k)dk$ is solved using the same coordinate transformation. Noting that $k^{2}={k'}^{2}+2k'q'+{q'}^{2}$, the cross-term $2k'q'$ vanishes due to parity. The integral splits into:
\begin{align}
	\mathcal{K} & =2\int dk'e^{-\frac{{k'}^{2}}{T}}\int dq'e^{-\frac{{q'}^{2}}{T}}\phi(2q')({k'}^{2}+{q'}^{2})\nonumber \\
	& =2\left[\left(\frac{\sqrt{\pi}T^{3/2}}{2}\right)J_{0}+\sqrt{\pi T}J_{2}\right]\nonumber \\
	& =\sqrt{\pi T}(TJ_{0}+2J_{2}),
\end{align}
where $J_{2}\equiv\int{q'}^{2}\phi(2q')e^{-{q'}^{2}/T}dq'$. Substituting $\mathcal{K}$ back into the energy density expression:
\begin{equation}
	\frac{E}{L}=\frac{T^{3/2}}{2\sqrt{2\pi}}\left[Z+\sqrt{2}Z^{2}\left(J_{0}+\frac{2}{T}J_{2}-\frac{1}{4}\right)\right].
\end{equation}
Finally, incorporating the pressure term P :
\begin{align}
	P & =\frac{T}{2\pi}\int dk\ln(1+X)\nonumber \\
	& =\frac{T}{2\pi}\int dk\left[-\ln(1-Ze^{-\frac{k^{2}}{2T}})+\ln\left(1+Z^{2}e^{-\frac{k^{2}}{2T}}I(k)-Z^{2}e^{-\frac{k^{2}}{T}}\right)\right]\nonumber \\
	& =P_{0}+\frac{T}{2\pi}\int dk\left(Z^{2}e^{-\frac{k^{2}}{2T}}I(k)-Z^{2}e^{-\frac{k^{2}}{T}}\right).
\end{align}
Here $P_{0}$ is the pressure of free bosons. In the high-temperature limit, $P_{0}\approx\frac{T^{3/2}Z}{\sqrt{2\pi}}$. Evaluating the integral in the second term:
\begin{align}
	\text{Integral} & =Z^{2}\left(\int dk,e^{-\frac{k^{2}}{2T}}I(k)-\int dk,e^{-\frac{k^{2}}{T}}\right)\nonumber \\
	& =Z^{2}\left(2\sqrt{\pi T}J_{0}-\sqrt{\pi T}\right)\nonumber \\
	& =\sqrt{\pi T}Z^{2}(2J_{0}-1).
\end{align}
Substituting this back:
\begin{align}
	P & \approx\frac{T^{3/2}Z}{\sqrt{2\pi}}+\frac{T}{2\pi}\sqrt{\pi T}Z^{2}(2J_{0}-1)\nonumber \\
	& =\frac{T^{3/2}}{\sqrt{2\pi}}Z+\frac{T^{3/2}}{2\sqrt{\pi}}Z^{2}(2J_{0}-1)\nonumber \\
	& =\frac{T^{3/2}}{\sqrt{2\pi}}\left[Z+\sqrt{2}Z^{2}\left(J_{0}-\frac{1}{2}\right)\right].
\end{align}
Using the previously defined $G_{1}=J_{0}-1/2$, we arrive at the final expression:
\begin{equation}
	P=\frac{T^{3/2}}{\sqrt{2\pi}}\left(Z+\sqrt{2}G_{1}Z^{2}\right).
\end{equation}
Combining these results:
\begin{align}
	D_{ne} & =\frac{E}{L}+P\nonumber \\
	& =\frac{3T^{3/2}}{2\sqrt{2\pi}}\left[Z+\frac{\sqrt{2}}{3}Z^{2}\left(3J_{0}+\frac{2}{T}J_{2}-\frac{5}{4}\right)\right]\nonumber \\
	& =\frac{3T^{3/2}}{2\sqrt{2\pi}}\left(Z+\frac{\sqrt{2}}{3}Z^{2}G_{2}\right).
\end{align}
Here, the second virial coefficient for the energy current is defined
as:
\begin{align}
	G_{2} & =3J_{0}+\frac{2}{T}J_{2}-\frac{5}{4} \nonumber \\
	& =3\int_{-\infty}^{\infty}\phi(2q')e^{-\frac{q'{}^{2}}{T}}dq'+\frac{2}{T}\int_{-\infty}^{\infty}q'{}^{2}\phi(2q')e^{-\frac{q'{}^{2}}{T}}dq'-\frac{5}{4}.
\end{align}

\section*{Appendix B: Drude weights in Bose-Fermi mixture model}

\def\thefigure{B\arabic{figure}}
\def\thetable{B\arabic{table}}
\def\theequation{B\arabic{equation}}

This section provides the definitions, Hamiltonian, and key Thermodynamic Bethe Ansatz (TBA) / Generalized Hydrodynamics (GHD) equations for the integrable Bose-Fermi (B-F) mixture model used in the main text.

\subsection{Bose-Fermi mixture Model}

\subsubsection*{Thermodynamic Bethe Ansatz Equations}

We consider a one-dimensional Bose-Fermi mixture with periodic boundary
conditions. The general Hamiltonian is given by~\cite{Phys.Rev.A2006ImambekovExactly}:
\begin{equation}
	H=\int_{0}^{L}dx\left(\frac{\hbar^{2}}{2m_{b}}\partial_{x}\Psi_{b}^{\dagger}\partial_{x}\Psi_{b}+\frac{\hbar^{2}}{2m_{f}}\partial_{x}\Psi_{f}^{\dagger}\partial_{x}\Psi_{f}+\frac{g_{bb}}{2}\Psi_{b}^{\dagger}\Psi_{b}^{\dagger}\Psi_{b}\Psi_{b}+g_{bf}\Psi_{b}^{\dagger}\Psi_{f}^{\dagger}\Psi_{f}\Psi_{b}-\mu_{{F}}\Psi_{f}^{\dagger}\Psi_{f}-\mu_{{B}}\Psi_{b}^{\dagger}\Psi_{b}\right).
\end{equation}
Here, $\Psi_{b(f)}$ are the field operators for bosons (fermions). 
The fermions are spin-polarized, so $g_{ff}=0$. The model is integrable when masses and interaction strengths are equal: $m_{b}=m_{f}=m_{0}$ and $g_{bb}=g_{bf}=g{=-(2\hbar^2/m_{0}a_{1D})}$, here ${a_{1D}}$ is an effective $1D$ scattering length. We adopt units where $m_0=\hbar=1$. The first-quantized Hamiltonian is:
\begin{equation}
	\hat{H}=-\sum_{i=1}^{N}{\frac{\hbar^{2}}{2m_0}}\frac{\partial^{2}}{\partial x_{i}^{2}}+{\frac{\hbar^2c}{m_{0}}}\sum_{i<j}^{N}\delta\left(x_{i}-x_{j}\right)-\mu N-\frac{\mathcal{H}}{2}\left(N_{f}-M\right),
\end{equation}
where $N=N_{f}+M$ is the total particle number ($M$ bosons, $N_{f}$
fermions) and $c=-2/a_{1D}$. 
We use the average chemical potential $\mu=(\mu_{F}+\mu_{B})/2$
and the effective field $\mathcal{H}=\mu_{F}-\mu_{B}$. 
This model is solvable via the nested Bethe Ansatz~\cite{J.Phys.A:Math.Theor.2020WangEmergent}. 
The energy spectrum is determined by $N$ quasi-momenta $\{k_{i}\}$ (representing the overall particle motion) and $M$ rapidities $\{\Lambda_{j}\}$ (associated with internal bosonic degrees of freedom).
In the thermodynamic limit, the macroscopic state at thermal equilibrium is described by the dressed energy $\epsilon(k),$ of the quasiparticle excitations. The dressed energy is the solution to the core TBA integral equations
\begin{align}
	\epsilon(k) & =-\mu_{F}+\frac{k^{2}}{2}-\frac{T}{2\pi}\int_{-\infty}^{\infty}\phi(k-\Lambda)\ln[1+\exp[-\varphi(\Lambda)/T]]d\Lambda, 
	\label{eq:BFTBA1}
	\\
	\phi(\Lambda) & =\mu_{F}-\mu_{B}-\frac{T}{2\pi}\int_{-\infty}^{\infty}\phi(\Lambda-k)\ln[1+\exp[-\epsilon(k)/T]]dk.
	\label{eq:BFTBA2}
\end{align}
In the thermodynamic limit ($N,L\to\infty$ with $n=N/L$ fixed), the state is described by the distribution densities for particles, $\rho(k)$ and $\sigma(\Lambda)$, and for holes, $\rho_{h}(k)$ and $\sigma_{h}(\Lambda)$
\begin{align}
	\rho(k)+\rho_{h}(k) & =\frac{1}{2\pi}+\frac{1}{2\pi}\int_{-\infty}^{\infty}\phi(\Lambda-k)\sigma(\Lambda)d\Lambda, \\
	\sigma(\Lambda)+\sigma_{h}(\Lambda) & =\frac{1}{2\pi}\int_{-\infty}^{\infty}\phi(\Lambda-k)\rho(k)dk.
\end{align}
The total particle density ($n$) and boson density ($m$) read 
\begin{equation}
	n=\frac{N}{L}=\int dk\rho(k),\quad m=\frac{M}{L}=\int d\Lambda\sigma(\Lambda).
\end{equation}
The momentum density ($P/L$) and energy density ($E/L$) are 
\begin{equation}
	\frac{P}{L}=\int dk\,k\rho(k),\quad\frac{E}{L}=\int dk\,\frac{k^{2}}{2}\rho(k). 
\end{equation}

\subsubsection*{Local conserved charges and currents}

Bose-Fermi mixture consists of two components, labeled by the index
$a\in\{\rho,\sigma\}$. The system possesses a set of conserved quantities
$Q_{i}$ and their associated currents $J_{i}$~\cite{J.Phys.A:Math.Theor.2012MosselGeneralized}.
These are composed of contributions from both species and the total
conserved charges and currents are integrals of their respective densities,
$q_{i}(x)$ and $j_{i}(x)$:
\begin{align}
	Q_{i} & =Q_{\rho_{i}}+Q_{\sigma_{i}}=\sum_{a}Q_{a_{i}}\nonumber \\
	& =\sum_{a}\sum_{x}q_{a_{i}}\left(x\right)=\sum_{x}q_{i}\left(x\right)=\int dxq_{i}\left(x\right), \\
	J_{i} & =J_{\rho_{i}}+J_{\sigma_{i}} =\sum_{a}J_{a_{i}} \nonumber\\
	&=\sum_{a}\sum_{x}j_{a_{i}}\left(x\right)=\sum_{x}j_{i}\left(x\right) =\int dxj_{i}\left(x\right).
\end{align}

We further introduce a unified momentum-like (rapidity) variable $u$, such
that $u=k$ for the $\rho$-species and $u=\Lambda$ for the $\sigma$-species.
Where the $\rho$-species represents the ``Total Component" and $\sigma$-species Represents the``Boson Component". 
This convention is used  in the sums like $\sum_{a}\int g_{a}(u)du\equiv\int g_{\rho}(k)dk+\int g_{\sigma}(\Lambda)d\Lambda$. 
In the thermodynamic limit, the charge and current densities are expressed in terms of the particle densities $\rho_{a}(u)$ and the single-particle eigenvalues $h_{a_{i}}(u)$ of the conserved charges~\cite{Phys.Rev.Lett.2016BertiniTransport,Phys.Rev.X2016Castro-AlvaredoEmergent}
\begin{align}
	q_{i}\left(x\right) & =\sum_{a}\int dk\rho_{a}\left(u,x\right)h_{a_{i}}\left(u\right),\\
	j_{i}\left(x\right) & =\sum_{a}\int duj_{a}\left(u,x\right)h_{a_{i}}\left(u\right).
\end{align}
Under the local equilibrium approximation, these densities obey the hydrodynamic conservation equation~\cite{Phys.Rev.Lett.2016BertiniTransport,Phys.Rev.X2016Castro-AlvaredoEmergent}
\begin{equation}
	\partial_{t}q\left(x,t\right)+\partial_{x}j\left(x,t\right)=0.
\end{equation}
For a spatially homogeneous (translationally invariant) steady state, the local densities $q_{i}(x)$ and $j_{i}(x)$ are constant and equal to their global averages
\begin{align}
	\frac{Q_{i}}{L}=q_{i} & =\sum_{a}\int du\rho_{a}(u)h_{a,i}(u), \\
	\frac{J_{i}}{L}=j_{i} & =\sum_{a}\int duj_{a}(u)h_{a,i}(u).
\end{align}

For our  convenience in calculation, we denote $h_{a,i}$ as $h_{a_i}$. As discussed in Refs.~\cite{Phys.Rev.B2017IlievskiBallistic,Phys.Rev.X2016Castro-AlvaredoEmergent}, these eigenvalues are explicitly given by
\begin{align}
	h_{n} & =\{h_{\rho_{n}},h_{\sigma_{n}}\}=\{1_{\rho},1_{\sigma}\}=\{1,0\}, & \text{for}\,i=n\\
	h_{f} & =\{h_{\rho_{f}},h_{\sigma_{f}}\}=\{f_{\rho},f_{\sigma}\}=\{1,-1\}, & \text{for}\,i=f\\
	h_{m} & =\{h_{\rho_{m}},h_{\sigma_{m}}\}=\{m_{\rho},m_{\sigma}\}=\{0,1\}, & \text{for}\,i=m\\
	h_{k} & =\{h_{\rho_{k}},h_{\sigma_{k}}\}=\{k_{\rho},k_{\sigma}\}=\{k,0\}, & \text{for}\,i=k\\
	h_{e} & =\{h_{\rho_{e}},h_{\sigma_{e}}\}=\{e_{\rho},e_{\sigma}\}=\{k^{2}/2,0\},  & \text{for}\,i=e\\
	h_{\epsilon} & =\{h_{\rho_{\epsilon}},h_{\sigma_{\epsilon}}\}=\{\epsilon_{\rho},\epsilon_{\sigma}\}=\{k^{2}/2-(\mu+\mathcal{H}/2),\mathcal{H}\} . & \text{for}\,i=\epsilon
\end{align}
Here, $h_{n}$, $h_{k}$, $h_{e}$, and $h_{\epsilon}$ are defined as in the Lieb-Liniger model, while $h_{f}$ and $h_{m}$ represent the single-particle eigenvalues for the Fermi particle number and Bose particle number, respectively.
The current density $j_{a}(u)$ is given by $j_{a}(u)=\rho_{a}(u)v_{a}^{\dr}(u)$,
where $v_{a}^{\dr} = e_{a}'/ k_{a}'$ is the effective velocity (with $\rho_{\rho} = \rho,\rho_{\sigma}=\sigma$).
The system is described by a Generalized Gibbs Ensemble (GGE) density matrix, which is constrained by all conserved quantities
\begin{align}
	\hat{\rho}_{GGE} & \simeq e^{-\sum_{i}\beta_{i}\hat{Q}_{i}} =e^{-\sum_{a}\int\mu_{a}(u)\rho_{a}(u)du}. 
\end{align}
Here, $\mu_{a}(u)=\sum_{i}\beta_{i}h_{a_{i}}(u)$ is the effective chemical potential for a particle of species $a$ with rapidity $u$ and $\beta_{i}$ are the associated potentials. The pseudoenergy $\epsilon(u)$ is defined by the following coupled integral equations~\cite{Phys.Rev.X2016Castro-AlvaredoEmergent}:
\begin{align}
	\epsilon(u) &= \mu_{a}(u) - \int \frac{1}{\pi}\frac{c}{c^{2}+(u'-u)^{2}}\ln\left(1 + e^{-\psi(u')}\right) du', 
	\label{eq:BFGTBA1}
	\\
	\psi(u') &= \mu_{a'}(u') - \int \frac{1}{\pi}\frac{c}{c^{2}+(u-u')^{2}}\ln\left(1 + e^{-\epsilon(u)}\right) du.
	\label{eq:BFGTBA2}
\end{align}
Specifically, in the grand canonical ensemble, the above equation takes the form of Eq.~(\ref{eq:BFTBA1}-\ref{eq:BFTBA2}).

\subsubsection*{The Dressing Operation}

Physical quantities are "dressed" by interactions. The dressing operation for an arbitrary function $\psi={\psi_{\rho},\psi_{\sigma}}$ is defined by the set of coupled linear integral equations
\begin{align}
	\psi_{\rho}^{\dr}(k) & =\psi_{\rho}(k)+\int d\Lambda\phi(k-\Lambda)\theta_{\sigma}(\Lambda)\psi_{\sigma}^{\dr}(\Lambda), \\
	\psi_{\sigma}^{\dr}(\Lambda) & =\psi_{\sigma}(\Lambda)+\int dk\phi(\Lambda-k)\theta_{\rho}(k)\psi_{\rho}^{\dr}(\Lambda). 
\end{align}
Here, $\phi(u-u')$ is the scattering kernel, and $\theta_{a}(u)$ is the occupation factor (e.g., $\theta_{\rho}(k)$). 
We can write this formalism more compactly. 
Let us define an integral operator $\tilde{T}_{a}$ acting on a function $\psi_{b}$: $\tilde{T}_{a}\theta_{b}\psi_{b}(u)=\int du^{'}\phi(u-u^{'})\theta_{b}(u^{'})\psi_{b}(u^{'})$.

The dressing equations can be solved formally by substitution. 
Substituting $\psi_{\sigma}^{\dr}$ into the $\psi_{\rho}^{\dr}$ equation, we obtain the following forms of dressed quantities 
\begin{align}
	\psi_{\rho}^{\dr}(k) & =\psi_{\rho}(k)+\int d\Lambda\phi(k-\Lambda)\theta_{\sigma}\left(\psi_{\sigma}(\Lambda)+\int dk\phi(\Lambda-k)\theta_{\rho}\psi_{\rho}^{\dr}(k)\right)\nonumber \\
	& =\psi_{\rho}(k)+\int d\Lambda\phi(k-\Lambda)\theta_{\sigma}\psi_{\sigma}(\Lambda)+\int d\Lambda\phi(k-\Lambda)\theta_{\sigma}\int dk\phi(\Lambda-k)\theta_{\rho}\psi_{\rho}^{\dr}(k)\nonumber \\
	& =\psi_{\rho}(k)+\tilde{T}_{\rho}\theta_{\sigma}\psi_{\sigma}(\Lambda)+T_{\rho}\theta_{\sigma}\tilde{T}_{\sigma}\theta_{\rho}\psi_{\rho}^{\dr}(k)\nonumber \\
	& =\left(1-\tilde{T}_{\rho}\theta_{\sigma}\tilde{T}_{\sigma}\theta_{\rho}\right)^{-1}\left[\psi_{\rho}(k)+\tilde{T}_{\rho}\theta_{\sigma}\psi_{\sigma}(\Lambda)\right], \\
	\psi_{\sigma}^{\dr}(\Lambda) & =\psi_{\sigma}(\Lambda)+\int dk\phi(\Lambda-k)\theta_{\rho}\left(\psi_{\rho}(k)+\int d\Lambda\phi(k-\Lambda)\theta_{\sigma}\psi_{\sigma}^{\dr}(\Lambda)\right)\nonumber \\
	& =\psi_{\sigma}(\Lambda)+\int dk\phi(\Lambda-k)\theta_{\rho}\psi_{\rho}(k)+\int dk\phi(\Lambda-k)\theta_{\rho}\int d\Lambda\phi(k-\Lambda)\theta_{\sigma}\psi_{\sigma}^{\dr}(\Lambda)\nonumber \\
	& =\psi_{\sigma}(\Lambda)+\tilde{T}_{\sigma}\theta_{\rho}\psi_{\rho}(k)+\tilde{T}_{\sigma}\theta_{\rho}\tilde{T}_{\rho}\theta_{\sigma}\psi_{\sigma}^{\dr}(\Lambda)\nonumber \\
	& =\left(1-\tilde{T}_{\sigma}\theta_{\rho}\tilde{T}_{\rho}\theta_{\sigma}\right)^{-1}\left[\psi_{\sigma}(\Lambda)+\tilde{T}_{\sigma}\theta_{\rho}\psi_{\rho}(k)\right]. 
\end{align}

Defining the resolvent operators $A\equiv(1-\tilde{T}_{\rho}\theta_{\sigma}\tilde{T}_{\sigma}\theta_{\rho})^{-1}$
and $A'\equiv(1-\tilde{T}_{\sigma}\theta_{\rho}\tilde{T}_{\rho}\theta_{\sigma})^{-1}$, the solution is
\begin{align}
	\psi_{\rho}^{\dr} & =A\left(\psi_{\rho}+\tilde{T}_{\rho}\theta_{\sigma}\psi_{\sigma}\right),\\
	\psi_{\sigma}^{\dr} & =A^{'}\left(\psi_{\sigma}+\tilde{T}_{\sigma}\theta_{\rho}\psi_{\rho}\right).
\end{align}
Alternatively, the dressing operation can be expressed in a compact matrix form
\begin{equation}
	\left(\begin{array}{c}
		\psi_{\rho}^{\dr}\\
		\psi_{\sigma}^{\dr}
	\end{array}\right)=\left(\begin{array}{c}
		\psi_{\rho}\\
		\psi_{\sigma}
	\end{array}\right)+\left(\begin{array}{cc}
		0 & \tilde{T}_{\rho}\theta_{\sigma}\\
		\tilde{T}_{\sigma}\theta_{\rho} & 0
	\end{array}\right)\left(\begin{array}{c}
		\psi_{\rho}^{\dr}\\
		\psi_{\sigma}^{\dr}
	\end{array}\right),
\end{equation}
or simply $\boldsymbol{\psi}^{\dr}=\boldsymbol{\psi}+\mathbf{\tilde{T}}\boldsymbol{\psi}^{\dr}$. 
The formal solution is obtained by inverting this matrix operator
\begin{equation}
	\left(\begin{array}{c}
		\psi_{\rho}^{\dr}\\
		\psi_{\sigma}^{\dr}
	\end{array}\right)=\left(1-\left(\begin{array}{cc}
		0 & \tilde{T}_{\rho}\theta_{\sigma}\\
		\tilde{T}_{\sigma}\theta_{\rho} & 0
	\end{array}\right)\right)^{-1}\left(\begin{array}{c}
		\psi_{\rho}\\
		\psi_{\sigma}
	\end{array}\right),
\end{equation}
or simply $\boldsymbol{\psi}^{\dr}=(\mathbf{1}-\mathbf{\tilde{T}})^{-1}\boldsymbol{\psi}$.

\subsection{Derivation of the Drude Weight Formalism}

In this section, we derive the formal expression for the Drude weight
matrix $D_{ij}$. 
The derivation connects the Drude weight to the
second derivatives of a generalized free energy functional $F_{g}$.
We define a functional $F_{g}$ based on a set of bare single-particle eigenvalues $g=(g_{\rho},g_{\sigma})$
\begin{equation}
	F_{g}=-\sum_{a\in\{\rho,\sigma\}}\frac{1}{2\pi}\int du_{a}\,g_{a}(u_{a})\ln\left(1+e^{-\epsilon_{a}(u_{a})}\right).
\end{equation}
If $g={h_{k}'}=(1,0)$, $F_{h_k'}$ is the free energy of the GGE. 
1f $g=e'=(k,0)$, $F_{e^{'}}$ is the current free energy~\cite{Phys.Rev.X2016Castro-AlvaredoEmergent}. 
Our goal is to compute the second derivative $\partial_{\beta_{i}}\partial_{\beta_{j}}F_{g}$.
We first compute the partial derivative of $F_{g}$ with respect to
a Lagrange multiplier $\beta_{i}$
\begin{align}
	\frac{\partial F_{g}}{\partial\beta_{i}} & =\frac{1}{2\pi}\int dkg_{\rho}(k)\theta_{\rho}h_{\rho,i}^{\dr}(k)+\frac{1}{2\pi}\int d\Lambda g_{\sigma}(\Lambda)\theta_{\sigma}h_{\sigma,i}^{\dr}(k).\label{eq:dFdb}
\end{align}
The resulting dressed charges are derived by taking the derivatives of Eqs.~\eqref{eq:BFGTBA1} and \eqref{eq:BFGTBA2} with respect to the Lagrange multipliers~\cite{Phys.Rev.B2017IlievskiBallistic}
	\begin{align}
		\partial_{\beta_{i}}\epsilon & =h_{\rho,i}^{\dr}, \,\,\,\partial_{\beta_{i}}\psi  =h_{\sigma,i}^{\dr}. 
	\end{align}

This relation Eq.~\ref{eq:dFdb} connects the thermodynamics to the GGE densities. By setting $g=h_k'$ and $g=e'$, we find
	\begin{align}
		\frac{\partial F_{h_k'}}{\partial\beta_{i}} & =\frac{1}{2\pi}\int dkk_{\rho}^{'}\theta_{\rho}h_{\rho_{i}}^{\dr}+\frac{1}{2\pi}\int d\Lambda k_{\sigma}^{'}\theta_{\sigma}h_{\sigma_{i}}^{\dr}\nonumber \\
		& =\frac{1}{2\pi}\int dk1_{\rho}^{\dr}\theta_{\rho}h_{\rho_{i}}+\frac{1}{2\pi}\int d\Lambda 1_{\sigma}^{\dr}\theta_{\sigma}h_{\sigma_{i}}\nonumber \\
		& =\frac{1}{2\pi}\int dk\frac{2\pi\rho}{\theta_{\rho}}\theta_{\rho}h_{\rho_{i}}+\frac{1}{2\pi}\int d\Lambda \frac{2\pi\sigma}{\theta_{\sigma}}\theta_{\sigma}h_{\sigma_{i}}\nonumber \\
		& =\int dk\rho h_{\rho,i}+\int d\Lambda\sigma h_{\sigma,i}\nonumber \\
		& =q_{\rho,i}+q_{\sigma,i}\nonumber \\
		& =q_{i}.
	\end{align}
	\begin{align}
		\frac{\partial F_{e^{'}}}{\partial\beta_{i}} & =\frac{1}{2\pi}\int dke_{\rho}^{'}\theta_{\rho}h_{\rho,i}^{\dr}+\frac{1}{2\pi}\int d\Lambda e_{\sigma}^{'}\theta_{\sigma}h_{\sigma,i}^{\dr}\nonumber \\
		& =\frac{1}{2\pi}\int dkk_{\rho}^{\dr}\theta_{\rho}h_{\rho,i}+\frac{1}{2\pi}\int d\Lambda k_{\sigma}^{\dr}\theta_{\sigma}h_{\sigma,i}\nonumber \\
		& =\frac{1}{2\pi}\int dkv^{\eff}_{\rho}1_{\rho}^{\dr}\theta_{\rho}h_{\rho,i}+\frac{1}{2\pi}\int d\Lambda v^{\eff}_{\sigma}1_{\sigma}^{\dr}\theta_{\sigma}h_{\sigma,i}\nonumber \\
		& =\frac{1}{2\pi}\int dkv^{\eff}_{\rho}\frac{2\pi\rho}{\theta_{\rho}}\theta_{\rho}h_{\rho,i}+\frac{1}{2\pi}\int d\Lambda v^{\eff}_{\sigma}\frac{2\pi\sigma}{\theta_{\sigma}}\theta_{\sigma}h_{\sigma,i}\nonumber \\
		& =\int dk\rho h_{\rho,i}v_{\rho}^{\eff}+\int d\Lambda\sigma h_{\sigma,i}v_{\sigma}^{\eff}\nonumber \\
		& =j_{\rho_{i}}+j_{\sigma_{i}}\nonumber \\
		& =j_{i}.
	\end{align}

\subsubsection*{Second Derivative: The $C_{ij}$ and $B_{ij}$ Matrices}

The Drude weight $D_{ij}$ is related to the static correlation matrices.
We define the charge-charge correlation matrix $C_{ij}$ and the current-charge
correlation matrix $B_{ij}$ \cite{Phys.Rev.X2016Castro-AlvaredoEmergent,SciPostPhys.2017DoyonNote}
\begin{align*}
	C_{ij} & =\int dx\left\langle q_{i}(x,0)q_{j}(0,0)\right\rangle ^{c}=-\partial_{\beta_{i}}\partial_{\beta_{j}}F_{h_k'},\\
	B_{ij} & =\int dx\left\langle j_{i}(x,0)q_{j}(0,0)\right\rangle ^{c}=-\partial_{\beta_{i}}\partial_{\beta_{j}}F_{e'}.
\end{align*}

As shown in \cite{Phys.Rev.B2017IlievskiBallistic,SciPostPhysics2017DoyonDrude}
and \cite{Phys.Rev.B2017IlievskiBallistic}, the Drude weight matrix
is given by $D=\beta B C^{-1} B^{T}$. 
Our task is   to find the general second derivative $\partial_{\beta_{i}}\partial_{\beta_{j}}F_{g}$.

We differentiate expression for $\frac{\partial F_{g}}{\partial\beta_{i}}$
Eq. (\ref{eq:dFdb}) with respect to $\beta_{j}$
\begin{equation}
	\frac{\partial^{2}F_{g}}{\partial\beta_{j}\partial\beta_{i}}=\sum_{a\in\{\rho,\sigma\}}\frac{1}{2\pi}\int du_{a}\,g_{a}\left[\frac{\partial \theta_{a}}{\partial\beta_{j}}h_{a}^{\dr}+\theta_{a}\frac{\partial h_{a_{i}}^{\dr}}{\partial\beta_{j}}\right].\label{eq:ddFgrho}
\end{equation}
This requires two terms. The first term reads 
\begin{equation}
	\ensuremath{\frac{\partial \theta_{a}}{\partial\beta_{j}}=\frac{\partial \theta_{a}}{\partial\epsilon_{a}}\frac{\partial\epsilon_{a}}{\partial\beta_{j}}=-\theta_{a}(1-\theta_{a})h_{a_{j}}^{\dr}}.
\end{equation}
The second term, $\frac{\partial h_{a_{i}}^{\dr}}{\partial\beta_{j}}$, is more complex. Differentiating the matrix form $\boldsymbol{h}_{i}^{\dr}=\boldsymbol{h}_{i}+\mathbf{\tilde{T}}\boldsymbol{h}_{i}^{\dr}$ (and noting $\mathbf{\tilde{T}}$ depends on $\beta_{j}$ through $\theta_{a}$) leads to:
\begin{align}
	\frac{\partial\boldsymbol{h}_{i}^{\dr}}{\partial\beta_{j}} & =\frac{\partial\mathbf{\tilde{T}}}{\partial\beta_{j}}\boldsymbol{h}_{i}^{\dr}+\mathbf{\tilde{T}}\frac{\partial\boldsymbol{h}_{i}^{\dr}}{\partial\beta_{j}}\nonumber, \\
	(\mathbf{1}-\mathbf{\tilde{T}})\frac{\partial\boldsymbol{h}_{i}^{\dr}}{\partial\beta_{j}} & =\left(\frac{\partial\mathbf{\tilde{T}}}{\partial\beta_{j}}\boldsymbol{h}_{i}^{\dr}\right)\nonumber \\
	\frac{\partial\boldsymbol{h}_{i}^{\dr}}{\partial\beta_{j}} & =\left(\frac{\partial\mathbf{\tilde{T}}}{\partial\beta_{j}}\boldsymbol{h}_{i}^{\dr}\right)^{\dr}.
\end{align}
Substituting these results into the expression for $\partial^{2}F_{g}$ [Eq.~(\ref{eq:ddFgrho})] and simplifying using the swap identity $\int \mathrm{d}k\, g_{\rho}\theta_{\rho}\psi_{\rho}^{\mathrm{\dr}}+\int \mathrm{d}\Lambda\, g_{\sigma}\theta_{\sigma}\psi_{\sigma}^{\mathrm{\dr}}=\int \mathrm{d}k\, g_{\rho}^{\mathrm{\dr}}\theta_{\rho}\psi_{\rho}+\int \mathrm{d}\Lambda\, g_{\sigma}^{\mathrm{\dr}}\theta_{\sigma}\psi_{\sigma}$, we obtain the final expression
\begin{equation}
	\partial_{\beta_{j}}\partial_{\beta_{i}}F_{g}=-\sum_{a\in\{\rho,\sigma\}}\frac{1}{2\pi}\int du_{a}\,g_{a}^{\dr}\theta_{a}(1-\theta_{a})h_{a_{i}}^{\dr}h_{a_{j}}^{\dr}.
\end{equation}
We can now compute $C_{ij}$ and $B_{ij}$ by substituting $g=h_k'$ and $g=e'$:
Charge-Charge Matrix $C_{ij}$:
\begin{align}
	C_{ij} & =\sum_{a}\frac{1}{2\pi}\int du_{a}\,1_{a}^{\dr}\theta_{a}(1-\theta_{a})h_{a_{i}}^{\dr}h_{a_{j}}^{\dr}\nonumber \\
	& =\sum_{a}\int du_{a}\,\rho_{a}(1-\theta_{a})h_{a_{i}}^{\dr}h_{a_{j}}^{\dr}\nonumber \\
	& \equiv\sum_{a}\int du_{a}h_{a_{i}}^{\dr}C_{a}h_{a_{j}}^{\dr}. \label{eq:Cij_BF}
\end{align}
where $1_{a}^{\dr}=2\pi\rho_{a}/\theta_{a}$, and we define the kernel $C_{a}=\rho_{a}(1-\theta_{a})$.

Current-Charge Matrix $B_{ij}$:
\begin{align}
	B_{ij} & =\sum_{a}\frac{1}{2\pi}\int du_{a}\,(e')_{a}^{\dr}\theta_{a}(1-\theta_{a})h_{a_{i}}^{\dr}h_{a_{j}}^{\dr}\nonumber \\
	& =\sum_{a}\int du_{a}\,\rho_{a}(1-\theta_{a})v_{a}^{\eff}h_{a_{i}}^{\dr}h_{a_{j}}^{\dr}\nonumber \\
	& \equiv\sum_{a}\int du_{a}h_{a_{i}}^{\dr}B_{a}h_{a_{j}}^{\dr}. \label{eq:Bij_BF}
\end{align}
Here $(e')_{a}^{\dr}=v_{a}^{\eff}1_{a}^{\dr}$ and we define the kernel $B_{a}=\rho_{a}(1-\theta_{a})v_{a}^{\eff}$.
The Drude Weight matrix $D_{ij}$ is then obtained by performing the matrix operation $D=\beta B C^{-1} B^{T}$ using these expressions for $B$ and $C$~\cite{Phys.Rev.B2017IlievskiBallistic,SciPostPhysics2017DoyonDrude}
\begin{align}
	D_{ij} & =\beta\sum_{a}\int du_{a}\rho_{a}(1-\theta_{a})\left(v_{a}^{\eff}\right)^{2}h_{a_{i}}^{\dr}h_{a_{j}}^{\dr}. \label{eq:Dij_BF}
\end{align}

\section*{Appendix C: Universal Relations and Approximate Solutions of the Drude Weight}
\def\thefigure{C\arabic{figure}}
\def\thetable{C\arabic{table}}
\def\theequation{C\arabic{equation}}

\subsection*{Universal Relations}

In Appendix B, we established the formal GHD framework for the Bose-Fermi mixture, culminating in expressions for the charge-charge matrix $C$ Eq.~(\ref{eq:Cij_BF}), the current-charge matrix $B$ Eq.~(\ref{eq:Bij_BF}) and the Drude weight matrix $D$ Eq.~(\ref{eq:Dij_BF}). In this section, we will use these expressions to derive the universal relations for the key Drude weight matrix elements.


To facilitate the calculations, we first transform the integrand. Taking the derivative with respect to the occupation number, we can obtain two relations:
\begin{align}
	d_{k}\theta_{\rho}(k) & =-\frac{2\pi}{T}\left[1-\theta_{\rho}(k)\right]v_{\rho}^{\eff}(k)\rho(k), \\
	d_{\Lambda}\theta_{\sigma}(\Lambda) & =-\frac{2\pi}{T}\left[1-\theta_{\sigma}(\Lambda)\right]v_{\sigma}^{\eff}(\Lambda)\sigma(\Lambda).
\end{align}
Substituting these relations into our starting formula for $D_{ij}$  Eq.~(\ref{eq:Dij_BF})
\begin{align}
	D_{ij}= & -\frac{1}{2\pi}\int-\frac{2\pi}{T}\rho_{P}(k)\left[1-\theta_{\rho}(k)\right]v_{\rho}^{\eff}(k)v_{\rho}^{\eff}(k)h_{\rho_{i}}^{\dr}(k)h_{\rho_{j}}^{\dr}(k)\,dk\nonumber \\
	& -\frac{1}{2\pi}\int-\frac{2\pi}{T}\sigma_{P}(\Lambda)[1-\theta_{\sigma}(\Lambda)]v_{\sigma}^{\eff}(\Lambda)v_{\sigma}^{\eff}(\Lambda)h_{\sigma_{i}}^{\dr}(\Lambda)h_{\sigma_{j}}^{\dr}(\Lambda)\,d\Lambda\nonumber \\
	= & -\frac{1}{2\pi}\int d_{k}\theta_{\rho}(k)v_{\rho}^{\eff}(k)h_{\rho_{i}}^{\dr}(k)h_{\rho_{j}}^{\dr}(k)\,dk\nonumber \\
	& -\frac{1}{2\pi}\int d_{\Lambda}\theta_{\sigma}(\Lambda)v_{\sigma}^{\eff}(\Lambda)h_{\sigma_{i}}^{\dr}(\Lambda)h_{\sigma_{j}}^{\dr}(\Lambda)\,d\Lambda. \label{eq:DW_dn_BF}
\end{align}
This form  is the key expression to compute  the Drude weights for the Bose-Fermi mixture. 

\subsubsection*{Universal Relation for $D_{nn}$}

We first calculate $D_{nn}$, the Drude weight for the total component number charge ($i=j=n$). The bare charge $h_{n}=\{1,0\}$ dresses to $h_{n}^{\dr}=1^{\dr}=\{1_{\rho}^{\dr},1_{\sigma}^{\dr}\}$.
Substituting this into Eq.~(\ref{eq:DW_dn_BF}), the expression becomes
\begin{equation}
	D_{nn}=-\frac{1}{2\pi}\int d_{k}\theta_{\rho}(k)k_{\rho}^{\dr}(k)1_{\rho}^{\dr}(k)\,dk-\frac{1}{2\pi}\int d_{\Lambda}\theta_{\sigma}(\Lambda)k_{\sigma}^{\dr}(\Lambda)1_{\sigma}^{\dr}(\Lambda)\,d\Lambda.\label{eq:Dnn_BF}
\end{equation}
For our convenience, we denote 
\begin{align}
	D_{nn}(1) & =-\frac{1}{2\pi}\int d_{k}\theta_{\rho}(k)k_{\rho}^{\dr}(k)1_{\rho}^{\dr}(k)\,dk\label{eq:Dnn1_BF},\\
	D_{nn}(2) & =-\frac{1}{2\pi}\int d_{\Lambda}\theta_{\sigma}(\Lambda)k_{\sigma}^{\dr}(\Lambda)1_{\sigma}^{\dr}(\Lambda)\,d\Lambda.\label{eq:Dnn2_BF}
\end{align}
Following the method  (\ref{Dnn-calculation}), we compute  the $D_{nn}(1)$ term with relating the term  $d_{k}\theta_{\rho}1_{\rho}^{\dr}$ to the TBA state densities $\rho$ and $\sigma$, namely, 
\begin{equation}
	d_{k}\theta_{\rho}1_{\rho}^{\dr}=2\pi(d_{k}\rho-\theta_{\rho}\tilde{T}_{\rho}d_{\Lambda}\sigma).
\end{equation}

Substituting this into $D_{nn}\left(1\right)$ Eq.~($\ref{eq:Dnn1_BF}$) and integrating by parts, we have 
\begin{align}
	D_{nn}\left(1\right) & =-\int\left(d_{k}\rho-\theta_{\rho}\tilde{T}_{\rho}d_{\Lambda}\sigma\right)k_{\rho}^{\dr}(k)\,dk\nonumber \\
	& =-\int d_{k}\rho k_{\rho}^{\dr}(k)\,dk+\int k_{\rho}^{\dr}(k)\theta_{\rho}\tilde{T}_{\rho}d_{\Lambda}\sigma\,dk\nonumber \\
	& =-\int d_{k}\rho k\,dk-\int dkd_{k}\rho \tilde{T}_{\rho}\left(\theta_{\sigma}k_{\sigma}^{\dr}\right)+\int dkk_{\rho}^{\dr}(k)\theta_{\rho}\tilde{T}_{\rho}d_{\Lambda}\sigma\nonumber \\
	& =-\rho k|_{-\infty}^{\infty}+\int\rho\,dk-\int dkd_{k}\rho \tilde{T}_{\rho}\theta_{\sigma}k_{\sigma}^{\dr}+\int dkk_{\rho}^{\dr}(k)\theta_{\rho}\tilde{T}_{\rho}d_{\Lambda}\sigma\nonumber \\
	& =n-\int dkd_{k}\rho \tilde{T}_{\rho}\theta_{\sigma}k_{\sigma}^{\dr}+\int dkk_{\rho}^{\dr}(k)\theta_{\rho}\tilde{T}_{\rho}d_{\Lambda}\sigma. \label{eq:Dnn1_c_BF}
\end{align}
Next, we analyze the third term using the derivative of the Boson state density $d_{\Lambda}\sigma$
\begin{align}
	d_{\Lambda}\sigma & =d_{\Lambda}\theta_{\sigma}\tilde{T}_{\sigma}\rho+\theta_{\sigma}\tilde{T}_{\sigma}d_{k}\rho\nonumber \\
	& =d_{\Lambda}\theta_{\sigma}\frac{\sigma}{\theta_{\sigma}}+\theta_{\sigma}\tilde{T}_{\sigma}d_{k}\rho.
	\label{eq:dlambdasigma}
\end{align}
Substituting this expression result back into the third term of Eq.~(\ref{eq:Dnn1_c_BF})
\begin{align}
	\int dkk_{\rho}^{\dr}(k)\theta_{\rho}\tilde{T}_{\rho}d_{\Lambda}\sigma & =\int dkk_{\rho}^{\dr}(k)\theta_{\rho}\int d\Lambda\phi\left(k-\Lambda\right)d_{\Lambda}\sigma\nonumber \\
	& =\int d\Lambda d_{\Lambda}\sigma \int dk\phi\left(k-\Lambda\right)\theta_{\rho}k_{\rho}^{\dr}(k)\nonumber \\
	& =\int d\Lambda d_{\Lambda}\sigma k_{\sigma}^{\dr}(k)\nonumber \\
	& =\int d\Lambda d_{\Lambda}\theta_{\sigma}\frac{\sigma}{\theta_{\sigma}}k_{\sigma}^{\dr}(k)+\int d\Lambda \theta_{\sigma}\tilde{T}_{\sigma}d_{k}\rho k_{\sigma}^{\dr}(k)\nonumber \\
	& =\frac{1}{2\pi}\int d\Lambda d_{\Lambda}\theta_{\sigma}1_{\sigma}^{\dr}k_{\sigma}^{\dr}(k)+\int d\Lambda \theta_{\sigma}k_{\sigma}^{\dr}(k)\int dk\phi\left(k-\Lambda\right)d_{k}\rho\nonumber \\
	& =\frac{1}{2\pi}\int d\Lambda d_{\Lambda}\theta_{\sigma}1_{\sigma}^{\dr}k_{\sigma}^{\dr}(k)+\int dkd_{k}\rho\int d\Lambda\phi\left(k-\Lambda\right)\theta_{\sigma}k_{\sigma}^{\dr}(k)\nonumber \\
	& =\frac{1}{2\pi}\int d\Lambda d_{\Lambda}\theta_{\sigma}1_{\sigma}^{\dr}k_{\sigma}^{\dr}(k)+\int dkd_{k}\rho \tilde{T}_{\rho}\theta_{\sigma}k_{\sigma}^{\dr}(k).
\end{align}
Here, in the third equal, the $ k_{\rho}^{\dr}(k)$ dressed equation was used. 
Substituting this result back into Eq.~(\ref{eq:Dnn1_c_BF}) yields
\begin{align}
	D_{nn}\left(1\right) & =-\frac{1}{2\pi}\int d_{k}n_{\rho}(k)k_{\rho}^{\dr}(k)1_{\rho}^{\dr}(k)\,dk\nonumber \\
	& =n-\int dkd_{k}\rho \tilde{T}_{\rho}\theta_{\sigma}k_{\sigma}^{\dr}+\int dkk_{\rho}^{\dr}(k)\theta_{\rho}\tilde{T}_{\rho}d_{\Lambda}\sigma\nonumber \\
	& =n-\int dkd_{k}\rho \tilde{T}_{\rho}\theta_{\sigma}k_{\sigma}^{\dr}+\frac{1}{2\pi}\int d\Lambda d_{\Lambda}\theta_{\sigma}1_{\sigma}^{\dr}k_{\sigma}^{\dr}(k)+\int dkd_{k}\rho \tilde{T}_{\rho}\theta_{\sigma}k_{\sigma}^{\dr}(k)\nonumber \\
	& =n+\frac{1}{2\pi}\int d\Lambda d_{\Lambda}\theta_{\sigma}1_{\sigma}^{\dr}k_{\sigma}^{\dr}(k).
\end{align}
Finally, substituting this expression into the $D_{nn}$ Eq.~(\ref{eq:Dnn_BF}) gives the universal relation
\begin{align}
	D_{nn} & =-\frac{1}{2\pi}\int d_{k}\theta_{\rho}(k)k_{\rho}^{\dr}(k)1_{\rho}^{\dr}(k)\,dk-\frac{1}{2\pi}\int d_{\Lambda}\theta_{\sigma}(\Lambda)k_{\sigma}^{\dr}(\Lambda)1_{\sigma}^{\dr}(\Lambda)\,d\Lambda\nonumber \\
	& =n+\frac{1}{2\pi}\int d\Lambda d_{\Lambda}\theta_{\sigma}1_{\sigma}^{\dr}k_{\sigma}^{\dr}(k)-\frac{1}{2\pi}\int d_{\Lambda}\theta_{\sigma}(\Lambda)k_{\sigma}^{\dr}(\Lambda)1_{\sigma}^{\dr}(\Lambda)\,d\Lambda\nonumber \\
	& =n.
\end{align}
We thus rigorously prove this universal relation for the Drude weight  $D_{nn}=n$.  A similar conclusion was also obtained for a single component Bose gas   in Ref.~\cite{arXiv2025GohmannBallistic}.

\subsubsection*{Universal Relation for $D_{ne}$}

We now calculate $D_{ne}$ ({particle-energy Drude weight}), corresponding to $h_{i}=h_{n}=\{1,0\}$ and $h_{j}=h_{e}=\{k^{2}/2,0\}$. Substituting into Eq.~(\ref{eq:DW_dn_BF}), the expression becomes
\begin{equation}
	D_{ne}= -\frac{1}{2\pi}\int d_{k}\theta_{\rho}(k)k_{\rho}^{\dr}(k)e_{\rho}^{\dr}(k)\,dk -\frac{1}{2\pi}\int d_{\Lambda}\theta_{\sigma}(\Lambda)k_{\sigma}^{\dr}(\Lambda)e_{\sigma}^{\dr}(\Lambda)\,d\Lambda \, .
	\label{eq:Dne_BF}
\end{equation}
We again adopt the method used for $D_{nn}$ by introducing the auxiliary function $u_{\rho}=e_{\rho}^{\dr}\theta_{\rho}=\theta_{\rho}\left(e_{\rho}+\tilde{T}_{\rho}u_{\sigma}\right)$. 
Following the similar method used in the Section A,  we take the derivative of $u_{\rho}$ 
\begin{align}
	\left(d_{k}\theta_{\rho}\right)e_{\rho}^{\dr} & =d_{k}u_{\rho}-\theta_{\rho}k-\theta_{\rho}\tilde{T}_{\rho}d_{\Lambda}u_{\sigma}. \label{eq:u_rho}
\end{align}
Similarly, we define $u_{\sigma}=\theta_{\sigma}e_{\sigma}^{\dr}$, and its derivative gives
\begin{equation}
	\left(d_{\Lambda}\theta_{\sigma}\right) e_{\sigma}^{\dr}=d_{\Lambda}u_{\sigma}-\theta_{\sigma}\tilde{T}_{\sigma}d_{k}u_{\rho}.
	\label{eq:u_sigma}
\end{equation}
Substituting Eq.~(\ref{eq:u_rho})  and Eq.~(\ref{eq:u_sigma}) into the $D_{ne}$ expression Eq.~(\ref{eq:Dne_BF}) and simplifying by a method similar to the $D_{nn}$ calculation, we obtain
\begin{align}
	D_{ne} & =\frac{1}{2\pi}\int dk\theta_{\rho}e_{\rho}^{\dr}+\frac{1}{2\pi}\int dkk_{\rho}^{\dr}(k)\theta_{\rho}k\nonumber \\
	& =\frac{1}{2\pi}\int dk\theta_{\rho}e_{\rho}(h_k')_{\rho}^{\dr}+\frac{1}{2\pi}\int dkk_{\rho}^{\dr}(k)\theta_{\rho}k\nonumber \\
	& =\frac{1}{2\pi}\int dk\theta_{\rho}e_{\rho}2\pi\frac{\rho}{\theta_{\rho}}+\frac{1}{2\pi}\int dkk_{\rho}^{\dr}(k)\theta_{\rho}k\nonumber \\
	& =\int dke_{\rho}\rho+\frac{1}{2\pi}\int dkk_{\rho}^{\dr}(k)\theta_{\rho}k\nonumber \\
	& =\frac{E}{L}+\frac{1}{2\pi}\int dkk_{\rho}^{\dr}(k)\theta_{\rho}k.
	\label{eq:Dne_ann_BF}
\end{align}

To calculate the second term, we need to calculate the pressure $p$
\begin{align}
	p & =\frac{T}{2\pi}\int_{-\infty}^{\infty}\ln\left(1+e^{-\epsilon(k)/T}\right)dk\nonumber \\
	& =\frac{T}{2\pi}\left\{ k\ln\left[1+e^{-\epsilon(k)/T}\right]\right\} _{-\infty}^{\infty}-\frac{T}{2\pi}\int_{-\infty}^{\infty}kd\ln\left(1+e^{-\epsilon(k)/T}\right)\nonumber \\
	& =\frac{T}{2\pi}\int_{-\infty}^{\infty}k\frac{1}{1+e^{\epsilon(k)/T}}\frac{1}{T}\frac{d\epsilon(k)}{dk}dk\nonumber \\
	& =\frac{1}{2\pi}\int_{-\infty}^{\infty}\frac{k}{1+e^{\epsilon(k)/T}}\frac{d\epsilon(k)}{dk}dk\nonumber \\
	& =\frac{1}{2\pi}\int_{-\infty}^{\infty}\theta (k)k\frac{d\epsilon(k)}{dk}dk. \label{eq:p_uni_BF}
\end{align}

Next, we identify $\frac{d\epsilon(k)}{dk}$
\begin{align}
	\frac{d\epsilon(k)}{dk} & =k-\frac{d}{dk}[T\int_{-\infty}^{\infty}\phi\left(\Lambda-k\right)\ln\left(1+e^{-\psi(\Lambda)/T}\right)d\Lambda]\nonumber \\
	& =k+\int_{-\infty}^{\infty}\phi\left(\Lambda-k\right)\theta_{\sigma}(\Lambda)\frac{d\psi(\Lambda)}{d\Lambda}d\Lambda. 
\end{align}

This relies on the identity $d_{k}[T\int\phi(\Lambda-k)\ln\left(1+e^{-\psi(\Lambda)/T}\right)\,d\Lambda]=-\int_{-\infty}^{\infty}\phi(\Lambda-k)\theta_{\sigma}(\Lambda)\frac{d\psi(\Lambda)}{d\Lambda}d\Lambda$,
which we now prove:
\begin{align}
	d_{k}[T\int\phi(\Lambda-k)\ln\left(1+e^{-\psi(\Lambda)/T}\right)\,d\Lambda] & =T\int\frac{d\phi(\Lambda-k)}{dk}\ln\left(1+e^{-\psi(\Lambda)/T}\right)\,d\Lambda\nonumber \\
	& =-T\int\frac{d\phi(\Lambda-k)}{d\Lambda}\ln\left(1+e^{-\psi(\Lambda)/T}\right)\,d\Lambda\nonumber \\
	& =-T\phi(\Lambda-k)\ln\left(1+e^{-\psi(\Lambda)/T}\right)\big|_{-\infty}^{\infty}+\frac{T}{2\pi}\int\phi(\Lambda-k)\frac{\ln\left(1+e^{-\psi(\Lambda)/T}\right)}{d\Lambda}d\Lambda\nonumber \\
	& =T\int\phi(\Lambda-k)\frac{\ln\left(1+e^{-\psi(\Lambda)/T}\right)}{d\Lambda}d\Lambda\nonumber \\
	& =-\int_{-\infty}^{\infty}\phi(\Lambda-k)\frac{1}{1+e^{\psi(\Lambda)/T}}\frac{d\psi(\Lambda)}{d\Lambda}d\Lambda\nonumber \\
	& =-\int_{-\infty}^{\infty}\phi(\Lambda-k)\theta{\sigma}(\Lambda)\frac{d\psi(\Lambda)}{d\Lambda}d\Lambda.
\end{align}
The derivative of the $\psi(\Lambda)$ is
\begin{align}
	\frac{d\psi(\Lambda)}{d\Lambda} & =\int_{-\infty}^{\infty}\phi\left(k-\Lambda\right)\theta_{\rho}(k)\frac{d\epsilon(k)}{dk}dk. 
\end{align}

We thus have a coupled system of equations for $\{\frac{d\epsilon}{dk},\frac{d\psi}{d\Lambda}\}$. We
compare this to the definition of the dressed momentum $\{k_{\rho}^{\dr},k_{\sigma}^{\dr}\}$,
which has the bare term $h_{k}=(k,0)$
\begin{align}
	\frac{d\epsilon(k)}{dk} & =k+\int_{-\infty}^{\infty}\phi\left(\Lambda-k\right)\theta_{\sigma}(\Lambda)\frac{d\psi(\Lambda)}{d\Lambda}d\Lambda,\\
	\frac{d\psi(\Lambda)}{d\Lambda} & =\int_{-\infty}^{\infty}\phi\left(k-\Lambda\right)\theta_{\rho}(k)\frac{d\epsilon(k)}{dk}dk,\\
	k_{\rho}^{\dr} & =k+\int_{-\infty}^{\infty}\phi\left(\Lambda-k\right)n(\Lambda)k_{\sigma}^{\dr}d\Lambda,\\
	k_{\sigma}^{\dr} & =\int_{-\infty}^{\infty}\phi\left(k-\Lambda\right)\theta_{\rho}(k)k_{\rho}^{\dr}dk.
\end{align}
By the uniqueness of the solution to this linear system, we make the identification
\begin{equation}
	\frac{d\epsilon(k)}{dk}=k_{\rho}^{\dr},\frac{d\psi(\Lambda)}{d\Lambda}=k_{\sigma}^{\dr}.
\end{equation}
Substituting this identity back into the pressure expression Eq.~(\ref{eq:p_uni_BF}) gives
\begin{align}
	p & =\frac{1}{2\pi}\int_{-\infty}^{\infty}\theta(k)k\frac{d\epsilon(k)}{dk}dk\nonumber \\
	& =\frac{1}{2\pi}\int_{-\infty}^{\infty}\theta(k)kk_{\rho}^{\dr}dk.
\end{align}

This exactly matches the second term in our $D_{ne}$ expression Eq.~(\ref{eq:Dne_ann_BF}). 
Thus, the universal relation for $D_{ne}$ is found to be
\begin{equation}
	D_{ne}=\frac{E}{L}+p. 
\end{equation}

\subsubsection*{For $D_{nk}$ and $D_{mm}$}

Following a similar procedure, we can also yields
\begin{align}
	D_{nk} & ={2P/L},\,\,\,\, D_{nm}=m. 
\end{align}

\subsubsection*{For $D_{n\epsilon}$}

Finally, we find the relation for $D_{n\epsilon}$, where $h_{n}=\{1,0\}$ and $h_{\epsilon}=\{k^{2}/2-(\mu+\mathcal{H}/2),\mathcal{H}\}$. We first establish a linear relation between the bare charges:
\begin{align}
	\epsilon_{\rho}^{\dr} & =\frac{k^{2}}{2}-\left(\mu+\frac{\mathcal{H}}{2}\right)+\int\phi\left(k-\Lambda\right)\theta_{\sigma}\epsilon_{\sigma}^{\dr}, \nonumber \\
	\epsilon_{\sigma}^{\dr} & =\mathcal{H}+\int\phi\left(\Lambda-k\right)\theta_{\rho}\epsilon_{\rho}^{\dr}, \nonumber \\
	e_{\rho}^{\dr} & =\frac{k^{2}}{2}+\int\phi\left(k-\Lambda\right)\theta_{\sigma}e_{\sigma}^{\dr}, \nonumber \\
	e_{\sigma}^{\dr} & =\int\phi\left(\Lambda-k\right)\theta_{\rho}e_{\rho}^{\dr}, \nonumber \\
	1_{\rho}^{\dr} & =1+\int\phi\left(k-\Lambda\right)\theta_{\sigma}1_{\sigma}^{\dr}, \nonumber \\
	1_{\sigma}^{\dr} & =\int\phi\left(\Lambda-k\right)\theta_{\rho}1_{\rho}^{\dr}, \nonumber \\
	m_{\rho}^{\dr} & =\int\phi\left(k-\Lambda\right)\theta_{\sigma}m_{\sigma}^{\dr}, \nonumber \\
	m_{\sigma}^{\dr} & =1+\int\phi\left(k-\Lambda\right)\theta_{\rho}m_{\rho}^{\dr}. 
\end{align}
By observing the bare terms, we construct the following identity:
\begin{align}
	e_{\rho}^{\dr}-\left(\mu+\frac{\mathcal{H}}{2}\right)1_{\rho}^{\dr}+\mathcal{H}m_{\rho}^{\dr} & =\frac{k^{2}}{2}+\int\phi\left(k-\Lambda\right)\theta_{\sigma}e_{\sigma}^{\dr}-\left(\mu+\frac{\mathcal{H}}{2}\right)\left[1+\int\phi\left(k-\Lambda\right)\theta_{\sigma}1_{\sigma}^{\dr}\right]+\mathcal{H}\int\phi\left(k-\Lambda\right)\theta_{\sigma}m_{\sigma}^{\dr}\nonumber \\
	& =\frac{k^{2}}{2}-\left(\mu+\frac{\mathcal{H}}{2}\right)+\int\phi\left(k-\Lambda\right)\theta_{\sigma}e_{\sigma}^{\dr}-\left(\mu+\frac{\mathcal{H}}{2}\right)\int\phi\left(k-\Lambda\right)\theta_{\sigma}1_{\sigma}^{\dr}+\mathcal{H}\int\phi\left(k-\Lambda\right)\theta_{\sigma}m_{\sigma}^{\dr}\nonumber \\
	& =\frac{k^{2}}{2}-\left(\mu+\frac{\mathcal{H}}{2}\right)+\int\phi\left(k-\Lambda\right)\theta_{\sigma}e_{\sigma}^{\dr}-\int\phi\left(k-\Lambda\right)\theta_{\sigma}\left(\mu+\frac{\mathcal{H}}{2}\right)1_{\sigma}^{\dr}+\int\phi\left(k-\Lambda\right)\theta_{\sigma}\mathcal{H}m_{\sigma}^{\dr}\nonumber \\
	& =\frac{k^{2}}{2}-\left(\mu+\frac{\mathcal{H}}{2}\right)+\int\phi\left(k-\Lambda\right)\theta_{\sigma}\left[e_{\sigma}^{\dr}-\left(\mu+\frac{\mathcal{H}}{2}\right)1_{\sigma}^{\dr}+\mathcal{H}m_{\sigma}^{\dr}\right],\\
	e_{\sigma}^{\dr}-\left(\mu+\frac{\mathcal{H}}{2}\right)1_{\sigma}^{\dr}+\mathcal{H}m_{\sigma}^{\dr} & =\int\phi\left(\Lambda-k\right)\theta_{\rho}e_{\rho}^{\dr}-\left(\mu+\frac{\mathcal{H}}{2}\right)\int\phi\left(\Lambda-k\right)\theta_{\rho}1_{\rho}^{\dr}+\mathcal{H}\left[1+\int\phi\left(k-\Lambda\right)n_{\rho}m_{\rho}^{\dr}\right]\nonumber \\
	& =\mathcal{H}+\int\phi\left(\Lambda-k\right)\theta_{\rho}e_{\rho}^{\dr}-\int\phi\left(\Lambda-k\right)\theta_{\rho}\left(\mu+\frac{\mathcal{H}}{2}\right)1_{\rho}^{\dr}+\int\phi\left(k-\Lambda\right)\theta_{\rho}\mathcal{H}m_{\rho}^{\dr}\nonumber \\
	& =\mathcal{H}+\int\phi\left(\Lambda-k\right)\theta_{\rho}\left[e_{\rho}^{\dr}-\left(\mu+\frac{\mathcal{H}}{2}\right)1_{\rho}^{\dr}+\mathcal{H}m_{\rho}^{\dr}\right].
\end{align}
We find that the combination $e^{\dr}-(\mu+\mathcal{H}/2)1^{\dr}+\mathcal{H}m^{\dr}$
obeys the same coupled integral equations as $\epsilon^{\dr}$. Due
to the uniqueness of the solution:
\begin{align}
	\epsilon^{\dr} & =e^{\dr}-\left(\mu+\frac{H}{2}\right)1^{\dr}+\mathcal{H}m^{dr}.
\end{align}

We substitute this identity into $D_{n\epsilon}$ and $D_{m\epsilon}$
yields:
\begin{align}
	D_{n\epsilon} & =D_{ne}-\left(\mu+\frac{\mathcal{H}}{2}\right)D_{nn}+\mathcal{H}D_{nm}\nonumber \\
	& =\frac{E}{L}+p-\left(\mu+\frac{\mathcal{H}}{2}\right)n+\mathcal{H}m\nonumber \\
	& {=\frac{E}{L}+p-\mu_{F}n_{f}-\mu_{B}m} \nonumber \\
	& =Ts\\
	D_{m\epsilon} & =D_{me}-\left(\mu+\frac{\mathcal{H}}{2}\right)D_{mn}+\mathcal{H}D_{mm}.
\end{align}
Here, $n_{f}$ is the fermion density.

\subsection{Ground state}

Based on Eq.~(\ref{eq:DW_dn_BF}), the Drude weights can be expressed as an integral involving the derivative of the occupation function
\begin{equation}D_{ij}=\sum_{a\in{\rho,\sigma}}\int_{-\infty}^{\infty} G_{a}(u) \frac{\partial \theta_{a}(u)}{\partial u} du,
\end{equation}
where the kernel is defined by $G_{a}(u)=v_{a}^{\eff}(u)h_{a_{i}}^{\dr}(u)h_{a_{j}}^{\dr}(u)$.
 Here, $\theta_a(u)$ follows the Fermi-Dirac distribution $\theta(\varepsilon) = 1/(1+e^{\varepsilon/T})$. 
 To evaluate this integral in the low-temperature limit, we employ the Sommerfeld expansion~\cite{Z.Physik1928SommerfeldZur}. 
 For a smooth function $H(\varepsilon)$, the integral involving the derivative of the Fermi function can be approximated as
\begin{equation}
	\int_{-\infty}^{\infty} H(\varepsilon) \left(-\frac{\partial \theta}{\partial \varepsilon}\right) d\varepsilon \approx H(0) + \frac{\pi^2 T^2}{6} H''(0) + \frac{7\pi^2 T^4}{360} H^{(4)}(0) + \cdots,\label{eq:Sommerfeld_Gen}
\end{equation}
where the expansion is performed around the Fermi surface ($\varepsilon=0$).
To apply this to the Drude weights, we utilize the parity symmetry of the integrands to restrict the integration domain to the positive axis (introducing a factor of 2) and change the integration variables from spectral parameters ($k, \Lambda$) to the dressed energies ($\epsilon, \psi$).
 This transformation yields
\begin{align}
	D_{ij} 
	& =\frac{1}{\pi}\int_{0}^{\infty} dk \left(-\frac{\partial \theta_{\rho}}{\partial k}\right) \left[v_{\rho}^{\eff}(k)h_{\rho_{i}}^{\dr}(k)h_{\rho_{j}}^{\dr}(k)\right] + \frac{1}{\pi}\int_{0}^{\infty} d\Lambda \left(-\frac{\partial \theta_{\sigma}}{\partial \Lambda}\right) \left[v_{\sigma}^{\eff}(\Lambda)h_{\sigma_{i}}^{\dr}(\Lambda)h_{\sigma_{j}}^{\dr}(\Lambda)\right],\nonumber \\
	& =\frac{1}{\pi}\int_{\epsilon_{0}}^{\infty} d\epsilon \left(-\frac{\partial f}{\partial \epsilon}\right) G_{\rho}(\epsilon) + \frac{1}{\pi}\int_{\psi_{0}}^{\infty} d\psi \left(-\frac{\partial \theta}{\partial \psi}\right) G_{\sigma}(\psi),
\end{align}
where $G_{a}(\varepsilon) \equiv \left[v_{a}^{\eff}h_{a_{i}}^{\dr}h_{a_{j}}^{\dr}\right]$ expressed as a function of energy. 
In the low-temperature regime, the lower integration limits $\epsilon_{0}$ and $\psi_{0}$ can be effectively extended to $-\infty$. Substituting $G_{\rho}(\epsilon)$ and $G_{\sigma}(\psi)$ into the expansion formula Eq.~\eqref{eq:Sommerfeld_Gen}, we obtain the low-temperature expression for the Drude weights:
\begin{equation}
	\begin{split}
		D_{ij} \approx 
		&\frac{1}{\pi}\left[ v_{\rho}^{\eff}h_{\rho_{i}}^{\dr}h_{\rho_{j}}^{\dr}+\frac{\pi^{2}T^{2}}{6}\frac{\partial^{2}}{\partial\epsilon^{2}}\left( v_{\rho}^{\eff}h_{\rho_{i}}^{\dr}h_{\rho_{j}}^{\dr}\right) +\frac{7\pi^{2}T^{4}}{360}\frac{\partial^{4}}{\partial\epsilon^{4}}\left( v_{\rho}^{\eff}h_{\rho_{i}}^{\dr}h_{\rho_{j}}^{\dr}\right) \right]_{\epsilon=0} \\
		& +\frac{1}{\pi}\left[ v_{\sigma}^{\eff}h_{\sigma_{i}}^{\dr}h_{\sigma_{j}}^{\dr}+\frac{\pi^{2}T^{2}}{6}\frac{\partial^{2}}{\partial\psi^{2}}\left( v_{\sigma}^{\eff}h_{\sigma_{i}}^{\dr}h_{\sigma_{j}}^{\dr}\right) +\frac{7\pi^{2}T^{4}}{360}\frac{\partial^{4}}{\partial\psi^{4}}\left( v_{\sigma}^{\eff}h_{\sigma_{i}}^{\dr}h_{\sigma_{j}}^{\dr}\right) \right]_{\psi=0}.
		\end{split}
		\label{eq:Dij_lT_BF_M}
\end{equation}
Consequently, in the zero-temperature limit ($T \to 0$), the higher-order corrections vanish. The expression for the Drude weights simplifies to the values evaluated strictly at the Fermi surfaces
\begin{equation}
	D_{ij} = \frac{1}{\pi} \left[ v_{\rho}^{\eff} h_{\rho_{i}}^{\dr} h_{\rho_{j}}^{\dr} \right]_{\epsilon=0} + \frac{1}{\pi} \left[ v_{\sigma}^{\eff} h_{\sigma_{i}}^{\dr} h_{\sigma_{j}}^{\dr} \right]_{\psi=0}.
\end{equation}
This result confirms that at zero temperature, transport properties are determined solely by the quasiparticle excitations at the Fermi edges.

\subsection{Zero-temperature strong interaction}

At $T=0$, the system is in the ground state. The Sommerfeld expansion
Eq. \eqref{eq:Dij_lT_BF_M} reduces to its leading term, evaluated
at the Fermi points ($k_{F},\Lambda_{F}$)
\begin{align}
	D_{ij} & =\frac{1}{\pi}\left[v_{\rho}^{\eff}(k)h_{\rho_{i}}^{\dr}(k)h_{\rho_{j}}^{\dr}(k)\right]_{k=k_{F}}+\frac{1}{\pi}\left[v_{\sigma}^{\eff}(\Lambda)h_{\sigma_{i}}^{\dr}(\Lambda)h_{\sigma_{j}}^{\dr}(\Lambda)\right]_{\Lambda=\Lambda_{F}}.
\end{align}
To solve this, we must calculate the dressed charges at the Fermi
points in the strong-interaction limit.

\subsubsection*{For $1^{\text{dr}}$}

At $T=0$, the occupation numbers are step functions:
\begin{align}
	\theta_{\rho}\left(k\right) & =\begin{cases}
		1 & -k_{F}\leq k\leq k_{F},\\
		0 & else,
	\end{cases}\\
	\theta_{\sigma}\left(\Lambda\right) & =\begin{cases}
		1 & -\Lambda_{F}\leq\Lambda\leq\Lambda_{F}, \\
		0 & else. 
	\end{cases}
\end{align}
The $T=0$ integral equations for $1^{\dr}$ become
\begin{align}
	1_{\rho}^{\dr}\left(k\right) & =1+\frac{c}{\pi}\int_{-\Lambda_{F}}^{\Lambda_{F}}\frac{1}{c^{2}+\left(k-\Lambda\right)^{2}}1_{\sigma}^{\dr}d\Lambda\label{eq:Idrrho_zTiC_BF},\\
	1_{\sigma}^{\dr}\left(\Lambda\right) & =\frac{c}{\pi}\int_{-k_{F}}^{k_{F}}\frac{1}{c^{2}+\left(\Lambda-q\right)^{2}}1_{\rho}^{\dr}dq.
	\label{eq:Idrsigma_zTiC_BF}
\end{align}
In the strong interaction limit ($c\to\infty$), the kernel is
\begin{align}
	\phi & =\frac{1}{2\pi}\frac{2c}{c^{2}+(k-q)^{2}}\approx\frac{1}{\pi c}.
\end{align}
Substituting this approximation into Eq. (\ref{eq:Idrrho_zTiC_BF})
and (\ref{eq:Idrsigma_zTiC_BF}) yields
\begin{align}
	1_{\rho}^{\dr} & \approx 1+\frac{1}{\pi c}\int_{-\Lambda_{F}}^{\Lambda_{F}}1_{\sigma}^{\dr}d\Lambda \nonumber \\
	& =1+\frac{1}{\pi c}P_{1\sigma}, \label{eq:Idrrho_zTiC_an_p_BF}
\end{align}
where $P_{1\sigma}=\int_{-\Lambda_{F}}^{\Lambda_{F}}1_{\sigma}^{\dr}d\Lambda$ We now solve this coupled algebraic system. Substituting $1_{\rho}^{\dr}$ (Eq.~\ref{eq:Idrrho_zTiC_an_p_BF}) into $1_{\sigma}^{\dr}$ (Eq.~\ref{eq:Idrsigma_zTiC_ann_BF})
\begin{align}
	1_{\sigma}^{\dr} & =\frac{c}{\pi}\int_{-k_{F}}^{k_{F}}\frac{1}{c^{2}+\left(\Lambda-q\right)^{2}}1_{\rho}^{\dr}dq \approx \frac{1}{\pi c}\int_{-k_{F}}^{k_{F}}1_{\rho}^{\dr}dq \nonumber\\
	& =\frac{1}{\pi c}\left(1+\frac{1}{\pi c}P_{1\sigma}\right)2k_{F}.
	\label{eq:Idrsigma_zTiC_ann_BF}
\end{align}
Now, we substitute the result Eq. (\ref{eq:Idrsigma_zTiC_ann_BF})
back into the definition of $P_{1\sigma}$.
It follows 
\begin{align}
	P_{1\sigma} & =\int_{-\Lambda_{F}}^{\Lambda_{F}}1_{\sigma}^{\dr}d\Lambda, \nonumber \\
	& =\frac{1}{\pi c}\left(1+\frac{1}{\pi c}P_{1\sigma}\right)2k_{F}2\Lambda_{F}.
\end{align}
Using the known expressions $k_{F}\approx \pi n\left(1-\frac{2}{\gamma}\alpha\right)$ 
and $\Lambda_{F} \approx \frac{\pi m\gamma}{2}$ (where $\gamma=\frac{n}{c},\alpha=\frac{m}{n}$) at $T=0$, 
we obtain $P_{1\sigma}$
\begin{equation}
	P_{1\sigma}=2\pi m.
\end{equation}
Substituting this result back into the expressions for $1^{\dr}$  (Eq.~\ref{eq:Idrrho_zTiC_BF}~\ref{eq:Idrsigma_zTiC_ann_BF})
\begin{align}
	1_{\rho}^{\dr}\left(k\right) & =1+\frac{2m}{c}, \\
	1_{\sigma}^{\dr}\left(\Lambda\right) & =\frac{2n}{c}.
\end{align}
These dressed charges are independent of momentum. Thus, the 
values at the Fermi points are given by 
\begin{align}
	1_{\rho}^{\dr}\left(k_{F}\right) & =1+\frac{2m}{c},\\
	1_{\sigma}^{\dr}\left(\Lambda_{F}\right) & =\frac{2n}{c}.
\end{align}
Following the same method, at $T=0$, the  strong-interaction expressions
for $m^{\dr},k^{\dr},e^{\dr}$ are given by 


\begin{align}
	m_{\rho}^{\dr}\left(k_{F}\right) & =\frac{m}{n}\left(1+\frac{2m}{c}\right),\\
	m_{\sigma}^{\dr}\left(\Lambda_{F}\right) & =1+\frac{2m}{c}
\end{align}
for  $m^{\text{dr}}$ and 
%
\begin{align}
	k_{\rho}^{\dr}\left(k_{F}\right) & =\pi n\left(1-\frac{2m}{c}\right),\\
	k_{\sigma}^{\dr}\left(\Lambda_{F}\right) & =0
\end{align}
for $k^{\text{dr}}$.  While for  $e^{\text{dr}}$, we have 
\begin{align}
	e_{\rho}^{\dr}\left(k_{F}\right) & =\frac{\pi^{2}n^{2}}{2}\left(1-\frac{10m}{3c}\right),\\
	e_{\sigma}^{\dr}\left(\Lambda_{F}\right) & =\frac{\pi^{3}n^{3}}{3\pi c}\left(1-\frac{2m}{c}\right)^{3}.
\end{align}

\subsubsection*{Calculation of  DWs}

We now use the DW expression at  $T=0$ 
\begin{align}
	D_{ij} & =\frac{1}{\pi}\left[v_{\rho}^{\eff}(k)h_{\rho_{i}}^{\dr}(k)h_{\rho_{j}}^{\dr}(k)\right]_{k=k_{F}}+\frac{1}{\pi}\left[v_{\sigma}^{\eff}(\Lambda)h_{\sigma_{i}}^{\dr}(\Lambda)h_{\sigma_{j}}^{\dr}(\Lambda)\right]_{\Lambda=\Lambda_{F}}.
\end{align}
This gives the specific matrix elements:
\begin{align}
	D_{nn} & =\frac{1}{\pi}\left[k_{\rho}^{\dr}(k)1_{\rho}^{\dr}(k)\right]_{k=k_{F}}+\frac{1}{\pi}\left[k_{\sigma}^{\dr}(\Lambda)1_{\sigma}^{\dr}(\Lambda)\right]_{\Lambda=\Lambda_{F}},\\
	D_{nm} & =\frac{1}{\pi}\left[k_{\rho}^{\dr}(k)m_{\rho}^{\dr}(k)\right]_{k=k_{F}}+\frac{1}{\pi}\left[k_{\sigma}^{\dr}(\Lambda)m_{\sigma}^{\dr}(\Lambda)\right]_{\Lambda=\Lambda_{F}},\\
	D_{mm} & =\frac{1}{\pi}\left[\frac{k_{\rho}^{\dr}(k)}{1_{\rho}^{\dr}(k)}m_{\rho}^{\dr}(k)m_{\rho}^{\dr}(k)\right]_{k=k_{F}}+\frac{1}{\pi}\left[\frac{k_{\sigma}^{\dr}(\Lambda)}{1_{\sigma}^{\dr}(\Lambda)}m_{\sigma}^{\dr}(\Lambda)m_{\sigma}^{\dr}(\Lambda)\right]_{\Lambda=\Lambda_{F}},\\
	D_{ne} & =\frac{1}{\pi}\left[k_{\rho}^{\dr}(k)e_{\rho}^{\dr}(k)\right]_{k=k_{F}}+\frac{1}{\pi}\left[k_{\sigma}^{\dr}(\Lambda)e_{\sigma}^{\dr}(\Lambda)\right]_{\Lambda=\Lambda_{F}},\\
	D_{me} & =\frac{1}{\pi}\left[\frac{k_{\rho}^{\dr}(k)}{1_{\rho}^{\dr}(k)}m_{\rho}^{\dr}(k)e_{\rho}^{\dr}(k)\right]_{k=k_{F}}+\frac{1}{\pi}\left[\frac{k_{\sigma}^{\dr}(\Lambda)}{1_{\sigma}^{\dr}(\Lambda)}m_{\sigma}^{\dr}(\Lambda)e_{\sigma}^{\dr}(\Lambda)\right]_{\Lambda=\Lambda_{F}}.
\end{align}
Substituting the $1/c$ expressions for all dressed charges at
the Fermi points into these formulas, we obtain the final results,
approximated to order $1/\gamma$:
\begin{align}
	D_{nn} & =n,\\
	D_{nm} & =m=n\alpha,\\
	D_{mm} & \approx\frac{m^{2}}{n}=n\alpha^{2},\\
	D_{ne} & \approx\frac{\pi^{2}n^{3}}{2}\left(1-\frac{16m}{3c}\right)=\frac{\pi^{2}n^{3}}{2}\left(1-\frac{16}{3\gamma}\alpha\right),\\
	D_{me} & \approx\frac{\pi^{2}mn^{2}}{2}\left(1-\frac{16m}{3c}\right)=\frac{\pi^{2}n^{3}}{2}\alpha\left(1-\frac{16}{3\gamma}\alpha\right).
\end{align}

To complement the results in the main text, Fig.~\ref{fig:BF_lowtep_strlim_app} presents the remaining Drude weight components ($D_{nm}$, $D_{mm}$, $D_{me}$) evaluated at the zero-temperature and strong-coupling limit. While $D_{nn}$ and $D_{ne}$ are detailed in Fig.~6, the cross and bosonic components shown here identically vanish in this limit, consistent with our analytical expansions.

\begin{figure}[H]
	\centering
	\includegraphics[width=1\linewidth]{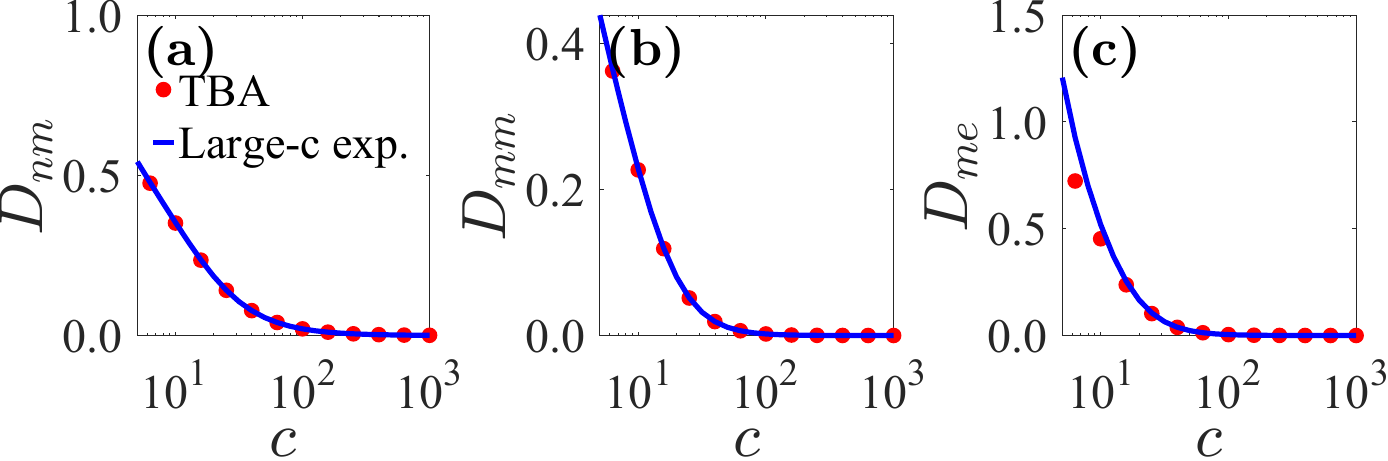}
	\caption{Supplemental plots for the Drude weights in the Bose-Fermi mixture at $T=0.001$, $\mu=1$, and $\mathcal{H}=-1$. Circles represent the exact TBA numerical solutions for the components: (a) $D_{nm}$, (b) $D_{mm}$, and (c) $D_{me}$. The blue lines denote the analytical large-$c$ expansions [Eqs.~\eqref{eq:Dnm_BF_zTiC_M}-\eqref{eq:Dme_BF_zTiC_M}].}
	\label{fig:BF_lowtep_strlim_app}
\end{figure}

\subsection{Finite-Temperature Weak Interaction}

Similar to the situation in the Lieb-Liniger model, the integral kernel
in the weak interaction approximation ($c\to0$) becomes a delta function:
\begin{align}
	\lim_{c\rightarrow0}\phi(k-\Lambda)=\lim_{c\rightarrow0}\frac{c}{\pi}\frac{1}{c^{2}+(k-\Lambda)^{2}} & \approx\delta\left(k-\Lambda\right).
\end{align}
This has a significant consequence for the integral operators
\begin{align}
	\int dkg\left(k\right)\int d\Lambda\phi\left(k-\Lambda\right)h\left(\Lambda\right) & =\int d\Lambda h\left(\Lambda\right)\int dk\phi\left(k-\Lambda\right)g\left(k\right)\nonumber, \\
	\int dkg\left(k\right)h\left(k\right) & =\int d\Lambda h\left(\Lambda\right)g\left(\Lambda\right).
\end{align}
This implies that the interaction term only couples states at the
same rapidity, effectively identifying the integration variables ($k=\Lambda$).
In this $c\to0$ limit, the dressed energy equations simplify to
\begin{align}
	\epsilon\left(k\right) & =\frac{k^{2}}{2}-\mu_{F}-T\ln\left(1+e^{-\psi(k)/T}\right),\\
	\psi\left(\Lambda\right) & =\mu_{F}-\mu_{B}-T\ln\left(1+e^{-\epsilon(\Lambda)/T}\right).
\end{align}
Substituting these into the occupation number definitions $\theta_{\rho}=\frac{1}{1+e^{\epsilon\left(k\right)/T}},\theta_{\sigma}=\frac{1}{1+e^{\psi\left(\Lambda\right)/T}}$ 
we have 
\begin{align}
	\theta_{\rho} & =\frac{\rho}{\rho+\rho_{h}}=\frac{\frac{1}{e^{\left(\mu_{F}-\mu_{B}\right)/T}}+1}{e^{\left(k^{2}/2-\mu_{F}\right)/T}+1}, \\
	\theta_{\sigma} & =\frac{\sigma}{\sigma+\sigma_{h}}=\frac{\frac{1}{e^{\left(k^{2}/2-\mu_{F}\right)/T}}+1}{e^{\left(\mu_{F}-\mu_{B}\right)/T}+1}.
	\label{eq:n_fTzC}
\end{align}
The momentum density distribution function also simplify
\begin{align}
	\rho\left(k\right)+\rho_{h}\left(k\right) & =\frac{1}{2\pi}+\sigma\left(k\right), \\
	\sigma\left(\Lambda\right)+\sigma_{h}\left(\Lambda\right) & =\rho\left(\Lambda\right).
	\label{eq:rho_n}
\end{align}
Substituting the occupation numbers Eq.~(\ref{eq:n_fTzC}) into these
momentum density distribution function Eq.~(\ref{eq:rho_n}) yields
\begin{align}
	\rho\left(k\right) & =\frac{1}{2\pi}\left[\frac{1}{e^{\left(k^{2}/2-\mu_{B}\right)/T}-1}+\frac{1}{e^{\left(k^{2}/2-\mu_{F}\right)/T}+1}\right],
	\label{eq:rho_fTzC_BF}\\
	\sigma\left(k\right) & =\frac{1}{2\pi}\left[\frac{1}{e^{\left(k^{2}/2-\mu_{B}\right)/T}-1}\right].
	\label{eq:sigma_fTzC_BF}
\end{align}
As seen from this expression, the  total density $\rho(k)$ decouples into a linear superposition of a free Boson component (with chemical potential $\mu_{B}$) and a free Fermion component (with chemical potential $\mu_{F}$). 

\subsubsection*{Dressed Charges in the $c\to 0$ Limit}

To calculate the DWs, we apply the $c\to0$ approximation ($\phi\to\delta$)
to the dressing equations
\begin{align}
	h_{\rho}^{\dr}(k) & =h_{\rho}(k)+\theta_{\sigma}h_{\sigma}^{\dr}\left(k\right), \label{h-rho}\\
	h_{\sigma}^{\dr}(\Lambda) & =h_{\sigma}(\Lambda)+\theta_{\rho}h_{\rho}^{\dr}(\Lambda).
\end{align}
Substituting the bare charges $h_{n},h_{k},h_{e},h_{m}$ and solving
gives:
\begin{align}
	1_{\rho}^{\dr}(k) & =\frac{1}{1-\theta_{\sigma}\theta_{\rho}},\\
	1_{\sigma}^{\dr}(\Lambda) & =\frac{\theta_{\rho}}{1-\theta_{\sigma}\theta_{\rho}}.\label{Equations-Dress-1}
\end{align}
\begin{align}
	k_{\rho}^{\dr}(k) & =\frac{k}{1-\theta_{\sigma}\theta_{\rho}},\\
	k_{\sigma}^{\dr}(\Lambda) & =\frac{\theta_{\rho}\Lambda}{1-\theta_{\sigma}\theta_{\rho}}.
\end{align}
\begin{align}
	e_{\rho}^{\dr}(k) & =\frac{k^{2}/2}{1-\theta_{\sigma}\theta_{\rho}},\\
	e_{\sigma}^{\dr}(\Lambda) & =\frac{\theta_{\rho}\Lambda^{2}/2}{1-\theta_{\sigma}\theta_{\rho}}.
\end{align}
\begin{align}
	m_{\rho}^{\dr}(k) & =\frac{\theta_{\sigma}}{1-\theta_{\sigma}\theta_{\rho}},\\
	m_{\sigma}^{\dr}(\Lambda) & =\frac{1}{1-\theta_{\sigma}\theta_{\rho}}.\label{Equations-Dress-8}
\end{align}
Initating the $m^{\dr}$ expression above, we also define auxiliary
charges for $i=k,e$ in the $m$-sector $h_{im}$($h_{km}=(0,\Lambda),h_{em}=(0,\Lambda^{2}/2)$),
The corresponding dressed charges are
\begin{align}
	km_{\rho}^{\dr}(k) & =\frac{\theta_{\sigma}k}{1-\theta_{\sigma}\theta_{\rho}},\\
	km_{\sigma}^{\dr}(\Lambda) & =\frac{\Lambda}{1-\theta_{\sigma}\theta_{\rho}},\\
	em_{\rho}^{\dr}(k) & =\frac{\theta_{\sigma}k^{2}/2}{1-\theta_{\sigma}\theta_{\rho}},\\
	em_{\sigma}^{\dr}(\Lambda) & =\frac{\Lambda^{2}/2}{1-\theta_{\sigma}\theta_{\rho}}.
\end{align}

\subsubsection*{Calculation of $D_{nn}$ and $D_{ne}$}

We calculate the $D_{ij}$ matrix for the Total Component charges ($i,j\in\{n,k,e\}$).
\begin{align}
	D_{ij} & =-\frac{1}{2\pi}\int d_{k}\theta_{\rho}(k)v_{\rho}^{\eff}(k)h_{\rho_{i}}^{\dr}(k)h_{\rho_{j}}^{\dr}(k)\,dk-\frac{1}{2\pi}\int d_{\Lambda}\theta_{\sigma}(\Lambda)v_{\sigma}^{\eff}(\Lambda)h_{\sigma_{i}}^{\dr}(\Lambda)h_{\sigma_{j}}^{\dr}(\Lambda)\,d\Lambda\nonumber \\
	& =-\frac{1}{2\pi}\int d_{k}\theta_{\rho}(k)\frac{e'{}_{\rho}^{\dr}}{p'{}_{\rho}^{\dr}}h_{\rho_{i}}^{\dr}(k)h_{\rho_{j}}^{\dr}(k)\,dk-\frac{1}{2\pi}\int d_{\Lambda}\theta_{\sigma}(\Lambda)\frac{e'{}_{\sigma}^{\dr}}{p'{}_{\sigma}^{\dr}}h_{\sigma_{i}}^{\dr}(\Lambda)h_{\sigma_{j}}^{\dr}(\Lambda)\,d\Lambda\nonumber \\
	& =-\frac{1}{2\pi}\int d_{k}\theta_{\rho}(k)\frac{k_{\rho}^{\dr}}{1_{\rho}^{\dr}}h_{\rho_{i}}^{\dr}(k)h_{\rho_{j}}^{\dr}(k)\,dk-\frac{1}{2\pi}\int d_{\Lambda}\theta_{\sigma}(\Lambda)\frac{k_{\sigma}^{\dr}}{1_{\sigma}^{\dr}}h_{\sigma_{i}}^{\dr}(\Lambda)h_{\sigma_{j}}^{\dr}(\Lambda)\,d\Lambda  \nonumber \\
	&=D_{ij}\left(1\right) +D_{ij}\left(2\right). \label{eq:Dij2_fTzC_BF}
\end{align}
Similar to the method in the Lieb-Liniger model, see Eq.~(\ref{relation-rho-theta}), we use the relation
for the momentum density distribution function derivative
\begin{align}
	\frac{d\theta_{\rho}}{dk}\frac{\rho}{\theta_{\rho}} & =\frac{1}{2\pi}1^{\dr}\frac{d\theta_{\rho}}{dk}=\frac{d\rho}{dk}-\theta_{\rho}\frac{d\sigma}{d\Lambda},\\
	\frac{d\theta_{\sigma}}{d\Lambda}\frac{\sigma}{\theta_{\sigma}} & =\frac{d\sigma}{d\Lambda}-\theta_{\sigma}\tilde{T}_{\sigma}\frac{d\rho}{dk}.
\end{align}
Substituting this into Eq.~ (\ref{eq:Dij2_fTzC_BF})
\begin{align}
	D_{ij}\left(1\right) & =-\frac{1}{2\pi}\int d_{k}\theta_{\rho}(k)k_{\rho}^{\dr}1_{\rho}^{\dr}\frac{h_{\rho_{i}}^{\dr}(k)}{1_{\rho}^{\dr}}\frac{h_{\rho_{j}}^{\dr}(k)}{1_{\rho}^{\dr}}\,dk\nonumber \\
	& =-\frac{1}{i!j!}\int\left(d_{k}\rho-\theta_{\rho}\tilde{T}_{\rho}d_{\Lambda}\sigma\right)k_{\rho}^{\dr}k^{i+j}\,dk\nonumber \\
	& =-\frac{1}{i!j!}\int\left(d_{k}\rho-\theta_{\rho}\tilde{T}_{\rho}d_{\Lambda}\sigma\right)\left(k^{i+j+1}\right){}_{\rho}^{\dr}\,dk\nonumber \\
	& =-\frac{1}{i!j!}\int d_{k}\rho\left(k^{i+j+1}\right){}_{\rho}^{\dr}\,dk+\frac{1}{i!j!}\int \theta_{\rho}\tilde{T}_{\rho}d_{\Lambda}\sigma\left(k^{i+j+1}\right){}_{\rho}^{\dr}\,dk\nonumber \\
	& =-\frac{1}{i!j!}\int d_{k}\rho k^{i+j+1}\,dk-\frac{1}{i!j!}\int dkd_{k}\rho \theta_{\sigma}\left(k^{i+j+1}\right){}_{\sigma}^{\dr}+\frac{1}{i!j!}\int dk\left(k^{i+j+1}\right){}_{\rho}^{\dr}\theta_{\rho}d_{\Lambda}\sigma\nonumber \\
	& =\frac{\left(i+j+1\right)}{i!j!}\int\rho k^{i+j}\,dk-\frac{1}{i!j!}\int dkd_{k}\rho\left(k^{i+j+1}\right){}_{\rho}^{\dr}+\frac{1}{i!j!}\int dk\left(k^{i+j+1}\right){}_{\sigma}^{\dr}d_{\Lambda}\sigma \nonumber \\
	& =\frac{\left(i+j+1\right)}{i!j!}\int\rho k^{i+j}\,dk-\frac{1}{i!j!}\int dkd_{k}\rho\left(k^{i+j+1}\right){}_{\rho}^{\dr}+\frac{1}{i!j!}\int dk\left(k^{i+j+1}\right){}_{\sigma}^{\dr}\left(\frac{d\theta_{\sigma}}{d\Lambda}\frac{\sigma}{\theta_{\sigma}}+\theta_{\sigma}\tilde{T}_{\sigma}\frac{d\rho}{dk}\right) \nonumber \\
	& =\frac{\left(i+j+1\right)}{i!j!}\int\rho k^{i+j}\,dk-\frac{1}{i!j!}\int dkd_{k}\rho\left(k^{i+j+1}\right){}_{\rho}^{\dr}  \nonumber \\
	&\quad +\frac{1}{i!j!}\frac{1}{2\pi}\int dk\frac{d\theta_{\sigma}}{d\Lambda}1_{\sigma}^{\dr}\left(k^{i+j+1}\right)_{\sigma}^{\dr}+\frac{1}{i!j!}\int dk\left(k^{i+j+1}\right){}_{\rho}^{\dr}\frac{d\rho}{dk} \nonumber \\
	& =\frac{\left(i+j+1\right)}{i!j!}\int\rho k^{i+j}\,dk+\frac{1}{2\pi}\frac{1}{i!j!}\int dk\frac{d\theta_{\sigma}}{d\Lambda}1_{\sigma}^{\dr}\left(k^{i+j+1}\right){}_{\sigma}^{\dr}.
	\label{eq:Dij1_fTzC_n_BF}
\end{align}
In the above equation, the  fifth equality  was used the Eq.~(\ref{h-rho}).
The second term $D_{ij}(2)$ Eq.~(\ref{eq:Dij2_fTzC_BF}) can be simplified 
\begin{align}
	D_{ij}\left(2\right) & =-\frac{1}{2\pi}\int d_{\Lambda}\theta_{\sigma}(\Lambda)k_{\sigma}^{\dr}1_{\sigma}^{\dr}\frac{h_{\sigma_{i}}^{\dr}(\Lambda)}{1_{\sigma}^{\dr}}\frac{h_{\sigma_{j}}^{\dr}(\Lambda)}{1_{\sigma}^{\dr}}\,d\Lambda\nonumber \\
	& =-\frac{1}{2\pi}\frac{1}{i!j!}\int\frac{d\theta_{\sigma}}{d\Lambda}1_{\sigma}^{\dr}\left(k^{i+j+1}\right)_{\sigma}^{\dr}\,d\Lambda.
\end{align}
From the sum of  $D_{ij}$, we have 
\begin{align}
	D_{ij} & =D_{ij}\left(1\right)+D_{ij}\left(2\right)\nonumber \\
	& =-\frac{1}{2\pi}\int d_{k}\theta_{\rho}(k)k_{\rho}^{\dr}(k)1_{\rho}^{\dr}(k)\,dk-\frac{1}{2\pi}\int d_{\Lambda}\theta_{\sigma}(\Lambda)k_{\sigma}^{\dr}(\Lambda)1_{\sigma}^{\dr}(\Lambda)\,d\Lambda\nonumber \\
	& =\frac{\left(i+j+1\right)}{i!j!}\int\rho k^{i+j}\,dk+\frac{1}{2\pi}\frac{1}{i!j!}\int d\Lambda d_{\Lambda}\theta_{\sigma}1_{\sigma}^{\dr}\left(k^{i+j+1}\right){}_{\sigma}^{\dr}-\frac{1}{2\pi}\frac{1}{i!j!}\int d_{\Lambda}\theta_{\sigma}(\Lambda)\left(k^{i+j+1}\right)_{\sigma}^{\dr}1_{\sigma}^{\dr}\,d\Lambda\nonumber \\
	& =\frac{\left(i+j+1\right)}{i!j!}\int\rho k^{i+j}\,dk.
	\label{eq:Dij_fTzC_rho_BF}
\end{align}
Substituting the specific Total Component charges $h_{n}$ and $h_{e}$ ($i=0,j=0/2$) into Eq.~(\ref{eq:Dij_fTzC_rho_BF}) yields
\begin{align}
	D_{nn} & =\int\rho\,dk=n,\\
	D_{ne} & =\frac{3}{2}\int\rho k^{2}\,dk=3\frac{E}{L}.
\end{align}
By substituting the expression of  $\rho$ Eq. (\ref{eq:rho_fTzC_BF}),  the simplifications of $D_{nn}$ and $D_{ne} $  reads
\begin{align}
	D_{nn} & =\int\rho\,dk\nonumber \\
	& =\frac{1}{2\pi}\int\left(\frac{1}{e^{\left(k^{2}/2-\mu_{B}\right)/T}-1}+\frac{1}{e^{\left(k^{2}/2-\mu_{F}\right)/T}+1}\right)\,dk\nonumber \\
	& =\frac{1}{2\pi}\int\frac{1}{e^{\left(k^{2}/2-\mu_{B}\right)/T}-1}\,dk+\frac{1}{2\pi}\int\frac{1}{e^{\left(k^{2}/2-\mu_{F}\right)/T}+1}\,dk\nonumber \\
	& =\frac{1}{\sqrt{2\pi}}T^{1/2}Li_{\frac{1}{2}}\left(e^{\mu_{B}/T}\right)+\frac{1}{\sqrt{2\pi}}T^{1/2}Li_{\frac{1}{2}}\left(e^{\mu_{F}/T}\right),\\
	D_{ne} & =\frac{3}{2}\int\rho k^{2}\,dk\nonumber \\
	& =\frac{3}{2}\int\left(\frac{1}{e^{\left(k^{2}/2-\mu_{B}\right)/T}-1}+\frac{1}{e^{\left(k^{2}/2-\mu_{F}\right)/T}+1}\right)k^{2}\,dk\nonumber \\
	& =\frac{3}{2}\frac{1}{2\pi}\int\frac{k^{2}}{e^{\left(k^{2}/2-\mu_{B}\right)/T}-1}dk+\frac{3}{2}\frac{1}{2\pi}\int\frac{k^{2}}{e^{\left(k^{2}/2-\mu_{F}\right)/T}+1}dk\nonumber \\
	& =\frac{3}{2\sqrt{2\pi}}T^{3/2}Li_{\frac{3}{2}}\left(e^{\mu_{B}/T}\right)+\frac{3}{2\sqrt{2\pi}}T^{3/2}Li_{\frac{3}{2}}\left(e^{\mu_{F}/T}\right).
\end{align}
Thus, the total  Drude weights $D_{nn}$ and $D_{ne} $ show a linear superposition of the free Bosons and free Fermions in the limit of $c\to 0$.

\subsubsection*{Calculation of the $D_{mn}$ and $D_{me}$}

For the cross Drude weights $D_{ji}$ ($j=m$=Boson, $i\in\{n,e\}$),
we write the expression as
\begin{align*}
	D_{mi} & =-\frac{1}{2\pi}\int d_{k}\theta_{\rho}(k)v_{\rho}^{\eff}(k)h_{\rho_i}^{\dr}m_{\rho}^{\dr}(k)\,dk-\frac{1}{2\pi}\int d_{\Lambda}\theta_{\sigma}(\Lambda)v_{\sigma}^{\eff}(\Lambda)h_{\sigma_i}^{\dr}(\Lambda)m_{\sigma}^{\dr}(\Lambda)\,d\Lambda\\
	& =-\frac{1}{2\pi}\int d_{k}\theta_{\rho}(k)\frac{e'{}_{\rho}^{\dr}} {k'{}_{\rho}^{\dr}}  h_{\rho_i}^{\dr}(k)m_{\rho}^{\dr}(k)\,dk-\frac{1}{2\pi}\int d_{\Lambda}\theta_{\sigma}(\Lambda)\frac{e'{}_{\sigma}^{\dr}}{k'{}_{\sigma}^{\dr}}h_{\sigma_i}^{\dr}(\Lambda)m_{\sigma}^{\dr}(\Lambda)\,d\Lambda\\
	& =-\frac{1}{2\pi}\int d_{k}\theta_{\rho}(k)\frac{k_{\rho}^{\dr}}{1_{\rho}^{\dr}}h_{\rho_i}^{\dr}(k)m_{\rho}^{\dr}(k)\,dk-\frac{1}{2\pi}\int d_{\Lambda}\theta_{\sigma}(\Lambda)\frac{k_{\sigma}^{\dr}}{1_{\sigma}^{\dr}}h_{\sigma_i}^{\dr}(\Lambda)m_{\sigma}^{\dr}(\Lambda)\,d\Lambda.
\end{align*}
We denote these two terms by 
\begin{align}
	D_{mi}(1) & =-\frac{1}{2\pi}\int d_{k}\theta_{\rho}(k)\frac{k_{\rho}^{\dr}}{1_{\rho}^{\dr}}h_{\rho_i}^{\dr}(k)m_{\rho}^{\dr}(k)\,dk,
	\label{eq:Dmi1}\\
	D_{mi}(2) & =-\frac{1}{2\pi}\int d_{\Lambda}\theta_{\sigma}(\Lambda)\frac{k_{\sigma}^{\dr}}{1_{\sigma}^{\dr}}h_{\sigma_i}^{\dr}(\Lambda)m_{\sigma}^{\dr}(\Lambda)\,d\Lambda.
	\label{eq:Dmi2}
\end{align}
We  calculate  the two parts separately.
By using the relations (\ref{Equations-Dress-1}-\ref{Equations-Dress-8}), $D_{mi}(1)$ is calculated as  following
\begin{align}
	D_{mi}(1) & =-\frac{1}{2\pi}\int d_{k}\theta_{\rho}(k)\frac{k_{\rho}^{\dr}}{1_{\rho}^{\dr}}h_{\rho_i}^{\dr}(k)m_{\rho}^{\dr}(k)\,dk\nonumber \\
	& =-\frac{1}{2\pi }\int\,dkd_{k}\theta_{\rho}(k)1_{\rho}^{\dr}k_{\rho}^{\dr}\frac{h_{\rho_i}^{\dr}(k)}{1_{\rho}^{\dr}}\frac{m_{\rho}^{\dr}(k)}{1_{\rho}^{\dr}}\nonumber \\
	& =-\frac{1}{2\pi i!}\int\,dkd_{k}\theta_{\rho}(k)1_{\rho}^{\dr}\frac{k}{1-\theta_{\sigma}\theta_{\rho}}\frac{k^{i}}{1-\theta_{\sigma}\theta_{\rho}}\frac{1-\theta_{\sigma}\theta_{\rho}}{1}\frac{\theta_{\sigma}}{1-\theta_{\rho}\theta_{\sigma}}\frac{1-\theta_{\sigma}\theta_{\rho}}{1}\nonumber \\
	& =-\frac{1}{2\pi i!}\int\,dkd_{k}\theta_{\rho}(k)1_{\rho}^{\dr}\frac{\theta_{\sigma}k^{i+1}}{1-\theta_{\sigma}\theta_{\rho}}\nonumber \\
	& =-\frac{1}{2\pi i!}\int dkd_{k}\theta_{\rho}1_{\rho}^{\dr}\left(hm\right)_{\rho}^{\dr},
\end{align}
where $hm$ is an auxiliary charge defined by$\left(hm\right)_{\rho}=0,\left(hm\right)_{\sigma}=\Lambda^{i+1}$.
Its dressed form is given by 
\begin{align}
	\left(hm\right)_{\rho}^{\dr}(k) & =\frac{k^{i+1}\theta_{\sigma}}{1-\theta_{\rho}\theta_{\sigma}},\\
	\left(hm\right)_{\sigma}^{\dr}(\Lambda) & =\frac{\Lambda^{i+1}}{1-\theta_{\rho}\theta_{\sigma}}.
\end{align}
The second part, $D_{mi}(2)$, is calculated by substituting the relation
for $d_{\Lambda}\theta_{\sigma}$:
\begin{align}
	D_{mi}(2) & =-\frac{1}{2\pi}\int d_{\Lambda}\theta_{\sigma}(\Lambda)\frac{k_{\sigma}^{\dr}}{1_{\sigma}^{\dr}}h_{\sigma_i}^{\dr}(\Lambda)m_{\sigma}^{\dr}(\Lambda)\,d\Lambda\nonumber \\
	& =-\frac{1}{2\pi}\int d_{\Lambda}\theta_{\sigma}(\Lambda)1_{\sigma}^{\dr}k_{\sigma}^{\dr}\frac{h_{\sigma_i}^{\dr}(\Lambda)}{1_{\sigma}^{\dr}}\frac{m_{\sigma}^{\dr}(\Lambda)}{1_{\sigma}^{\dr}}\,d\Lambda\nonumber \\
	& =-\frac{1}{2\pi}\int d_{\Lambda}\theta_{\sigma}(\Lambda)1_{\sigma}^{\dr}k_{\sigma}^{\dr}\frac{h_{\sigma_i}^{\dr}(\Lambda)}{1_{\sigma}^{\dr}}\frac{m_{\sigma}^{\dr}(\Lambda)}{1_{\sigma}^{\dr}}\,d\Lambda\nonumber \\
	& =-\frac{1}{2\pi}\int d_{\Lambda}\theta_{\sigma}(\Lambda)1_{\sigma}^{\dr}\frac{\theta_{\rho}\Lambda}{1-\theta_{\sigma}\theta_{\rho}}\frac{\theta_{\rho}h(\Lambda)}{1-\theta_{\sigma}\theta_{\rho}}\frac{1-\theta_{\sigma}\theta_{\rho}}{\theta_{\rho}}\frac{1}{1-\theta_{\rho}\theta_{\sigma}}\frac{1-\theta_{\sigma}\theta_{\rho}}{\theta_{\rho}}\,d\Lambda\nonumber \\
	& =-\frac{1}{2\pi}\int d_{\Lambda}\theta_{\sigma}(\Lambda)1_{\sigma}^{\dr}\frac{\theta_{\rho}\Lambda}{1-\theta_{\sigma}\theta_{\rho}}h(\Lambda)\frac{1}{\theta_{\rho}}\,d\Lambda\nonumber \\
	& =-\frac{1}{2\pi i!}\int d_{\Lambda}\theta_{\sigma}(\Lambda)1_{\sigma}^{\dr}\frac{\Lambda^{i+1}}{1-\theta_{\sigma}\theta_{\rho}}\,d\Lambda\nonumber \\
	& =-\frac{1}{2\pi i!}\int d_{\Lambda}\theta_{\sigma}(\Lambda)1_{\sigma}^{\dr}\left(hm\right)_{\sigma}^{\dr}\,d\Lambda\nonumber \\
	& =-\frac{1}{2\pi}\frac{1}{i!}\int\left(d_{\Lambda}\sigma-\theta_{\sigma}\tilde{T}_{\sigma}d_{k}\rho\right)\left(hm\right)_{\sigma}^{\dr}\,d\Lambda\nonumber \\
	& =-\frac{1}{i!}\int d_{\Lambda}\sigma\left(hm\right)_{\sigma}^{\dr}\,d\Lambda+\frac{1}{i!}\int \theta_{\sigma}\tilde{T}_{\sigma}d_{k}\rho\left(hm\right)_{\sigma}^{\dr}\,d\Lambda\nonumber \\
	& =-\frac{1}{i!}\int d_{\Lambda}\sigma\left(\Lambda^{i+1}+\tilde{T}_{\sigma}\theta_{\rho}\left(hm\right)_{\rho}^{\dr}\right)\,d\Lambda+\frac{1}{i!}\int \theta_{\sigma}\tilde{T}_{\sigma}d_{k}\rho\left(hm\right)_{\sigma}^{\dr}\,d\Lambda\nonumber \\
	& =-\frac{1}{i!}\int d_{\Lambda}\sigma\Lambda^{i+1}\,d\Lambda-\frac{1}{i!}\int d_{\Lambda}\sigma \tilde{T}_{\sigma}\theta_{\rho}\left(hm\right)_{\rho}^{\dr}\,d\Lambda+\frac{1}{i!}\int \theta_{\sigma}\tilde{T}_{\sigma}d_{k}\rho\left(hm\right)_{\sigma}^{\dr}\,d\Lambda\nonumber \\
	& =-\sigma\Lambda^{i+1}|_{-\infty}^{\infty}+\frac{i+1}{i!}\int\sigma\Lambda^{i}\,d\Lambda-\frac{1}{i!}\int d_{\Lambda}\sigma \tilde{T}_{\sigma}\theta_{\rho}\left(hm\right)_{\rho}^{\dr}\,d\Lambda+\frac{1}{i!}\int \theta_{\sigma}\tilde{T}_{\sigma}d_{k}\rho\left(hm\right)_{\sigma}^{\dr}\,d\Lambda\nonumber \\
	& =\frac{1+i}{i!}\int\sigma\Lambda^{i}\,d\Lambda-\frac{1}{i!}\int\,d\Lambda d_{\Lambda}\sigma \tilde{T}_{\sigma}\theta_{\rho}\left(hm\right)_{\rho}^{\dr}+\frac{1}{i!}\int\,d\Lambda\left(hm\right)_{\sigma}^{\dr}\theta_{\sigma}\tilde{T}_{\sigma}d_{k}\rho.
	\label{eq:Dmi2_sigma}
\end{align}
Substituting $d_{k}\rho=d_{k}\theta_{\rho}\frac{\rho}{\theta_{\rho}}+\theta_{\rho}\tilde{T}_{\rho}d_{\Lambda}\sigma$ into the third term  of  the Eq.~\ref{eq:Dmi2_sigma}, we get 
\begin{align}
	D_{mi}(2) 
	& =\frac{1+i}{i!}\int\sigma\Lambda^{i}\,d\Lambda-\int\,d\Lambda d_{\Lambda}\sigma \tilde{T}_{\sigma}\theta_{\rho}\left(hm\right)_{\rho}^{\dr}+\frac{1}{i!}\frac{1}{2\pi}\int dkd_{k}\theta_{\rho}1_{\rho}^{\dr}\left(hm\right)_{\rho}^{\dr}+\frac{1}{i!}\int d\Lambda\left(hm\right)_{\rho}^{\dr}\tilde{T}_{\sigma}\theta_{\rho}d_{\Lambda}\sigma\nonumber \\
	& =\frac{1+i}{i!}\int\sigma\Lambda^{i}\,d\Lambda+\frac{1}{i!}\frac{1}{2\pi}\int dkd_{k}\theta_{\rho}1_{\rho}^{\dr}\left(hm\right)_{\rho}^{\dr}.
\end{align}
Adding up $D_{mi}(1)$ and $D_{mi}(2)$, we finally obtain 
\begin{align}
	D_{im} & =D_{mi}(1)+D_{mi}(2)\nonumber \\
	& =\frac{i+1}{i!}\int\sigma\Lambda^{i}\,d\Lambda+\frac{1}{2\pi}\frac{1}{i!}\int dkd_{k}\theta_{\rho}1_{\rho}^{\dr}\left(hm\right)_{\rho}^{\dr}-\frac{1}{2\pi}\frac{1}{i!}\int dkd_{k}\theta_{\rho}1_{\rho}^{\dr}\left(hm\right)_{\rho}^{\dr}\nonumber \\
	& =\frac{i+1}{i!}\int\sigma\Lambda^{i}\,d\Lambda.
\end{align}
Substituting the specific charges $i=n$ ($i=0$) and $i=e$ ($i=2$)
into this result, we get:
\begin{align}
	D_{mn} & =\int\sigma\,d\Lambda=m,\\
	D_{me} & =\frac{3}{2}\int\sigma\Lambda^{2}\,d\Lambda=3E_{\sigma}.
\end{align}
Where $E_{\sigma}=\int\sigma\frac{\Lambda^{2}}{2}\,d\Lambda$ is the
boson energy density.

Substituting the expression of $\sigma$ Eq~(\ref{eq:sigma_fTzC_BF}), we compute the Drude weighst $D_{mn}$ and $D_{me}$
\begin{align}
	D_{mn} & =\frac{1}{2\pi}\int_{\infty}^{\infty}\frac{1}{e^{\left(k^{2}/2-\mu_{B}\right)/T}-1}\,dk\nonumber \\
	& =\frac{1}{2\pi}\int_{\infty}^{\infty}\frac{1}{e^{\left(\epsilon_{0}-\mu_{B}\right)/T}-1}\frac{d\epsilon_{0}}{\sqrt{2\epsilon_{0}}}\nonumber \\
	& =\frac{1}{\sqrt{2}\pi}\int_{0}^{\infty}\frac{\epsilon_{0}^{-1/2}}{e^{\left(\epsilon_{0}-\mu_{B}\right)/T}-1}d\epsilon_{0}\nonumber \\
	& =\frac{1}{\sqrt{2}\pi}\int_{0}^{\infty}\frac{\epsilon_{0}^{-1/2}T^{-1/2}}{e^{\left(\epsilon_{0}-\mu_{B}\right)/T}-1}d\frac{\epsilon_{0}}{T}\nonumber \\
	& =\frac{1}{\sqrt{2}\pi}T^{1/2}\int_{0}^{\infty}\frac{\epsilon_{0}^{-1/2}/T^{-1/2}}{e^{\left(\epsilon_{0}-\mu_{B}\right)/T}-1}d\frac{\epsilon_{0}}{T}\nonumber \\
	& =\frac{1}{\sqrt{2}\pi}T^{1/2}\sqrt{\pi}Li_{\frac{1}{2}}\left(e^{\mu_{B}/T}\right)\nonumber \\
	& =\frac{1}{\sqrt{2\pi}}T^{1/2}Li_{\frac{1}{2}}\left(e^{\mu_{B}/T}\right).
\end{align}
\begin{align}
	D_{me} & =\frac{3}{2}\int\sigma k^{2}\,dk\nonumber \\
	& =\frac{3}{2}\frac{1}{2\pi}\int\frac{k^{2}}{e^{\left(k^{2}/2-\mu_{B}\right)/T}-1}dk\nonumber \\
	& =\frac{3}{2\pi}\int_{0}^{\infty}\frac{2\epsilon_{0}}{e^{\left(k^{2}/2-\mu_{B}\right)/T}-1}\frac{d\epsilon_{0}}{\sqrt{2\epsilon_{0}}}\nonumber \\
	& =\frac{3}{2\pi}\int\frac{\sqrt{2\epsilon_{0}}}{e^{\left(k^{2}/2-\mu_{B}\right)/T}-1}d\epsilon_{0}\nonumber \\
	& =\frac{3}{\sqrt{2}\pi}T^{3/2}\int\frac{\sqrt{\epsilon_{0}/T}}{e^{\left(\epsilon_{0}-\mu_{B}\right)/T}-1}d\frac{\epsilon_{0}}{T}\nonumber \\
	& =\frac{3}{\sqrt{2}\pi}T^{3/2}\frac{\sqrt{\pi}}{2}Li_{\frac{3}{2}}\left(e^{\mu_{B}/T}\right)\nonumber \\
	& =\frac{3}{2\sqrt{2\pi}}T^{3/2}Li_{\frac{3}{2}}\left(e^{\mu_{B}/T}\right).
\end{align}
Thus, the cross-Drude weights $D_{mi}$ (correlating the Boson component
$m$ with a Total component $i$) are equal to the isolated Boson
component's integrals.

\subsubsection*{Calculation of $D_{mm}$}

For calculation of  the DW of bosonic  component $D_{mm}$, we write the expression as 
\begin{align}
	D_{mm} & =-\frac{1}{2\pi}\int d_{k}[\theta_{\rho}(k)]\frac{k_{\rho}^{\dr}}{1_{\rho}^{\dr}}m_{\rho}^{\dr}(k)m_{\rho}^{\dr}(k)\,dk-\frac{1}{2\pi}\int d_{\Lambda}[\theta_{\sigma}(\Lambda)]\frac{k_{\sigma}^{\dr}}{1_{\sigma}^{\dr}}m_{\sigma}^{\dr}(\Lambda)m_{\sigma}^{\dr}(\Lambda)\,d\Lambda\nonumber \\
	& =D_{mm}(1)+D_{mm}(2),
\end{align}
where
\begin{align}
	D_{mm}(1) & =-\frac{1}{2\pi}\int d_{k}[\theta_{\rho}(k)]\frac{k_{\rho}^{\dr}}{1_{\rho}^{\dr}}m_{\rho}^{\dr}(k)m_{\rho}^{\dr}(k)\,dk,\label{eq:Dmm1_fTzC}\\
	D_{mm}(2) & =-\frac{1}{2\pi}\int d_{\Lambda}[\theta_{\sigma}(\Lambda)]\frac{k_{\sigma}^{\dr}}{1_{\sigma}^{\dr}}m_{\sigma}^{\dr}(\Lambda)m_{\sigma}^{\dr}(\Lambda)\,d\Lambda.
	\label{eq:Dmm2_fTzC}
\end{align}
We define the auxiliary functions
\begin{align}
	u_{\rho} & =m_{\rho}^{\dr}\theta_{\rho},\\
	u_{\sigma} & =m_{\sigma}^{\dr}\theta_{\sigma}.
\end{align}
Their derivatives are:
\begin{align}
	d_{k}\theta_{\rho}m_{\rho}^{\dr} & =d_{k}u_{\rho}-\theta_{\rho}\tilde{T}_{\rho}d_{\Lambda}u_{\sigma}\label{eq:d_k_n_rho},\\
	d_{\Lambda}\theta_{\sigma}m_{\sigma}^{\dr} & =d_{\Lambda}u_{\sigma}-\theta_{\sigma}\tilde{T}_{\sigma}d_{k}u_{\rho}.
	\label{eq:d_Lambda_n_sigma}
\end{align}
Substituting these into $D_{mm}(1)$ and $D_{mm}(2)$ leads to:
\begin{align}
	D_{mm}(2) & =-\frac{1}{2\pi}\int d\Lambda\Lambda d_{\Lambda}u_{\sigma}-\frac{1}{2\pi}\int d\Lambda k_{\sigma}^{\dr}\theta_{\rho}d_{\Lambda}u_{\sigma}+\frac{1}{2\pi}\int dkd_{k}u_{\rho}\left(km\right)_{\rho}^{\dr},\\
	D_{mm}(1) & =\frac{1}{2\pi}\int d_{k}u_{\rho}\left(km\right)_{\rho}^{\dr}\,dk-\frac{1}{2\pi}\int k_{\sigma}^{\dr}d_{\Lambda}u_{\sigma}\,dk.
\end{align}
Adding up  the two terms, we have 
\begin{align}
	D_{mm} & =-\frac{1}{2\pi}\int d_{\Lambda}\theta_{\sigma}(\Lambda)\frac{k_{\sigma}^{\dr}}{1_{\sigma}^{\dr}}m_{\sigma}^{\dr}(\Lambda)m_{\sigma}^{\dr}(\Lambda)\,d\Lambda-\frac{1}{2\pi}\int d_{k}\theta_{\rho}(k)\frac{k_{\rho}^{\dr}}{1_{\rho}^{\dr}}m_{\rho}^{\dr}(k)m_{\rho}^{\dr}(k)\,dk\nonumber \\
	& =-\frac{1}{2\pi}\int d\Lambda\Lambda d_{\Lambda}u_{\sigma}-\frac{1}{2\pi}\int d\Lambda k_{\sigma}^{\dr}d_{\Lambda}u_{\sigma}+\frac{1}{2\pi}\int d\Lambda\left(km\right)_{\rho}^{\dr}d_{k}u_{\rho}-\frac{1}{2\pi}\int d_{k}u_{\rho}\left(km\right)_{\rho}^{\dr}\,dk+\frac{1}{2\pi}\int k_{\sigma}^{\dr}d_{\Lambda}u_{\sigma}\,dk\nonumber \\
	& =-\frac{1}{2\pi}\int d\Lambda\Lambda d_{\Lambda}\left(m_{\sigma}^{\dr}\theta_{\sigma}\right)\nonumber \\
	& =-\frac{1}{2\pi}\int dkkd_{k}\left[m_{\sigma}^{\dr}\left(k\right)\theta_{\sigma}\right].
	\label{eq:Dmm_fTzC_ann}
\end{align}
We now need the relation $d_{k}\left[m_{\sigma}^{\dr}\theta_{\sigma}\right]$. By definition, we further have 
\begin{align}
	\theta_{\sigma}m_{\sigma}^{\dr} & =\frac{\theta_{\sigma}}{1-\theta_{\sigma}\theta_{\rho}}\nonumber \\
	& =\frac{\frac{1}{e^{\left(k^{2}/2-\mu_{F}\right)/T}}+1}{e^{\left(\mu_{F}-\mu_{B}\right)/T}+1}\frac{e^{\left(k^{2}/2-\mu_{B}\right)/T}}{e^{\left(k^{2}/2-\mu_{B}\right)/T}-1}\nonumber \\
	& =\frac{e^{\left(\mu_{F}-\mu_{B}\right)/T}+e^{\left(k^{2}/2-\mu_{B}\right)/T}}{\left(e^{\left(\mu_{F}-\mu_{B}\right)/T}+1\right)\left(e^{\left(k^{2}/2-\mu_{B}\right)/T}-1\right)}\nonumber \\
	& =\frac{1}{e^{\left(\mu_{F}-\mu_{B}\right)/T}+1}+\frac{1}{e^{\left(k^{2}/2-\mu_{B}\right)/T}-1}.
\end{align}
The first term is a constant with respect to $k$. Taking the derivative:
\begin{align}
	d_{k}\left(m_{\sigma}^{\dr}\theta_{\sigma}\right) & =d_{k}\left(\frac{1}{e^{\left(\mu_{F}-\mu_{B}\right)/T}+1}+\frac{1}{e^{\left(k^{2}/2-\mu_{B}\right)/T}-1}\right)\nonumber \\
	& =d_{k}\left(\frac{1}{e^{\left(k^{2}/2-\mu_{B}\right)/T}-1}\right)\nonumber \\
	& =-\frac{k}{T}\left(\frac{e^{\left(k^{2}/2-\mu_{B}\right)/T}}{\left(e^{\left(k^{2}/2-\mu_{B}\right)/T}-1\right)^{2}}\right)\nonumber \\
	& =-\frac{k}{T}\left(\frac{e^{\left(k^{2}/2-\mu_{B}\right)/T}}{\left(e^{\left(k^{2}/2-\mu_{B}\right)/T}-1\right)^{2}}\right)\nonumber \\
	& =-kd_{\mu_{B}}\left(\frac{1}{e^{\left(k^{2}/2-\mu_{B}\right)/T}-1}\right)\nonumber \\
	& =-d_{\mu_{B}}\left(\frac{k}{e^{\left(k^{2}/2-\mu_{B}\right)/T}-1}\right).
\end{align}
Substituting this elegant identity back into the expression for $D_{mm}$
Eq.~(\ref{eq:Dmm_fTzC_ann}), we have 
\begin{align}
	D_{mm} & =-\frac{1}{2\pi}\int dkkd_{k}\left(m_{\sigma}^{\dr}\theta_{\sigma}\right)\nonumber \\
	& =\frac{1}{2\pi}\int dkkd_{\mu_{B}}\left(\frac{k}{e^{\left(k^{2}/2-\mu_{B}\right)/T}-1}\right)\nonumber \\
	& =\frac{1}{2\pi}d_{\mu_{B}}\int dk\frac{k^{2}}{e^{\left(k^{2}/2-\mu_{B}\right)/T}-1}\nonumber \\
	& =\frac{1}{2\pi}d_{\mu_{B}}\int\frac{d\epsilon_{0}}{\sqrt{2\epsilon_{0}}}\frac{2\epsilon_{0}}{e^{\left(\epsilon_{0}-\mu_{B}\right)/T}-1}\nonumber \\
	& =\frac{\sqrt{2}}{\pi}d_{\mu_{B}}\int_{0}^{\infty}d\epsilon_{0}\frac{\sqrt{\epsilon_{0}}}{e^{\left(\epsilon_{0}-\mu_{B}\right)/T}-1}\nonumber \\
	& =\frac{\sqrt{2}}{\pi}T^{3/2}d_{\mu_{B}}\int_{0}^{\infty}d\frac{\epsilon_{0}}{T}\frac{\sqrt{\epsilon_{0}/T}}{e^{\left(\epsilon_{0}-\mu_{B}\right)/T}-1}\nonumber \\
	& =\frac{\sqrt{2}}{\pi}T^{3/2}d_{\mu_{B}}\frac{\sqrt{\pi}}{2}Li_{\frac{3}{2}}\left(e^{\mu_{B}/T}\right)\nonumber \\
	& =\frac{1}{\sqrt{2\pi}}T^{1/2}Li_{\frac{1}{2}}\left(e^{\mu_{B}/T}\right).
\end{align}
The last step uses the polylogarithm property $\frac{d}{dz}\mathbf{Li}_{n}\left(z\right)=\frac{\mathbf{Li}_{n-1}\left(z\right)}{z}$.
The  Boson component Drude weight, $D_{mm}$, is identical
to the cross-Drude weight $D_{mn}$. 
This is an expected result, since in the weak-interaction ($c\to0$) limit, the components decouple,
and the correlations $D_{mn}$ (total Charge-Boson) and $D_{mm}$ (Boson-Boson)
are both sourced purely by the free Boson component.

\subsection{Strong Interaction Expansion at Finite-Temperature (Low-T) }

\subsubsection*{Phase Boundaries}
The thermodynamic Bethe ansatz (TBA) equations for the Bose-Fermi mixture are given by:
\begin{align}
	\epsilon(k) & =\frac{k^{2}}{2}-\mu_{F}-\frac{Tc}{\pi}\int\frac{1}{c^{2}+(k-\Lambda)^{2}}\ln\left(1+e^{-\psi(\Lambda)/T}\right)\,d\Lambda,\\
	\psi(\Lambda) & =\mu_{F}-\mu_{B}-\frac{Tc}{\pi}\int\frac{1}{c^{2}+(\Lambda-k)^{2}}\ln\left(1+e^{-\epsilon(k)/T}\right)\,dk,
\end{align}
where $\mu_{F}=\mu+\mathcal{H}/2, \, \mu_{B}=\mu-\mathcal{H}/2$. 
At the V-F (Vacuum-Fermi) phase boundary, the conditions are $\epsilon\left(k\right)=k^{2}/2-\mu_{F},\epsilon\left(0\right)=0,\psi\left(0\right)=0$.
We find the relation $\mu_{c}=-\mathcal{H}_{c}/2$ \cite{Phys.Rev.A2012YinQuantum}. 
At the F-BF (Fermi - Bose-Fermi) phase boundary, the conditions are given by $\psi\left(0\right)=0,\epsilon\left(k\right)=k^{2}/2-\mu_{F}$.
The relation at the critical point in the strong interaction limit
is $\mathcal{H}_{c}=\frac{4\sqrt{2}}{3\pi c}\left(\mu_{c}+\frac{\mathcal{H}_{c}}{2}\right)^{3/2}$.

\subsubsection*{Dressed Energy (Low-T, Strong-Interaction Approximation)}

Under strong interaction conditions ($c \to \infty$), the integral kernel is expanded up to order $1/c^3$:
\begin{align}
	\frac{c}{\pi}\frac{1}{c^{2}+(k-\Lambda)^{2}} & =\frac{c}{\pi}\frac{1}{c^{2}+k^{2}-2k\Lambda+\Lambda^{2}} \nonumber\\
	& \approx \frac{c}{\pi}\frac{1}{c^{2}+k^{2}}\left[1-\frac{\Lambda^{2}-2k\Lambda}{c^{2}+k^{2}}\right] \nonumber\\
	& =\frac{1}{\pi}\left(\frac{1}{c}-\frac{\Lambda^{2}-2k\Lambda+k^{2}}{c^{3}}\right), \nonumber\\
	\frac{c}{\pi}\frac{1}{c^{2}+(\Lambda-k)^{2}} & =\frac{c}{\pi}\frac{1}{c^{2}+\Lambda^{2}-2k\Lambda+k^{2}} \nonumber\\
	& \approx \frac{c}{\pi}\frac{1}{c^{2}+\Lambda^{2}}\left[1-\frac{k^{2}-2k\Lambda}{c^{2}+\Lambda^{2}}\right] \nonumber\\
	& =\frac{1}{\pi}\left(\frac{1}{c}-\frac{k^{2}-2k\Lambda+\Lambda^{2}}{c^{3}}\right).
\end{align}
At low temperatures, substituting this approximated kernel into the
dressed energy TBA equations yields
\begin{align}
	\epsilon(k) & =\frac{k^{2}}{2}-\mu_{F}-\frac{T}{\pi}\int\frac{c}{c^{2}+(k-\Lambda)^{2}}\ln\left(1+e^{-\psi(\Lambda)/T}\right)\,d\Lambda\nonumber \\
	& =\frac{k^{2}}{2}-\mu_{F}-\frac{Tc}{\pi}\int\frac{1}{c^{2}+k^{2}}\left[1-\frac{\Lambda^{2}-2k\Lambda}{c^{2}+k^{2}}\right]\ln\left(1+e^{-\psi(\Lambda)/T}\right)\,d\Lambda\nonumber \\
	& =\frac{k^{2}}{2}-\mu_{F}-\frac{2c}{c^{2}+k^{2}}\frac{T}{2\pi}\int\ln\left(1+e^{-\psi(\Lambda)/T}\right)\,d\Lambda+\frac{T}{\pi c^{3}}\int\Lambda^{2}\ln\left(1+e^{-\psi(\Lambda)/T}\right)\,d\Lambda\nonumber \\
	& =\frac{k^{2}}{2}-\mu_{F}-\frac{2p_{\psi}}{c}+\frac{2p_{\psi}}{c^{3}}k^{2}+\frac{T}{\pi c^{3}}\int\Lambda^{2}\ln\left(1+e^{-\psi(\Lambda)/T}\right)\,d\Lambda\nonumber \\
	& =k^{2}\left(\frac{1}{2}+\frac{2p_{\psi}}{c^{3}}\right)-\mu_{F}-\frac{2p_{\psi}}{c}+\frac{T}{\pi c^{3}}\int\Lambda^{2}\ln\left(1+e^{-\psi(\Lambda)/T}\right)\,d\Lambda\nonumber \\
	& =k^{2}\beta_{\epsilon}-A_{\epsilon},\\
	\psi(\Lambda) & =\mu_{F}-\mu_{B}-\frac{Tc}{\pi}\int\frac{1}{c^{2}+(\Lambda-k)^{2}}\ln\left(1+e^{-\epsilon(k)/T}\right)\,dk\nonumber \\
	& =\mu_{F}-\mu_{B}-\frac{cT}{\pi}\int\frac{1}{c^{2}+\Lambda^{2}}\left[1-\frac{k^{2}-2k\Lambda}{c^{2}+\Lambda^{2}}\right]\ln\left(1+e^{-\epsilon(k)/T}\right)\,dk\nonumber \\
	& =\mu_{F}-\mu_{B}-\frac{2c}{c^{2}+\Lambda^{2}}\frac{T}{2\pi}\int\ln\left(1+e^{-\epsilon(k)/T}\right)\,dk+\frac{c}{\pi}\frac{T}{\left(c^{2}+\Lambda^{2}\right)^{2}}\int k^{2}\ln\left(1+e^{-\epsilon(k)/T}\right)\,dk\nonumber \\
	& =\mu_{F}-\mu_{B}-\frac{2p}{c}+\frac{2p}{c^{3}}\Lambda^{2}+\frac{T}{\pi c^{3}}\int k^{2}\ln\left(1+e^{-\epsilon(k)/T}\right)\,dk\nonumber \\
	& =\frac{2p}{c^{3}}\Lambda^{2}+\mu_{F}-\mu_{B}-\frac{2p}{c}+\frac{T}{\pi c^{3}}\int k^{2}\ln\left(1+e^{-\epsilon(k)/T}\right)\,dk\nonumber \\
	& =\Lambda^{2}\beta_{\psi}-A_{\psi},
\end{align}
where $p=\frac{T}{2\pi}\int\ln\left(1+e^{-\epsilon(k)/T}\right)\,dk$ is the pressure.
 We denote the coefficients
\begin{align}
	\beta_{\epsilon} & =\frac{1}{2}+\frac{2p_{\psi}}{c^{3}}, \\
	p_{\psi} & =\frac{T}{2\pi}\int\ln\left(1+e^{-\psi(\Lambda)/T}\right)\,d\Lambda,\\
	A_{\epsilon} & =\mu_{F}+\frac{2p_{\psi}}{c}-\frac{T}{\pi c^{3}}\int\Lambda^{2}\ln\left(1+e^{-\psi(\Lambda)/T}\right)\,d\Lambda,\\
	\beta_{\psi} & =\frac{2p}{c^{3}},\\
	A_{\psi} & =-\mu_{F}+\mu_{B}+\frac{2p}{c}-\frac{T}{\pi c^{3}}\int k^{2}\ln\left(1+e^{-\epsilon(k)/T}\right)\,dk. 
\end{align}
Using these notations and  substituting the quadratic forms $\epsilon(k)=k^{2}\beta_{\epsilon}-A_{\epsilon},\psi(\Lambda)=\Lambda^{2}\beta_{\psi}-A_{\psi}$
 into the integrals, we may evaluate the integrals in terms of the  polylogarithm functions
\begin{align}
	\beta_{\epsilon} & =\frac{1}{2}-\frac{T^{\frac{3}{2}}}{\pi^{\frac{1}{2}}\beta_{\psi}^{\frac{1}{2}}c^{3}}Li_{\frac{3}{2}}\left(-e^{A_{\psi}/T}\right) ,\\
	p_{\psi} & =-\frac{T^{\frac{3}{2}}}{2\pi^{\frac{1}{2}}\beta_{\psi}^{\frac{1}{2}}}Li_{\frac{3}{2}}\left(-e^{A_{\psi}/T}\right) ,\\
	A_{\epsilon} & =\mu+\frac{\mathcal{H}}{2}-\frac{T^{\frac{3}{2}}}{\pi^{\frac{1}{2}}\beta_{\psi}^{\frac{1}{2}}c}Li_{\frac{3}{2}}\left(-e^{A_{\psi}/T}\right)+\frac{T^{\frac{5}{2}}}{2\sqrt{\pi}\beta_{\psi}^{\frac{3}{2}}c^{3}}Li_{\frac{5}{2}}\left(-e^{A_{\psi}/T}\right),
	\label{eq:Aepsilon}\\
	\beta_{\psi} & =\frac{2p}{c^{3}},\\
	A_{\psi} & =-\mathcal{H}+\frac{2p}{c}+\frac{T^{\frac{5}{2}}}{2\sqrt{\pi}\beta_{\epsilon}^{\frac{3}{2}}c^{3}}Li_{\frac{5}{2}}\left(-e^{A_{\epsilon}/T}\right),
	\label{eq:Apsi}\\
	p & =-\frac{T^{\frac{3}{2}}}{\sqrt{2\pi}}Li_{\frac{3}{2}}\left(-e^{A_{\epsilon}/T}\right)\left[1+\frac{T^{\frac{3}{2}}}{\pi^{\frac{1}{2}}\beta_{\psi}^{\frac{1}{2}}c^{3}}Li_{\frac{3}{2}}\left(-e^{A_{\psi}/T}\right)\right].
\end{align}

\subsubsection*{Calculation of the Drude weight  $D_{ij}$}

We first integrate the DWs  (Eq.~\ref{eq:DW_dn_BF}) by part
\begin{align}
	D_{ij} & =\frac{1}{2\pi}\int \theta_{\rho}(k)d_{k}\left[v_{\rho}^{\eff}(k)h_{\rho_{i}}^{\dr}(k)h_{\rho_{j}}^{\dr}(k)\right]+\frac{1}{2\pi}\int \theta_{\sigma}(\Lambda)d_{\Lambda}\left[v_{\sigma}^{\eff}(\Lambda)h_{\sigma_{i}}^{\dr}(\Lambda)h_{\sigma_{j}}^{\dr}(\Lambda)\right].
	\label{eq:Dij_fTiC_BF}
\end{align}
In order to carry out  the following thermal integrals, we use the quadratic forms
$\epsilon=k^{2}\beta_{\epsilon}-A_{\epsilon}$ and $\psi=\Lambda^{2}\beta_{\psi}-A_{\psi}$
\begin{align}
	B_{\psi_{\frac{1}{2}}} & =\int \theta_{\sigma}\,d\Lambda=-\beta_{\psi}^{-\frac{1}{2}}T^{\frac{1}{2}}\pi^{\frac{1}{2}}Li_{\frac{1}{2}}\left(-e^{A_{\psi}/T}\right),\\
	B_{\psi_{\frac{3}{2}}} & =\int\Lambda^{2}\theta_{\sigma}\,d\Lambda=-\beta_{\psi}^{-\frac{3}{2}}T^{\frac{3}{2}}\frac{\pi^{\frac{1}{2}}}{2}Li_{\frac{3}{2}}\left(-e^{A_{\psi}/T}\right),\\
	B_{\psi_{\frac{5}{2}}} & =\int\Lambda^{4}\theta_{\sigma}\,d\Lambda=-\beta_{\psi}^{-\frac{5}{2}}T^{\frac{5}{2}}\frac{3\pi^{\frac{1}{2}}}{4}Li_{\frac{5}{2}}\left(-e^{A_{\psi}/T}\right),\\
	B_{\epsilon_{\frac{1}{2}}} & =\int \theta_{\rho}\,dk=-\beta_{\epsilon}^{-\frac{1}{2}}T^{\frac{1}{2}}\pi^{\frac{1}{2}}Li_{\frac{1}{2}}\left(-e^{A_{\epsilon}/T}\right),\\
	B_{\epsilon_{\frac{3}{2}}} & =\int k^{2}\theta_{\rho}\,dk=-\beta_{\epsilon}^{-\frac{3}{2}}T^{\frac{3}{2}}\frac{\pi^{\frac{1}{2}}}{2}Li_{\frac{3}{2}}\left(-e^{A_{\epsilon}/T}\right),\\
	B_{\epsilon_{\frac{5}{2}}} & =\int k^{4}\theta_{\rho}\,dk=-\beta_{\epsilon}^{-\frac{5}{2}}T^{\frac{5}{2}}\frac{3\pi^{\frac{1}{2}}}{4}Li_{\frac{5}{2}}\left(-e^{A_{\epsilon}/T}\right).
\end{align}

\paragraph*{Calculating Dressed Charges.}
In order to solve the DWs, we should find the expressions for the dressed charges in this approximation. 
We calculate dressed charge $1^{\dr}$ ($h_{n}=(1,0)$)  up to the order of $1/c^{3}$
\begin{align}
	1_{\rho}^{\dr}(k) & =1+\frac{1}{\pi}\int\frac{c}{c^{2}+(k-\Lambda)^{2}}\theta_{\sigma}1_{\sigma}^{\dr}\,d\Lambda\nonumber \\
	& =1+\frac{c}{\pi}\int\frac{1}{c^{2}+k^{2}}\theta_{\sigma}h_{\sigma}^{\dr}\,d\Lambda-\frac{c}{\pi}\int\frac{\Lambda^{2}-2k\Lambda}{\left(c^{2}+k^{2}\right)^{2}}\theta_{\sigma}1_{\sigma}^{\dr}\,d\Lambda\nonumber \\
	& =1+\frac{c}{\pi\left(c^{2}+k^{2}\right)}\int \theta_{\sigma}h_{\sigma}^{\dr}\,d\Lambda-\frac{1}{\pi c^{3}}\int\Lambda^{2}\theta_{\sigma}1_{\sigma}^{\dr}\,d\Lambda\nonumber \\
	& =1+\frac{c}{\pi\left(c^{2}+k^{2}\right)}p_{1_{\sigma}}-\frac{1}{\pi c^{3}}\int\Lambda^{2}\theta_{\sigma}1_{\sigma}^{\dr}\,d\Lambda\nonumber \\
	& =-\frac{p_{1_{\sigma}}}{\pi c^{3}}k^{2}+1+\frac{p_{1_{\sigma}}}{\pi c}-\frac{1}{\pi c^{3}}\int\Lambda^{2}\theta_{\sigma}1_{\sigma}^{\dr}\,d\Lambda\nonumber \\
	& =\beta_{1_{\rho}}k^{2}+A_{1_{\rho}},\\
	1_{\sigma}^{\dr}(\Lambda) & =\frac{1}{\pi}\int\frac{c}{c^{2}+(\Lambda-k)^{2}}\theta_{\rho}1_{\rho}^{\dr}\,dk\nonumber \\
	& =\frac{c}{\pi\left(c^{2}+\Lambda^{2}\right)}\int \theta_{\rho}1_{\rho}^{\dr}\,dk-\frac{1}{\pi c^{3}}\int k^{2}\theta_{\rho}1_{\rho}^{\dr}\,dk\nonumber \\
	& =\frac{c}{\pi\left(c^{2}+\Lambda^{2}\right)}p_{1_{\rho}}-\frac{1}{\pi c^{3}}\int k^{2}\theta_{\rho}1_{\rho}^{\dr}\,dk\nonumber \\
	& =-\frac{p_{1_{\rho}}}{\pi c^{3}}\Lambda^{2}+\frac{p_{1_{\rho}}}{\pi c}-\frac{1}{\pi c^{3}}\int k^{2}\theta_{\rho}1_{\rho}^{\dr}\,dk\nonumber \\
	& =\beta_{1_{\sigma}}\Lambda^{2}+A_{1_{\sigma}}.
\end{align}
We define the corresponding coefficients:
\begin{align}
	p_{1_{\sigma}} & =\int \theta_{\sigma}h_{\sigma}^{\dr}\,d\Lambda,\\
	\beta_{1_{\rho}} & =-\frac{p_{1_{\sigma}}}{\pi c^{3}},\\
	A_{1_{\rho}} & =1+\frac{p_{1_{\sigma}}}{\pi c}-\frac{1}{\pi c^{3}}\int\Lambda^{2}\theta_{\sigma}1_{\sigma}^{\dr}\,d\Lambda\label{eq:AIrho},\\
	p_{1_{\rho}} & =\int \theta_{\rho}1_{\rho}^{\dr}\,dk,\\
	\beta_{1_{\sigma}} & =-\frac{p_{1_{\rho}}}{\pi c^{3}},\\
	A_{1_{\sigma}} & =\frac{p_{1_{\rho}}}{\pi c}-\frac{1}{\pi c^{3}}\int k^{2}\theta_{\rho}1_{\rho}^{\dr}\,dk,
	\label{eq:AIsigma}
\end{align}
providing  nested structure in  the coefficients. We solve  that   iteratively  up to order $O(1/c^{3})$
\begin{align}
	\int\Lambda^{2}\theta_{\sigma}1_{\sigma}^{\dr}\,d\Lambda & =\int\Lambda^{2}\theta_{\sigma}\left(\beta_{1_{\sigma}}\Lambda^{2}+A_{1_{\sigma}}\right)\,d\Lambda\nonumber \\
	& =\beta_{1_{\sigma}}\int\Lambda^{4}\theta_{\sigma}\,d\Lambda+A_{1_{\sigma}}\int\Lambda^{2}\theta_{\sigma}\,d\Lambda\nonumber \\
	& =\beta_{1_{\sigma}}B_{\psi_{\frac{5}{2}}}+A_{1_{\sigma}}B_{\psi_{\frac{3}{2}}}\nonumber \\
	& =-\frac{p_{1_{\rho}}}{\pi c^{3}}B_{\psi_{\frac{5}{2}}}+\left(\frac{p_{1_{\rho}}}{\pi c}-\frac{1}{\pi c^{3}}\int k^{2}\theta_{\rho}1_{\rho}^{\dr}\,dk\right)B_{\psi_{\frac{3}{2}}}\nonumber ,\\
	& =\frac{p_{1_{\rho}}}{\pi c}B_{\psi_{\frac{3}{2}}}-\frac{p_{1_{\rho}}}{\pi c^{3}}B_{\psi_{\frac{5}{2}}}-\frac{1}{\pi c^{3}}B_{\epsilon_{\frac{3}{2}}}B_{\psi_{\frac{3}{2}}},
\end{align}
\begin{align}
	\int k^{2}\theta_{\rho}1_{\rho}^{\dr}\,dk & =\int k^{2}n_{\rho}\left(\beta_{1_{\rho}}k^{2}+A_{1_{\rho}}\right)\,dk\nonumber \\
	& =\beta_{1_{\rho}}\int k^{4}\theta_{\rho}\,dk+A_{1_{\rho}}\int k^{2}n_{\rho}\,dk\nonumber \\
	& =\beta_{1_{\rho}}B_{\epsilon_{\frac{5}{2}}}+A_{1_{\rho}}B_{\epsilon_{\frac{3}{2}}}\nonumber \\
	& =-\frac{p_{1_{\rho}}}{\pi c^{3}}B_{\epsilon_{\frac{5}{2}}}+\left(1+\frac{p_{1_{\sigma}}}{\pi c}-\frac{1}{\pi c^{3}}\int\Lambda^{2}\theta_{\sigma}1_{\sigma}^{\dr}\,d\Lambda\right)B_{\epsilon_{\frac{3}{2}}}\nonumber \\
	& =B_{\epsilon_{\frac{3}{2}}}+\frac{p_{1_{\sigma}}}{\pi c}B_{\epsilon_{\frac{3}{2}}}-\frac{p_{1_{\rho}}}{\pi c^{3}}B_{\epsilon_{\frac{5}{2}}}-\frac{1}{\pi c^{3}}B_{\epsilon_{\frac{3}{2}}}\int\Lambda^{2}\theta_{\sigma}1_{\sigma}^{\dr}\,d\Lambda \nonumber\\
	& =B_{\epsilon_{\frac{3}{2}}}+\frac{p_{1_{\sigma}}}{\pi c}B_{\epsilon_{\frac{3}{2}}}-\frac{p_{1_{\rho}}}{\pi c^{3}}B_{\epsilon_{\frac{5}{2}}}.
\end{align}
Substituting these back into the definitions Eq.~\eqref{eq:AIrho} and \eqref{eq:AIsigma}, we finally obtain 
\begin{align}
	\beta_{1_{\rho}} & =-\frac{p_{1_{\sigma}}}{\pi c^{3}},\\
	A_{1_{\rho}} & =1+\frac{p_{1_{\sigma}}}{\pi c},\\
	\beta_{1_{\sigma}} & =-\frac{p_{1_{\rho}}}{\pi c^{3}},\\
	A_{1_{\sigma}} & =\frac{p_{1_{\rho}}}{\pi c}-\frac{1}{\pi c^{3}}B_{\epsilon_{\frac{3}{2}}}.
\end{align}
Next, we further solve for $p_{1\sigma}$ and $p_{1\rho}$
\begin{align}
	p_{1_{\sigma}} & =\int \theta_{\sigma}1_{\sigma}^{\dr}\,d\Lambda\nonumber \\
	& =\int\frac{\beta_{1_{\sigma}}\Lambda^{2}+A_{1_{\sigma}}}{1+e^{\psi(\Lambda)/T}}\,d\Lambda\nonumber \\
	& =\beta_{1_{\sigma}}\int\frac{\Lambda^{2}}{1+e^{\psi(\Lambda)/T}}\,d\Lambda+A_{1_{\sigma}}\int\frac{1}{1+e^{\psi(\Lambda)/T}}\,d\Lambda\nonumber \\
	& =\beta_{1_{\sigma}}B_{\psi_{\frac{3}{2}}}+A_{1_{\sigma}}B_{\psi_{\frac{1}{2}}}\nonumber \\
	& =-\frac{p_{1_{\rho}}}{\pi c^{3}}B_{\psi_{\frac{3}{2}}}+\frac{p_{1_{\rho}}}{\pi c}B_{\psi_{\frac{1}{2}}}-\frac{1}{\pi c^{3}}B_{\epsilon_{\frac{3}{2}}}B_{\psi_{\frac{1}{2}}},
\end{align}
\begin{align}
	p_{1_{\rho}} & =\int \theta_{\rho}1_{\rho}^{\dr}\,dk\nonumber \\
	& =\int\frac{\beta_{1_{\rho}}k^{2}+A_{1_{\rho}}}{1+e^{\epsilon(k)/T}}\,dk\nonumber \\
	& =\beta_{1_{\rho}}\int\frac{k^{2}}{1+e^{\epsilon(k)/T}}\,dk+A_{1_{\rho}}\int\frac{1}{1+e^{\epsilon(k)/T}}\,dk\nonumber \\
	& =\beta_{1_{\rho}}B_{\epsilon_{\frac{3}{2}}}+A_{1_{\rho}}B_{\epsilon_{\frac{1}{2}}}\nonumber \\
	& =-\frac{p_{1_{\sigma}}}{\pi c^{3}}B_{\epsilon_{\frac{3}{2}}}+B_{\epsilon_{\frac{1}{2}}}+\frac{p_{1_{\sigma}}}{\pi c}B_{\epsilon_{\frac{1}{2}}},
\end{align}
 We iterate term by term to reach the solutions to order of  $1/c^{3}$
\begin{align}
	p_{1_{\sigma}} & =-\frac{p_{1_{\rho}}}{\pi c^{3}}B_{\psi_{\frac{3}{2}}}+\frac{p_{1_{\rho}}}{\pi c}B_{\psi_{\frac{1}{2}}}-\frac{1}{\pi c^{3}}B_{\epsilon_{\frac{3}{2}}}B_{\psi_{\frac{1}{2}}}\nonumber \\
	& =-\frac{1}{\pi c^{3}}\left(-\frac{p_{1_{\sigma}}}{\pi c^{3}}B_{\epsilon_{\frac{3}{2}}}+B_{\epsilon_{\frac{1}{2}}}+\frac{p_{1_{\sigma}}}{\pi c}B_{\epsilon_{\frac{1}{2}}}\right)B_{\psi_{\frac{3}{2}}}+\frac{1}{\pi c}\left(-\frac{p_{1_{\sigma}}}{\pi c^{3}}B_{\epsilon_{\frac{3}{2}}}+B_{\epsilon_{\frac{1}{2}}}+\frac{p_{1_{\sigma}}}{\pi c}B_{\epsilon_{\frac{1}{2}}}\right)B_{\psi_{\frac{1}{2}}}-\frac{1}{\pi c^{3}}B_{\psi_{\frac{1}{2}}}B_{\epsilon_{\frac{3}{2}}}\nonumber \\
	& =\frac{p_{1_{\sigma}}}{\pi^{2}c^{6}}B_{\epsilon_{\frac{3}{2}}}B_{\psi_{\frac{3}{2}}}-\frac{1}{\pi c^{3}}B_{\epsilon_{\frac{1}{2}}}B_{\psi_{\frac{3}{2}}}-\frac{p_{1_{\sigma}}}{\pi^{2}c^{4}}B_{\epsilon_{\frac{1}{2}}}B_{\psi_{\frac{3}{2}}}-\frac{p_{1_{\sigma}}}{\pi^{2}c^{4}}B_{\epsilon_{\frac{3}{2}}}B_{\psi_{\frac{1}{2}}}+\frac{1}{\pi c}B_{\epsilon_{\frac{1}{2}}}B_{\psi_{\frac{1}{2}}}+\frac{p_{1_{\sigma}}}{\pi^{2}c^{2}}B_{\epsilon_{\frac{1}{2}}}B_{\psi_{\frac{1}{2}}}-\frac{1}{\pi c^{3}}B_{\psi_{\frac{1}{2}}}B_{\epsilon_{\frac{3}{2}}}\nonumber \\
	& =\frac{1}{\pi c}B_{\epsilon_{\frac{1}{2}}}B_{\psi_{\frac{1}{2}}}-\frac{1}{\pi c^{3}}B_{\epsilon_{\frac{1}{2}}}B_{\psi_{\frac{3}{2}}}-\frac{1}{\pi c^{3}}B_{\psi_{\frac{1}{2}}}B_{\epsilon_{\frac{3}{2}}}+\frac{p_{1_{\sigma}}}{\pi^{2}c^{2}}B_{\epsilon_{\frac{1}{2}}}B_{\psi_{\frac{1}{2}}}-\frac{p_{1_{\sigma}}}{\pi^{2}c^{4}}B_{\epsilon_{\frac{1}{2}}}B_{\psi_{\frac{3}{2}}}-\frac{p_{1_{\sigma}}}{\pi^{2}c^{4}}B_{\epsilon_{\frac{3}{2}}}B_{\psi_{\frac{1}{2}}}+\frac{p_{1_{\sigma}}}{\pi^{2}c^{6}}B_{\epsilon_{\frac{3}{2}}}B_{\psi_{\frac{3}{2}}}\nonumber \\
	& \approx \frac{1}{\pi c}B_{\epsilon_{\frac{1}{2}}}B_{\psi_{\frac{1}{2}}}-\frac{1}{\pi c^{3}}B_{\epsilon_{\frac{1}{2}}}B_{\psi_{\frac{3}{2}}}-\frac{1}{\pi c^{3}}B_{\epsilon_{\frac{3}{2}}}B_{\psi_{\frac{1}{2}}}+\frac{1}{\pi^{3}c^{3}}B_{\epsilon_{\frac{1}{2}}}^{2}B_{\psi_{\frac{1}{2}}}^{2},
\end{align}
\begin{align}
	p_{1_{\rho}} & =-\frac{p_{1_{\sigma}}}{\pi c^{3}}B_{\epsilon_{\frac{3}{2}}}+B_{\epsilon_{\frac{1}{2}}}+\frac{p_{1_{\sigma}}}{\pi c}B_{\epsilon_{\frac{1}{2}}}\nonumber \\
	& =-\frac{p_{1_{\sigma}}}{\pi c^{3}}B_{\epsilon_{\frac{3}{2}}}+B_{\epsilon_{\frac{1}{2}}}+\frac{p_{1_{\sigma}}}{\pi c}B_{\epsilon_{\frac{1}{2}}}\nonumber \\
	& \approx B_{\epsilon_{\frac{1}{2}}}+\frac{1}{\pi^{2}c^{2}}B_{\psi_{\frac{1}{2}}}B_{\epsilon_{\frac{1}{2}}}^{2}.
\end{align}
Substituting these final forms back into the $A_{1},\beta_{1}$ coefficients, we have 
\begin{align}
	\beta_{1_{\rho}} & =-\frac{p_{1_{\sigma}}}{\pi c^{3}},\\
	A_{1_{\rho}} & =1+\frac{p_{1_{\sigma}}}{\pi c},\\
	\beta_{1_{\sigma}} & =-\frac{p_{1_{\rho}}}{\pi c^{3}},\\
	A_{1_{\sigma}} & =\frac{p_{1_{\rho}}}{\pi c}-\frac{1}{\pi c^{3}}B_{\epsilon_{\frac{3}{2}}}.
\end{align}
It follows that the final expressions of  the dressed charge $1^{\dr}$ reads
\begin{align}
	1_{\rho}^{\dr}(k) & =1+\frac{1}{\pi^{2}c^{2}}B_{\psi_{\frac{1}{2}}}B_{\epsilon_{\frac{1}{2}}},\\
	1_{\sigma}^{\dr}(\Lambda) & =-\frac{1}{\pi c^{3}}B_{\epsilon_{\frac{1}{2}}}\Lambda^{2}+\frac{1}{\pi c}B_{\epsilon_{\frac{1}{2}}}+\frac{1}{\pi^{3}c^{3}}B_{\psi_{\frac{1}{2}}}B_{\epsilon_{\frac{1}{2}}}^{2}-\frac{1}{\pi c^{3}}B_{\epsilon_{\frac{3}{2}}}.\label{1-sigam-dress}
\end{align}
Using the same method, we can calculate  other dressed charges ($m^{\dr},k^{\dr},e^{\dr}$).
The wanted  results are given by 
\begin{align}
	m_{\rho}^{\dr}(k) & =-\frac{1}{\pi c^{3}}k^{2}B_{\psi_{\frac{1}{2}}}+\frac{1}{\pi c}B_{\psi_{\frac{1}{2}}}-\frac{1}{\pi c^{3}}B_{\psi_{\frac{3}{2}}},\\
	m_{\sigma}^{\dr}(\Lambda) & =1+\frac{1}{\pi^{2}c^{2}}B_{\epsilon_{\frac{1}{2}}}B_{\psi_{\frac{1}{2}}},\\
	k_{\rho}^{\dr}(k) & =k,\\
	k_{\sigma}^{\dr}(\Lambda) & =\frac{2\Lambda}{\pi c^{3}}B_{\epsilon_{\frac{3}{2}}}, \label{k-sigma-dress}\\
	e_{\rho}^{\dr}(k) & =\frac{k^{2}}{2}+\frac{1}{2}\frac{1}{\pi^{2}c^{2}}B_{\epsilon_{\frac{3}{2}}}B_{\psi_{\frac{1}{2}}},\\
	e_{\sigma}^{\dr}(\Lambda) & =-\frac{1}{2}\frac{1}{\pi c^{3}}B_{\epsilon_{\frac{3}{2}}}\Lambda^{2}+\frac{1}{2}\frac{1}{\pi c}B_{\epsilon_{\frac{3}{2}}}+\frac{1}{2}\frac{1}{\pi^{3}c^{3}}B_{\epsilon_{\frac{3}{2}}}B_{\psi_{\frac{1}{2}}}B_{\epsilon_{\frac{1}{2}}}-\frac{1}{2\pi c^{3}}B_{\epsilon_{\frac{5}{2}}}.
\end{align}


We now substitute the above  expressions of the dressed charges  into the integrated form of the Drude weight 
formula of Eq. (\ref{eq:Dij_fTiC_BF}). 
First, we calculate the Drude weight $D_{nn}$
\begin{alignat}{1}
	D_{nn}&=\frac{1}{2\pi}\int \theta_{\rho}(k)d_{k}\left(v_{\rm eff} 1_{\rho}^{\dr} 1_{\rho}^{\dr}\right)+\frac{1}{2\pi}\int \theta_{\sigma}(k)d_{\Lambda}\left(v_{\rm eff} 1_{\sigma}^{\dr}1_{\sigma}^{\dr}\right)\nonumber\\
	 & =\frac{1}{2\pi}\int \theta_{\rho}(k)d_{k}\left(k_{\rho}^{\dr}1_{\rho}^{\dr}\right)+\frac{1}{2\pi}\int \theta_{\sigma}(k)d_{\Lambda}\left(k_{\sigma}^{\dr}1_{\sigma}^{\dr}\right)\nonumber \\
	& =D_{nn}\left(1\right)+D_{nn}\left(2\right).
	\label{eq:Dnn_fTiC_BF}
\end{alignat}
The derivative of the integrand for the first term $D_{nn}(1)$ reads
\begin{align}
	d_{k}(k_{\rho}^{\dr}1_{\rho}^{\dr}) & =d_{k}\left[\left(1+\frac{1}{\pi^{2}c^{2}}B_{\psi_{\frac{1}{2}}}B_{\epsilon_{\frac{1}{2}}}\right)k\right]\nonumber \\
	& =1+\frac{1}{\pi^{2}c^{2}}B_{\psi_{\frac{1}{2}}}B_{\epsilon_{\frac{1}{2}}}.
\end{align}
Substituting this back into Eq. (\ref{eq:Dnn_fTiC_BF}) 
\begin{align}
	D_{nn}\left(1\right) & =\frac{1}{2\pi}\int dk\theta_{\rho}(k)d_{k}\left[\left(1+\frac{1}{\pi^{2}c^{2}}B_{\psi_{\frac{1}{2}}}B_{\epsilon_{\frac{1}{2}}}\right)k\right]\nonumber \\
	& =\frac{1}{2\pi}\int dk\theta_{\rho}(k)\left(1+\frac{1}{\pi^{2}c^{2}}B_{\psi_{\frac{1}{2}}}B_{\epsilon_{\frac{1}{2}}}\right)\nonumber \\
	& =\frac{1}{2\pi}\left(1+\frac{1}{\pi^{2}c^{2}}B_{\psi_{\frac{1}{2}}}B_{\epsilon_{\frac{1}{2}}}\right)B_{\epsilon_{\frac{1}{2}}}\nonumber \\
	& =\frac{1}{2\pi}\left(B_{\epsilon_{\frac{1}{2}}}+\frac{1}{\pi^{2}c^{2}}B_{\psi_{\frac{1}{2}}}B_{\epsilon_{\frac{1}{2}}}^{2}\right),\\
	D_{nn}\left(2\right) & =\frac{1}{2\pi}\int \theta_{\sigma}(k)d_{\Lambda}\left(\frac{2\Lambda}{\pi c^{3}}B_{\epsilon_{\frac{3}{2}}}\right)\left(-\frac{B_{\epsilon_{\frac{1}{2}}}}{\pi c^{3}}\Lambda^{2}+\frac{1}{\pi c}B_{\epsilon_{\frac{1}{2}}}+\frac{1}{\pi^{3}c^{3}}B_{\psi_{\frac{1}{2}}}B_{\epsilon_{\frac{1}{2}}}^{2}-\frac{B_{\epsilon_{\frac{3}{2}}}}{\pi c^{3}}\right)\nonumber \\
	& \approx 0.
\end{align}
In the calculation of $D_{nn}\left(2\right) $, the results (\ref{1-sigam-dress}) and (\ref{k-sigma-dress}) were used. 
Thus, the final result for $D_{nn}$ is given by 
\begin{align}
	D_{nn} & =\frac{1}{2\pi}\left(B_{\epsilon_{\frac{1}{2}}}+\frac{1}{\pi^{2}c^{2}}B_{\psi_{\frac{1}{2}}}B_{\epsilon_{\frac{1}{2}}}^{2}\right)\nonumber \\
	& =\frac{1}{2\pi}B_{\epsilon_{\frac{1}{2}}}+\frac{1}{2\pi^{3}c^{2}}B_{\psi_{\frac{1}{2}}}B_{\epsilon_{\frac{1}{2}}}^{2}.
\end{align}
Following the same procedure, we expand the expressions up to order $1/c^3$ to obtain the remaining Drude weights:
\begin{alignat}{1}
	D_{ne}  =&\frac{1}{2\pi}\int \theta_{\rho}(k)d_{k}\left(k_{\rho}^{\dr}e_{\rho}^{\dr}\right)+\frac{1}{2\pi}\int d\Lambda \theta_{\sigma}(k)d_{\Lambda}\left(k_{\sigma}^{\dr}e_{\sigma}^{\dr}\right)\nonumber \\
	\approx&\frac{3}{4\pi}B_{\epsilon_{\frac{3}{2}}}+\frac{1}{4\pi^{3}c^{2}}B_{\epsilon_{\frac{3}{2}}}B_{\psi_{\frac{1}{2}}}B_{\epsilon_{\frac{1}{2}}},\\
	D_{nm}  
	=&\frac{1}{2\pi}\int dk\theta_{\rho}(k)d_{k}\left(k_{\rho}^{\dr}m_{\rho}^{\dr}\right)+\frac{1}{2\pi}\int d\Lambda \theta_{\sigma}(k)d_{\Lambda}\left(k_{\sigma}^{\dr}m_{\sigma}^{\dr}\right)\nonumber \\
	\approx&\frac{1}{2\pi^{2}c}B_{\psi_{\frac{1}{2}}}B_{\epsilon_{\frac{1}{2}}}-\frac{1}{2\pi^{2}c^{3}}B_{\psi_{\frac{3}{2}}}B_{\epsilon_{\frac{1}{2}}}-\frac{1}{2\pi^{2}c^{3}}B_{\psi_{\frac{1}{2}}}B_{\epsilon_{\frac{3}{2}}},\\
	D_{mm} =&\frac{1}{2\pi}\int \theta_{\rho}(k)d_{k}\left(\frac{k_{\rho}^{\dr}}{1_{\rho}^{\dr}}m_{\rho}^{\dr}m_{\rho}^{\dr}\right)+\frac{1}{2\pi}\int d\Lambda \theta_{\sigma}(k)d_{\Lambda}\left(\frac{k_{\sigma}^{\dr}}{1_{\sigma}^{\dr}}m_{\sigma}^{\dr}m_{\sigma}^{\dr}\right)\nonumber \\
	\approx&\frac{1}{2\pi^{3}c^{2}}B_{\psi_{\frac{1}{2}}}^{2}B_{\epsilon_{\frac{1}{2}}}+\frac{1}{\pi c^{2}}B_{\epsilon_{\frac{3}{2}}}B_{\epsilon_{\frac{1}{2}}}^{-1}B_{\psi_{\frac{1}{2}}},\\
	D_{me} =&\frac{1}{2\pi}\int \theta_{\rho}(k)d_{k}\left(\frac{k_{\rho}^{\dr}}{1_{\rho}^{\dr}}m_{\rho}^{\dr}e_{\rho}^{\dr}\right)+\frac{1}{2\pi}\int \theta_{\sigma}(k)d_{\Lambda}\left(\frac{k_{\sigma}^{\dr}}{1_{\sigma}^{\dr}}m_{\sigma}^{\dr}e_{\sigma}^{\dr}\right)\nonumber \\
	\approx&\frac{3}{4\pi^{2}c}B_{\psi_{\frac{1}{2}}}B_{\epsilon_{\frac{3}{2}}}-\frac{5}{4\pi^{2}c^{3}}B_{\epsilon_{\frac{5}{2}}}B_{\psi_{\frac{1}{2}}}-\frac{3}{4\pi^{2}c^{3}}B_{\epsilon_{\frac{3}{2}}}B_{\psi_{\frac{3}{2}}}-\frac{1}{2\pi^{2}c^{3}}B_{\epsilon_{\frac{1}{2}}}^{-1}B_{\epsilon_{1}}B_{\epsilon_{\frac{3}{2}}}B_{\psi_{1}}\nonumber \\
	&+\frac{1}{2\pi^{2}c^{3}}B_{\epsilon_{\frac{1}{2}}}^{-1}B_{\epsilon_{\frac{3}{2}}}^{2}B_{\psi_{\frac{1}{2}}}-\frac{1}{\pi^{4}c^{3}}B_{\epsilon_{\frac{1}{2}}}B_{\epsilon_{\frac{3}{2}}}B_{\psi_{\frac{1}{2}}}^{2}.
\end{alignat}

The finite-temperature strong-coupling expansions are numerically verified in Fig.~\ref{fig:BF_fintep_strlim_app}. Here, we plot the cross and bosonic sector components ($D_{nm}$, $D_{mm}$, $D_{me}$), supplementing the primary transport channels ($D_{nn}$ and $D_{ne}$) discussed in Fig.~7.

\begin{figure}[H]
	\centering
	\includegraphics[width=1\linewidth]{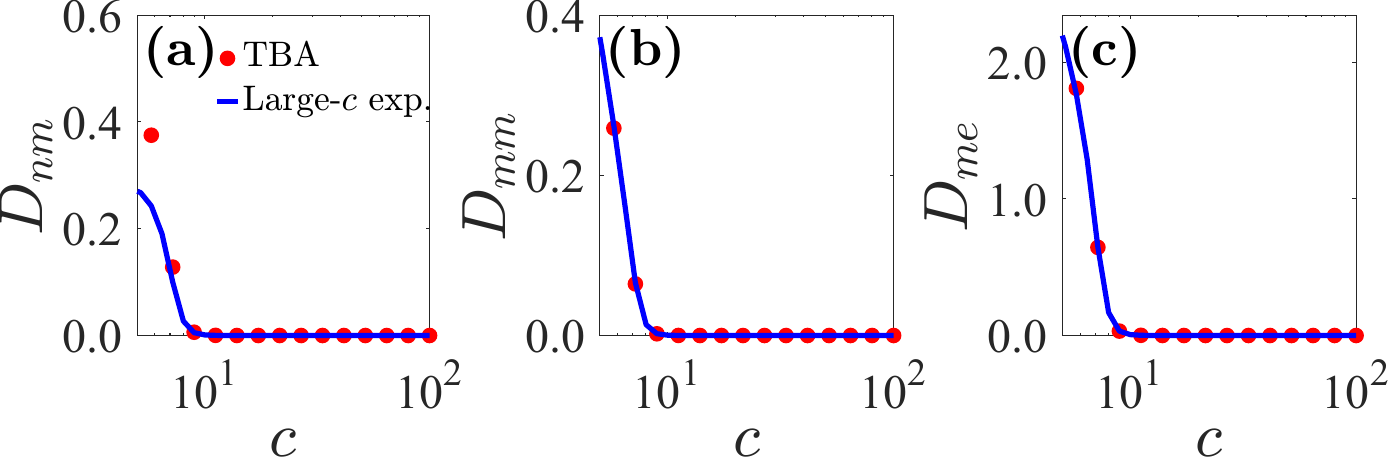}
	\caption{Supplemental plots for the Drude weights in the Bose-Fermi mixture at finite temperature ($T=0.1$, $\mu=5$, and $\mathcal{H}=1$). Circles represent the exact TBA numerical solutions for the components: (a) $D_{nm}$, (b) $D_{mm}$, and (c) $D_{me}$. The blue lines denote the analytical large-$c$ expansions. The primary components $D_{nn}$ and $D_{ne}$ are detailed in Fig.~7 of the main text.}
	\label{fig:BF_fintep_strlim_app}
\end{figure}

\subsection{Drude Weights at quantum criticality }

The quantum critical (QC) state describes the strong coupling of thermal and quantum fluctuations when a quantum phase transition (QPT) occurs at $T\rightarrow0$. A key feature of quantum criticality  is manifested by that scaling functions of physical properties, from which  critical exponents are conceived. 
 Here, we show that the universal scaling functions for the Drude weights can be derived from the low-temperature thermodynamic properties which we calculated in the previous section. 
 In the QC region, defined by $T\gg \mu-\mu_{c}|$ and $T\gg \mathcal{H}-\mathcal{H}_{c}|$, we expand our  expressions of $D_{ij}$ to get their universal scaling forms.

As a preliminary step, we substitute the Polylogarithm forms of the coefficients $\beta_{\epsilon},\beta_{\psi},A_{\epsilon},A_{\psi}$ and the integrals $B_{s}$ into the final $D_{ij}$ expressions. This yields complex expressions in terms of $T$ and the $Li_{s}(z)$ functions, For the Drude weight $D_{nn}$
\begin{align}
	D_{nn}  =&\frac{1}{2\pi}B_{\epsilon_{\frac{1}{2}}}+\frac{1}{2\pi^{3}c^{2}}B_{\psi_{\frac{1}{2}}}B_{\epsilon_{\frac{1}{2}}}^{2}\nonumber \\
	=&-\frac{1}{2\pi}\beta_{\epsilon}^{-\frac{1}{2}}T^{\frac{1}{2}}\pi^{\frac{1}{2}}Li_{\frac{1}{2}}\left(-e^{A_{\epsilon}/T}\right)+\frac{1}{2\pi^{3}c^{2}}\left[-\beta_{\psi}^{-\frac{1}{2}}T^{\frac{1}{2}}\pi^{\frac{1}{2}}Li_{\frac{1}{2}}\left(-e^{A_{\psi}/T}\right)\right]\left[-\beta_{\epsilon}^{-\frac{1}{2}}T^{\frac{1}{2}}\pi^{\frac{1}{2}}Li_{\frac{1}{2}}\left(-e^{A_{\epsilon}/T}\right)\right]^{2}\nonumber \\
	=&-\frac{1}{2\pi}\sqrt{2}\pi^{\frac{1}{2}}T^{\frac{1}{2}}Li_{\frac{1}{2}}\left(-e^{A_{\epsilon}/T}\right)\left[1+\frac{T^{\frac{3}{2}}}{\pi^{\frac{1}{2}}\beta_{\psi}^{\frac{1}{2}}c^{3}}Li_{\frac{3}{2}}\left(-e^{A_{\psi}/T}\right)\right] \nonumber\\
	&+\frac{1}{2\pi^{3}c^{2}}\left\{ -\left[1+\frac{2T^{\frac{3}{2}}}{\pi^{\frac{1}{2}}\beta_{\psi}^{\frac{1}{2}}c^{3}}Li_{\frac{3}{2}}\left(-e^{A_{\psi}/T}\right)\right]\beta_{\psi}^{-\frac{1}{2}}T^{\frac{3}{2}}\pi^{\frac{3}{2}}Li_{\frac{1}{2}}\left(-e^{A_{\psi}/T}\right)Li_{\frac{1}{2}}^{2}\left(-e^{A_{\epsilon}/T}\right)\right\} \nonumber \\
	=&-\frac{T^{\frac{1}{2}}}{\sqrt{2}\pi^{\frac{1}{2}}}Li_{\frac{1}{2}}\left(-e^{A_{\epsilon}/T}\right)-\frac{T^{\frac{3}{2}}}{\pi^{\frac{3}{2}}\beta_{\psi}^{\frac{1}{2}}c^{2}}Li_{\frac{1}{2}}\left(-e^{A_{\psi}/T}\right)Li_{\frac{1}{2}}^{2}\left(-e^{A_{\epsilon}/T}\right)\nonumber \\
	&-\frac{T^{2}}{\sqrt{2}\pi\beta_{\psi}^{\frac{1}{2}}c^{3}}Li_{\frac{1}{2}}\left(-e^{A_{\epsilon}/T}\right)Li_{\frac{3}{2}}\left(-e^{A_{\psi}/T}\right)-\frac{2T^{3}}{\pi^{2}\beta_{\psi}c^{5}}Li_{\frac{1}{2}}\left(-e^{A_{\psi}/T}\right)Li_{\frac{1}{2}}^{2}\left(-e^{A_{\epsilon}/T}\right)Li_{\frac{3}{2}}\left(-e^{A_{\psi}/T}\right)\nonumber \\
	\approx&-\frac{T^{\frac{1}{2}}}{\sqrt{2\pi}}Li_{\frac{1}{2}}\left(-e^{A_{\epsilon}/T}\right)-\frac{T^{\frac{3}{2}}}{\pi^{\frac{3}{2}}\beta_{\psi}^{\frac{1}{2}}c^{2}}Li_{\frac{1}{2}}\left(-e^{A_{\psi}/T}\right)Li_{\frac{1}{2}}^{2}\left(-e^{A_{\epsilon}/T}\right).
	\label{eq:Dnn_fTiC_an_BF}
\end{align}
Following the same approximation procedure for the other matrix elements:
\begin{align}
	D_{ne} & \approx 
	-\frac{3T^{\frac{3}{2}}}{2\sqrt{2\pi}}Li_{\frac{3}{2}}\left(-e^{A_{\epsilon}/T}\right)\left[1+\frac{3T^{\frac{3}{2}}}{\pi^{\frac{1}{2}}\beta_{\psi}^{\frac{1}{2}}c^{3}}Li_{\frac{3}{2}}\left(-e^{A_{\psi}/T}\right)\right]
		-\frac{T^{\frac{5}{2}}\beta_{\psi}^{-\frac{1}{2}}}{2\pi^{\frac{3}{2}}c^{2}}Li_{\frac{3}{2}}\left(-e^{A_{\epsilon}/T}\right)Li_{\frac{1}{2}}\left(-e^{A_{\epsilon}/T}\right)Li_{\frac{1}{2}}\left(-e^{A_{\psi}/T}\right)
	\label{eq:Dne_fTiC_an_BF}\\
	D_{nm} & \approx \frac{1}{\sqrt{2}\pi}\frac{T}{\beta_{\psi}^{\frac{1}{2}}c}Li_{\frac{1}{2}}\left(-e^{A_{\epsilon}/T}\right)Li_{\frac{1}{2}}\left(-e^{A_{\psi}/T}\right),
	\label{eq:Dnm_fTiC_an_BF}\\
	D_{mm} & \approx -\frac{T^{\frac{3}{2}}}{\sqrt{2}\pi^{\frac{3}{2}}c^{2}}\beta_{\psi}^{-1}Li_{\frac{1}{2}}^{2}\left(-e^{A_{\psi}/T}\right)Li_{\frac{1}{2}}\left(-e^{A_{\epsilon}/T}\right)-\frac{T^{\frac{3}{2}}}{\pi^{\frac{1}{2}}c^{2}}\frac{Li_{\frac{3}{2}}\left(-e^{A_{\epsilon}/T}\right)}{Li_{\frac{1}{2}}\left(-e^{A_{\epsilon}/T}\right)}\beta_{\psi}^{-\frac{1}{2}}Li_{\frac{1}{2}}\left(-e^{A_{\psi}/T}\right),
	\label{eq:Dmm_fTiC_an_BF}\\
	D_{me} & \approx \frac{3T^{2}}{2^{3/2}\pi c}\beta_{\psi}^{-\frac{1}{2}}Li_{\frac{1}{2}}\left(-e^{A_{\psi}/T}\right)Li_{\frac{3}{2}}\left(-e^{A_{\epsilon}/T}\right).
	\label{eq:Dme_fTiC_an_BF}
\end{align}
We now analyze these expressions at the two QPT points.

\subsubsection*{Quantum Phase Transition from Vacuum to Free Fermi Gas}

At the Vacuum-to-Fermi boundary ($\mu_{c}=-\mathcal{H}_{c}/2$), the $\sigma$-component (Boson) is fully gapped ($A_{\psi}\to-\infty$), causing $Li_{n}(-e^{A_{\psi}/T})=0$.
Consequently, $A_{\epsilon}$  Eq.~(\ref{eq:Aepsilon}) becomes $A_{\epsilon}\approx\mu+\mathcal{H}/2$.   
In the critical region
\begin{align}
	A_{\epsilon} & =(\mu_{c}+\Delta\mu)+(\mathcal{H}_{c}+\Delta\mathcal{H})/2 \nonumber \\
	& =(\mu_{c}+\mathcal{H}_{c}/2)+\Delta\mu+\Delta\mathcal{H}/2 \nonumber \\
	& =\Delta\mu+\Delta\mathcal{H}/2,
\end{align}
where $\Delta\mu=\mu-\mu_{c},\Delta\mathcal{H}=\mathcal{H}-\mathcal{H}_{c}$. Substituting $Li_{n}(-e^{A_{\psi}/T})=0$ and $A_{\epsilon}=\Delta\mu+\Delta\mathcal{H}/2$ into the approximated $D_{ij}$ formulas Eq.~(\ref{eq:Dnn_fTiC_an_BF}-\ref{eq:Dme_fTiC_an_BF}):
\begin{align}
	D_{nn} & \approx-\frac{1}{\sqrt{2\pi}}T^{\frac{1}{2}}Li_{\frac{1}{2}}\left(-e^{\frac{\Delta\mu+\Delta\mathcal{H}/2}{T}}\right),\\
	D_{ne} & \approx-\frac{3}{2\sqrt{2\pi}}T^{\frac{3}{2}}Li_{\frac{3}{2}}\left(-e^{\frac{\Delta\mu+\Delta\mathcal{H}/2}{T}}\right),\\
	D_{nm} & =0,\\
	D_{mm} & =0,\\
	D_{me} & =0.
\end{align}
These equations describe the scaling of transport near the V-F phase transition.

To reveal the universal scaling behavior independent of the interaction strength $c$, we introduce the natural binding energy scale $\epsilon_{b} = c^2/2$ (setting $\hbar=m=1$) and define the dimensionless thermodynamic variables $\tilde{\mu} \equiv \mu/\epsilon_{b}$, $\tilde{\mathcal{H}} \equiv \mathcal{H}/\epsilon_{b}$, and $\mathcal{T} \equiv T/\epsilon_{b}$. The Drude weights are correspondingly rescaled by their macroscopic dimensional limits, i.e., $\mathcal{D}_{nn} \equiv D_{nn}/c$ and $\mathcal{D}_{ne} \equiv D_{ne}/|c\epsilon_{b}|$. 

Substituting the scaling relations $\Delta\mu+\Delta\mathcal{H}/2 = \epsilon_{b}(\Delta\tilde{\mu}+\Delta\tilde{\mathcal{H}}/2)$ and $T = \epsilon_{b}\mathcal{T}$ into the above equations, the interaction parameter $c$ factorizes perfectly. Taking the particle Drude weight as an example:
\begin{align}
	D_{nn} &= -\frac{1}{\sqrt{2\pi}} \left( \frac{c^2}{2}\mathcal{T} \right)^{\frac{1}{2}} \Li_{\frac{1}{2}}\left(-e^{\frac{\epsilon_{b}(\Delta\tilde{\mu}+\Delta\tilde{\mathcal{H}}/2)}{\epsilon_{b}\mathcal{T}}}\right) \nonumber \\
	&= -\frac{c}{2\sqrt{\pi}}\mathcal{T}^{\frac{1}{2}}\Li_{\frac{1}{2}}\left(-e^{\frac{\Delta\tilde{\mu}+\Delta\tilde{\mathcal{H}}/2}{\mathcal{T}}}\right).
\end{align}
Dividing by $c$, we arrive at the purely dimensionless universal scaling function:
\begin{equation}
	\mathcal{D}_{nn} = -\frac{1}{2\sqrt{\pi}}\mathcal{T}^{\frac{1}{2}}\Li_{\frac{1}{2}}\left(-e^{\frac{\Delta\tilde{\mu}+\Delta\tilde{\mathcal{H}}/2}{\mathcal{T}}}\right).
\end{equation}
Following the exact same procedure with the energy characteristic scale $|c\epsilon_{b}|$, the dimensionless energy-current Drude weight $\mathcal{D}_{ne}$ is obtained as:
\begin{equation}
	\mathcal{D}_{ne} = -\frac{3}{4\sqrt{\pi}}\mathcal{T}^{\frac{3}{2}}\Li_{\frac{3}{2}}\left(-e^{\frac{\Delta\tilde{\mu}+\Delta\tilde{\mathcal{H}}/2}{\mathcal{T}}}\right).
\end{equation}

\subsubsection*{Quantum Phase Transition from Free  Fermi gas  to Bose-Fermi Mixture}

At the second phase transition point  from the free Fermi gas  phase to the Bose-Fermi mixed phase, $\mu_{c}$ and $\mathcal{H}_{c}$ satisfy the condition  $\mathcal{H}_{c}\approx\frac{10\sqrt{2}}{3\pi c}\left(\mu_{c}+\frac{\mathcal{H}_{c}}{2}\right)^{3/2}$. 
The physical condition is $\mathcal{H}>0$, $\mu_{F}/T\to\infty$ or $(\mu+\mathcal{H}/2)/T\to\infty$. 
In this limit, $A_{\epsilon}\approx\mu_{F}$ is large, while $A_{\psi}$ is the small parameter governing the transition
\begin{align}
	A_{\epsilon} & \approx\mu_{F},\\
	A_{\psi} & \approx-\mathcal{H}-\frac{\sqrt{2}T^{\frac{3}{2}}}{\sqrt{\pi}c}Li_{\frac{3}{2}}\left(-e^{\frac{\mu+\mathcal{H}/2}{T}}\right).
\end{align}
We use the asymptotic expansion $\lim_{z\rightarrow\infty}Li_{s}\left(-e^{z}\right)=-\frac{z^{s}}{\Gamma\left(s+1\right)},\Gamma\left(n+\frac{1}{2}\right)=\frac{\sqrt{\pi}}{2^{n}}\left(2n-1\right),\Gamma\left(\frac{1}{2}\right)=\sqrt{\pi}$.
\begin{align}
	Li_{\frac{1}{2}}\left(-e^{\frac{\mu_{F}}{T}}\right) & =-\frac{\left(\frac{\mu+\mathcal{H}/2}{T}\right)^{1/2}}{\Gamma\left(\frac{3}{2}\right)}=-\frac{2\left(\frac{\mu+\mathcal{H}/2}{T}\right)^{1/2}}{\sqrt{\pi}}.
\end{align}
Substituting this into the $A_{\psi}$ expression, we have 
\begin{align}
	A_{\psi} & =-\mathcal{H}-\frac{\sqrt{2}T^{\frac{3}{2}}}{\sqrt{\pi}c}Li_{\frac{3}{2}}\left(-e^{\frac{\mu+\mathcal{H}/2}{T}}\right)\nonumber \\
	& \approx-\mathcal{H}+\frac{\sqrt{2}T^{\frac{3}{2}}}{\sqrt{\pi}c}\frac{\left(\frac{\mu+\mathcal{H}/2}{T}\right)^{3/2}}{\Gamma\left(\frac{5}{2}\right)}\nonumber \\
	& =-\mathcal{H}+\frac{4\sqrt{2}}{3\pi c}\left(\mu+\mathcal{H}/2\right)^{3/2}.
\end{align}
At the phase transition point, we expand $(\mu+\mathcal{H}/2)^{3/2}$
around $(\mu_{c},\mathcal{H}_{c})$:
\begin{align}
	\left(\mu+\mathcal{H}/2\right)^{3/2} & =\left(\mu_{c}+\mathcal{H}_{c}/2+\Delta\mu+\Delta\mathcal{H}/2\right)^{3/2}\nonumber \\
	& \approx\left(\mu_{c}+\mathcal{H}_{c}/2\right)^{3/2}+\frac{3}{2}\left(\mu_{c}+\mathcal{H}_{c}/2\right)^{1/2}\Delta\mu+\frac{3}{4}\left(\mu_{c}+\mathcal{H}_{c}/2\right)^{1/2}\Delta\mathcal{H}.
\end{align}
Using the boundary condition $\mathcal{H}_{c}=\frac{4\sqrt{2}}{3\pi c}(\mu_{c}+\mathcal{H}_{c}/2)^{3/2}$,
the $\mathcal{H}_{c}$ terms are canceled 
\begin{align}
	A_{\psi} & =-\mathcal{H}+\frac{4\sqrt{2}}{3\pi c}\left[\left(\mu_{c}+\mathcal{H}_{c}/2\right)^{3/2}+\frac{3}{2}\left(\mu_{c}+\mathcal{H}_{c}/2\right)^{1/2}\Delta\mu+\frac{3}{4}\left(\mu_{c}+\mathcal{H}_{c}/2\right)^{1/2}\Delta\mathcal{H}\right]\nonumber \\
	& =-\mathcal{H}+\frac{4\sqrt{2}}{3\pi c}\left(\mu_{c}+\mathcal{H}_{c}/2\right)^{3/2}+\frac{2\sqrt{2}}{\pi c}\left(\mu_{c}+\mathcal{H}_{c}/2\right)^{1/2}\Delta\mu+\frac{\sqrt{2}}{\pi c}\left(\mu_{c}+\mathcal{H}_{c}/2\right)^{1/2}\Delta\mathcal{H}\nonumber \\
	& =-\Delta\mathcal{H}+\frac{2\sqrt{2}}{\pi c}\left(\mu_{c}+\mathcal{H}_{c}/2\right)^{1/2}\Delta\mu+\frac{\sqrt{2}}{\pi c}\left(\mu_{c}+\mathcal{H}_{c}/2\right)^{1/2}\Delta\mathcal{H}\nonumber \\
	& =\frac{2\sqrt{2}}{\pi c}\left(\mu_{c}+\mathcal{H}_{c}/2\right)^{1/2}\Delta\mu+\left[-1+\frac{\sqrt{2}}{\pi c}\left(\mu_{c}+\mathcal{H}_{c}/2\right)^{1/2}\right]\Delta\mathcal{H}\nonumber \\
	& =R_{0}\Delta\mu+S_{0}\Delta\mathcal{H},
\end{align}
where the scaling coefficients are:
\begin{align}
	R_{0} & =\frac{2\sqrt{2}}{\pi c}\left(\mu_{c}+\mathcal{H}_{c}/2\right)^{1/2},\\
	S_{0} & =-1+\frac{\sqrt{2}}{\pi c}\left(\mu_{c}+\mathcal{H}_{c}/2\right)^{1/2}.
\end{align}
For $\beta_{\psi}$, we also expand the polylog function
\begin{align}
	\beta_{\psi} & =\frac{2p}{c^{3}}=-\frac{\sqrt{2}T^{\frac{3}{2}}}{\sqrt{\pi}c^{3}}Li_{\frac{3}{2}}\left(-e^{\frac{\mu_{F}}{T}}\right)\nonumber \\
	& \approx \frac{\sqrt{2}T^{\frac{3}{2}}}{\sqrt{\pi}c^{3}}\frac{\left(\frac{\mu+\mathcal{H}/2}{T}\right)^{3/2}}{\Gamma\left(\frac{5}{2}\right)} =\frac{4\sqrt{2}}{3\pi c^{3}}\left(\mu+\mathcal{H}/2\right)^{3/2}.
\end{align}

Substituting these expansions of $\beta_{\psi}$, $A_{\epsilon}\to\mu_{F}$,
and $A_{\psi}\to R_{0}\Delta\mu+S_{0}\Delta\mathcal{H}$ into the DWs Eq.~(\ref{eq:Dnn_fTiC_an_BF}-\ref{eq:Dme_fTiC_an_BF}), we have the scaling forms near the critical point. As an explicit example, the asymptotic expansion for the particle Drude weight $D_{nn}$ is carried out step-by-step as follows:
\begin{align}
	D_{nn} & =-\frac{T^{\frac{1}{2}}}{\sqrt{2\pi}}Li_{\frac{1}{2}}\left(-e^{A_{\epsilon}/T}\right)-\frac{T^{\frac{3}{2}}}{\pi^{\frac{3}{2}}\beta_{\psi}^{\frac{1}{2}}c^{2}}Li_{\frac{1}{2}}\left(-e^{A_{\psi}/T}\right)Li_{\frac{1}{2}}^{2}\left(-e^{A_{\epsilon}/T}\right)\nonumber \\
	& \approx-\frac{T^{\frac{1}{2}}}{\sqrt{2\pi}}Li_{\frac{1}{2}}\left(-e^{\mu_{F}/T}\right)-\frac{T^{\frac{3}{2}}}{\pi^{\frac{3}{2}}\beta_{\psi}^{\frac{1}{2}}c^{2}}Li_{\frac{1}{2}}\left(-e^{\frac{R_{0}\Delta\mu+S_{0}\Delta\mathcal{H}}{T}}\right)Li_{\frac{1}{2}}^{2}\left(-e^{\mu_{F}/T}\right)\nonumber \\
	& \approx\frac{T^{\frac{1}{2}}}{\sqrt{2\pi}}\frac{\left(\frac{\mu_{c}+\mathcal{H}_{c}/2}{T}\right)^{1/2}}{\Gamma\left(\frac{3}{2}\right)}-\frac{T^{\frac{3}{2}}}{\pi^{\frac{3}{2}}\beta_{\psi}^{\frac{1}{2}}c^{2}}Li_{\frac{1}{2}}\left(-e^{\frac{R_{0}\Delta\mu+S_{0}\Delta\mathcal{H}}{T}}\right)\left[-\frac{\left(\frac{\mu_{c}+\mathcal{H}_{c}/2}{T}\right)^{1/2}}{\Gamma\left(\frac{3}{2}\right)}\right]^{2}\nonumber \\
	& =\frac{T^{\frac{1}{2}}}{\sqrt{2\pi}}\frac{2\left(\frac{\mu_{c}+\mathcal{H}_{c}/2}{T}\right)^{1/2}}{\sqrt{\pi}}-\frac{T^{\frac{3}{2}}}{\pi^{\frac{3}{2}}\beta_{\psi}^{\frac{1}{2}}c^{2}}Li_{\frac{1}{2}}\left(-e^{\frac{R_{0}\Delta\mu+S_{0}\Delta\mathcal{H}}{T}}\right)\frac{4\left(\frac{\mu_{c}+\mathcal{H}_{c}/2}{T}\right)}{\pi}\nonumber \\
	& =\frac{\sqrt{2}\left(\mu_{c}+\mathcal{H}_{c}/2\right)^{1/2}}{\pi}-\frac{4T^{\frac{1}{2}}\left(\mu_{c}+\mathcal{H}_{c}/2\right)}{\pi^{\frac{5}{2}}\beta_{\psi}^{\frac{1}{2}}c^{2}}Li_{\frac{1}{2}}\left(-e^{\frac{R_{0}\Delta\mu+S_{0}\Delta\mathcal{H}}{T}}\right)\nonumber \\
	& =\frac{\sqrt{2}\left(\mu_{c}+\mathcal{H}_{c}/2\right)^{1/2}}{\pi}-\frac{4T^{\frac{1}{2}}\left(\mu_{c}+\mathcal{H}_{c}/2\right)}{\pi^{\frac{5}{2}}\sqrt{\frac{4\sqrt{2}}{3\pi c^{3}}\left(\mu_{c}+\mathcal{H}_{c}/2\right)^{3/2}}c^{2}}Li_{\frac{1}{2}}\left(-e^{\frac{R_{0}\Delta\mu+S_{0}\Delta\mathcal{H}}{T}}\right)\nonumber \\
	& =\frac{\sqrt{2}\left(\mu_{c}+\mathcal{H}_{c}/2\right)^{1/2}}{\pi}-\frac{\sqrt{6\sqrt{2}}T^{\frac{1}{2}}\left(\mu_{c}+\mathcal{H}_{c}/2\right)^{1/4}}{\pi^{2}\sqrt{c}}Li_{\frac{1}{2}}\left(-e^{\frac{R_{0}\Delta\mu+S_{0}\Delta\mathcal{H}}{T}}\right).
\end{align}
Following a similar procedure for the remaining components, we obtain the scaling expressions:
\begin{align}
	D_{ne} & \approx \frac{\sqrt{2}}{\pi}\left(\mu+\frac{\mathcal{H}}{2}\right)^{3/2}
	-\frac{2T^{\frac{1}{2}}}{\pi^{2}\sqrt{3\sqrt{2}c}}\left(\mu+\frac{\mathcal{H}}{2}\right)^{\frac{5}{4}}Li_{\frac{1}{2}}\left(-e^{\frac{R_{0}\Delta\mu+S_{0}\Delta\mathcal{H}}{T}}\right) \nonumber \\
	& \quad \,\,  -\frac{3T^{\frac{3}{2}}}{2\pi\sqrt{\frac{2\sqrt{2}}{3c}}\left(\mu+\frac{\mathcal{H}}{2}\right)^{1/4}}Li_{\frac{3}{2}}\left(-e^{\frac{R_{0}\Delta\mu+S_{0}\Delta\mathcal{H}}{T}}\right), \\
	D_{nm} & \approx -\frac{T^{\frac{1}{2}}}{\pi\sqrt{\frac{2\sqrt{2}}{3c}}\left(\mu+\mathcal{H}/2\right)^{1/4}}Li_{\frac{1}{2}}\left(-e^{\frac{R_{0}\Delta\mu+S_{0}\Delta\mathcal{H}}{T}}\right), \\
	D_{mm} &  \approx -\frac{T^{\frac{1}{2}}\left(\mu+\frac{\mathcal{H}}{2}\right)^{1/4}}{\sqrt{3\sqrt{2}c}}Li_{\frac{1}{2}}\left(-e^{\frac{R_{0}\Delta\mu+S_{0}\Delta\mathcal{H}}{T}}\right)+\frac{3T}{4\pi\left(\mu+\mathcal{H}/2\right)\frac{1}{c}}Li_{\frac{1}{2}}^{2}\left(-e^{\frac{R_{0}\Delta\mu+S_{0}\Delta\mathcal{H}}{T}}\right), \\
	D_{me} & \approx -\frac{T^{\frac{1}{2}}\left(\mu+\frac{\mathcal{H}}{2}\right)^{3/4}}{\pi\sqrt{\frac{2\sqrt{2}}{3c}}}Li_{\frac{1}{2}}\left(-e^{\frac{R_{0}\Delta\mu+S_{0}\Delta\mathcal{H}}{T}}\right).
\end{align}

Following the same non-dimensionalization procedure established for the first quantum critical point, we introduce the dimensionless units $\tilde{\mu} \equiv \mu/\epsilon_{b}$, $\tilde{\mathcal{H}} \equiv \mathcal{H}/\epsilon_{b}$, and $\mathcal{T} \equiv T/\epsilon_{b}$. By rescaling the Drude weights with their respective dimensional factors and retaining only the leading-order terms,, we obtain the completely dimensionless universal scaling functions for the second critical point:
\begin{align}
	\mathcal{D}_{nn} & \equiv \frac{D_{nn}}{c} = \frac{\left(\tilde{\mu}_{c}+\tilde{\mathcal{H}}_{c}/2\right)^{1/2}}{\pi}-\frac{\sqrt{3}\mathcal{T}^{1/2}\left(\tilde{\mu}_{c}+\tilde{\mathcal{H}}_{c}/2\right)^{1/4}}{\pi^{2}}\text{Li}_{\frac{1}{2}}\left(-e^{\frac{\tilde{R}_{0}\Delta\tilde{\mu}+\tilde{S}_{0}\Delta\tilde{\mathcal{H}}}{\mathcal{T}}}\right), \\
	\mathcal{D}_{ne} & \equiv \frac{D_{ne}}{c\epsilon_{b}} = \frac{1}{\pi}\left(\tilde{\mu}+\frac{\tilde{\mathcal{H}}}{2}\right)^{3/2}
	-\frac{\mathcal{T}^{1/2}\left(\tilde{\mu}+\frac{\tilde{\mathcal{H}}}{2}\right)^{5/4}}{\sqrt{3}\pi^{2}}\text{Li}_{\frac{1}{2}}\left(-e^{\frac{\tilde{R}_{0}\Delta\tilde{\mu}+\tilde{S}_{0}\Delta\tilde{\mathcal{H}}}{\mathcal{T}}}\right) \\
	\mathcal{D}_{nm} & \equiv \frac{D_{nm}}{c} = -\frac{\sqrt{3}\mathcal{T}^{1/2}}{2\pi\left(\tilde{\mu}+\frac{\tilde{\mathcal{H}}}{2}\right)^{1/4}}\text{Li}_{\frac{1}{2}}\left(-e^{\frac{\tilde{R}_{0}\Delta\tilde{\mu}+\tilde{S}_{0}\Delta\tilde{\mathcal{H}}}{\mathcal{T}}}\right), \\
	\mathcal{D}_{mm} & \equiv \frac{D_{mm}}{c} = -\frac{\mathcal{T}^{1/2}\left(\tilde{\mu}+\frac{\tilde{\mathcal{H}}}{2}\right)^{1/4}}{2\sqrt{3}}\text{Li}_{\frac{1}{2}}\left(-e^{\frac{\tilde{R}_{0}\Delta\tilde{\mu}+\tilde{S}_{0}\Delta\tilde{\mathcal{H}}}{\mathcal{T}}}\right) ,\\
	\mathcal{D}_{me} & \equiv \frac{D_{me}}{c\epsilon_{b}} = -\frac{\sqrt{3}\mathcal{T}^{1/2}\left(\tilde{\mu}+\frac{\tilde{\mathcal{H}}}{2}\right)^{3/4}}{2\pi}\text{Li}_{\frac{1}{2}}\left(-e^{\frac{\tilde{R}_{0}\Delta\tilde{\mu}+\tilde{S}_{0}\Delta\tilde{\mathcal{H}}}{\mathcal{T}}}\right).
\end{align}

The universal scaling behavior across the Fermi to Bose-Fermi (F-BF) quantum phase transition is shown in Fig.~\ref{fig:BF_criticality-2_app}. This figure details the scaling of the cross and bosonic components ($\mathcal{D}_{nm}$, $\mathcal{D}_{mm}$, $\mathcal{D}_{me}$), complementing the analysis of the diagonal components ($\mathcal{D}_{nn}$, $\mathcal{D}_{ne}$) presented in Fig.~9.
\begin{figure}[H]
	\centering
	\includegraphics[width=1\linewidth]{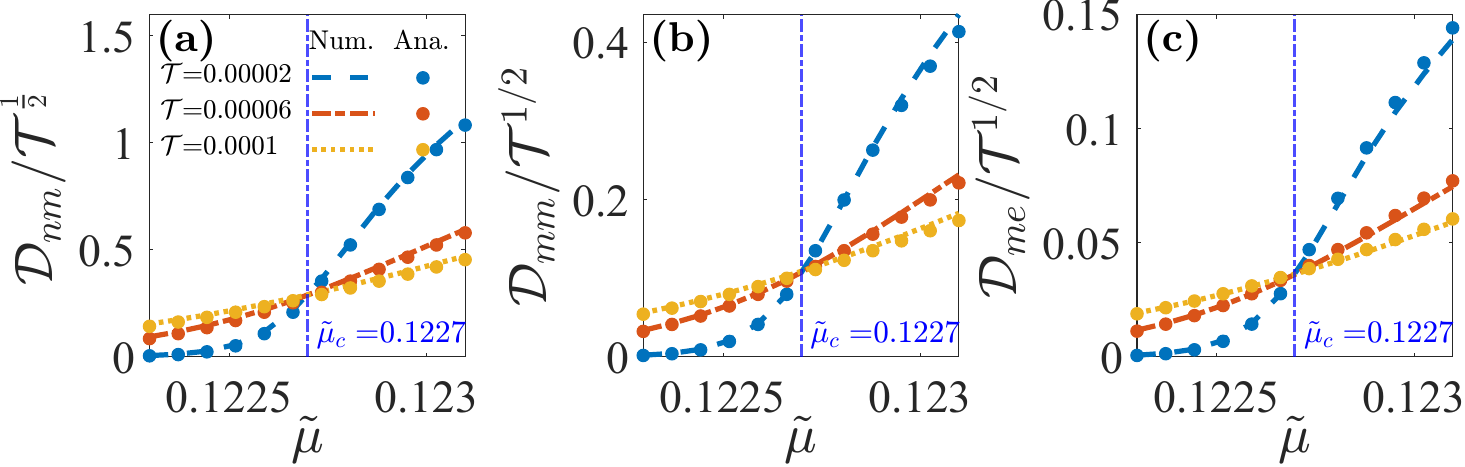}
	\caption{
	Supplemental scaling behavior of the Drude weights across the F-BF phase transition for $\mathcal{H}=1$. Circles represent the exact TBA numerical solutions for (a) $\mathcal{D}_{nm}$, (b) $\mathcal{D}_{mm}$, and (c) $\mathcal{D}_{me}$ versus chemical potential $\tilde{\mu}$. The blue lines denote the analytical scaling forms. The vertical dashed line marks the critical point $\tilde{\mu}_{c} = 0.1227$. The scaling of $D_{nn}$ and $D_{ne}$ is presented in Fig.~9 of the main text.
	}
	\label{fig:BF_criticality-2_app}
\end{figure}

\section*{Appendix D: Drude Weights from Non-Equilibrium Protocols}
\def\thefigure{D\arabic{figure}}
\def\thetable{D\arabic{table}}
\def\theequation{D\arabic{equation}}

In this section, we provide alternative derivations of the Drude weight
formula that are more directly connected to experimental protocols,
confirming the results obtained from the thermodynamic GHD formalism.

\subsubsection*{Constant Potential Gradient}
To circumvent the computational challenges associated with the Kubo formula for dynamic auto-correlation functions, we employ a direct linear response protocol to extract the Drude weight. The system is initialized in a global thermal equilibrium state. At $t=0$, a linear chemical potential gradient $V(x) = -\phi x$ is quenched into the system, generating a constant thermodynamic force. The Drude weight $D_{ij}$ is identified as the asymptotic growth rate of the macroscopic current in the limit of a vanishingly small force~\cite{Commun.Math.Phys.2013IlievskiThermodyamic}
\begin{equation}
	D_{ij}=\lim_{\phi\rightarrow0}\lim_{t\rightarrow\infty}\frac{j_{j}(t)}{t\phi}.
\end{equation}
Experimentally, this is equivalent to measuring the imbalance of atom numbers $\Delta N(t)$ between two reservoirs, linked to the current via $j(0,t) = \frac{1}{2} d\Delta N/dt$.
The non-equilibrium dynamics of this protocol are rigorously described by the inhomogeneous Generalized Hydrodynamics (iGHD) framework~\cite{Phys.Rev.B2022ScopaGeneralized}. The time evolution of the quasi-particle filling functions $\theta_a(x,u,t)$ is governed by the continuity equation in phase space:
\begin{equation}
	\partial_{t}\theta_{a} + v_{a}^{\eff}\partial_{x}\theta_{a} + a_{a}^{\eff}\partial_{u}\theta_{a} = 0,
\end{equation}
where $v_a^{\eff}$ is the effective velocity and $a_a^{\eff}$ is the effective acceleration induced by external fields. The external potential enters the GHD equations through the one-particle potential vector $\vec{w}(x)$, defined as the single-particle eigenvalue function of the external Hamiltonian operator $\hat{H}_{\text{ext}}$. Specifically, the external part of the Hamiltonian involves the spatially dependent chemical potential $\mu(x)=\mu_{0}+\phi x$ and the magnetic field $\mathcal{H}$~\cite{SciPostPhys.2017DoyonNote,Phys.Rev.B2022ScopaGeneralized}:
\begin{equation}
	\hat{H}_{\text{ext}} = \int dx \left[ -\left(\mu(x)+\frac{\mathcal{H}}{2}\right)\hat{n}(x) + \mathcal{H}\hat{m}(x) \right].
\end{equation}
In the basis of elementary excitations, the particle number density operator $\hat{n}$ and magnetization density operator $\hat{m}$ correspond to the single-particle charge vectors $h_n = \{1, 0\}$ and $h_m = \{0, 1\}$, respectively. Thus, the pseudopotential vector $\vec{w}(x)$ for each species is constructed as~\cite{SciPostPhys.2017DoyonNote,Phys.Rev.B2022ScopaGeneralized}
\begin{equation}
	\vec{w}(x) = -\left(\mu(x)+\frac{\mathcal{H}}{2}\right) h_{n} + \mathcal{H} h_{m} = \left\{-\left(\mu(x)+\frac{\mathcal{H}}{2}\right) , \mathcal{H}\right\}.
\end{equation}
In the present protocol, we apply a linear gradient to the chemical potential, $\mu(x) = \mu_{0} + \phi x$, while maintaining a constant magnetic field ($\partial_x \mathcal{H} = 0$). 
The ``bare" force vector, defined as the negative spatial gradient of the pseudopotential, acts explicitly only on the charge sector
\begin{equation}
	-\partial_x \vec{w}(x) = \{\phi , 0 \} = \phi h_{n}.
\end{equation}
However, in an interacting integrable system, the physical acceleration experienced by the quasi-particles is determined by the ``dressed" force. The effective acceleration is given by the ratio of the dressed force gradient to the dressed momentum derivative~\cite{SciPostPhys.2017DoyonNote,Phys.Rev.B2022ScopaGeneralized}
\begin{equation}
	\vec{a}^{\eff}(x,t) = \frac{(-\partial_x \vec{w})^{\dr}}{(\partial_k h_{k})^{\dr}}.
\end{equation}Utilizing the linearity of the dressing operation and the relation $-\partial_x \vec{w} = \phi h_n$, the numerator simplifies directly to $(-\partial_x \vec{w})^{\dr} = \phi h_n^{\dr}$. Since the dressed charge for particle number $h_n^{\dr}$ is identical to the dressed momentum derivative $(\partial_k h_{k})^{\dr}$, we obtain
\begin{equation}
	\vec{a}^{\eff} = \frac{\phi h_{n}^{\dr}}{h_{n}^{\dr}} = \frac{\phi \{1_{\rho}^{\dr},1_{\sigma}^{\dr}\}}{\{1_{\rho}^{\dr},1_{\sigma}^{\dr}\}} = \phi \{1, 1\}.
\end{equation}
This derivation yields a remarkable result: despite the bare force acting solely on the charge species, the interactions induce a uniform effective acceleration for both charge and spin degrees of freedom
\begin{equation}
	a_{\rho}^{\eff}(x,t) = a_{\sigma}^{\eff}(x,t) = \phi.
\end{equation}
The numerical solution of the iGHD equations is implemented using the method of characteristics. 

This approach transforms the partial differential equation governing the filling functions into a set of ordinary differential equations describing the flow of fluid cells. 
We begin by considering the total time derivative of the filling function $\theta_a(t, x(t), u(t))$ along a specific trajectory in phase space
\begin{equation}
	\frac{d\theta_{a}}{dt} = \partial_{t}\theta_{a} + \frac{dx(t)}{dt}\partial_{x}\theta_{a} + \frac{du(t)}{dt}\partial_{u}\theta_{a}.
\end{equation}
Comparing this expression with the iGHD continuity equation, $\partial_{t}\theta_{a} + v_{a}^{\eff}\partial_{x}\theta_{a} + a_{a}^{\eff}\partial_{u}\theta_{a} = 0$, we observe that the total derivative vanishes
\begin{equation}
	\frac{d\theta_{a}}{dt} = 0,
\end{equation}
provided that the trajectory is defined by the characteristic flow equations
\begin{equation}
	\frac{dx(t)}{dt} = v_{a}^{\eff}(x, u, t), \quad \frac{du(t)}{dt} = a_{a}^{\eff}(x, u, t).
\end{equation}
The condition $d\theta_a/dt = 0$ implies that the filling function is an invariant of motion along these characteristic curves. Mathematically, integrating a zero derivative with respect to time yields a constant. Therefore, the value of the filling function carried by a fluid cell remains unchanged as it propagates from an initial time $t_1$ to a later time $t_2$~\cite{SciPostPhys.2017DoyonNote,Phys.Rev.B2022ScopaGeneralized}
\begin{equation}
	\theta_{a}(t_{1}, x(t_{1}), u(t_{1})) = \theta_{a}(t_{2}, x(t_{2}), u(t_{2})).
\end{equation}
This conservation law is the essence of the method of characteristics, allowing us to compute the system's evolution by simply updating the coordinates of the fluid cells rather than solving the PDE on a fixed grid. Numerically, we discretize the time evolution with a step $\Delta t$. Given the constant effective acceleration $\phi$ derived in our protocol, the phase-space coordinates of the fluid cells are updated as follows
\begin{align}
	x_{a}(t+\Delta t) & = x_{a}(t) + v_{a}^{\eff}(x, t)\Delta t, \\
	u(t+\Delta t) & = u(t) + \phi \Delta t, \\
	\Lambda(t+\Delta t) & = \Lambda(t) + \phi \Delta t.
\end{align}
Iterating these updates allows for the reconstruction of the full non-equilibrium state. The instantaneous current density $j_i(x,t)$ is then computed by integrating the quasi-particle contributions
\begin{align}
	j_{i}(x,t) & =\int du \rho(u,x,t)h_{\rho_{i}}(u)v_{\rho}^{\eff}(u,x,t) + \int d\Lambda \sigma(\Lambda,x,t)h_{\sigma_{i}}(\Lambda)v_{\sigma}^{\eff}(\Lambda,x,t),
\end{align}
where the root densities are related to the filling functions via $\rho = \frac{1}{2\pi}\theta_{\rho} 1_{\rho}^{\dr}$ and $\sigma = \frac{1}{2\pi}\theta_{\sigma} 1_{\sigma}^{\dr}$. The linear growth of the total current in the long-time limit provides the measure for extracting the Drude weight.

\subsubsection*{Bipartite Quench Protocol}

We now provide the rigorous derivation connecting the dynamical GHD bipartitioning protocol (Eq.~(110) in the main text) to the analytical TBA expression for the Drude weight. To align with Protocol 2, we consider a bipartite system prepared at a uniform temperature $T = 1/\beta$, but with a small chemical potential bias $\delta\mu$ across the junction at $x=0$. The initial chemical potentials are $\mu_L = \mu + \delta\mu/2$ for $x<0$ and $\mu_R = \mu - \delta\mu/2$ for $x>0$.
	
In the long-time limit, the non-equilibrium dynamics are governed by the Euler-scale GHD equations, and the system develops a quasi-stationary state that depends exclusively on the ray $\xi = x/t$. The local filling fraction $\theta_a(u,\xi)$ is given by the solution of the continuity equation
\begin{equation}
	 \theta_{a}(u,\xi) = \theta_{a,L}(u)\Theta(v_{a}^{\eff}(u)-\xi) + \theta_{a,R}(u)\Theta(\xi-v_{a}^{\eff}(u)),
\end{equation}
where $\Theta(x)$ is the Heaviside step function. In the linear response regime ($\delta\mu \to 0$), the difference between the left and right reservoirs can be expanded as $\theta_{a,L}(u) - \theta_{a,R}(u) \approx \frac{\partial \theta_{a}}{\partial \mu} \delta\mu$. The variation of the filling fraction from the homogeneous equilibrium background is thus
\begin{equation}
	\delta\theta_{a}(u,\xi) = \frac{1}{2}\delta\mu \frac{\partial \theta_{a}}{\partial \mu} \left[\Theta(v_{a}^{\eff}-\xi) - \Theta(\xi-v_{a}^{\eff})\right].
	\label{eq:app_delta_theta}
\end{equation}

Within the TBA framework, the derivative of the filling fraction with respect to the chemical potential introduces the thermodynamic fluctuation factor. Since the pseudo-energy $\epsilon_a$ depends on $\mu$ as $\partial \epsilon_a/\partial \mu = -\beta h_{a,n}^{\dr}$ (where $h_{a,n}^{\dr}$ is the dressed particle charge), we have
\begin{equation}
	\frac{\partial \theta_{a}}{\partial \mu} = -\theta_{a}(1-\theta_{a}) \frac{\partial \epsilon_{a}}{\partial \mu} = \beta \theta_{a}(1-\theta_{a}) h_{a,n}^{\dr}.
	\label{eq:app_dtheta_dmu}
\end{equation}

The macroscopic quasi-stationary current for the generic charge $j$ along the ray $\xi$ is evaluated as $j_{j}(\xi) = \frac{1}{2\pi} \sum_{a} \int du\, (e_{a}')^{\dr} h_{a,j}^{\dr} \delta\theta_{a}(u,\xi)$. Substituting Eq.~\eqref{eq:app_delta_theta} and Eq.~\eqref{eq:app_dtheta_dmu} into the current expression yields
\begin{equation}
	j_{j}(\xi) = \frac{\delta\mu \beta}{4\pi} \sum_{a} \int du\, (e_{a}')^{\dr} \theta_{a}(1-\theta_{a}) h_{a,n}^{\dr} h_{a,j}^{\dr} \left[\Theta(v_{a}^{\eff}-\xi) - \Theta(\xi-v_{a}^{\eff})\right].
\end{equation}

To extract the Drude weight, we integrate this steady-state current over all rays $\xi$. The integration over the Heaviside functions strictly evaluates to $\int d\xi \left[\Theta(v_{a}^{\eff}-\xi) - \Theta(\xi-v_{a}^{\eff})\right] = 2v_{a}^{\eff}$. Using the fundamental TBA identity $(e_{a}')^{\dr} v_{a}^{\eff} = (p_{a}')^{\dr} (v_{a}^{\eff})^2 = 2\pi \rho_{a}^{tot} (v_{a}^{\eff})^2$, the integrated current becomes
\begin{align}
	\int d\xi\, j_{j}(\xi) &= \delta\mu \left[ \beta \sum_{a} \int du\, \rho_{a}^{tot} \theta_{a}(1-\theta_{a}) (v_{a}^{\eff})^2 h_{a,n}^{\dr} h_{a,j}^{\dr} \right] \nonumber \\
	&= \delta\mu \left[ \beta \sum_{a} \int du\, \rho_{a} (1-\theta_{a}) (v_{a}^{\eff})^2 h_{a,n}^{\dr} h_{a,j}^{\dr} \right].
\end{align}

The term enclosed in the brackets is identically the exact analytical TBA expression for the Drude weight $D_{nj}$. We therefore arrive at the exact relationship
\begin{equation}
	\int d\xi\, j_{j}(\xi) = \delta\mu D_{nj},
\end{equation}
which immediately recovers Eq.~(110) of the main text, $D_{nj} = \lim_{\delta\mu \to 0} \frac{1}{\delta\mu} \int d\xi\, j_{j}(\xi)$. This confirms that the dynamical integration of the steady-state GHD currents mathematically reproduces the static linear-response Drude weights.

\end{widetext}

\end{document}